\documentclass[12pt]{article}
 \pdfoutput=1 
\usepackage{geometry}
\usepackage{amsthm,amsmath,amssymb}
\usepackage{natbib}
\usepackage{multirow}
\usepackage{graphicx}
\usepackage{subfigure}
\usepackage{makecell}
\usepackage{booktabs}
\usepackage{array}
\usepackage{fullpage}
\usepackage{url}
\usepackage{algorithm}
\usepackage{algorithmic}
\usepackage{bm}
\usepackage{mathtools}
\usepackage{wrapfig}
\usepackage{lipsum}
\usepackage{mathrsfs}
\usepackage{dsfont}
\usepackage{verbatim}
\usepackage{graphicx}
\usepackage{enumitem}
\usepackage{hyperref}
\usepackage[utf8]{inputenc}
\usepackage[english]{babel}
\usepackage{multirow}
\usepackage{rotating}
\usepackage[toc,page]{appendix}
\usepackage{adjustbox}

\usepackage{appendix}
\allowdisplaybreaks

\title{Estimating and Forecasting the Smoking-Attributable Mortality Fraction
	for Both Genders Jointly in Over 60 Countries\thanks{Yicheng Li is Graduate 
Research Assistant and Adrian E. Raftery is Boeing International Professor
of Statistics and Sociology, both at the Department of Statistics,
Box 354322, University of Washington, Seattle, WA 98195-4322.
This research was supported by NIH grants R01 HD054511 and R01 HD070936,
and by the Center for Advanced Research in the Behavioral Sciences at
Stanford University. The authors are grateful to John Bongaarts for 
helpful discussions.} }
\date{\today}
\author{Yicheng Li,  Adrian E. Raftery \\
University of Washington}

\begin{document}
	\maketitle
	\begin{abstract}
	Smoking is one of the leading preventable threats to human health and a major risk factor for lung cancer, upper aero-digestive cancer, and chronic obstructive pulmonary disease. Estimating and forecasting the smoking attributable fraction (SAF) of mortality can yield insights into smoking epidemics and also provide a basis for more accurate mortality and life expectancy projection. \citet{peto1992mortality} proposed a method to estimate the SAF using the lung cancer mortality rate as an indicator of exposure to smoking in the population of interest. Here we use the same method to estimate the all-age SAF (ASAF) for both genders for over 60 countries. We document a strong and cross-nationally consistent pattern of the evolution of the SAF over time. We use this as the basis for a new Bayesian hierarchical model to project future male and female ASAF from over 60 countries simultaneously. This gives forecasts as well as predictive 
	distributions that can be used to find uncertainty intervals for any quantity of interest. We assess the model using out-of-sample predictive validation, and find that it provides good forecasts and well calibrated
	forecast intervals, comparing favorably with other methods.
	\end{abstract}

\baselineskip=18pt

\section{Introduction}\label{sec:intro}
Smoking is known to have adverse impacts on health and is one of the leading preventable causes of death \citep{peto1992mortality, bongaarts2014trends, mons2017demographic}. It is a major risk factor for lung cancer, chronic obstructive pulmonary disease (COPD), respiratory diseases, and vascular diseases, and tobacco use causes approximately 6 million deaths per year \citep{britton2017death}. For instance, tobacco use causes more than 480,000 deaths per year in the United States, accounting for about $20\%$ of the total deaths of US adults, even though smoking prevalence in United States has declined from $42\%$ in the 1960s to $14\%$ in 2018 \citep{mons2017demographic}. 

%As a result, estimating the smoking attributable fraction (SAF) of mortality is important for disease control and intervention policy planning. 
%It can also help to improve mortality and life expectancy projections, 
%because smoking can account for nonlinear trends, cohort effects and  
%differentials between countries and between genders in mortality 
%\citep{rogers2005mortality, bongaarts2006long, wang2009forecasting,preston2011contribution, janssen2013including,stoeldraijer2013impact,  bongaarts2014trends, preston2014projecting, stoeldraijer2015future,peters2016life}.

The smoking attributable fraction (SAF) is the proportion by 
which mortality would be reduced if the population were not exposed to smoking.
It is defined as 
\begin{align*}
\mbox{SAF} = \frac{n_{S}}{n_{D}},
\end{align*} 
where $n_{S}$ is the number of smokers who died because of their smoking habit and $n_{D}$ is the total number of people who died.
It can be shown that this is equivalent to
\begin{align}\label{eq:SAF}
\mbox{SAF} = \frac{p(r-1)}{p(r-1) + 1},
\end{align} 
where $p$ is the underlying prevalence of smoking in the population and $r$ is the risk of dying of smokers divided by the risk of dying of nonsmokers in the population \citep{rosen2013intuitive}. 

Estimating and forecasting the SAF of mortality is essential for assessing how the smoking epidemic influences mortality measures from the past to the future.
First of all, nonlinear patterns of increase in life expectancy over time are partially due to the smoking epidemic. \citet{bongaarts2006long} used the SAF to calculate the non-smoking life expectancy, which turned out to evolve in a more linear fashion than overall life expectancy (including smoking effects). \citet{janssen2013including} used a similar technique to calculate the non-smoking attributable mortality, and showed that its decline is more linear than that of overall mortality. 

Second, smoking partly accounts for regional variations in mortality. 
In most developed regions in the world including
Western Europe, North America and some East Asian countries,
the smoking epidemic among males started earlier than elsewhere,
in the first half of the 20th century.
The adverse effect of the smoking epidemic accumulated for several decades, 
leading to SAF peaking in these countries around the 1980s. With the continuous decline of male smoking prevalence in these countries due to anti-smoking movements and  tobacco control, years of life lost due to smoking began to decrease in recent decades. 
In contrast, many developing countries are currently in the early stage of the smoking epidemic, with high and increasing smoking prevalence among males,
even though tobacco control policies are in place. 
%Hence, for example,  \citet{chen2015contrasting} anticipated a continous increase in SAF in China for the next decades. 

Smoking also accounts for some subnational differences in mortality. For example, \citet{fenelon2012estimating} found that smoking accounts for the southern mortality disadvantage relative to other regions of the United States. They showed that smoking explained $65\%$ of the subnational variation in male mortality in 2004. 

\begin{figure}[tbp]
	\begin{center}
		\includegraphics[scale=0.3]{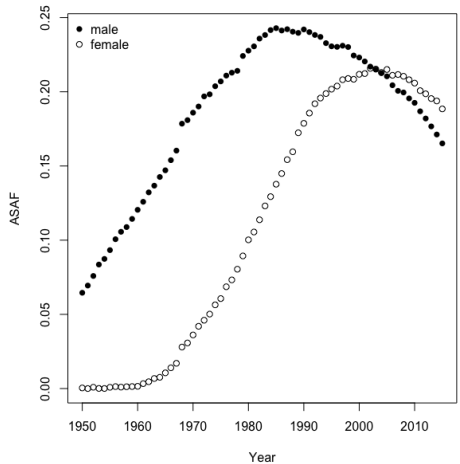}
	\end{center}	
	\caption{United States: All-age smoking attributable fractions of mortality for males and females from 1950 to 2015, estimated using the Peto-Lopez method.}
	\label{fg:usa}
\end{figure}

Third, changes in smoking mortality largely account for 
changes in the between-gender differences in mortality.
The gap in mortality between males and females has tended to first widen
and then narrow in most developed countries, 
and  reduced between-gender differences in smoking largely explain the current closing of the between-gender mortality gap \citep{pampel2006global, preston2006sex}. 
Indeed, in these countries the female smoking epidemic usually started one or two decades later than the male epidemic, and thereafter followed a similar pattern.
In mid- to low-income countries, female smoking-related mortality remains low but still follows a similar rising-peaking-falling trend to the male one. The SAF for males and females clearly follows the same general increasing-peaking-decreasing trend but with different times of onset, times-to-peak and maximum values 
(see Figure \ref{fg:usa}). 

Therefore, estimating and forecasting the SAF can help to improve mortality forecasts by taking the nonlinearity of mortality decline together with between-country and between-gender differentials into account \citep{bongaarts2006long,janssen2013including,stoeldraijer2015future}.
Here we propose a new Bayesian hierarchical model to project SAF that captures the observed increasing-peaking-declining trend so that it could be used for making better mortality forecasts.

Estimating the SAF is not easy for several reasons \citep{bongaarts2014trends, tachfouti2014mortality}. First, the smoking habits of individuals can differ in terms of smoking intensity, smoking history, types of tobacco used, as well as first-hand or second-hand smoking, so that estimating the prevalence of smoking ($p$ in Eq. \ref{eq:SAF}) based on smoking behavior data is not straightforward. 
Secondly, to estimate the relative risk of smoking ($r$ in Eq. \ref{eq:SAF}) requires accurate cohort data. 
Such data are challenging to collect because smoking is not a direct killer but rather has a lifelong impact, with deaths occurring mostly at older ages.
The American Cancer Society's Cancer Prevention Study II (CPS-II), which began in 1982, is so far the largest study that collects such data \citep{tachfouti2014mortality}. Thirdly, the quality of registration and survey data varies across countries and between genders, which makes estimation and comparison of SAF across countries difficult.

Three categories of methods have been proposed to estimate SAF. The first is prevalence-based analysis in cohort studies (SAMMEC) \citep{levin1953occurrence}. This uses estimated smoking prevalence from surveys and relative risk from CPS-II. The second method is prevalence-based analysis in case-control studies. This method is similar to the first one, except that the relative risk is estimated from a case-control study. It has been used for India \citep{gajalakshmi2003smoking}, Hong Kong \citep{lam2001mortality}, and China \citep{niu1998emerging}. The main drawback of prevalence-based methods is the scarcity of reliable historical data on smoking prevalence, especially for developing countries. 

The third method, which overcomes this limitation, is an indirect method. It is called the Peto-Lopez method and was first proposed by \citet{peto1992mortality}. This method estimates the proportion of the population exposed to smoking using lung cancer mortality data, since most lung cancer deaths are due to smoking in developed countries. According to \citet{CDC2019}, cigarette smoking is associated with more than $80\%$ of lung cancer deaths in the United States. \citet{simonato2001lung} also concluded by case-control studies in 6 developed European countries that smoking is associated with over $90\%$ of lung cancer cases. We use this method to estimate the SAF and we describe the procedure in Section \ref{subsec:plest}. 

Another indirect method, the PGW method of \citet{preston2009new}, also uses lung cancer mortality rate as an indicator of the cumulative hazard of smoking. Instead of using relative risks from the CPS-II as the Peto-Lopez method does, the PGW method adopts a regression-based procedure. We discuss these two methods in Section \ref{subsec:plestdif}. More comparisons among different estimation methods of SAF can be found in \citet{perez2008methodologies}, \citet{tachfouti2014mortality}, \citet{kong2016comparison}, and \citet{peters2016life}. 

Figure \ref{fg:usa} plots the estimated all-age SAF (ASAF) of males and females for the United States from 1950 to 2015. It can be seen that the evolution of SAF
over time follows a remarkably strong pattern, first rising and then falling.
Qualitatively very similar patterns were found in most countries that we 
studied, although in countries with less good data, higher levels of measurement
error can be seen. It seems intuitive to expect that such a regular pattern could be
used to obtain good forecasts. Here we describe our method for doing this. 
It turns out that, indeed, good forecasts can be obtained, thanks to the 
strong and consistent pattern of SAF over time. Here we propose a new probabilistic projection method for the SAF using a Bayesian hierarchical model. Our method will provide estimates and projections of the SAF for both genders jointly for more than 60 countries. 

%Since ignoring smoking will potential distort the projection of mortality, a modern paradigm for mortality projection is proposed by separately projecting smoking-attributable mortality (SAM), which is the mortality multiplied by SAF, and the non-smoking-attributable mortality \citep{janssen2013including, peters2016life}. Thus projection of SAF provides one way to project SAM. \citet{janssen2013including} and \citet{stoeldraijer2015future} provided two methods for projecting SAF based on the age-period-cohort (APC) analysis, and used the methods for the cases of England $\&$ Wales, Denmark, and the Netherlands. 

The paper is organized as follows. The data, the detailed SAF calculation based on the Peto-Lopez method, and the proposed Bayesian hierarchical model are described in Section \ref{sec:method}. An out-of-sample validation experiment is reported in Section \ref{sec:result}. We then discuss general estimation and forecasting results for all the countries
considered in this work, with detailed case studies for four countries chosen from North America, South America, Asia, and Europe in Section \ref{sec:casestudy}. We conclude with a discussion in Section \ref{sec:disc}.

\section{Method} \label{sec:method}
\subsection{Notation}
We use the symbol $y$ to denote the estimated (observed) all-age smoking attributable fraction (ASAF), which is defined as the smoking attributable fraction for all age groups combined, and we use the symbol $h$ to denote the true (unobserved) ASAF. All of these quantities are indexed by country $c$, gender $s$, and year $t$. The quantities of interest are the unobserved true past and present ASAF together with their future projections. 
Here the estimation time period is 1950--2015  and the projection time period 
is 2015--2050.
Section \ref{subsec:plest} describes the estimation procedure for ASAF using the Peto-Lopez method for all available countries. A Bayesian hierarchical model will be used to model the estimated ASAF. In the Bayesian hierarchical model, the country-specific parameter vector determining the time evolution pattern of ASAF for country $c$ and gender $s$ is denoted by $\theta_{c,s}$, and the global parameters by $\psi$.

\subsection{Data}\label{subsec:data}
We use the annual death counts by country, age group, gender, and cause of death from the WHO Mortality Database \citep{WHO2017} which covers data from 1950 to 2015 for more than 130 countries and regions around the world. This dataset comprises death counts registered in national vital registration systems and is coded under the rules of the International Classification of Diseases (ICD). There are 5 raw datasets available by the most recent update on 11 April 2018. The first three datasets are labeled as ICD versions 7, 8, and 9 respectively, and the last two are labeled as ICD version 10. 

Each version of ICD codes causes of death differently and a summary of the codes used for estimating ASAF in Section \ref{subsec:plest} is given in Table \ref{tb:ICD}. For each country, the death counts data can differ by geographical coverage, number of years available and age group breakdown. Some countries such as China only have data from selected regions, and these countries will not be included here.

We use the quinquennial population by five-year age groups from the 2017 Revision of the World Population Prospects \citep{WPP2017} for each country, gender and age group. Since this dataset provides population estimates at five-year intervals, we use linear interpolation to obtain annual population estimates for each five-year age group.

\begin{table}	
	\caption{ICD codes for different cause of death categories across versions.}
	\begin{center}\resizebox{\textwidth}{!}{
			\begin{tabular}{ c| c c c }
				\hline
				Causes	& ICD-7 (A-list) & ICD-8 (A-list) & ICD-9 (09A, 09B)\\
				\hline
				Lung Cancer & A050 & A051 & B101\\
				Upper Aero-digestive Cancer & A044, A045, A040& A045, A046, A050& B08, B090, B100   \\
				Other Cancer &rest of A044-A059 & rest of A045-A060 & rest of B08-B14  \\
				COPD & A092, A093 & A093 & B323, B324, B325 \\
				Other Respiratory & rest of A087-A097 & rest of A089-A096 & rest of B31-B32  \\
				Vascular Disease & A079-A086 & A080-A088 & B25-B30\\
				Liver Cirrhosis & A105 & A102 & A347 \\
				Other non-med & A138-A150 & A138-A150 & B47-B56 \\
				Other medical &rest & rest &rest \\
				All causes & A000 & A000 & B00 \\	
				\hline\hline
				Causes& ICD-9 (09N) & ICD-10 (101) &ICD-10 (103, 104, 10M)\\
				\hline
				Lung Cancer &   B101 & 1034& C33-C34\\
				Upper Aero-digestive Cancer & B08, B090, B100 &1027, 1028, 1033 & C00-C15, C32  \\
				Other Cancer  & rest of CH02 &rest of 1027-1046&rest of C00-C97 \\
				COPD  & B323, B324, B325 &1076&J40-J47 \\
				Other Respiratory &rest of CH08 & rest of 1072 & J00-J99 \\
				Vascular Disease & CH07 &1064& I00-I99\\
				Liver Cirrhosis & S347&1080& K74, K70 \\
				Other non-med  & CH17&1095&V00-Y89\\
				Other medical  &rest & rest &rest\\
				All causes &  B00 &1000& AAA \\	\hline
		\end{tabular}}
	\end{center}
	\label{tb:ICD}
\end{table}

\subsection{ASAF Estimation}\label{subsec:plest}
We apply the original Peto-Lopez indirect method to estimate ASAF for male and female separately. This method uses the lung cancer mortality rate as an indicator of the accumulated hazard of smoking to estimate the proportion of population exposed to smoking. As commented in \citet{peto1992mortality}, it is very rare to observe lung cancer cases among non-smokers in developed countries, even in areas with pollution sources such as radon and asbestos. The original papers \citep{peto1992mortality, peto1994mortality, peto2006mortality} applied the method to developed countries only, especially in Western Europe and North America. With the shift of global smoking pattern, and diffusion of smoking in middle- and low-income countries, this method has been extended to less developed countries \citep{ezzati2003estimates,ezzati2004regional, pampel2006global}. 

For estimating ASAF using the Peto-Lopez method, we need first to estimate age- and cause-of-death-specific SAF. The age groups used for estimation are 0-34, 35-59, 60-64, 65-69, 70-74, 75-79, and 80+. For each age group, annual death counts of the following nine categories of causes of death are obtained from the five raw datasets of WHO Mortality Database: lung cancer, upper aero-digestive cancer, other cancers, COPD, other respiratory diseases, vascular diseases, liver cirrhosis, non-medical causes, and all other medical causes. A detailed list of codes from ICD 7, 8, 9, and 10 for these nine categories is provided in Table \ref{tb:ICD}.

The ICD categorizes death count data according to availability using so-called sublists, which can be one of A-list or several others; see Table \ref{tb:ICD}.
The sublists we use are those satisfying the minimum requirements for ASAF calculation.
More specifically, for ICD 7 and 8, only countries whose ICD sublist is A-list are used. For ICD 9, only those countries whose ICD sublist is 09A-, 09B-, or 09N-list are used. For ICD 10, countries whose ICD sublist is one of 101-, 103-, 104-, 10M-list are used. 
In addition, we only calculate age-specific SAF for countries whose age group breakdown is finer than the following age group breakdown: 0-34, 35-39, 40-44, 45-49, 50-54, 55-59, 60-64, 65-69, 70-74, 75+. This corresponds to the age group format number 00, 01, 02, 03, 04 in the raw datasets. 

To estimate the proportion of a population exposed to smoking, i.e., $p$ in Eq. \ref{eq:SAF}, the method compares the observed lung cancer mortality rate with the lung cancer mortality rate of smokers estimated from CPS-II. The estimated proportion, indexed by country $c$, age group $a$, gender $s$, and year $t$, is estimated by
$$p_{c,a,s,t} = \frac{d_{c,a,s,t} - d_{a,s}^S}{d_{a,s}^S - d_{a,s}^{NS}},$$
where $d_{c,a,s,t}$ is the observed country-age-gender-year-specific lung cancer mortality rate, and $d_{a,s}^S$ and $d_{a,s}^{NS}$ are age-gender-specific lung cancer mortality rates for smokers and nonsmokers from the CPS-II respectively. Here the observed lung cancer mortality rate $d_{c,a,s,t}$ is the observed lung cancer death count divided by the population estimated from the 2017 Revision of the World Population Prospects for country $c$, age group $a$, gender $s$, and year $t$. 

The Peto-Lopez method uses the CPS-II to estimate the relative risk of dying for each cause of death for smokers and nonsmokers, i.e., $r$ in Eq. \ref{eq:SAF}. Specifically, the Cochran-Mantel-Haenszel method is used to estimate the relative risk for age group 35-59 by combining five sub-age groups (35-39, 40-44, 45-49, 50-54, 55-59). The relative risk is indexed by cause-of-death $k$, age group $a$, and gender $s$. Here $k$ takes integer values 1-9 corresponding to the nine categories mentioned above. 

The excess mortality rate attributable to smoking is denoted by $er_{k,a,s}$ for cause-of-death $k$, age group $a$, and gender $s$. For lung cancer, the excess mortality rate attributable to smoking is calculated as $er_{1,a,s} = r_{1,a,s} -1$. For all other categories except liver cirrhosis ($k = 7$) and non-medical causes ($k = 8$), the excess risk is discounted by $50\%$, i.e., $er_{k,a,s} = 0.5(r_{k,a,s} -1)$ for $k = 2,3,4,5,6,9$, so as to control for confounding factors. The excess risks for liver cirrhosis and non-medical causes are set to 0, i.e., $er_{7,a,s} = er_{8,a,s} = 0$. The country-cause-age-gender-year-specific SAF, denoted by $y_{c,k,a,s,t}$, is then
$$y_{c,k,a,s,t} = \frac{p_{c,a,s,t}\times er_{k,a,s}}{p_{c,a,s,t}\times er_{k,a,s}+1}.$$
Any estimated negative values are set to zero.

Since the hazard due to smoking is accumulated across years and mostly causes deaths at older ages, the fraction of deaths due to smoking for ages 0-34 is typically very small and is set to 0. In addition, the SAF for ages 80+ is set to the same value as that for ages 75-79 since smoking data are unreliable for very old ages. Finally, the country-gender-year-specific ASAF, denoted by $y_{c,s,t}$, is a weighted average of the age-specific smoking attributable fractions $y_{c,k,a,s,t}$. Thus
$$y_{c,s,t} = \sum_{a}\sum_{k} y_{c,k,a,s,t}\times d_{c,k,a,s,t},$$
where $d_{c,k,a,s,t}$ is the country-cause-age-gender-year-specific mortality rate.

We chose the Peto-Lopez method to estimate the ASAF because it has been validated and widely used \citep{preston2009new,bongaarts2014trends, tachfouti2014mortality,kong2016comparison}. 
Also, the data required for the estimation are cause- and age-specific death counts and population, which are provided with high quality by the WHO Mortality Database and the 2017 Revision of the World Population Prospects.

There are some variants of the Peto-Lopez method, which also assume that the lung cancer mortality rate is a good indicator for measuring smoking exposure. Some of the modifications include using different relative risk estimation instead of the CPS-II to extend the method to developing countries \citep{ezzati2003estimates} or using a regression-based approach \citep{preston2009new}. Section \ref{subsec:plestdif} contains more detailed discussion and comparison of these methods.

\subsection{Model}\label{subsec:model}
We develop a four-level Bayesian hierarchical framework to model male and female ASAF jointly for multiple regions simultaneously.

\paragraph{Random walk with drift for the true ASAF} The observed ASAF data show a strong and consistent pattern of increasing, then leveling, and then declining again for both genders \citep{stoeldraijer2015future} (see Figure \ref{fg:usa} for the example of United States). This pattern can be captured by the following five-parameter double logistic curve: 
\begin{align}\label{eq:dlc}
&g(t|\theta) = \frac{k}{1+\exp\{-a_1(t - 1950- a_2)\}} - \frac{k}{1+\exp\{-a_3(t - 1950- a_2 - a_4)\}},
\end{align}
where $t$ is the year of observation and $\theta$ is the double-logistic parameter vector, $\theta = (a_1, a_2, a_3, a_4, k)$. 

Models based on the double logistic curve have been used quite widely for human population measures such as life expectancy and total fertility rates \citep{marchetti1996human, raftery2013bayesian, alkema2011probabilistic}). 
Due to its natural scientific interpretability, the double logistic curve has also been used in other scientific fields such as hematology \citep{head1987vagal, head2004non}, phenology \citep{yang2012regional}, and agricultural science \citep{shabani2018using}. 
This function has also been used to describe social change, diffusion, and substitution processes
\citep{grubler1999dynamics,fokas2007growth,kucharavy2011logistic}.

Most developed countries have had male smoking prevalence that started before 1950, and peaked around the 1950s or 1960s when the adverse impacts of smoking on health became known and tobacco control measures started being put in place.
This led to a peak in smoking-related mortality a generation or so later,
followed by a continuous decline since then. \citet{pampel2005forecasting} argued that the smoking epidemic involves diffusion from males to females, and from more developed countries to less developed ones. Hence, the strong increasing-peaking-decreasing trend of ASAF observed in most countries is a consequence of the smoking epidemic diffusion process, and the double logistic curve can naturally describe its dynamics.

\begin{figure}[tbp]
	\begin{center}
		\includegraphics[scale=0.41]{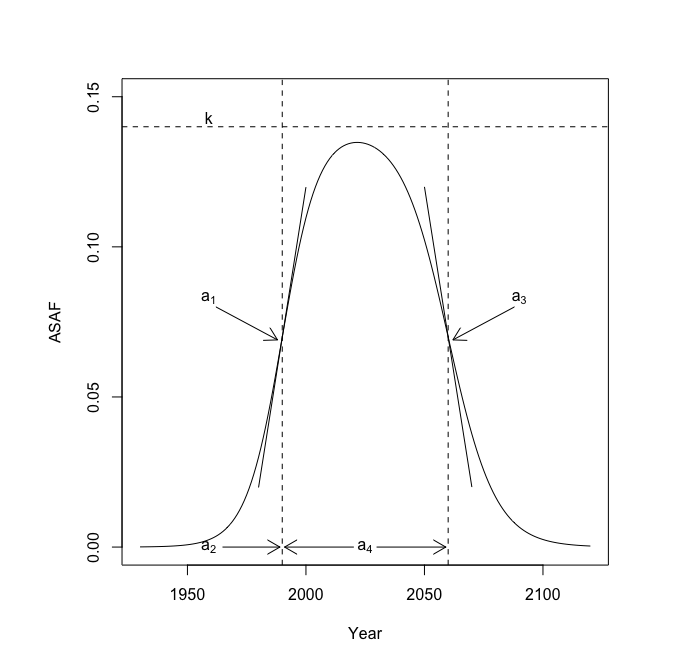}
		\includegraphics[scale=0.35]{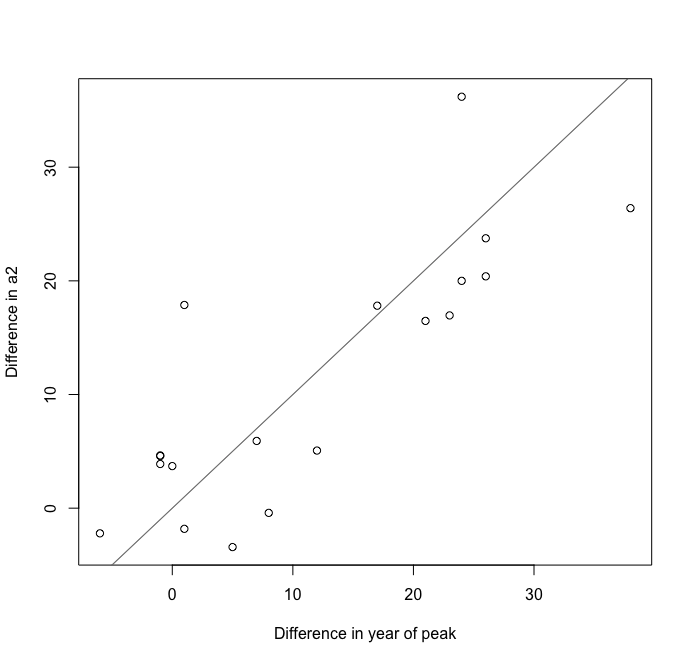}
	\end{center}
	\caption{Upper: The five-parameter double logistic curve. $a_2$ controls the left inflection point, $a_4$ controls the distance between left and right inflections points, $a_1, a_3$ determine the rate of change at left and right inflection points, and $k$ approximates the maximum value. Lower: The difference of country-specific $a_2^m$ and $a_2^f$ plotted against the difference between the country-specific peaks for males and females. The peak and $a_2$ are estimated from the countries whose male and female ASAF have all passed the maximum by 2015, according to the results of the non-linear least squares estimation. The solid line is the 45 degree line.}\label{fg:dlc}
\end{figure}

For the five-parameter double logistic function in Eq. \ref{eq:dlc}, $a_2$ controls the first (left) inflection point of the curve and $a_4$ controls the distance between the first (left) and the second (right) inflection points. The rates of change at these inflection points are controlled by $a_1$ and $a_3$ respectively. The parameter $k$ is an upper bound for the maximum value of the curve. See the upper panel of Figure \ref{fg:dlc} for an illustration.

To represent this and also take account of the observed pattern of variability,
we model changes in the true ASAF between adjacent time points using a random walk with drift given by the difference between the double logistic curve at the two points. This takes the form
\begin{align}\label{eq:rwd}
h_{c,s,t} = h_{c, s, t-1}  + g(t|\theta_{c,s}) - g(t-1|\theta_{c,s}) + \varepsilon^h_{c,s,t},
\end{align}
where $g(\cdot|\theta_{c,s})$ (i.e., Eq. \ref{eq:dlc}) quantifies the expected change of the true ASAF governed by the country- and gender-specific parameters $\theta_{c,s} = (a_1^{c,s}, a_2^{c,s}, a_3^{c,s},\allowbreak a_4^{c,s}, k^{c,s})$, and $\varepsilon^h_{c,s,t}$ are independent Gaussian noises. This random walk with drift model is designed to capture the variability of the true ASAF and allows the uncertainty of the forecast to increase when projecting further into the future. 
%Validation results in Section \ref{sec:result} show that this model is well calibrated.

\paragraph{Male-female joint model} Since the female smoking epidemic usually starts one to two decades after the male one, the start of the increase in the female ASAF is also later than that of the male ASAF. For most countries, the observed female ASAF is still in the increasing or leveling phase up to 2015. 
However, as the smoking epidemic diffuses from the male to the female population, it is reasonable to assume that the female ASAF will follow the same trend of increasing-leveling-declining as that of the male ASAF. This has already been observed for several countries with early smoking epidemics, such as the United Kingdom, Denmark, and Japan \citep{ pampel2005forecasting,peto2006mortality, janssen2013including, bongaarts2014trends,stoeldraijer2015future}. For these countries, the female ASAF follows the same trend as that of the male ASAF, but differs mainly in terms of the rate of increase or decrease, the number of years taken to reach the peak, and the peak ASAF value.

For males, we need only estimate the rate of decline of the ASAF. For females, especially for those countries whose observed ASAF data have not levelled yet, one needs first to determine the time and value of leveling. By modeling male and female data jointly, the lower panel of Figure \ref{fg:dlc} shows that for countries whose male and female ASAF both passed the leveling period, the difference between the years of maximum of male and female is approximately the same as the difference in the $a_2$ parameter estimated from Eq. \ref{eq:dlc}. The $a_2$ parameter represents the time point where the speed of the increasing part of the double logistic curve begins to slow down. 

The difference between the times-to-peak of male and female ASAF also differs among countries. For example, the time-to-peak of the female ASAF in the United States is about 15 years later than that of the male ASAF, while the time-to-peak of the ASAF happened at about the same time for both genders in Hong Kong. To incorporate these observations, we model the difference between male and female country-specific $a_2^c$ using a Gaussian distribution:
\begin{align}\label{eq:female}
a_2^{c,f} = a_2^{c,m} + \bigtriangleup^c_{a_2}, \hspace{1cm} \bigtriangleup^c_{a_2}|\bigtriangleup_{a_2}, \sigma^2_{\bigtriangleup_{a_2}} \sim \mathcal{N}(\bigtriangleup_{a_2}, \sigma^2_{\bigtriangleup_{a_2}}),
\end{align}
where $a_2^{c,m}$ and $a_2^{c,f}$ are the country- and gender-specific values of $a_2$, and $\bigtriangleup^c_{a_2}$ is the country-specific difference between these two parameters with prior mean $\bigtriangleup_{a_2}$ and variance $\sigma^2_{\bigtriangleup_{a_2}}$.

Moreover, since there are very few countries whose female ASAF have begun to decline by 2015, while the male ASAF has been declining for many years in most countries, we set the same global parameters for the gender-specific parameters $a_4^{c,m}$ and $a_4^{c,f}$ for each country, namely,
\begin{align}\label{eq:a4}
a_4^{c,m}, a_4^{c,f}|a_4, \sigma^2_{a_4} \stackrel{\rm ind}{\sim}
\mathcal{N}(a_4, \sigma^2_{a_4}).
\end{align}
Except for $a_4^{c}$, the other four country-specific parameters of the double logistic curve are conditioned on their own gender-specific global parameters.

\paragraph{Measurement error model for observed ASAF} 
The observed country-gender-year-specific ASAF $y_{c,s,t}$ are modeled based on the true (unobserved) ASAF $h_{c,s,t}$ by incorporating measurement error due to the variability of data quality across different countries:
\begin{align}\label{eq:measure}
y_{c,s,t}|h_{c,s,t},\sigma_c^2 \sim_{ind} \mathcal{N}(h_{c,t,s}, \sigma_c^2).
\end{align}
We assume that the variance of the observed ASAF for each country is time- and gender-invariant based on exploratory analyses that indicate that the data quality is consistent across time and between genders within the same country.

\paragraph{Summary of model} We combine the Bayesian hierarchical model and measurement error model into a four-level Bayesian hierarchical model. 
We model the observed ASAF estimates using the measurement error model in Level 1, conditional on the true (unobserved) ASAF data which are modeled with a random walk with drift in Level 2, conditional on the country-specific parameters. Country-specific parameters are modeled in Level 3, where parameters for male and female ASAF are modelled jointly conditional on the global parameters, whose prior distributions are specified in Level 4. 

The overall model is specified as follows:
\begin{align*}
\mbox{Level 1:}\ &y_{c,s,t}|h_{c,s,t} \sim \mathcal{N}(h_{c,s,t}, \sigma_c^2);\\
\mbox{Level 2:}\ & h_{c,s, t_{0,c}} = g(t_{0,c}|\theta_{c,s}) + \varepsilon^h_{c, s, t_{0,c}},\\
&h_{c,s,t} = h_{c,s,t-1}  + g(t|\theta_{c,s}) - g(t-1|\theta_{c,s}) + \varepsilon^h_{c,s,t}\ \mbox{for}\ t>t_{0,c},\\
&\varepsilon_{c,s,t}^h \stackrel{\rm ind}{\sim} \mathcal{N}(0, \sigma_h^2);\\
\mbox{Level 3:}\ & \theta_{c,s} \sim f(\cdot|\psi),\\
&\sigma_c^2 \sim \mbox{Lognormal} (\nu, \rho^2);\\
\mbox{Level 4:}\ & \psi, \nu, \rho^2, \sigma_h^2 \sim \pi(\cdot).
\end{align*}

Here, $t_{0,c}$ is the year of the first available ASAF data for country $c$, $g$ denotes the five-parameter double logistic curve in Eq. \ref{eq:dlc}, $f$ denotes the conditional distribution of the country-specific parameters $\theta_{c,s}$, and $\pi$ denotes the hyperpriors for the global parameters $\psi, \nu, \rho^2, \sigma_h^2$. The country-specific parameters $\theta_{c,s} = (a_1^{c,s}, a_2^{c,s}, a_3^{c,s}, a_4^{c,s}, k^{c,s})$ are gender-specific and the interaction between male and female parameters are governed by Eq. \ref{eq:female} and \ref{eq:a4}. The global parameters $\psi = (a_1^m, a_2^m, a_3^m, a_4, k^m, a_1^f, a_3^f, k^f, \bigtriangleup_{a_2}, \sigma^2_{a_2^m}, \allowbreak \sigma^2_{a_4}, \sigma^2_{k^m},\sigma^2_{k^f}, \sigma^2_{\bigtriangleup_{a_2}})$ are also gender-specific except for $\bigtriangleup_{a_2}, \sigma^2_{\bigtriangleup_{a_2}}, a_4, \sigma^2_{a_4}$. More information about the specification of the full model is given in the Appendix \ref{app1}.

\paragraph{Estimation and prediction} Statistical analysis of the model
is carried out in two phases, estimation and prediction.
The goal of the estimation phase is to obtain the joint posterior distribution of the true ASAF $h_{c,s,t}$ during the estimation period 1950--2015 and the country-specific parameters for the underlying double-logistic curve. 
The aim of the prediction phase is to forecast the future ASAF of both genders for the prediction period 2015--2050  based on the observed ASAF for over 60 countries whose male ASAF data are classifed as clear-pattern (see Section \ref{subsec:safcl} for the definition of clear-pattern). 

The functional form of the prior distribution $\pi(\cdot)$ is assessed using results from non-linear least squares estimation based on clear-pattern countries (see Section \ref{subsec:safcl} for details). Specifically, the priors for $(a_1^m, a_2^m, a_3^m, a_4, k^m, \sigma^2_{a_2^m}, \allowbreak \sigma^2_{a_4}, \sigma^2_{k^m}, \sigma^2_{a_2^m}, \sigma^2_{a_4}, \sigma^2_{k^m})$ are based on non-linear least squares results from the male ASAF of over 60 clear-pattern countries, the prior for $a_1^f$ is estimated based on non-linear least squares results from the female ASAF of 52 clear-pattern countries, the priors for $(a_3^f, k^f, \sigma^2_{a_3^f})$ are set to the same priors as their counterparts for males, while the priors for $(\bigtriangleup_{a_2}, \sigma^2_{\bigtriangleup_{a_2}})$ are estimated based on 19 countries for which both male and female ASAF have passed the leveling stage by 2015. The priors for $\nu, \rho^2, \sigma_h^2$ are estimated by pooling male and female ASAF from all clear-pattern countries. A complete specification of the model is given in the Appendix \ref{app1}.

\subsection{ASAF Categorization} \label{subsec:safcl}
We categorize estimated ASAF for 127 countries and regions into two categories according to the data availability and quality: clear-pattern and non-clear-pattern. On one hand, the Peto-Lopez method is not guaranteed to produce reliable ASAF estimates for some less developed countries because of poor data quality. On the other hand, modeling only with clear-pattern countries can improve estimation and projection accuracy without introducing too much random noise. 

The classification is based on non-linear least squares estimation of the following model for each country and gender separately:
\begin{align*}
&y_t  = g(t|\theta) + \varepsilon_t,
\end{align*}
where $g(t|\theta)$ is as in Eq. \ref{eq:dlc} and $\varepsilon_t$ are independent standard Gaussian errors. Its fit to the data in a given country provides an indication of data quality for that country.

Our categorization is based on the number of observations, maximum of observed values, and the $R^2$ value of the non-linear least squares fit. Due to the differences between the diffusion processes of smoking in the male and female populations \citep{pampel2006global}, we use different criteria for male and female data. For male data, we require that (1) the number of available annual observations up to 2015 be greater than 10; 
(2) at least one of the observations be  greater than $0.05$; and (3) that the $R^2$ value be greater than 0.5. 

For female data, since the smoking epidemic in general started one to two decades later than the male one, the onset and the value of the ASAF is later and smaller than that of the male epidemic \citep{pampel2005forecasting,preston2006sex}. The criteria for female data are that (1) the number of observations up to 2015 be greater than 10; (2) at least one of the observations be greater than $0.01$; and  (3) that the $R^2$ value be greater than 0.6. 

By these rules, there are over 60 countries whose male data are classifed as clear-pattern (2 in Africa, 16 in the Americas, 9 in Asia, 40 in Europe and 2 in Oceania), and 52 countries whose female data are classified as clear-pattern (12 in the Americas, 7 in Asia, 31 in Europe and 2 in Oceania).

\subsection{Estimation}
Estimation is based on the male and female ASAF data from over 60 countries whose male ASAF is classified as clear-pattern for the period 1950--2015. The reason why we chose clear-pattern ASAF data is that non-clear-pattern data either have too few observations, very low values, or their shapes are not identifiable. 

We used the \texttt{Rstan} package (Version 2.18.2) in R to obtain the joint posterior distributions of the parameters of interest \citep{carpenter2017stan}. Rstan uses a No-U-turn sampler, which is an adaptive variant of Hamiltonian Monte Carlo \citep{neal2011mcmc,hoffman2014no}. We ran 3 chains with different initial values, each of length 10,000 iterations with a burn-in of 2,000 without thining. This yielded a final, approximately independent sample of size 8,000 for each chain. We monitored convergence by inspecting trace plots and using standard convergence diagnostics. 

We also conducted a sensitivity analysis on the hyperparameters that specify the priors $\pi(\cdot)$ for the global parameters $\psi$, and concluded that the proposed model is not sensitive to the choice of hyperparameters. More information about the convergence diagnostics and the sensitivity analysis is given in the Appendices \ref{app2} and \ref{app3}.

\subsection{Projection}
We produce projections of future ASAF for the period 2015--2050 for over 60 countries whose male ASAF is classified as clear-pattern. The prediction of future ASAF for each country is based on past and present ASAF. We sample from the joint posterior distribution of the country-specific parameters $\theta_{c,s}$ and of the past, and present true ASAF $h_{c,s,t}$. We then use Eq. \ref{eq:rwd} and \ref{eq:measure} to generate a sample of trajectories of future true and observed ASAF respectively from their joint posterior predictive distribution. It is possible that the quantity generated by Eq. \ref{eq:rwd} and Eq. \ref{eq:measure} is negative, and we set such values to zero. This yields a sample from the joint posterior predictive distribution of the future ASAF for over 60 countries, for both genders, 
taking account of uncertainty about the past observations as well as the future
evolution. We include the plots of ASAF projections for both genders and all countries considered in this work in the Appendix \ref{app4}.

\section{Results}\label{sec:result}
We assess the predictive performance of our model using out-of-sample predictive validation.

\subsection{Study Design}
The data we used for out-of-sample validation cover the period 1950--2015. We assess the quality of our model based on different choices of estimation and validation data from the observed data. Since the trend of increasing-leveling-declining pattern plays an important role for estimation and projection, assessing how the model works when only part of the trend has been observed is crucial. We consider different choices for estimation and validation periods, namely  (1) 1950-- 2000 for estimation and 2000--2015 for validation; (2) 1950--2005 for estimation and 2005--2015 and for validation; and  (3) 1950--2010 for estimation, 2010--2015 for validation. The countries used for validation in each time-split scenario are required to be clear-pattern countries based on the male ASAF, to contain more than 10 observations in the estimation period, and to have at least one observation in the prediction period. This results in 63, 66 and 66 countries used for validation under choices (1), (2) and (3), respectively.

Since we are making probabilistic projections, our evaluation is based on both accuracy of point prediction and calibration of prediction intervals. Our goal is not only to produce accurate point predictions, but also to account for variability of future predictions based on historic data, especially for those countries whose data in the estimation period reveal only part of the pattern. If the proposed model works well, we would expect the point predictor to have small gender-specific mean absolute error (MAE), which is defined as
\begin{align}\label{eq:mae}
\mbox{MAE}_s = \frac{1}{N}\sum_{c \in \mathcal{C}} \sum_{t \in \mathcal{T}_c} |\hat{y}_{c,s,t} - y_{c,s,t}|,
\end{align}
where $\mathcal{C}$ is the set of countries considered in the validation, $\mathcal{T}_c$ is the set of country-year combinations used for validation, $\hat{y}_{c,s,t}$ is the posterior median of the predictive distribution of ASAF at year $t$ for country $c$ and gender $s$, and $N$ is the total number of data used for validation. 

We wish the  prediction to be well calibrated and sharp, i.e., the coverage of the prediction interval to be close to the nominal level with its half-width as short as possible. Thus, we include the empirical coverage and the half-width of the prediction interval in the validation. To assess the overall predictive performance, we also calculate the gender-specific continuous ranked probability score (CRPS) \citep{gneiting2007strictly}, which is defined as
\begin{align}\label{eq:crps}
\mbox{CRPS}_s = \frac{1}{|\mathcal{C}|} \sum_{c \in \mathcal{C}} \bigg[\frac{1}{|\mathcal{T}_c|}\sum_{t \in \mathcal{T}_c} \int_{-\infty}^{\infty} \{F_{c,s,t}(y) - \mathbf{1}(y_{c,s,t} \leq y)\}^2 dy\bigg],
\end{align}
where $F_{c,s,t}(y)$ is the predictive distribution of the future ASAF for country $c$, gender $s$, and time $t$, and $\mathbf{1}(\cdot)$ is equal to 1 if the condition in the parenthesis is satisfied and 0 otherwise. CRPS is a summary statistic measuring the quality of the probabilistic forecast, which evaluates model calibration and sharpness simultaneously. The smaller the CRPS, the closer the predictive distribution to the true data-generating distribution.

\subsection{Out-of-sample Validation Results}
To our knowledge, no other method is available in the literature to produce probabilistic forecasts for male and female ASAF for developed and developing countries jointly. \citet{janssen2013including} and \citet{ stoeldraijer2015future} developed methods for projection of age-specific SAF and age-standardized SAF, and their methods are based on age-period-cohort analysis, which cannot be trivially extended to ASAF. See Section \ref{subsec:plstatproj} for more discussion of their procedures and comparison to the present ones. 

As benchmarks against which to compare our method, we consider  four other forecast procedures. The first one is the persistence forecast, which takes the last observed value as the forecast
for the prediction period. The second method is the Bayesian thin plate regression spline method \citep{wood2003thin}, implemented in the \texttt{mgcv} package (Version 1.8-27) in R. The third method is the Bayesian structural time series model \citep{harvey1990forecasting,durbin2012time}, implemented in the \texttt{bsts} package (Version 0.8.0) in R. Here we choose to use two state components --- local linear trend and autocorrelation with lag 1 --- to build the structural time series model. 
Our fourth comparison method is a non-hierarchical version of our proposed model,
namely our proposed model without Level 4 (i.e., the global parameters). 
This is included to see whether the hierarchical structure is necessary. 

We summarize the validation results in Table \ref{tb:valid} for males and females separately. This shows the MAE, the coverage and half-width of the prediction intervals, and the continuous ranked probability score (CRPS). For males, our method improved the prediction accuracy for all three scenarios over the persistence forecast. For forecasting one and two five-year periods ahead, our method improved the MAE by $30\%$ and $21\%$ respectively. Since most male ASAF series had passed the peak by 2005 and had experienced declines for  several years, the double logistic curve captures this trend well. For predictions three five-year periods into the future, during which the male ASAF series for some countries were just reaching the peak, our method still improved the MAE by $6\%$. For females, we observed similar improvements. Our method decreased the MAE by $22\%$, $17\%$, and $27\%$ for predictions one, two, and, three five-year periods ahead compared to those of the persistence forecast.

Also, compared with other probabilistic forecast methods, our method produced shorter prediction intervals with empirical coverages close to the nominal level for one and two five-year predictions, while it produced prediction intervals with reasonably close to nomial for the three five-year predictions for the male ASAF. On the other hand, since most female ASAF series have not yet reached the peak, capturing the variability of future female ASAF is essential. The coverage of our method is close to the nominal level, indicating that our method is well calibrated. 

\begin{figure}[tbp]
	\begin{center}
		\includegraphics[scale=0.25]{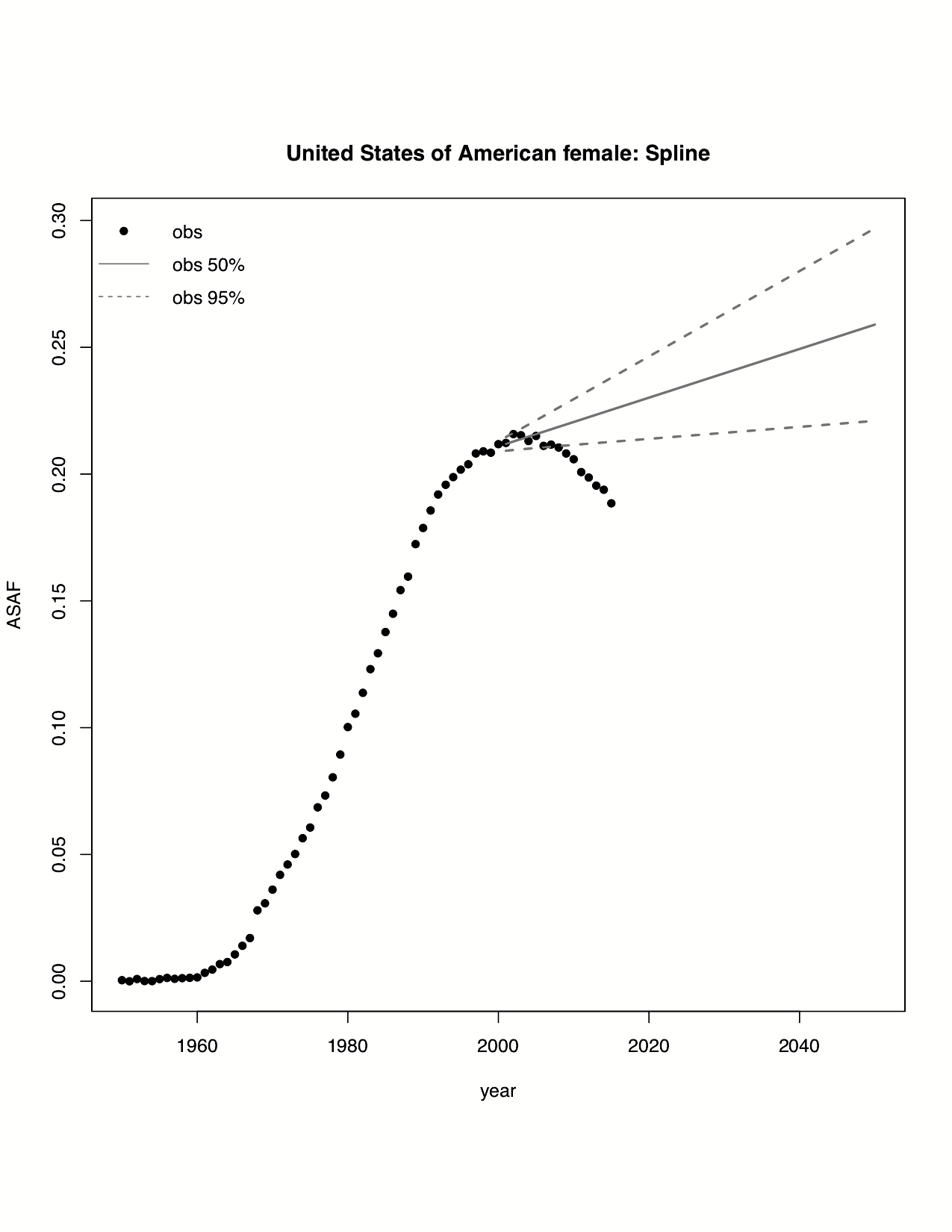}
		\includegraphics[scale=0.25]{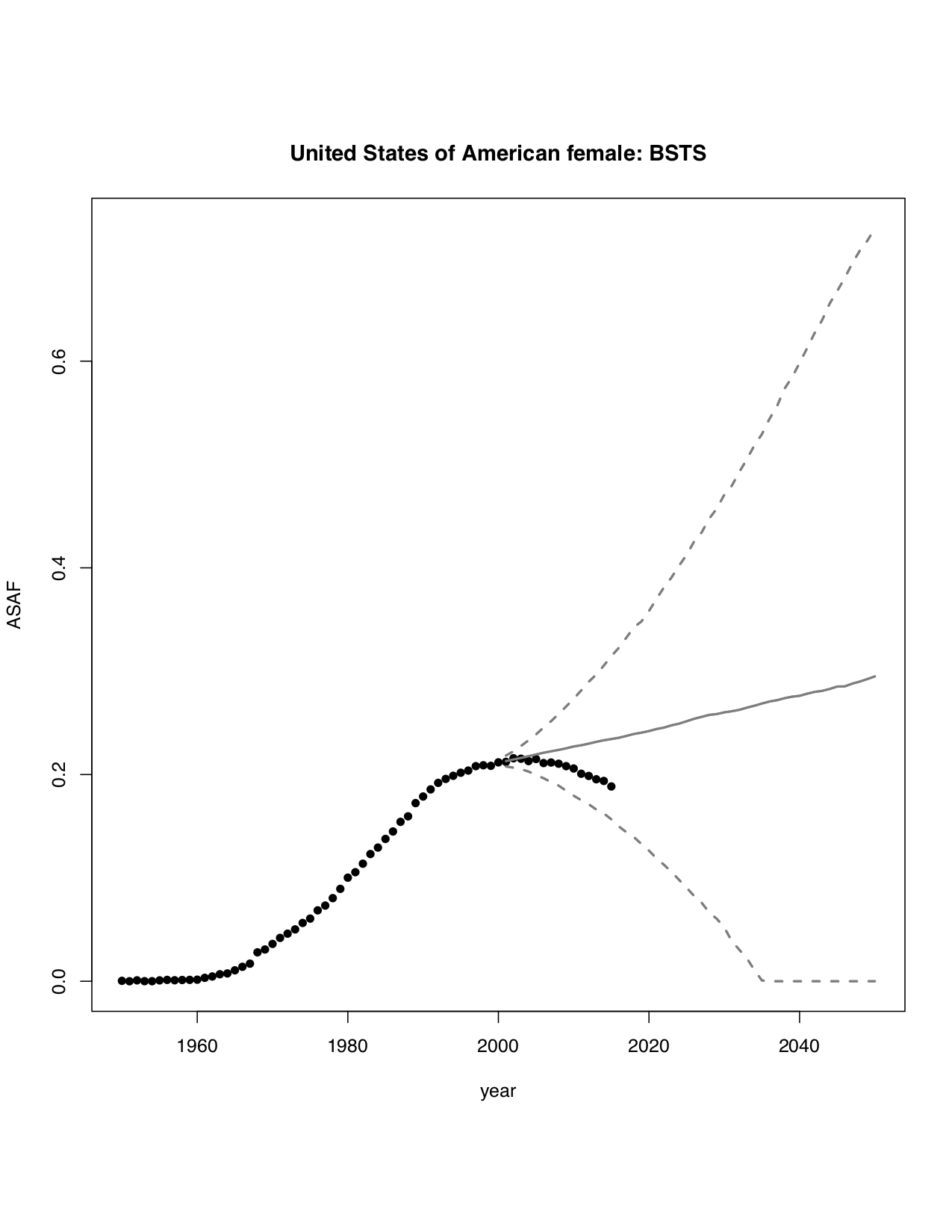}
	\end{center}
	\caption{Forecast of US female ASAF based on data before 2000 using Bayesian spline method (left) and Bayesian structural time series method (right). Observed ASAF values are represented by black dots. The solid lines and dashed lines represent the posterior median and the $95\%$ prediction interval, respectively.}\label{fg:bsts}
\end{figure}

Overall, our proposed BHM yielded the smallest CRPS among all methods in most cases for both the male and female epidemics. Among all five methods compared in the validation exercise, the Bayesian spline method was worst in terms of forecast accuracy, and tended to underestimate the variability of future values. 
The Bayesian structural time series model produced prediction interval close to the nominal level with slightly larger average half-width than our method. However, a significant drawback of the persistence forecast, the Bayesian spline method, and the Bayesian structural time series model is that they tend to produce unrealistic forecasts when all the observed data are before the peak, since they do not incorporate the increasing-peaking-decreasing information in the model. The left panel of Figure \ref{fg:bsts} indicates that the Bayesian thin plate spline method projected a monotonically increasing ASAF for United States female based on data before 2000, where the entire prediction interval missed the observed data after 2000. The right panel of Figure \ref{fg:bsts} shows that the Bayesian structural time series model did cover the data but with an unrealistically wide prediction interval. 

The Bayesian model without the global level parameters produced results similar to those from our BHM for projecting short term male ASAF. When forecasting three five-year periods ahead, or the female ASAF, in both of which cases the peak has often not been reached, the Bayesian model without the global level parameters was worse in accuracy and CRPS. 
This indicates that the hierarchical structure did indeed improve the
overall forecast when only part of the trend has been observed, 
by sharing information among all the countries.

Table \ref{tb:validoecd} gives validation results for subgroups of countries,
categorized  by membership of the Organization for Economic Cooperation and Development (OECD). Most of the countries in the OECD are regarded as developed countries with high GDP and human development index (HDI). For male ASAF, our BHM improved most of the forecasts for OECD countries, especially the longer term projections. For OECD countries, the increasing-peaking-decreasing pattern is clearer and stronger, which fits with our modeling well.  In contrast, our BHM performed less well among non-OECD countries. 

%For non-OECD countries, the observed ASAF has more deviations from the main increasing-peaking-decreasing trend. But after inspecting the validation plots, we found that the validation for most non-OECD countries were better than the summarizing quantities suggested. For forecasting female ASAF, our BHM works well under almost all cases for both OECD and non-OECD countries. Since most of them have only partially revealed trend of ASAF, our BHM can borrow information from the male series and from other countries to enhance the forecast. In general, our BHM substantially improved the forecasting results for countries with clear pattern ASAF series, while it did not hurt on those with less clear patterns by producing unrealistic projections. }

\begin{table}
	\caption{Predictive validation results for all-age smoking attributable fraction (ASAF). The first and second columns indicate the estimation and validation periods . The ``Gender" and ``$n$" columns indicate the gender and the number of countries used for the validation. In the ``Model" column, ``Bayes" represents the Bayesian hierarchcial model with measurement error and random walk with drift, ``Bayes(S)" represents the same model as ``Bayes" without the global parameters, ``Persistence" represents the persistence forecast, ``Spline" represents the Bayesian thin plate regression spline method, and ``BSTS" represents the Bayesian structural time series method. The ``MAE" column contains the mean absolute prediction error defined by Eq. \protect\ref{eq:mae}. The ``Coverage" columns show the proportion of validation observations contained in the $80\%$, $90\%$, $95\%$ prediction intervals with their average half-widths in parentheses. The ``CRPS" column contains the continuous ranked probability score defined by Eq. \protect\ref{eq:crps}.}
	\begin{center}
		\resizebox{\textwidth}{!}{	\begin{tabular}{ |c|c|c|c|c|c|ccc|c|}
				\hline
				\multirow{2}{*}{Training} & \multirow{2}{*}{Test}& \multirow{2}{*}{$n$}  & \multirow{2}{*}{Gender}  & \multirow{2}{*}{Model} & \multirow{2}{*}{MAE} & \multicolumn{3}{c|}{Coverage} & \multirow{2}{*}{CRPS}\\\cline{7-9}
				& & & & & & $80\%$ & $90\%$ & $95\%$ & \\
				\hline
				
				\multirow{10}{*}{1950--2010} &  \multirow{10}{*}{2010--2015} &\multirow{10}{*}{66} & 
				\multirow{5}{*}{Male}& Persistence &0.010 & - & - & - & - \\ 
				& & && Bayes &0.007& 0.78 (0.011) & 0.86 (0.014) & 0.90 (0.017) & 0.00523\\
				& & && Bayes(S) & 0.007 & 0.86 (0.014) & 0.94 (0.018) & 0.97 (0.022)  & 0.00505\\
				& & && Spline &0.008& 0.58 (0.009) & 0.65 (0.011) & 0.72 (0.013) & 0.00648\\
				& & && BSTS & 0.008 & 0.85 (0.015) & 0.94 (0.020) & 0.94 (0.025) & 0.00570\\
				
				\cline{4-10}
				
				& &  &\multirow{5}{*}{Female} &Persistence &0.009 & - & - & - & -\\
				& & && Bayes & 0.007& 0.83 (0.012) & 0.93 (0.015) & 0.96 (0.018) & 0.00507\\ 
				& & && Bayes(S) & 0.008 & 0.88 (0.014) & 0.94 (0.018) & 0.97 (0.022) & 0.00538\\
				& & && Spline & 0.010 & 0.42 (0.007) & 0.52 (0.009) & 0.61 (0.011) & 0.00763\\
				& & && BSTS & 0.008 & 0.80 (0.013) & 0.89 (0.016) & 0.94 (0.020) & 0.00562\\
				
				\hline
				
				\multirow{10}{*}{1950--2005} &  \multirow{10}{*}{2005--2015} & \multirow{10}{*}{66} & \multirow{5}{*}{Male}& Persistence &0.014 & - & - & - & - \\ 
				& & && Bayes & 0.011& 0.72 (0.014) & 0.83 (0.018) & 0.89 (0.022) & 0.00797 \\
				&&&& Bayes(S) & 0.010 & 0.85 (0.020) & 0.93 (0.027) & 0.97 (0.033) & 0.00795\\
				& & && Spline & 0.014 & 0.54 (0.014) & 0.65 (0.018) & 0.72 (0.021) & 0.01096\\
				& & && BSTS & 0.013 & 0.83 (0.026) & 0.90 (0.035) & 0.95 (0.043) & 0.00989\\

				\cline{4-10}
				& &  &\multirow{5}{*}{Female} & Persistence &0.012 & - & - & - &-  \\
				& & && Bayes & 0.010& 0.80 (0.015) & 0.90 (0.020)& 0.92 (0.025) & 0.00721\\ 
				& & && Bayes(S) & 0.011 & 0.88 (0.021) & 0.93 (0.028) & 0.95 (0.035) & 0.00808\\
				& & && Spline & 0.014 & 0.44 (0.011) & 0.51 (0.014) & 0.58 (0.016) & 0.01133\\
				& & && BSTS & 0.011 & 0.77 (0.017) & 0.88 (0.023) & 0.93 (0.029) & 0.00802\\
				
				\hline
				
				\multirow{10}{*}{1950--2000} &  \multirow{10}{*}{2000--2015}& \multirow{10}{*}{63}&  \multirow{5}{*}{Male} & Persistence & 0.017 & - & - & -  & -\\ 
				& & && Bayes & 0.016 & 0.65 (0.020) & 0.76 (0.026) & 0.84 (0.031)  &0.01214 \\
				& & && Bayes(S) & 0.018 & 0.84 (0.031) & 0.92 (0.042) & 0.95 (0.052) & 0.01278\\
				& & && Spline & 0.018 & 0.59 (0.019) & 0.69 (0.024) & 0.76 (0.029) & 0.01335\\
				& & && BSTS & 0.016 & 0.85 (0.039) & 0.93 (0.053) & 0.98 (0.068) & 0.01281\\
				
				\cline{4-10}
				& & & \multirow{5}{*}{Female} &Persistence & 0.015 & - & - & - & -\\
				& & && Bayes & 0.011 & 0.81 (0.021) & 0.90 (0.029)& 0.95 (0.037) & 0.00817\\ 
				& & && Bayes(S) & 0.012 & 0.88 (0.027) & 0.96 (0.039) & 0.98 (0.050) & 0.00887\\
				& & && Spline & 0.016 & 0.48 (0.014) & 0.59 (0.018) & 0.70 (0.022) & 0.01151\\
				& & && BSTS & 0.012 & 0.79 (0.022) & 0.89 (0.030) & 0.94 (0.039) & 0.00831\\ 
				
				\hline
		\end{tabular}}
		\label{tb:valid}
	\end{center}
\end{table}

\begin{table}
	\footnotesize
	\caption{Predictive validation results for all-age smoking attributable fraction (ASAF) for categories of countries. The ``OECD" column represents whether the countries in the subgroup belong to the OECD. The number of countries in the subgroup used for the validation is in parentheses. All the other columns are the same as those in Table \protect\ref{tb:valid}.}\label{tb:validoecd}
	\begin{center}
		\resizebox{\textwidth}{0.45\textheight}{	\begin{tabular}{ |c|c|c|c|c|c|ccc|c|}
				\hline
				\multirow{2}{*}{Training} & \multirow{2}{*}{Test}& \multirow{2}{*}{Gender} & \multirow{2}{*}{OECD}   & \multirow{2}{*}{Model} & \multirow{2}{*}{MAE} & \multicolumn{3}{c|}{Coverage} & \multirow{2}{*}{CRPS}\\\cline{7-9}
				& & &  & & & $80\%$ & $90\%$ & $95\%$ & \\
				\hline
				
				\multirow{20}{*}{1950--2010} &  \multirow{20}{*}{2010--2015} & \multirow{10}{*}{Male}& \multirow{5}{*}{Y(34)} & Persistence &0.011 & - & - & - & - \\ 
				& & && Bayes & 0.006  &  0.81 (0.011) & 0.90 (0.014)  & 0.95 (0.016)   & 0.00448 \\
				& & && Bayes(S) & 0.006  &  0.88 (0.013)& 0.94 (0.017) & 0.99 (0.021) &  0.00459 \\
				& & && Spline &  0.007 & 0.60 (0.008) & 0.67 (0.010)  & 0.73 (0.012) & 0.00565 \\
				& & && BSTS &  0.007 &    0.86 (0.014)  & 0.95 (0.018)  & 0.98 (0.022)  & 0.00529 \\
				\cline{4-10}
				
				&  & & \multirow{5}{*}{N(32)} & Persistence &0.008& - & - & - & - \\ 
				& & && Bayes & 0.009   & 0.75 (0.011)  & 0.81 (0.015)  & 0.84 (0.018)   & 0.00601 \\
				& & && Bayes(S) & 0.008  & 0.85 (0.015) & 0.92 (0.019) & 0.94 (0.023)  & 0.00554\\ 
				& & && Spline &0.010   &  0.56(0.010) &  0.63 (0.012)&   0.70 (0.015)   & 0.00736  \\
				& & && BSTS & 0.009    &  0.86 (0.017)  &  0.95 (0.023) & 0.98 (0.028) & 0.00629\\
				\cline{3-10}
				
				&  & \multirow{10}{*}{Female}& \multirow{5}{*}{Y(34)} & Persistence &0.009& - & - & - & - \\ 
				& & && Bayes & 0.007 &  0.82 (0.011)& 0.92 (0.015)  &0.94 (0.018)   & 0.00505 \\
				& & && Bayes(S) & 0.008  &  0.86 (0.013) &0.93 (0.017)  & 0.96 (0.021) & 0.00560  \\
				& & && Spline &  0.010& 0.42 (0.007) & 0.51 (0.008)  & 0.58 (0.010)& 0.00762 \\
				& & && BSTS &  0.009 &  0.78 (0.012)  & 0.85 (0.015) & 0.91 (0.019) & 0.00616 \\
				\cline{4-10}
				
				&  & & \multirow{5}{*}{N(32)} & Persistence &0.008& - & - & - & - \\ 
				& & && Bayes & 0.008   & 0.83 (0.012)  & 0.95 (0.015)  & 0.95 (0.018)   & 0.00507 \\
				& & && Bayes(S) & 0.007 & 0.89 (0.015) & 0.95 (0.019) & 0.98 (0.023)  & 0.00516\\ 
				& & && Spline &0.011   &  0.42(0.008) &  0.54 (0.010)&   0.63 (0.012)   & 0.00764  \\
				& & && BSTS & 0.007    &  0.82 (0.013)  &  0.89 (0.017) & 0.94 (0.021) & 0.00506\\
				
				\hline
				
				\multirow{20}{*}{1950--2005} &  \multirow{20}{*}{2005--2015} & \multirow{10}{*}{Male}& \multirow{5}{*}{Y(34)} & Persistence &0.016 & - & - & - & - \\ 
				& & && Bayes & 0.010  &  0.73 (0.014)  & 0.85 (0.018)   &  0.90 (0.021) & 0.00676 \\
				& & && Bayes(S) & 0.010  & 0.84 (0.019) &  0.93 (0.025) &0.97 (0.032)  & 0.00717  \\
				& & && Spline & 0.013  & 0.52 (0.012) & 0.61 (0.015)   & 0.69 (0.018) & 0.01008  \\
				& & && BSTS &  0.012 &   0.85 (0.028)   & 0.91 (0.039)  & 0.97 (0.049)   &  0.01000 \\
				\cline{4-10}
				&   && \multirow{5}{*}{N(32)} & Persistence &0.011 & - & - & - & - \\ 
				& & && Bayes & 0.012  &  0.70 (0.014)  & 0.81 (0.019) &  0.88 (0.022)&  0.00928  \\
				& & && Bayes(S) & 0.011 & 0.87 (0.021)& 0.93 (0.029) & 0.96 (0.035)  &  0.00879  \\
				& & && Spline & 0.015  &0.57 (0.016)  &0.68 (0.020)   & 0.76 (0.024) &0.01189  \\
				& & && BSTS &0.013   & 0.83 (0.026)     &  0.90 (0.035) & 0.95 (0.043)  & 0.00989 \\
				\cline{3-10}
				&   & \multirow{10}{*}{Female}& \multirow{5}{*}{Y(34)} & Persistence &0.012 & - & - & - & - \\ 
				& & && Bayes & 0.009   &  0.82 (0.015) &  0.92 (0.020)  &  0.95 (0.025)  & 0.00669 \\
				& & && Bayes(S) & 0.010 & 0.88 (0.019)&  0.95 (0.025) &  0.96 (0.032) &  0.00736  \\
				& & && Spline & 0.012  & 0.38 (0.008) & 0.45 (0.011)   & 0.52 (0.013) & 0.00945  \\
				& & && BSTS &  0.012 &  0.82 (0.019)    &  0.90 (0.026) & 0.92 (0.033)  & 0.00851 \\
				\cline{4-10}
				&  & & \multirow{5}{*}{N(32)} & Persistence & 0.013& - & - & - & - \\ 
				& & && Bayes & 0.011  & 0.78 (0.015)  &0.88 (0.020)   &  0.90 (0.025) & 0.00780  \\
				& & && Bayes(S) &0.012  & 0.88 (0.023)&  0.91 (0.031)& 0.93 (0.039) & 0.00885   \\
				& & && Spline &0.017   & 0.51 (0.013) &0.58 (0.017)   & 0.66 (0.020) & 0.01333  \\
				& & && BSTS & 0.011  &  0.77 (0.017)    & 0.88 (0.023)  & 0.93 (0.029)  & 0.00802  \\
				\cline{4-10}
				
				\hline
				
				\multirow{20}{*}{1950--2000} &  \multirow{20}{*}{2000--2015} & \multirow{10}{*}{Male}& \multirow{5}{*}{Y(33)} & Persistence & 0.018& - & - & - & - \\ 
				& & && Bayes & 0.014  &  0.67 (0.020) &  0.79 (0.026)  & 0.88 (0.032)  & 0.01063  \\
				& & && Bayes(S) & 0.017 & 0.83 (0.030)& 0.90 (0.040)  & 0.95 (0.050)&  0.01221  \\
				& & && Spline & 0.017  & 0.58 (0.015) & 0.68 (0.020)  &0.74 (0.023)  & 0.01338 \\
				& & && BSTS & 0.018   & 0.88 (0.035)     & 0.93 (0.047)  & 0.97 (0.060)  & 0.01308 \\
				\cline{4-10}
				&   && \multirow{5}{*}{N(30)} & Persistence &0.017 & - & - & - & - \\ 
				& & && Bayes & 0.019  & 0.63 (0.020)   & 0.72 (0.026)  &0.80 (0.031)   & 0.01377 \\
				& & && Bayes(S) &0.018  &0.86 (0.032) & 0.93 (0.042) & 0.95 (0.053)  &  0.01341 \\
				& & && Spline &0.018   &  0.60 (0.022)& 0.71 (0.029)   & 0.79 (0.034)  &  0.01331\\
				& & && BSTS & 0.016   &   0.85 (0.045)   & 0.94 (0.063)  & 0.98 (0.082)  & 0.01308  \\
				\cline{3-10}
				&   & \multirow{10}{*}{Female}& \multirow{5}{*}{Y(33)} & Persistence & 0.016 & - & - & - & - \\ 
				& & && Bayes & 0.011   & 0.80 (0.021)  & 0.89 (0.029)  & 0.95 (0.037)  & 0.00817 \\
				& & && Bayes(S) & 0.013 & 0.84 (0.028)& 0.94 (0.038)  &  0.97 (0.048)& 0.00981  \\
				& & && Spline &0.016   & 0.41 (0.012) &0.54 (0.015)   &0.62 (0.018)  & 0.01230 \\
				& & && BSTS & 0.011  &  0.73 (0.016)    &  0.86 (0.022) & 0.92 (0.027)  & 0.00777 \\
				\cline{4-10}
				&  & & \multirow{5}{*}{N(30)} & Persistence & 0.013& - & - & - & - \\ 
				& & && Bayes & 0.010 & 0.82 (0.017) & 0.92 (0.023) & 0.95 (0.030) & 0.00699 \\
				& & && Bayes(S) &  0.010  &  0.93 (0.028) & 0.98 (0.040)  & 0.99 (0.052)  & 0.00784  \\  
				& & && Spline & 0.016  &  0.56 (0.017) & 0.66 (0.022)   & 0.79 (0.026)  & 0.01066   \\
				& & && BSTS &0.010   & 0.86 (0.022)     &  0.94 (0.030)  & 0.95 (0.039)  & 0.00735  \\
				\cline{4-10}
				
				\hline
		\end{tabular}}

	\end{center}
\end{table}

\begin{figure}[tbp]
	\begin{center}
		\includegraphics[scale=0.32]{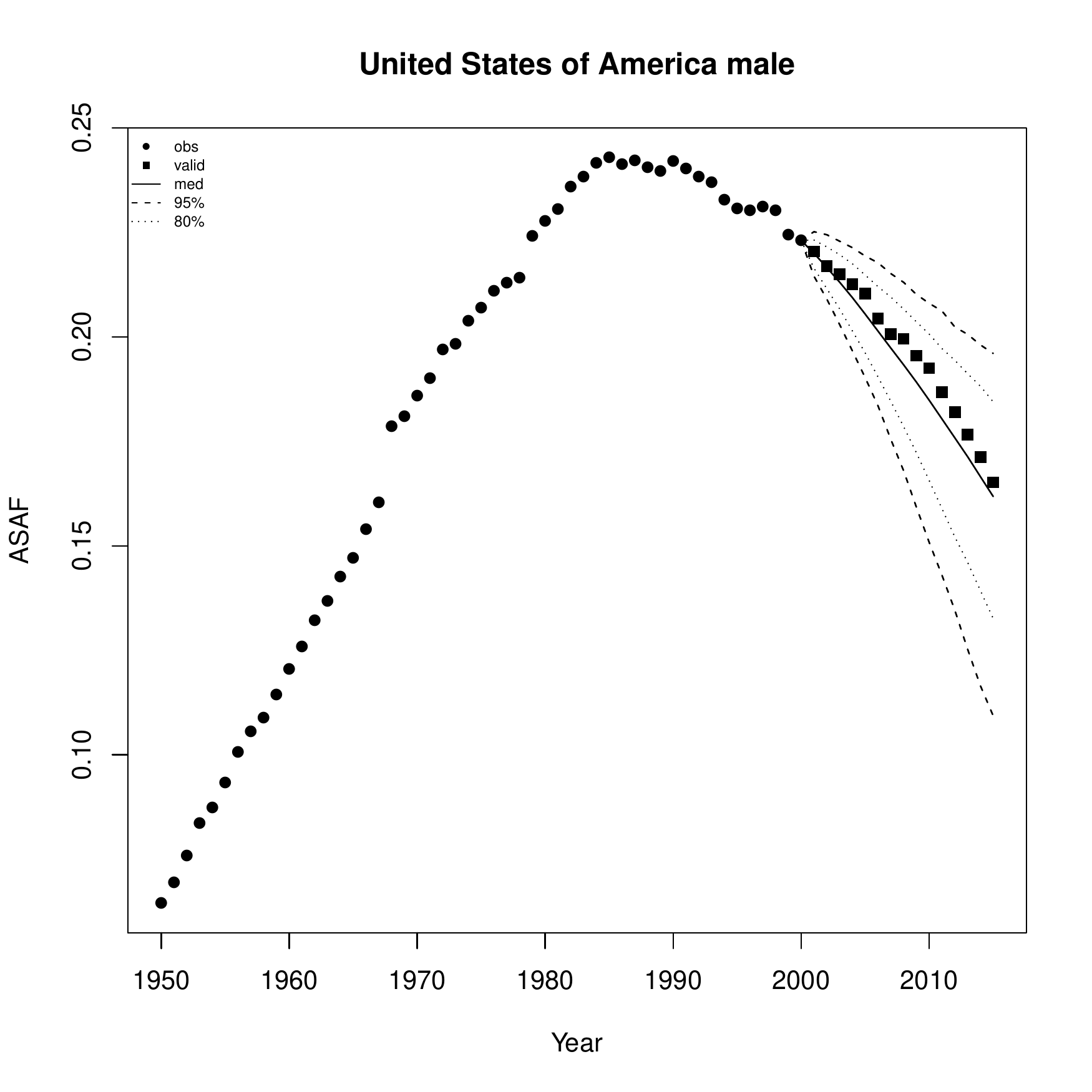}
		\includegraphics[scale=0.32]{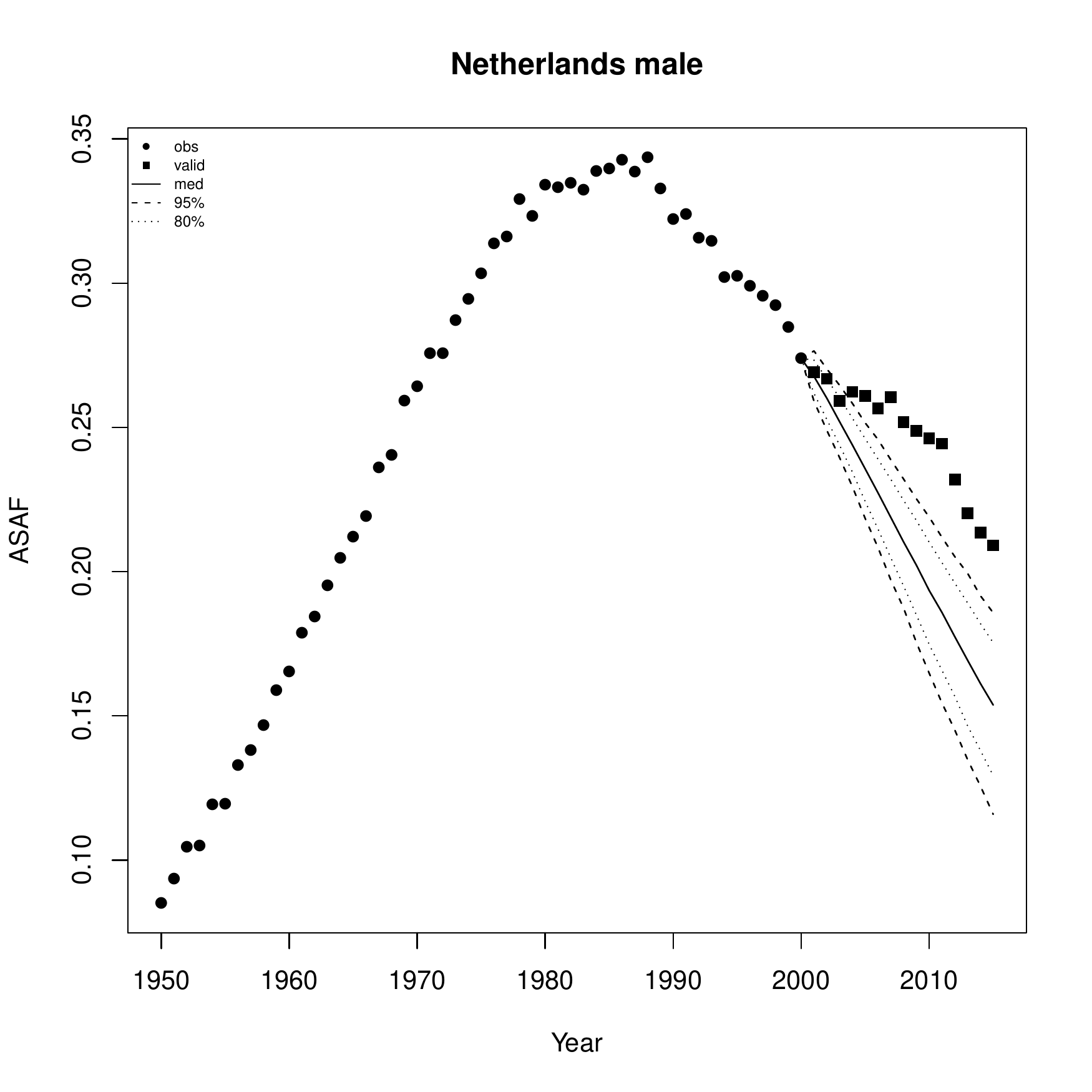}
		\includegraphics[scale=0.32]{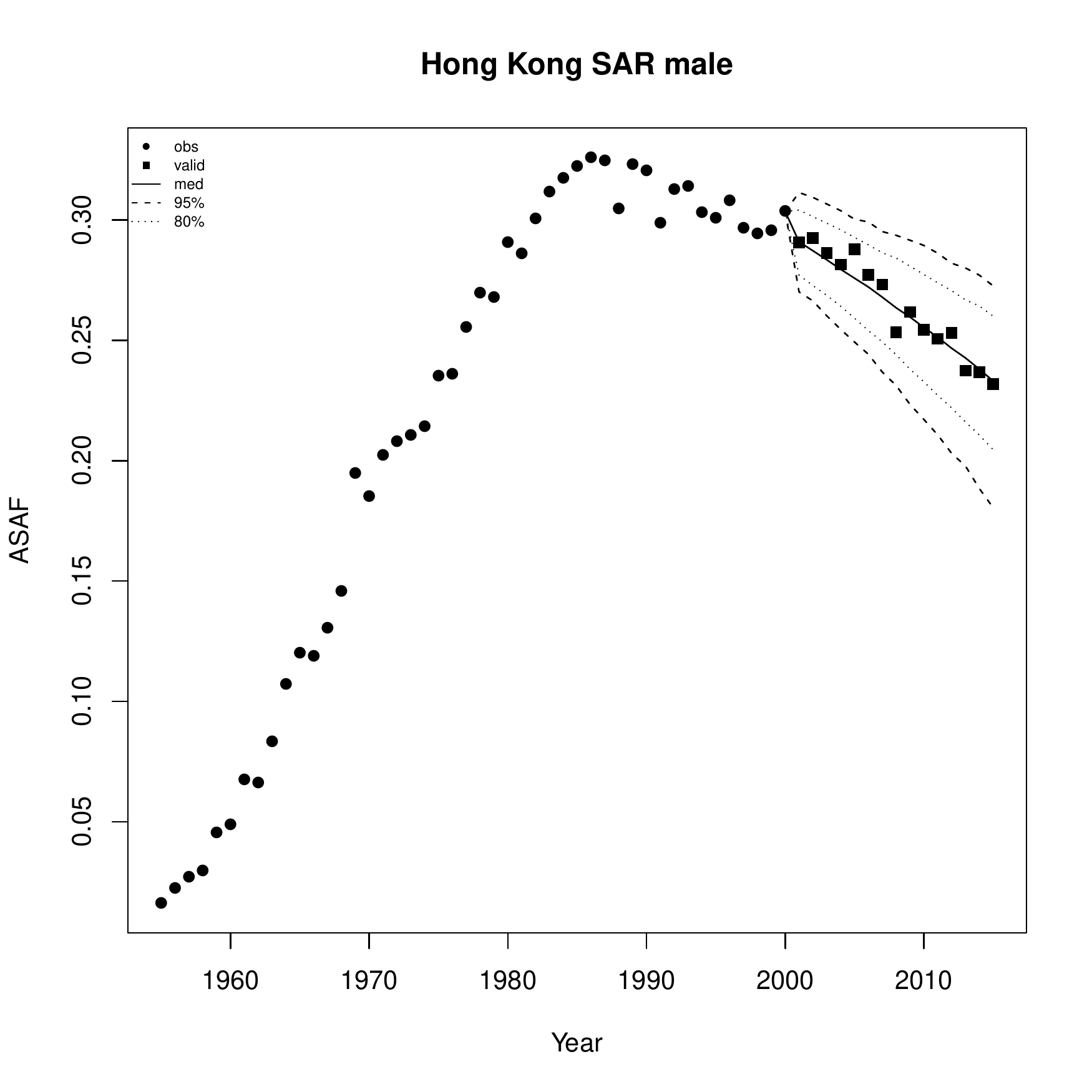}
		\includegraphics[scale=0.32]{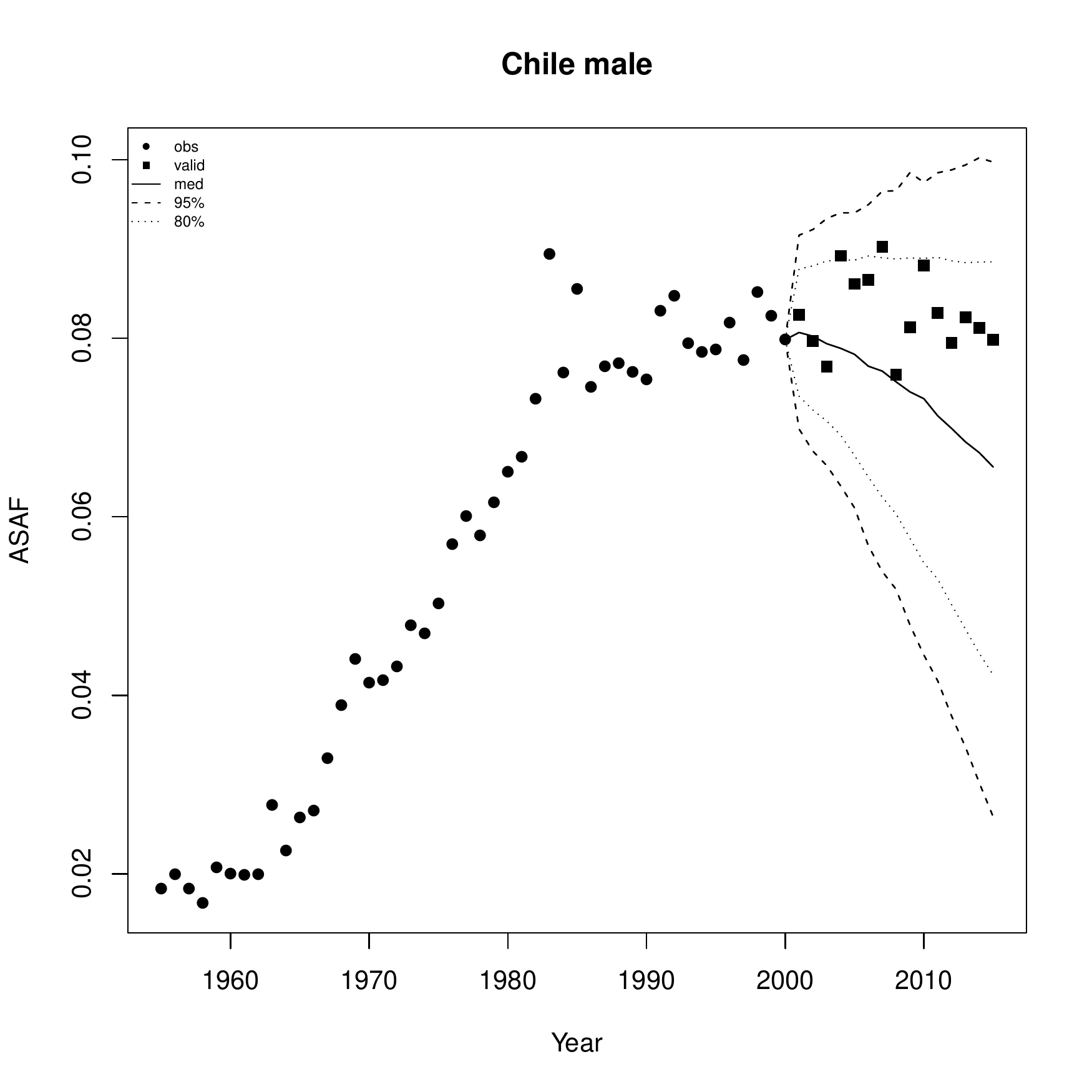}
	\end{center}
	\caption{Validation of male all-age smoking attributable fraction for the United States, Netherlands, Hong Kong, and Chile. Past observed ASAF values are shown by black dots for 1950--2000 and by black squares for 2000--2015. The posterior median for 2000--2015 is shown by the solid line, and the $80\%$ and $95\%$ prediction intervals are shown by the dotted and dashed lines respectively.}	\label{fg:validm}
\end{figure}

\begin{figure}[tbp] 
	\begin{center}
		\includegraphics[scale=0.32]{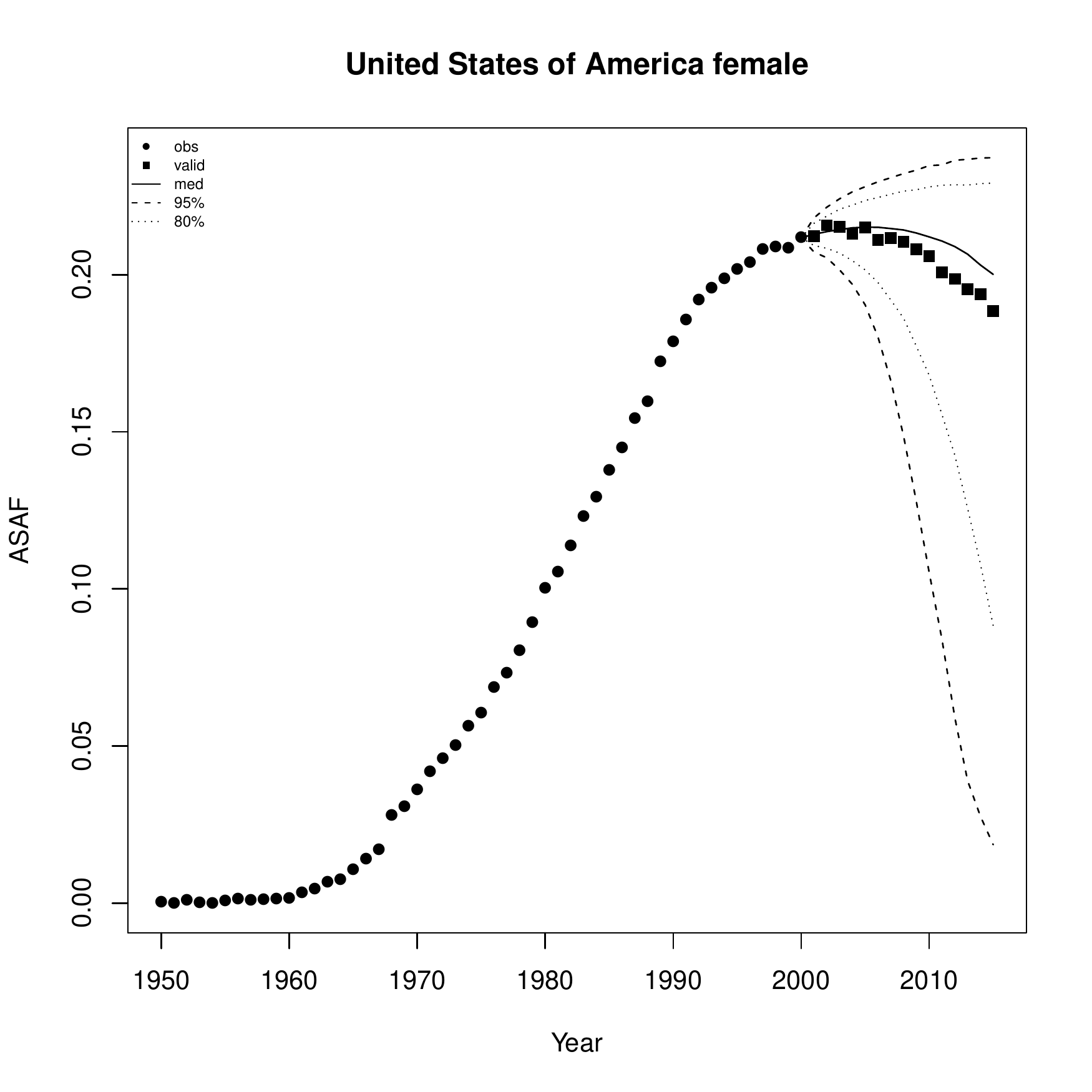}
		\includegraphics[scale=0.32]{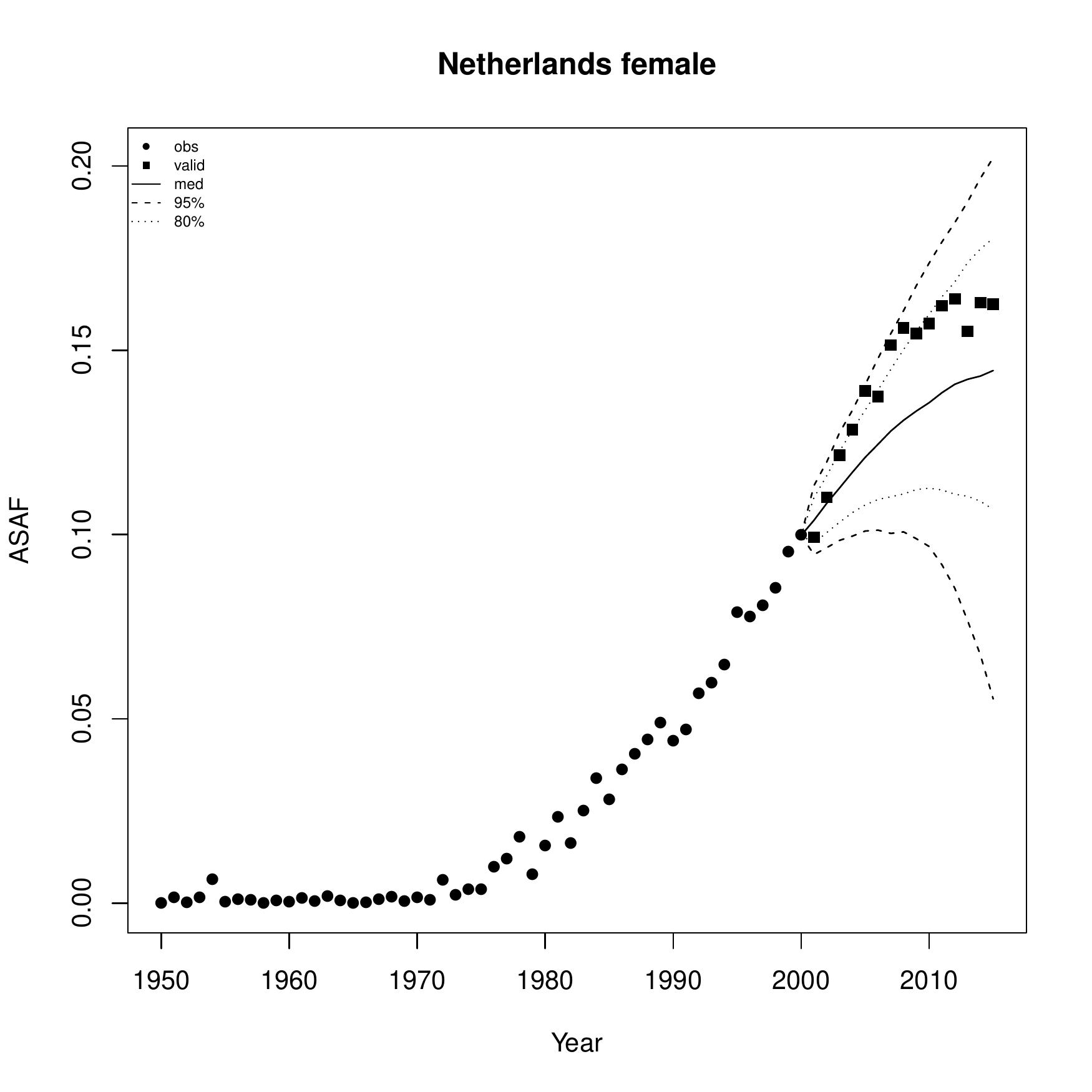}
		\includegraphics[scale=0.32]{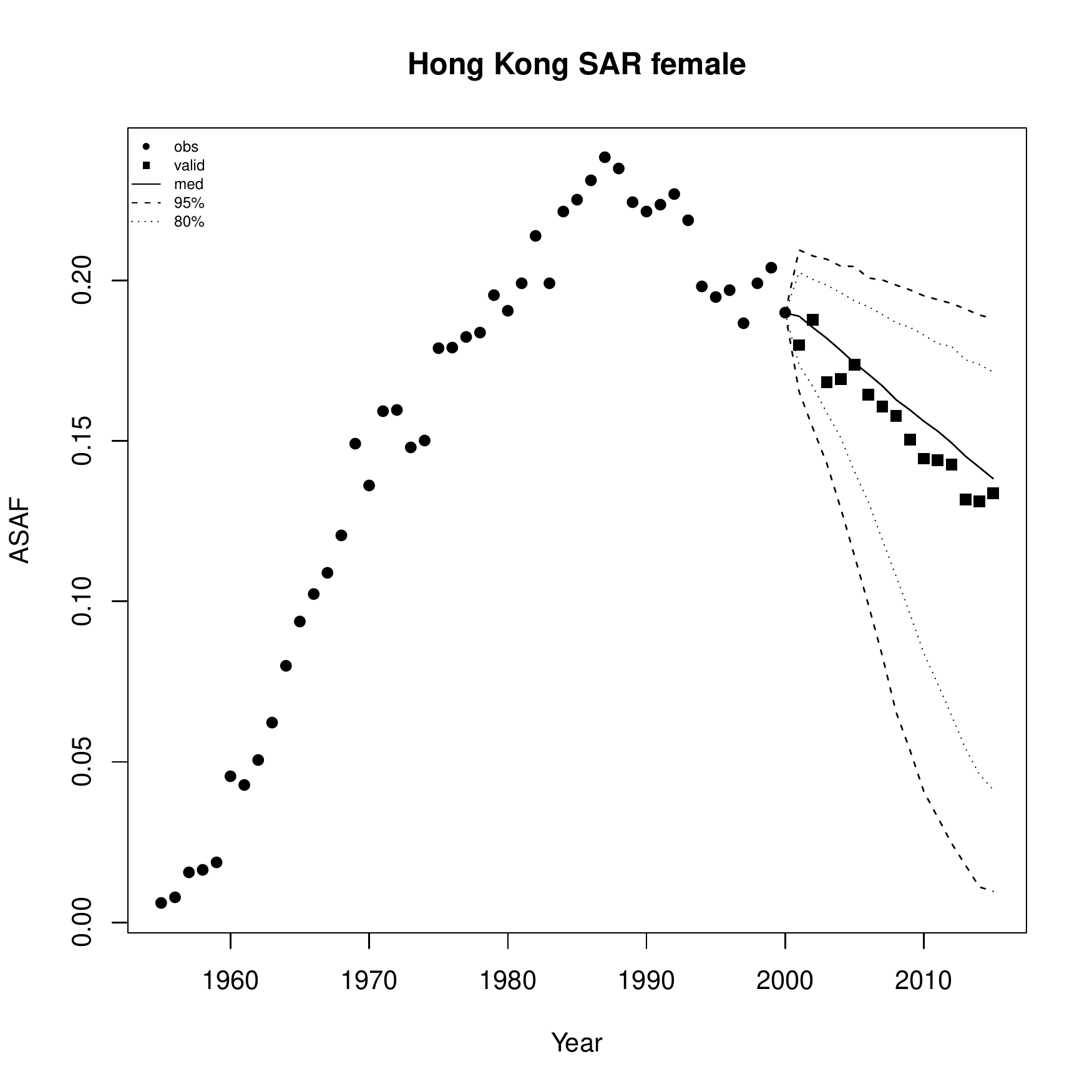}
		\includegraphics[scale=0.32]{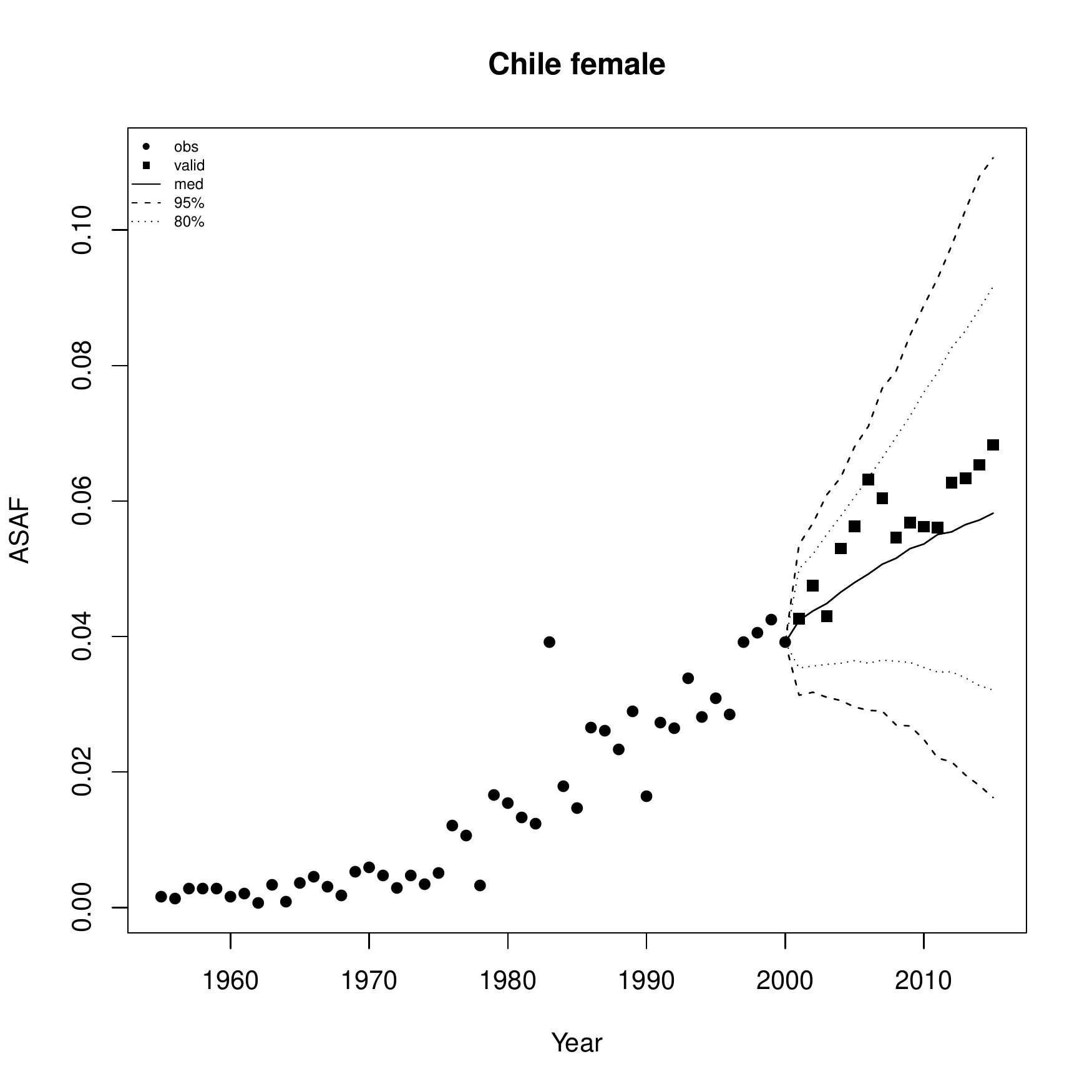}
	\end{center}
	\caption{Validation of female all-age smoking attributable fraction for the United States, Netherlands, Hong Kong, and Chile. Past observed ASAF values are shown by black dots for 1950--2000 and by black squares for 2000--2015. 
		The posterior median for 2000--2015 is shown by the solid line, and the $80\%$ and $95\%$ prediction intervals are shown by the dotted and dashed lines respectively.}\label{fg:validf}
\end{figure}

Figure \ref{fg:validm} shows validation results for the male ASAF of four countries or regions for predictions three five-year periods ahead. We see that our method works quite well for the United States and Hong Kong, and the prediction interval captures the variability of the male ASAF of Chile. Figure \ref{fg:validf} shows the results from Scenario (1) where most female ASAF of countries among the examples have not reached the peak by the year 2000. We see that the posterior median of the predictive distribution captures the general trend of future female ASAF of the United States, the Netherlands, and Chile reasonably well. For countries or regions like Hong Kong whose female ASAF already passed the peak, our method also accurately estimates the rate of decline.

\section{Case Studies}\label{sec:casestudy}

Probabilistic forecasts of ASAF to 2050 are given in the Appendix \ref{app4} for over 60 countries. Broadly, the patterns in the OECD countries are similar, with male ASAF having declined from about $30\%$ in the 1990s to around $15\%$ in 2015, with further declines projected to 2050, reaching around $5\%$. The patterns vary more for females in OECD, and for both males and females in non-OECD countries because they are currently  at different stages of the epidemic.

We now give four cases studies which illustrate various aspects of the proposed method for estimating and forecasting ASAF.

\subsection{United States}
The annual ASAF for both male and female for the time period 1950--2015 is shown in Figure \ref{fg:usa}. The very clear pattern is due to the high quality of the data, reflecting the fact that the United States has one of the the best vital registration systems in the world.

\begin{figure}[tbp]
	\begin{center}
		\includegraphics[scale=0.35]{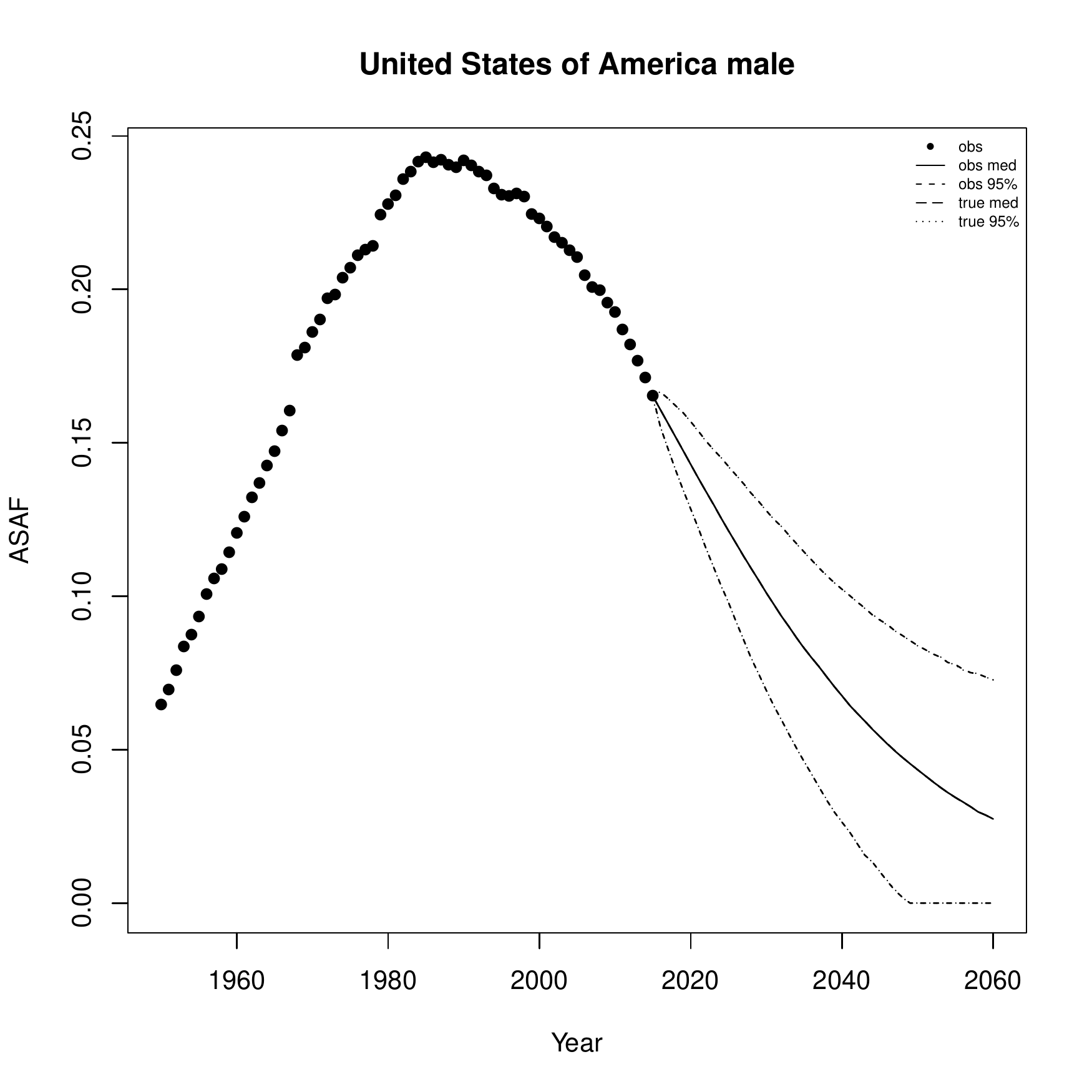}
		\includegraphics[scale=0.35]{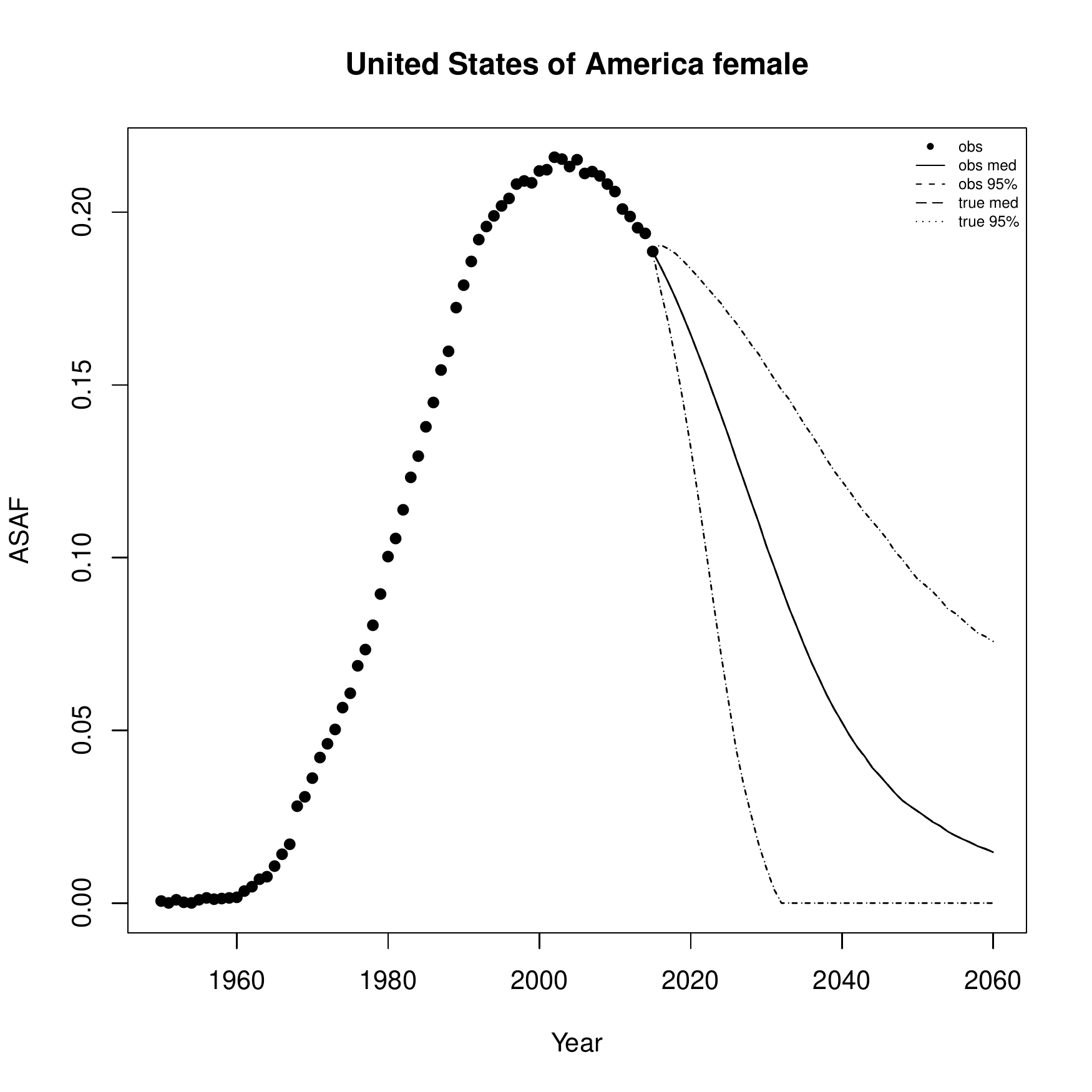}
	\end{center}
	\caption{United States: The left and right panels show the projection of ASAF up to 2050 under the proposed model for male and female respectively. The solid and long dashed lines show the posterior median of projected observed ASAF and true ASAF respectively. The dashed and dotted lines represent $95\%$ prediction intervals for observed ASAF and true ASAF respectively.}\label{fg:USAproj}
\end{figure}

The smoking epidemic in the male population in the United States started in the earlier 1900s, and there was a substantial decrease of smoking prevalence and lung cancer mortality rate after the 1950s. Smoking prevalence among US male adults was approximately $60\%$ in 1950s, and went down to about $20\%$ in the 1990s, and the general decline is still continuing \citep{burns1997cigarette, islami2015global}. The observed ASAF levelled around the 1990s and declined afterwards. We forecast that by 2050, the median observed ASAF for US males will be around $4.3\%$ (with $95\%$ prediction interval [$0.0\%$, $8.3\%$]). Because the measurement error for the US is tiny, the projected true ASAF (long dashed line for posterior mean and dotted line for $95\%$ prediction interval in Figure \ref{fg:USAproj}) for US males is almost equal to that of the observed ASAF.

The female smoking epidemic started two decades later than the male one and the maximum prevalence was around $30\%$ in the 1960s, and then declined to about $20\%$ in the 1990s \citep{burns1997cigarette}. The pattern of smoking prevalence among US females is similar to that for males, but around 20 years behind \citep{burns1997cigarette, islami2015global}. The female ASAF started to rise around the 1960s and reached its peak of $23\%$ around 2005. We forecast that by 2050, the median observed ASAF for US females will be around $2.7\%$ (with $95\%$ prediction interval [$0.0\%$, $9.3\%$]). Similarly, the projected US female true ASAF follows closely with that of the observed ASAF. Figure \ref{fg:USAproj} shows the historical records of the observed male and female ASAF during the time period 1950--2015, along with projections up to 2050 with posterior median and prediction intervals. 

\subsection{The Netherlands}
The Netherlands is a high-income western Europe country whose smoking epidemic started relatively early. 
Smoking prevalence reached $90\%$ in the 1950s and dropped to $30\%$ in the 2010s. The male observed ASAF in Netherlands passed its maximum ASAF around the 1990s and we project that it will go down to around $5.7\%$ (with $95\%$ prediction interval [$1.4\%$, $9.7\%$]) in 2050. 

\begin{figure}[tbp]
	\begin{center}
		\includegraphics[scale=0.35]{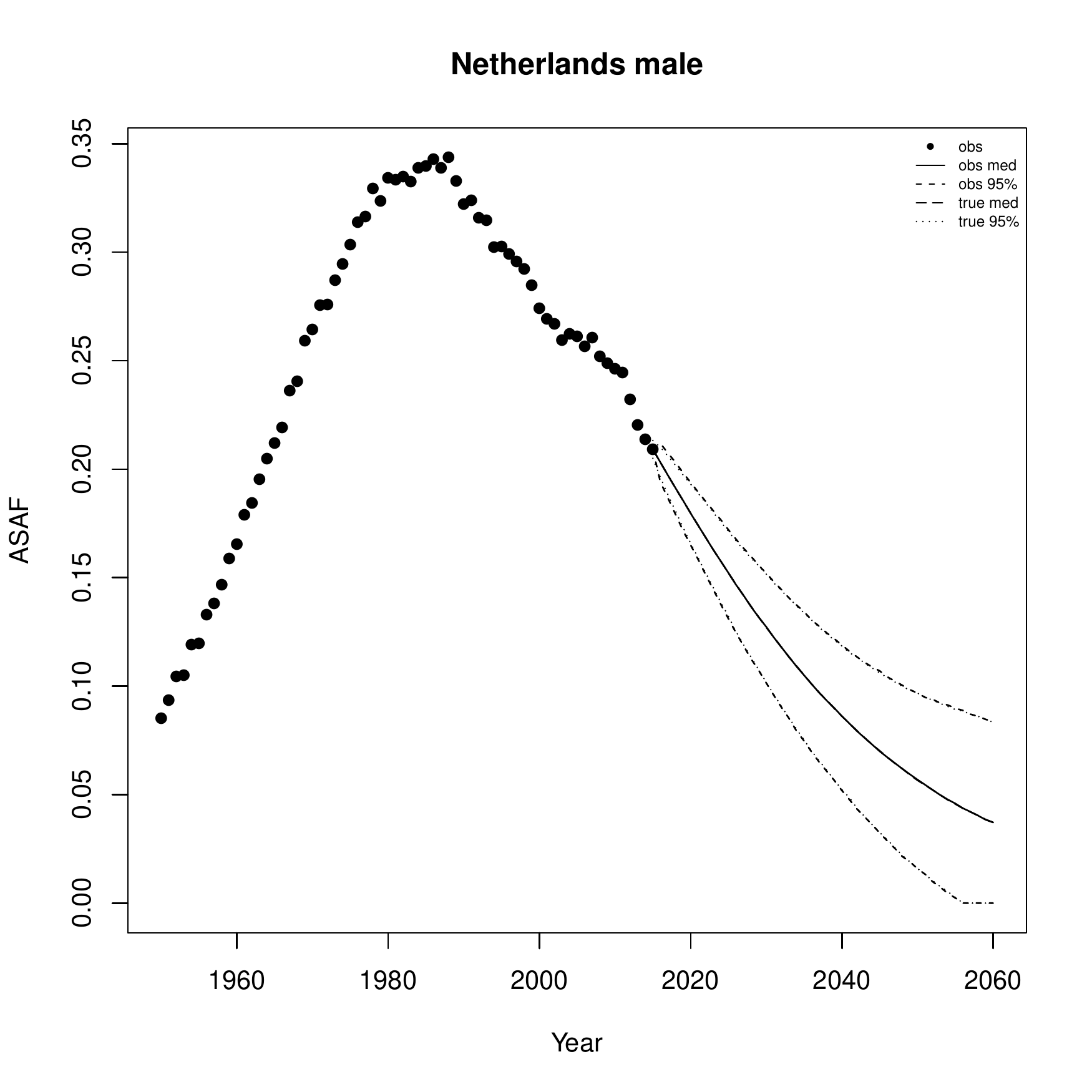}
		\includegraphics[scale=0.35]{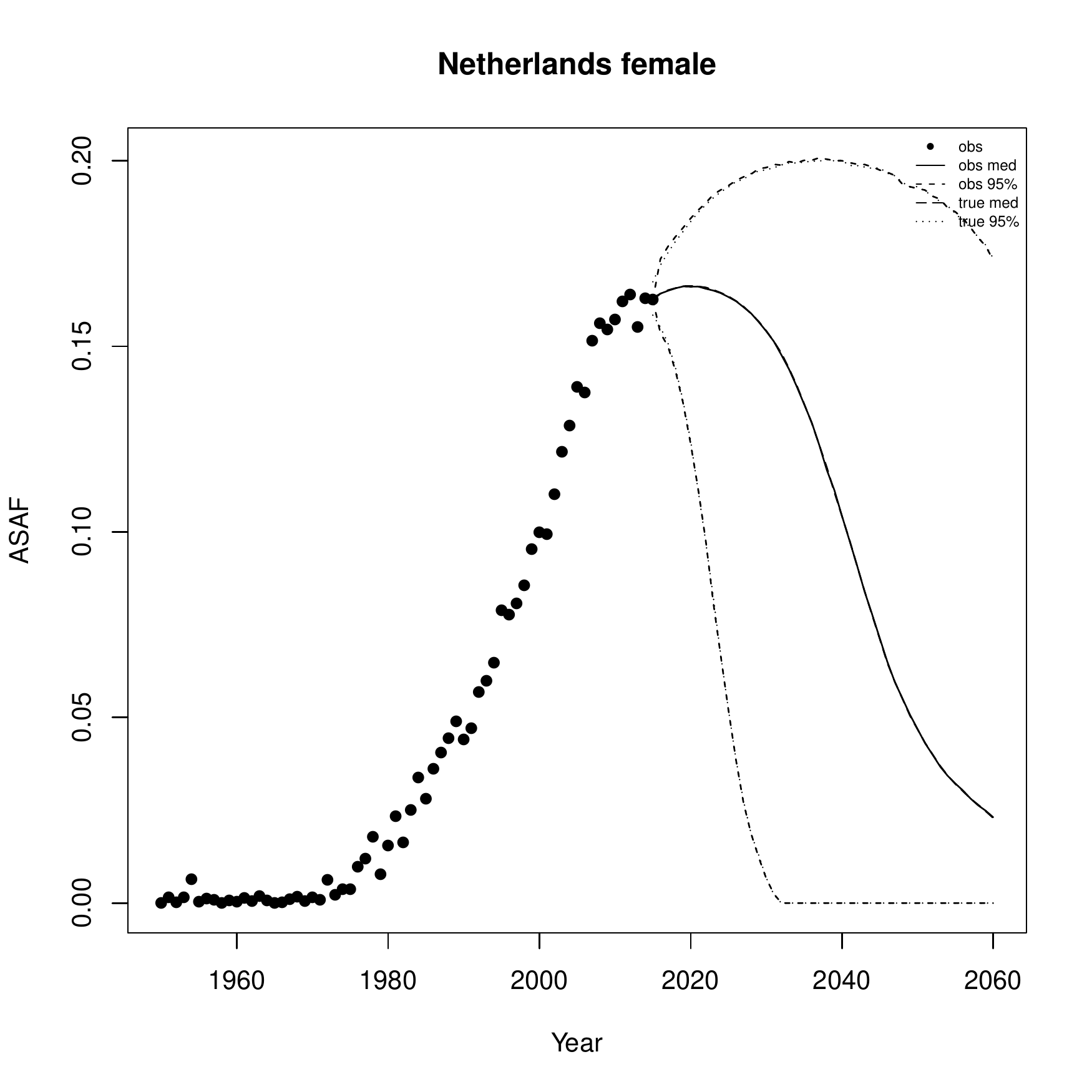}
	\end{center}
	\caption{Netherlands: The left and right panels show the projection of ASAF up to 2050 under the proposed model for male and female respectively. The solid and long dashed lines show the posterior median of projected observed ASAF and true ASAF respectively. The dashed and dotted lines represent $95\%$ prediction intervals for observed ASAF and true ASAF respectively.}\label{fg:Netherlandsproj}
\end{figure}

For females, smoking prevalence is also relatively high, and reached its peak of about $40\%$ in the 1970s  and dropped to $24\%$ in the 2010s \citep{stoeldraijer2015future}. The female ASAF in Netherlands is among the few that is already experiencing the leveling stage. By our projection, the median year-to-peak for the female ASAF will be around 2020, which is about 30 years after the male peak, and will reach $16.6\%$ (with $95\%$ prediction interval [$12.4\%$, $18.5\%$]). By 2050, the median observed female ASAF will be $4.7\%$ (with $95\%$ prediction interval [$0.0\%$, $19.3\%$]). Similarly to the case of US, the projected true ASAF follows that of the observed ASAF closely, due to the small measurement error. Figure \ref{fg:Netherlandsproj} shows the historical records of the observed male and female ASAF during time period 1950--2015, and projections are given up to 2050 with posterior median and prediction intervals for both observed and true ASAF. 

%\subsection{Denmark}

%\begin{figure}[tbp][!h]
%	\begin{center}
%		\includegraphics[scale=0.4]{Denmarkmale}
%		\includegraphics[scale=0.4]{Denmarkfemale}
%	\end{center}
%	\caption{The left and right panel show projection of ASAF in Denmark under proposed model for male and female respectively. Red circles represent posterior median of projected ASAF, colored dashed lines represent the prediction intervals with different nominal coverage.}\label{fg:denproj}
%\end{figure}

\subsection{Hong Kong}
Hong Kong has an advanced smoking epidemic, but
had a decrease in male smoking prevalence from about $40\%$ in the 1980s to $22\%$ in 2000. A decline has also been observed in female smoking prevalence, from $5.6\%$ to $3.3\%$ \citep{au2004time}. Like Japan, Singapore, and South Korea, both male and female ASAF have passed the leveling stage and have been declining for two decades. Unlike in most western developed countries, the time trend of the ASAF has been almost identical for males and females in Hong Kong, with similar times of onset and times-to-peak. \citet{au2004time} showed that the time trends of lung cancer incidence were similar for both genders. 

By our projection, the observed ASAF will reach $9.7\%$ for males (with $95\%$ prediction interval [$4.9\%$, $14.3\%$]) and $4.1\%$ for females (with $95\%$ prediction interval [$0.0\%$, $8.1\%$]) by 2050. Compared with US and the Netherlands, the projected true ASAF of Hong Kong will have narrower prediction intervals than those of the observed ASAF due to larger measurement error exhibited in the historical data. However, the difference becomes less and less since the majority uncertainty of the future ASAF will be account mainly by the variance from the random walk model of the true ASAF. 

\begin{figure}[tbp]
	\begin{center}
		\includegraphics[scale=0.35]{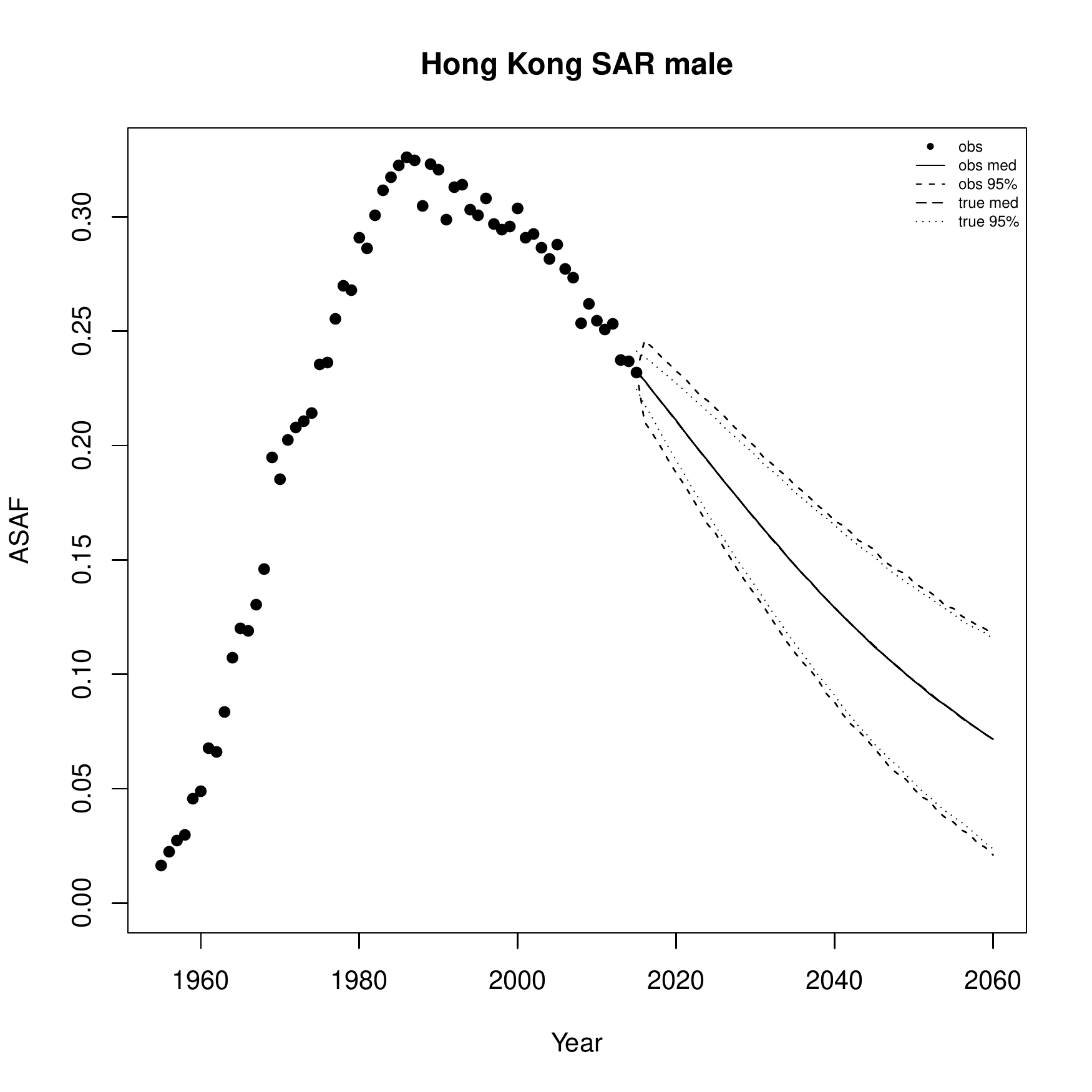}
		\includegraphics[scale=0.35]{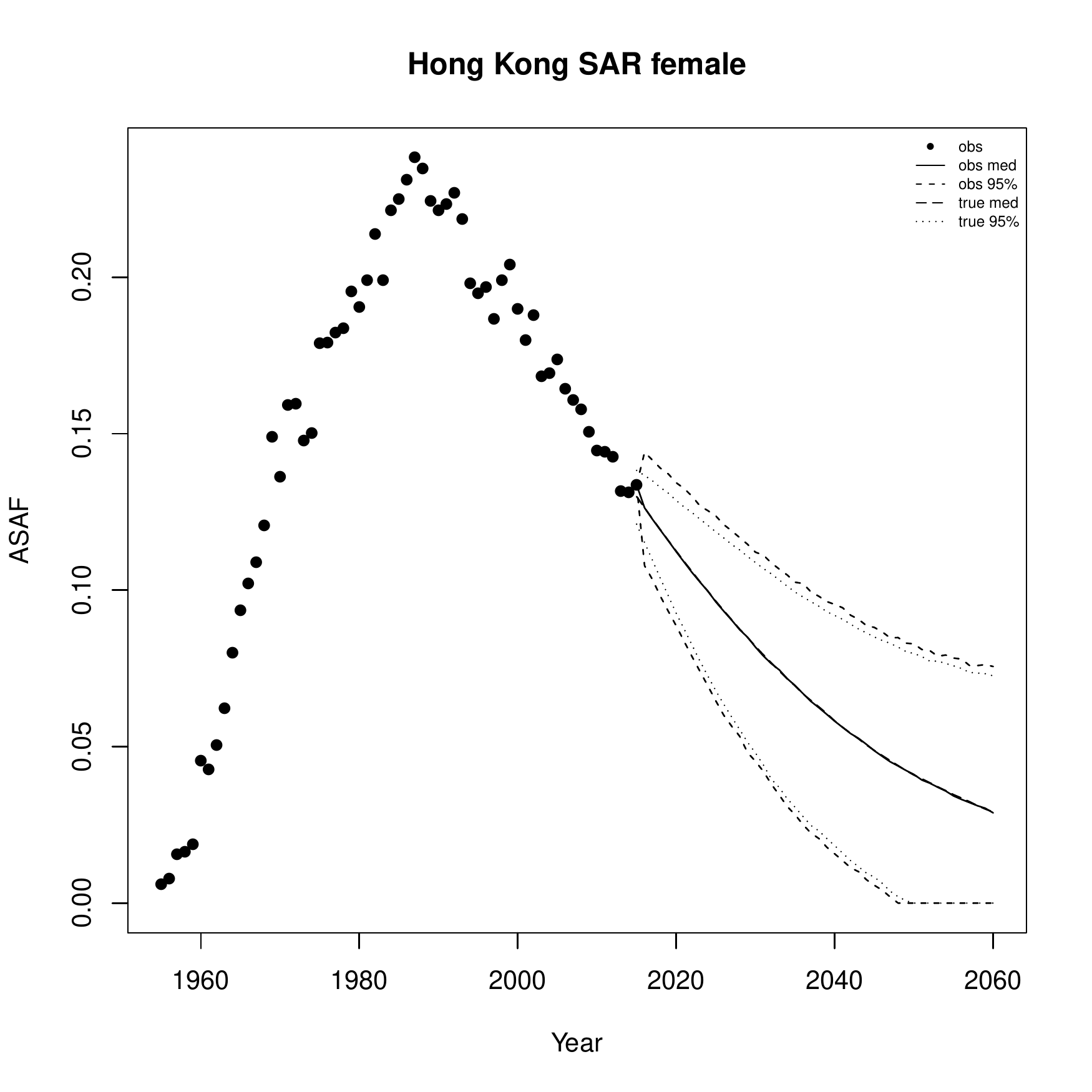}
	\end{center}
	\caption{Hong Kong: The left and right panels show the projection of ASAF up to 2050 under the proposed model for male and female respectively. The solid and long dashed lines show the posterior median of projected observed ASAF and true ASAF respectively. The dashed and dotted lines represent $95\%$ prediction intervals for observed ASAF and true ASAF respectively.}\label{fg:HKproj}
\end{figure}

As discussed by \citet{lam2001mortality}, Hong Kong may be a good indicator for the future development of the smoking epidemic and its impact on mortality in mainland China and other developing countries. Figure \ref{fg:HKproj} shows the historical records of the observed male and female ASAF during time period 1950--2015, along with projections up to 2050 with posterior median and prediction intervals.

\subsection{Chile}
Chile is one of the South America countries that have clear-pattern ASAF data for both males and females. It also has relatively high smoking prevalence.  A decline in prevalence among males and females has been observed in recent years but is modest compared to the decline in the United States \citep{islami2015global}. Also, female smoking prevalence is far behind that of males. 

\begin{figure}[tbp]
	\begin{center}
		\includegraphics[scale=0.35]{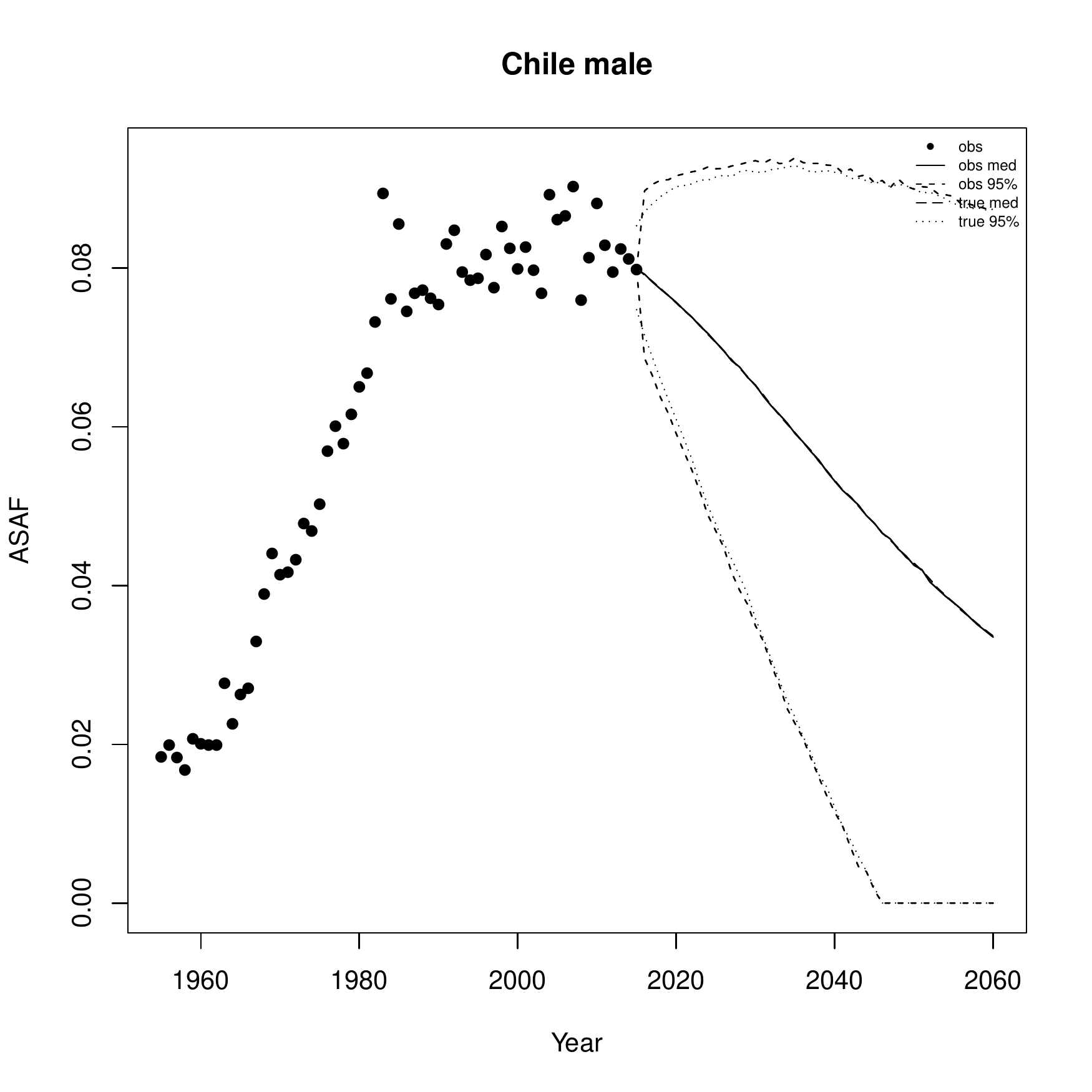}
		\includegraphics[scale=0.35]{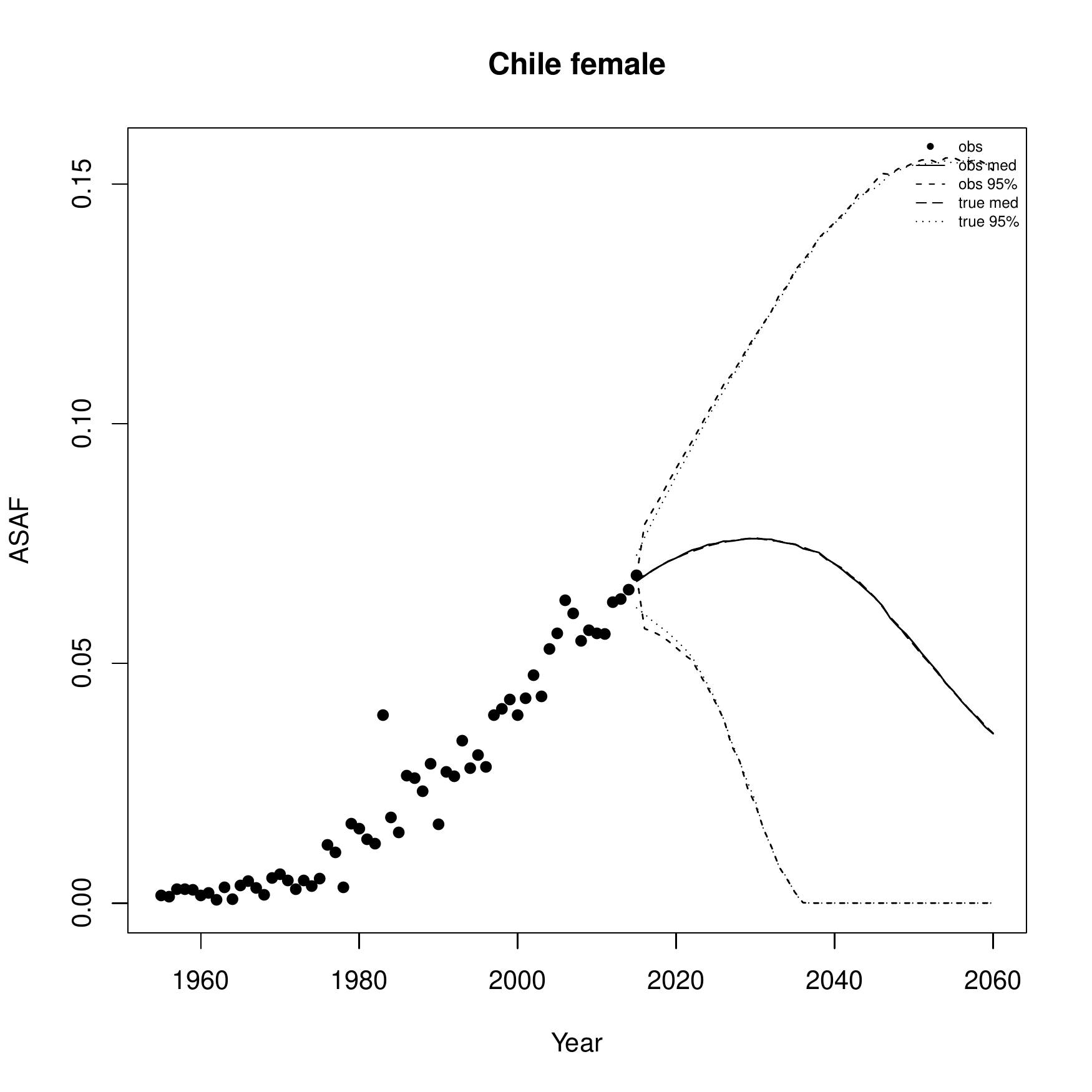}
	\end{center}
	\caption{Chile: The left and right panels show the projection of ASAF up to 2050 under the proposed model for male and female respectively. The solid and long dashed lines show the posterior median of projected observed ASAF and true ASAF respectively. The dashed and dotted lines represent $95\%$ prediction intervals for observed ASAF and true ASAF respectively.}\label{fg:Chileproj}
\end{figure}

Our method projects that the male ASAF will decline gradually. By 2050, the projected median observed ASAF for the male population will be $4.3\%$ (with $95\%$ prediction interval [$0.0\%$, $9.1\%$]). For females, we expect an increase for another 10 years with the median observed ASAF reaching the maximum $7.6\%$ (with $95\%$ prediction interval [$2.0\%$, $11.8\%$]) by 2030. By 2050, the median observed female ASAF be $5.36\%$ (with $95\%$ prediction interval [$0.0\%$, $15.2\%$]); see Figure \ref{fg:Chileproj}. Similarly to Hong Kong, Chile also has larger measurement error and the pattern is less clear, so that the projected true ASAF has wider prediction intervals compared with previous cases and the difference between true and observed projections also appears in the short term.

\section{Discussion}\label{sec:disc}

\subsection{Comparison between SAF Estimation Methods}\label{subsec:plestdif}
In Section \ref{sec:intro}, we briefly described three categories of estimation methods for SAF. Prevalence-based methods depend heavily on smoking prevalence history. Since the lag between smoking prevalence and SAF is usually around two or three decades, in order to use smoking prevalence to estimate and predict SAF, especially for those countries whose onset of SAF is before 1950, one needs data at least back to the 1920s or 1930s. However, such smoking prevalence history is not available for most countries, and reconstruction of such data is challenging.  \citet{ng2014smoking} provided estimates of smoking prevalence for many countries only from 1980 onwards. 

Insufficient historical data is a major obstacle to using smoking prevalence for estimation and projection of SAF, and with currently available historical data, the predictive power using smoking prevalence data is not very high. In addition, smoking prevalence only reveals one aspect of the smoking epidemic, which cannot capture other aspects such as smoking intensity and duration and thus has been argued to be a poor indicator of the smoking exposure of the population \citep{ shibuya2005statistical,luo2018lung}. Prevalence-based estimation and projection have generally been applied only to specific countries on an individual basis, and examples include Taiwan \citep{wen2005smoking} and the United States \citep{ma2018smoking}. 

There are two main indirect methods used widely in the literature, which both use the lung cancer mortality rate as an indicator for the accumulated hazard of smoking. The first one is the Peto-Lopez method which we have used here. 
This has been widely used in the demographic literature, in part because 
its data requirements are relatively modest.
It has been validated in many studies \citep{preston2009new, kong2016comparison}. 

One drawback of the Peto-Lopez method is that it uses the CPS-II to estimate the relative risk. Since the CPS-II was conducted in 1982 with volunteer participants only from the United States and most of them were middle-class, the CPS-II might not be fully representative and may potentially underestimate lung cancer mortality in nonsmokers \citep{tachfouti2014mortality}. Moreover, the Peto-Lopez method assumes that the relative risk is constant over time and homogeneous across nations. \citet{mehta2012continued}, \citet{teng2017changing}, and \citet{lariscy2018cigarette} have shown that the risks from smoking are changing over time. 
Also, in China and India, the lung cancer mortality rate among nonsmokers is higher than that of the developed countries such as that in the CPS-II \citep{liu1998emerging, gajalakshmi2003smoking}. 
Another issue is that the original Peto-Lopez paper reduced the smoking excess risk of each cause-of-death except lung cancer by $50\%$ to control for other confounders. As stated in their paper, this reduction is somewhat arbitrary. To avoid some of these issues, here we have used only data from clear-pattern
countries, which avoids some countries for which the Peto-Lopez method
may not give good estimates. 

Some variants of the Peto-Lopez method have been proposed. For example, \citet{ezzati2003estimates} reduced the correction factor for excess risk from $50\%$ to $30\%$ for all countries and extended this method to less developed countries by estimating the non-smoker lung cancer mortality rate based on household use of coal in poorly-vented stoves. They also provided an analysis of uncertainty. \citet{mackenbach2004inequalities} used a simplified version which only used the all-cause relative risk in the CPS-II study and avoided calculations for the nine disease categories separately. \cite{janssen2013including} used this version to calculate age-specific SAF to partition mortality into smoking and non-smoking attributable parts, and projected them separately. 

\cite{muszynska2014role} and \cite{stoeldraijer2015future} used the same method to calculate an age-standardized SAF, whose purposes are to compare the role of smoking in different regions of Poland, and to estimate and compare smoking attributable fraction of mortality among England $\&$ Wales, Denmark and the Netherlands, respectively. While age-standardization is used mainly to compare SAF among different populations, ASAF provides the all-cause SAF with all age-groups aggregated and is the main quantity reported in the iterature, e.g., \cite{peto1992mortality, peto1994mortality, peto2006mortality, preston2009new}.

Based on these concerns about the Peto-Lopez method, \citet{preston2009new} and \citet{preston2011contribution} came up with the PGW method, which used a regression-based method to connect lung cancer mortality rate with other causes of death mortality rate instead of using the CPS-II. The PGW method avoids the relative risk problem faced by the Peto-Lopez method and provides estimates of uncentainty. However, its authors stated that the Peto-Lopez method might work better for countries where the cause-of-death structure is very different from that observed in developed countries, such as tropical African countries. They also pointed out that both methods would not work well for countries whose lung cancer mortality rate is also influenced largely by some other factors such as air pollution. 
As discussed by \citet{preston2009new}, the PGW method produces similar estimates to the Peto-Lopez method in general for both males and females. 

%Authors of this work fully understand the advantages as well as drawbacks of using original Peto-Lopez method to calculate ASAF for all countries with lung cancer data available. First of all, we only use data from clear-pattern countries as defined in Section \ref{subsec:safcl} to avoid some countries that Peto-Lopez method do not estimate well. Secondly, estimating ASAF is not the main contribution of this work. Our main purpose of this work is to propose a valid method for projecting ASAF based on historical data. If better estimation of ASAF is available, our method could still apply to the updated data.

\subsection{Projection Methodology}\label{subsec:plstatproj}
To our knowledge, there are only two other methods available for projecting SAF based on the Peto-Lopez method. \citet{janssen2013including} proposed the first method to forecast age-specific SAF and to our knowledge it has so far been applied only to the Netherlands. For projecting male age-specific SAF, a constant decline rate ($-1.5\%$) based on the current trend of all-age combined SAF is applied for each age group. For females, it first estimates the time-to-peak and value of peak of female SAF. It uses age-period-cohort (APC) analysis to find the cohort with the highest lung cancer mortality rate and then adds 68, which is the average age of dying from lung cancer, to that cohort to estimate the year which the all-age combined female SAF would reach the maximum. Then the difference between year-to-peak of male and female SAF with all ages combined is estimated and applied to get the time-to-peak and thus the age-specific female SAF. Finally, the rate of decline of female age-specific SAF is set to the same as that of the male. 

The other method proposed for projecting SAF is to first estimate and project lung cancer mortality rate by considering the cohort effect, and use it to calculate the age-specific SAF. 
\citet{stoeldraijer2015future} used an APC model to estimate and forecast the lung cancer mortality rate of three countries: England $\&$ Wales, Denmark, and the Netherlands. For female data, they first estimated the time-to-peak for each age group by assuming that the time-to-peak of age-specific lung cancer mortality rate for females is when it reaches the corresponding rate for males. 
By assuming that the female lung cancer mortality will follow the same increasing-leveling-declining time trend as that for males for each age group, the authors argued that their method could provide long-term projections of lung cancer mortality rate, while previous work which only used historic trends in APC analysis could only provide short-term projections. 

APC analysis is widely used, but it is also plagued by the unidentifiability issue resulting from the perfect linear relationship between the three effects. 
To resolve this requires extra constraints on the parameter space, many of which are not desirable \citep{luo2013,smith2016review}. 
Also, projection of the future lung cancer mortality rate also requires the projection of age, period, and cohort effects, which introduces additional projection error, even more so for young
cohorts for which historical data are not available.

Another way to resolve the unidentifiability issue in APC analysis is by introducing cohort explanatory variables \citep{smith2016review}. Cohort smoking history is one such powerful tool for estimating and projecting mortality. \citet{preston2006sex} and \citet{wang2009forecasting} used the average year of smoking before 40 of a cohort as a covariate to explain the mortality differences between genders and forecasted mortality of United States for both genders up to 2035. \citet{shibuya2005statistical} and \citet{luo2018lung} used APC anlaysis with selected smoking covariates such as cigarette tar exposure to estimate and project the lung cancer mortality rate. Cohort smoking history is a powerful tool, but 
it requires additional data \citep{burns1997cigarette} that are not available
for many of the countries we considered.

%Besides, we cannot directly compare these two methods with ours. First of all, \citet{janssen2013including} and \citet{stoeldraijer2015future} used a simplifed version of Peto-Lopez method to calculated the ``age-standardized SAF" based on estimated age-specific SAF, which only requires estimation and projection of the lung cancer mortality rate while we use mortality rate for all nine categories of cause-of-death to calculate the all-age SAF (ASAF).  In addition, to extend their methods under our definition, we need to perform APC analysis for the mortality rate for all nine categories of cause-of-death, which requires making many constrains and assumptions. Thirdly, \citet{janssen2013including} and \citet{stoeldraijer2015future} only used data from individual country for estimation with only point forecast provided while our method models 69 countries jointly with uncertainty considered.

\subsection{China and India}
According to \citet{reitsma2017smoking}, China and India are the two countries that have seen the largest percentage increase in smoking prevalence. 
As a result, the ASAF for these two countries is important for 
understanding and projecting the world trend of the effect of smoking on 
mortality since the diffusion of the smoking epidemic from developed countries to developing countries has already started.  

\citet{parascandola2019tobacco} found that smoking-related health issues in China have increased over the past two decades, and the trend resembles the early pattern observed in high income countries such as the US and Japan. Smoking prevalence among Chinese men has remained high (around $60\%$) since the 1980s, with a modest decrease to $52\%$ by 2015. 
Smokers born after 1970 tended to start smoking earlier and more intensely than those born before 1970.

\citet{chen2015contrasting} analyzed two nationwide prospective cohort studies on smoking conducted in China during 1991-99 and  2006-14. 
They found that the excess risk among smokers almost doubled over the 15-year period. 
They reported that the SAF of males aged 40-79 
increased from $11\%$ in the first study to $18\%$ in the second study, and 
they predicted that it would be over $20\%$ in the mid-2010s.

In contrast, female smoking prevalence decreased from $7\%$ in the 1980s to 
$3\%$ in 2015 \citep{parascandola2019tobacco}. 
However, second-hand smoking remains high among Chinese females. \citet{zheng2018prevalence} estimated that $65\%$ of Chinese female non-smokers were exposed to second hand smoking in 2012. Nonetheless, the SAF for Chinese females aged 40-79 years was around $3\%$ in 2006-14. 
%and will continue to decrease to $1\%$ by the 2030s. 

There are also substantial geographic differences in smoking prevalence.
In big cities like Beijing and Shanghai, smoking control measures have developed more rapidly than in other areas.

India has become the country with the second largest cigarette consumption 
in the world, after China.
Smoking, including manufactured cigarettes, bidis, and chewing tobacco is one of the major causes of death for middle-aged Indians. \citet{mishra2016trends} estimated that smoking prevalence among male Indians aged 15-69 years declined modestly from $27\%$ in 1998 to $24\%$ in 2010, 
while smoking prevalence among young adults aged 15-29 years rose.

%In contrast, smoking prevalence of female born after 1960 at ages 30-69 years is only half of the prevalence of female born before 1960, even though the age-standardized prevalence of women aged 15-69 years increased from $1.4\%$ to $2.7\%$. \citet{lal2012attributable} using the manufactured cigarettes and bidis to estimate the SAF in India from 1910 to 2010. The smoking attributable deaths increased in the historical period 1910-1990s largely due to the use of bidis. After 1990s, the attributable deaths from the use of bidis began to decrease, though it still took up 2/3 of the attributable deaths in 2010s, and the attributable deaths from the use of cigarettes are still steadily increasing. }

We have not included these two countries in our analysis for the following two reasons. Firstly, we do not have enough data to estimate the ASAF for China and India. Even though there are some records of lung cancer death count data in the WHO Mortality Database for China \citep{WHO2017}, these are only regional data and so could be biased. India has a reasonably good vital registration system but it also has lung cancer mortality data only for selected regions and locations.

Secondly, as pointed out by \citet{preston2009new}, neither the Peto-Lopez original method nor the PGW method will provide reliable estimates of SAF for countries like China since smoking is not the only major factor that can cause lung cancer. The main assumptions of the Peto-Lopez and PGW methods are that lung cancer mortality is primarily caused by smoking and that the lung cancer mortailty rate is very low among nonsmokers. Therefore, as proposed by \citet{ezzati2003estimates} and others, some extra covariates such as household use of coal in poorly-vented stoves are used to adjust the estimates. Incorporating China and India in the joint model could be feasible in the future if better ASAF estimation methods and related data become available.

\subsection{Decision-making and covariates}
A main purpose of our method is to help improve mortality forecasts.
One could also ask whether our approach could be used directly for policy-making.  One possible use would be to provide
a baseline forecast of what would happen with a continuation of current 
trends in general health, development and tobacco control measures. 
This could help to assess the effectiveness of additional policies 
in accelerating the decline of smoking-related mortality.
This could be done retrospectively, by considering a time point in the past at
which a new tobacco control policy was introduced, and then comparing the probabilistic forecast based on data up to that point with what actually happened.

To do this prospectively would require the addition of covariates to the model.
This is challenging, and would be a good topic for further research.
A difficulty with forecasting using covariates is that the covariates themselves need to be forecast, and the covariates can be harder to forecast than the quantity being forecast. This is especially the case when, as here, the quantity being forecast has a strong time trend, and thus may well itself be easier to forecast than the covariates. In this situation, adding covariates can lead to forecasts that are noisier.  This is one reason why, after decades of research, the majority of demographic studies do not use covariates in forecasting demographic quantities.

\bibliographystyle{apalike}
\bibliography{smokingbib}
\pagebreak

\begin{appendices}
\section{Full Bayesian Hierarchical  Model}\label{app1}
The details of the four-layer Bayesian Hierarchical model described in Section \ref{subsec:model} are as follows. Here $\mathcal{N}_{l}^{u}(a,b)$ represents a normal distribution with mean $a$ and variance $b$ truncated at interval $[l,u]$ ($l$($u$) is omitted if it takes value $-\infty$ ($\infty$)). $\mbox{Gamma}(a,b)$ represents a Gamma distribution with shape $a$ and rate $b$. $\mbox{Lognormal}(a,b)$ represents a log-normal distribution with parameters $a,b$. $\mbox{InvGamma}(a,b)$ represents a inverse-Gamma distribution with shape $a$ and scale $b$.
\begin{alignat*}{2}
\mbox{Level 1:}\ &y_{c,s,t}|h_{c,s,t} \sim \mathcal{N}(h_{c,s,t}, \sigma_c^2); \\
\mbox{Level 2:}\ & h_{c,s, t_0} = g(t_0|\theta_{c,s}) + \varepsilon^h_{c, s, t_0}, \\
&h_{c,s,t} = h_{c,s,t-1}  + g(t|\theta_{c,s}) - g(t-1|\theta_{c,s}) &&+ \varepsilon^h_{c,s,t}\ \mbox{for}\ t>t_0, \\
&\varepsilon_{c,s,t}^h \sim_{ind} \mathcal{N}(0, \sigma_h^2);  \\
\mbox{Level 3:}\ & a_1^{c,m} \sim \mbox{Gamma}(2, 2/a_1^m), && a_1^{c,f} \sim  \mbox{Gamma}(2, 2/a_1^f),\\
& a_2^{c,m} \sim \mathcal{N}^{65}(a_2^m, \sigma_{a_2^m}^2), && a_2^{c,f} = a_2^{c,m} +  \bigtriangleup^c_{a_2},\\
& a_3^{c,m} \sim \mbox{Gamma}(2, 2/a_3^m), && \bigtriangleup^c_{a_2} \sim \mathcal{N}(\bigtriangleup_{a_2},\sigma^2_{\bigtriangleup_{a_2}}),\\
& a_4^{c,m} \sim \mathcal{N}^{100}_0(a_4, \sigma_{a_4}^2), &&a_3^{c,f} \sim \mbox{Gamma}(2, 2/a_3^f),\\
& k^{c,m} \sim \mathcal{N}_{0} (k^m, \sigma_{k^m}^2), && a_4^{c,f} \sim \mathcal{N}^{100}_0(a_4, \sigma_{a_4}^2),\\
&\sigma_c^2 \sim \mbox{Lognormal} (\nu, \rho^2), && k^{c,f} \sim  \mathcal{N}_0(k^f, \sigma_{k^f}^2);\\
%\mbox{Level 4:}\ & a_1^m \sim \mbox{Gamma} (1.477, 9.423), && a_1^f \sim \mbox{Gamma} (2.093, 16.302),\\
%&a_2^m \sim \mathcal{N}(24.362, 12.488), && \bigtriangleup_{a_2} \sim \mathcal{N}(12.080, 11.140),\\
%&a_3^m \sim \mathcal{N}(1.031, 7.378), && a_3^f \sim \mbox{Gamma}(1.031, 7.378),\\
%&a_4 \sim \mathcal{N}(38.362, 19.058), && k^f \sim \mathcal{N}(0.362, 0.255),\\
%&k^m \sim \mathcal{N}(0.362, 0.255), &&\sigma^2_{\bigtriangleup_{a_2}} \sim \mbox{InvGamma}(2, 11^2),\\
%&\sigma_{a_2^m}^2 \sim \mbox{InvGamma}(2, 12.488^2), &&\sigma^2_{k^f} \sim \mbox{InvGamma}(2, 0.255^2),\\
%&\sigma_{a_4}^2 \sim \mbox{InvGamma}(2, 19.058^2), &&\nu \sim \mathcal{N}(-10.414, 1.186),\\
%&\sigma_{k^m}^2 \sim \mbox{InvGamma}(2, 0.255^2), &&\rho^2 \sim \mbox{InvGamma}(2, 1.186^2),\\
%&\sigma^2_h \sim \mbox{InvGamma}(2, 0.01^2).
\mbox{Level 4:}\ & a_1^m \sim \mbox{Gamma} (\alpha_{a_1^m}, \beta_{a_1^m}), && a_1^f \sim \mbox{Gamma} (\alpha_{a_1^f}, \beta_{a_1^f}),\\
&a_2^m \sim \mathcal{N}(\alpha_{a_2^m}, \beta_{a_1^m}), && \bigtriangleup_{a_2} \sim \mathcal{N}(\alpha_{\bigtriangleup_{a_2}}, \beta_{\bigtriangleup_{a_2}}),\\
&a_3^m \sim \mathcal{N}(\alpha_{a_3^m}, \beta_{a_3^m}), && a_3^f \sim \mbox{Gamma}(\alpha_{a_3^f}, \beta_{a_3^f}),\\
&a_4 \sim \mathcal{N}(\alpha_{a_4}, \beta_{a_4}), && k^f \sim \mathcal{N}(\alpha_{k^f}, \beta_{k^f}),\\
&k^m \sim \mathcal{N}(\alpha_{k^m}, \beta_{k^m}), &&\sigma^2_{\bigtriangleup_{a_2}} \sim \mbox{InvGamma}(\alpha_{\sigma^2_{\bigtriangleup_{a_2}}}, \beta_{\sigma^2_{\bigtriangleup_{a_2}}}),\\
&\sigma_{a_2^m}^2 \sim \mbox{InvGamma}(\alpha_{\sigma_{a_2^m}^2 }, \beta_{\sigma_{a_2^m}^2 }), &&\sigma^2_{k^f} \sim \mbox{InvGamma}(\alpha_{\sigma^2_{k^f} }, \beta_{\sigma^2_{k^f} }),\\
&\sigma_{a_4}^2 \sim \mbox{InvGamma}(\alpha_{\sigma_{a_4}^2 }, \beta_{\sigma_{a_4}^2 }), &&\nu \sim \mathcal{N}(\alpha_{\nu }, \beta_{\nu }),\\
&\sigma_{k^m}^2 \sim \mbox{InvGamma}(\alpha_{\sigma_{k^m}^2}, \beta_{\sigma_{k^m}^2}), &&\rho^2 \sim \mbox{InvGamma}(\alpha_{\rho^2 }, \beta_{\rho^2 }),\\
&\sigma^2_h \sim \mbox{InvGamma}(\alpha_{\sigma^2_h }, \beta_{\sigma^2_h}),
\end{alignat*}
where $\alpha_{a_1^m} = 1.477, \beta_{a_1^m} = 9.423, \alpha_{a_2^m} = 24.362, \beta_{a_2^m} = 12.488,
\alpha_{a_3^m} = 1.031,  \beta_{a_3^m} = 7.378, \alpha_{a_4} = 38.362, \beta_{a_4} = 19.058, 
\alpha_{k^m} = 0.362, \beta_{k^m} = 0.255, \alpha_{\sigma_{a_2^m}^2 } = 2,  \beta_{\sigma_{a_2^m}^2 } = 12.488^2, \alpha_{\sigma_{a_4}^2 } = 2, \beta_{\sigma_{a_4}^2 } = 19.058^2, \alpha_{\sigma_{k^m}^2} = 2,
\beta_{\sigma_{k^m}^2} = 0.255^2, \alpha_{a_1^f} = 2.093, \beta_{a_1^f} = 16.302, \alpha_{\bigtriangleup_{a_2}} = 12.080,  \beta_{\bigtriangleup_{a_2}} = 11.140, \alpha_{a_3^f} = 1.031,
\beta_{a_3^f} = 7.378, \alpha_{k^f} = 0.362,  \beta_{k^f}=0.255, \alpha_{\sigma^2_{\bigtriangleup_{a_2}}} = 2,
\beta_{\sigma^2_{\bigtriangleup_{a_2}}} = 11^2, \alpha_{\sigma^2_{k^f} } = 2, \beta_{\sigma^2_{k^f} } = 0.255^2,
\alpha_{\nu } = -10.414, \beta_{\nu } = 1.186^2, \alpha_{\rho^2 } = 2, \beta_{\rho^2 } = 1.186^2, \alpha_{\sigma^2_h } = 2, \beta_{\sigma^2_h} = 0.01^2$.

\section{MCMC Convergence Diagnostics} \label{app2}
\subsection{Hyperparameter Diagnostics}
\begin{figure}[H]
	\begin{center}
		\includegraphics[scale=0.45]{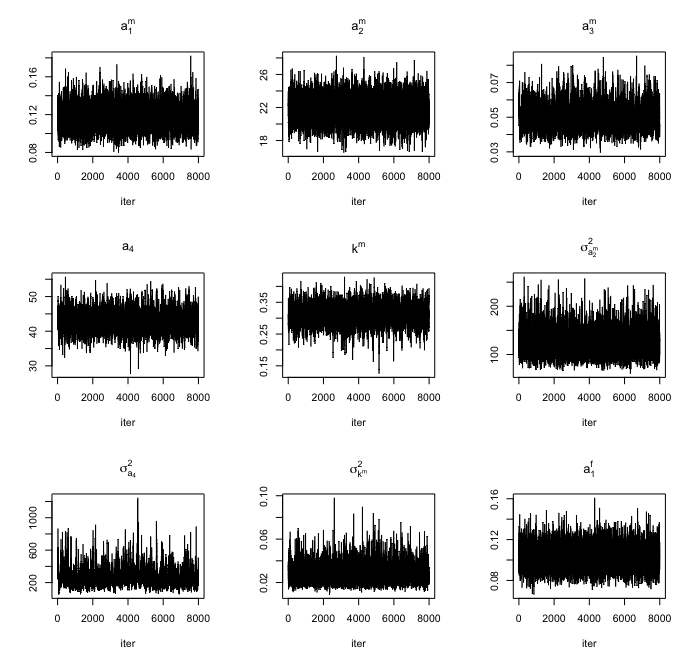}
		\includegraphics[scale=0.45]{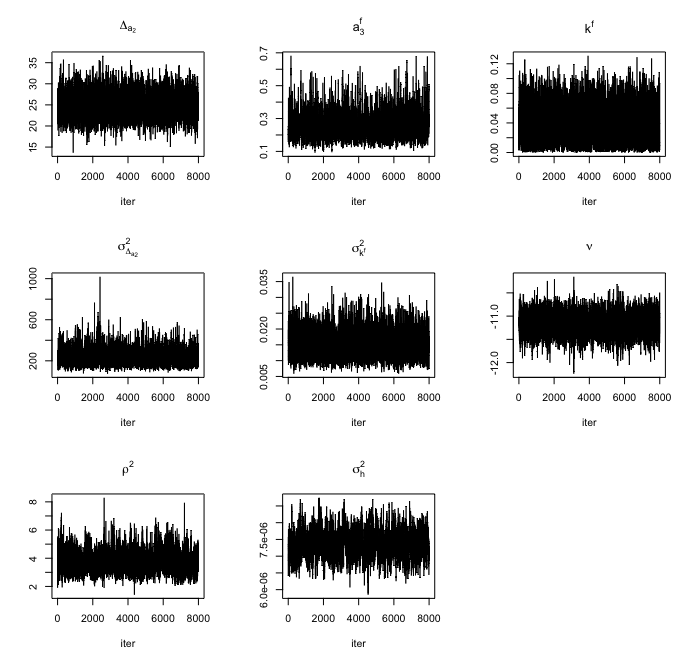}
	\end{center}
	\caption{Traceplots for the hyperparameters.}
	\label{fg:hypertrace}
\end{figure}

In this section, we present the MCMC convergence diagnostics of the hyperparameters in Level 4 of the model in terms of traceplots, Raftery diagnostic statistics \citep{raftery1992mcmc}, and Gelman diagnostic statistics \citep{gelman1992inference}. Table \ref{tb:hyperdiag} provides the Gelman and Raftery diagnostic statistics of all hyperparameters. We use 3 chains with 2000 burnin and 8000 samples without thinning for the Gelman diagnostics, and randomly choose one of the chain to perform the Raftery diagnostics. Figure \ref{fg:hypertrace} shows the traceplots of all 8000 samples of hyperparameters.

\begin{table}[H]
	\caption{Diagnostic statistics for hyperparameters. PSRF and $95\%$ UCI are the point estimator and upper bound of the $95\%$ CI of the Gelman potential scale reduction factor. Burn1, Size1, and DF1 are the length of burn-in, requied sample size, and dependent factor of Raftery diagnostics with parameters $q = 0.025, r=  0.0125, s = 0.95$. Burn2, Size2, and DF2 are the length of burn-in, requied sample size, and dependent factor of Raftery diagnostics with parameters $q = 0.975, r=  0.0125, s = 0.95$. }
	\begin{center}
		\begin{tabular}{ |c|cc|ccc|ccc|}
			\hline
			\multirow{2}{*}{Parameters} & \multicolumn{2}{c|}{Gelman Diag} & \multicolumn{6}{c|}{Raftery Diag} \\\cline{2-9}
			& PSRF & $95\%$ UCI & Burn1 & Size1 & DF1 & Burn2 & Size2 & DF2\\ \hline
			$a_1^m$ &1 & 1.00 & 6& 1318 & 2.20 & 6 &1164 & 1.94\\
			$a_2^m$ &1 & 1.00 & 3& 710 & 1.18 & 6 &1504 & 2.51 \\
			$a_3^m$ &1& 1.00 & 6& 1584 & 2.64 & 6& 1424 & 2.37 \\
			$a_4$ &1& 1.01 & 8&1750 & 2.92 &9 & 2028 & 3.38 \\
			$k^m$ &1& 1.01 & 10& 1952 & 3.25 & 6 & 1236 & 2.06 \\
			$\sigma_{a_2^m}^2$ &1& 1.00 &2&640 &1.07 &6 & 1730 & 2.88\\
			$\sigma_{a_4}^2$ &1& 1.00 &21& 4410 & 7.35& 12 & 2132 & 3.55\\
			$\sigma_{k^m}^2$ &1& 1.00 &6& 1334 & 2.22 & 8& 1448 & 2.41\\
			$a_1^f$ &1& 1.00 &2& 640 & 1.07 & 6 &1318 & 2.20 \\
			$\bigtriangleup_{a_2}$ &1& 1.00 &4& 756 & 1.26& 6 & 688 & 1.15\\
			$a_3^f$ &1& 1.00 &8& 1504 &2.51 & 12 & 1852 & 3.09\\
			$k^f$ &1& 1.00 &5& 895 &1.49 &2 & 640 & 1.07\\
			$\sigma^2_{\bigtriangleup_{a_2}}$&1& 1.00 &3&696& 1.16 &4 &839 & 1.40\\
			$\sigma^2_{k^f}$ &1& 1.00 &2&640 & 1.07 &6 & 1376 & 2.29\\
			$\nu$ &1& 1.00 &8&1518 & 2.53 & 8 & 1934 & 3.22\\
			$\rho^2$ &1& 1.00 &6& 1270 &2.21 & 6 & 1392 & 2.32\\
			$\sigma_h^2$ &1& 1.00 &10& 1872 & 3.12 & 12 & 2337 & 3.90\\
			\hline
		\end{tabular}
		\label{tb:hyperdiag}
	\end{center}
\end{table}

\subsection{Country-specific Parameter Diagnostics}
\begin{figure}[H]
	\begin{center}
		\includegraphics[scale=0.45]{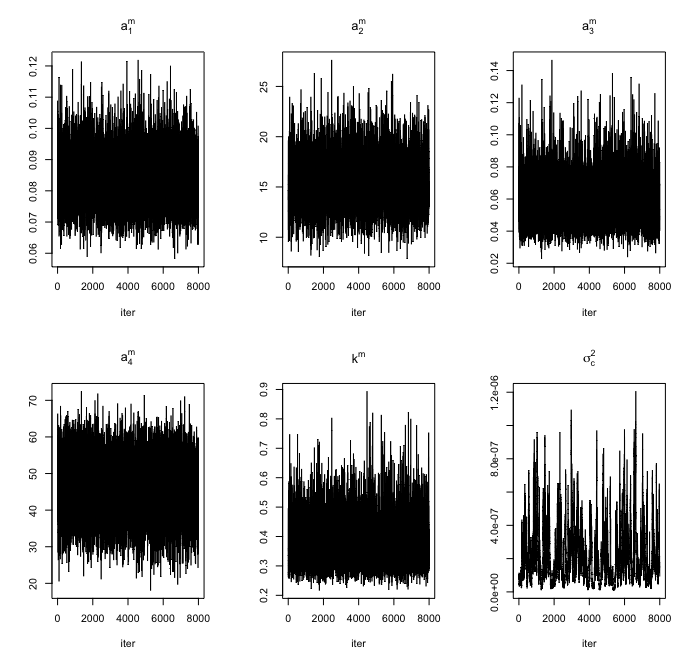}
		\includegraphics[scale=0.45]{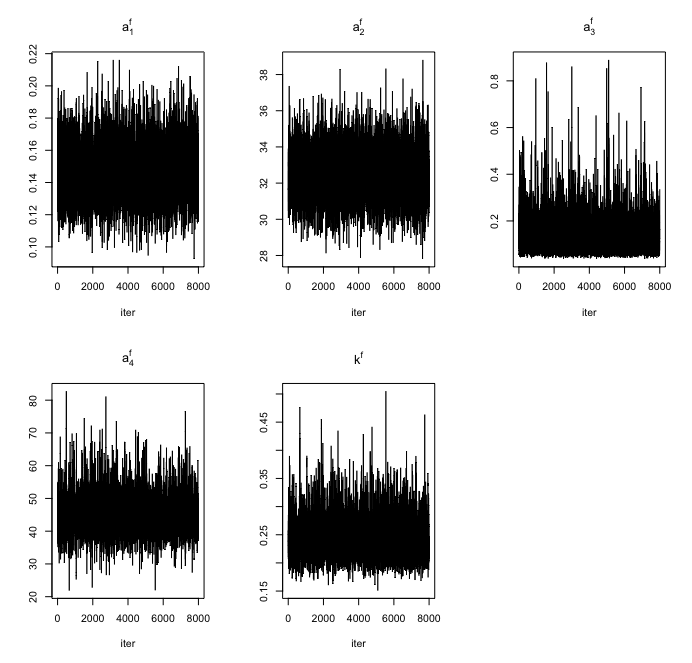}
	\end{center}
	\caption{Traceplots for the country-specific parameters of the United States.}
	\label{fg:trace}
\end{figure}

In this section, we present the MCMC convergence diagnostics of country-specific parameters of the model in terms of traceplots, Raftery diagnostic statistics, and Gelman diagnostic statistics. Table \ref{tb:conparadiag} provides the Gelman and Raftery diagnostic statistics of  country-specific parameters of the United States for male and female. The chains are the same as in the previous section. Figure \ref{fg:trace} shows the traceplots of all 8000 samples of country-specific paremeters for male and female of the United States.

\begin{table}[H]
	\caption{Diagnostic statistics for country-specific parameters for the United States. PSRF and $95\%$ UCI are the point estimator and upper bound of the $95\%$ CI of the Gelman potential scale reduction factor. Burn1, Size1, and DF1 are the length of burn-in, requied sample size, and dependent factor of Raftery diagnostics with parameters $q = 0.025, r=  0.0125, s = 0.95$. Burn2, Size2, and DF2 are the length of burn-in, requied sample size, and dependent factor of Raftery diagnostics with parameters $q = 0.975, r=  0.0125, s = 0.95$. }
	\begin{center}
		\begin{tabular}{ |c|cc|ccc|ccc|}
			\hline
			\multirow{2}{*}{Parameters} & \multicolumn{2}{c|}{Gelman Diag} & \multicolumn{6}{c|}{Raftery Diag} \\\cline{2-9}
			& PSRF & $95\%$ UCI & Burn1 & Size1 & DF1 & Burn2 & Size2 & DF2\\ \hline
			$a_1^m$ &1 & 1.00 & 4& 830 & 1.38 & 2 &640 & 1.07\\
			$a_2^m$ &1 & 1.00 & 6& 1326 & 2.21 & 4 &756 & 1.26 \\
			$a_3^m$ &1& 1.00 & 4& 822 & 1.37 & 2& 633 & 1.06 \\
			$a_4^m$ &1& 1.00 & 2& 614 & 1.02&4 & 772 & 1.29 \\
			$k^m$ &1& 1.00 & 6& 1106 & 1.10 & 2 & 621 & 1.03 \\
			$a_1^f$ &1 & 1.00 & 3& 661 & 1.10 & 2&614 & 1.02\\
			$a_2^f$ &1 & 1.00 & 2& 627 & 1.04 & 2 &614 & 1.02 \\
			$a_3^f$ &1& 1.00 & 2& 627 & 1.04 & 6& 1314 & 2.19 \\
			$a_4^f$ &1& 1.00 & 3&661 & 1.10 &4 & 848 & 1.41 \\
			$k^f$ &1& 1.00 & 3& 668 & 1.11 & 6 & 1444 & 2.41 \\
			$\sigma^2_c$ &1.01& 1.03& 96& 15198 & 25.30 & 24 & 3834 & 6.39 \\
			\hline
		\end{tabular}
		\label{tb:conparadiag}
	\end{center}
\end{table}

\section{Hyperparameter Sensitivity Analysis}\label{app3}
In this section, we present the sensitivity analysis for the hyperparameters set in $\pi(\cdot)$ on the posterior distributions of the global parameters $\psi$ in Level 4 of our model. We use \texttt{rstansensitivity} package \citep{Giordano2019}  in R to perform the sensitivity analysis. The local sensitivity of the posterior mean of parameter $\theta$ under hyperparameters $\zeta$ (i.e., $\mathbb{E}(\theta|x, \zeta)$) to $\zeta$ at $\zeta_0$ is defined as
\[ S_{\zeta_0}:= \frac{d \mathbb{E}(\theta|x, \zeta)}{d\zeta}\Big|_{\zeta_0},\]
where $x$ is the observed data (cf. \citet{basu1996local}, \citet{gustafson1996local}, and \citet{giordano2018covariances} for more discussions on local sensitivity in Bayesian analysis). By scaling the local sensitivity to be comparable with the possible range of the posterior distribution of $\theta$, the normalized local sensitivity is defined as
\[ \tilde{S}_{\zeta_0}:=\bigg| \frac{S_{\zeta_0}}{\mbox{sd}(\theta|x, \zeta_0)}\bigg|.\]
As commented in \citet{giordano2018covariances}, if the quantity $\tilde{S}_{\zeta_0}$ is less than 1 or if $\tilde{S}_{\zeta_0}$ is greater than 1 but the final results barely change when modifying the hyperparameters, then the model is robust. First of all, Table \ref{tb:sen} investigates the normalized local sensitivity of the hyperparameters set in $\pi(\cdot)$ on posterior distributions of the global parameters $\psi$.  For most hyperparameters, the normalized local sensitivity are much smaller than 1. For those whose normalized local sensitivity are greater than 1, we conduct out-of-sample validations for three five-year period prediction with the hyperparameters changed to evaluate the actual changes on the validation results. Table \ref{tb:sigma2km}, \ref{tb:sigma2kf}, \ref{tb:sigmah2}, and \ref{tb:mua4} show the out-of-sample validtion results after modifying $\beta_{\sigma^2_{k_m}}$ ($0.255^2$ to $1$), $\beta_{\sigma^2_{k_f}}$ ($0.255^2$ to $1$), $\beta_{\sigma^2_h}$ ($0.01^2$ to $0.02^2$), and $\alpha_{a_4}$ (38.362 to 20), respectively. All four cases show that the validation results barely change, and we conclude that model is  robust under the current choices of  hyperparameters.

\begin{sidewaystable}
	\caption{Normalized local sensitivity of hyperparameters on the global parameters. The bold numbers are local sensitivity $\tilde{S}_{\zeta_0}$ with absolute value greater than 1.}\label{tb:sen}
	\resizebox{\textwidth}{!}{	
		\begin{tabular}{rrrrrrrrrrrrrrrrrr}
			\hline
			& $a_1^m$ & $a_2^m$ & $a_3^m$& $a_4$ & $k_m$ & $\sigma^2_{a_2^m}$ & $\sigma^2_{a_4}$ & $\sigma^2_{km}$ & $a_1^f$ & $\triangle_{a_2}$ & $a_3^f$ & $k_f$ & $\sigma^2_{\triangle{a_2}}$ & $\sigma^2_{k_f}$ & $\nu$ & $\rho^2$ & $\sigma^2_h$ \\ 
			\hline
			$\alpha_{a_1^m}$ & 0.014 & 0.001 & 0.001 & 0.046 & -0.006 & -0.031 & 0.198 & 0.000 & -0.000 & -0.018 & 0.000 & -0.001 & -0.096 & -0.000 & 0.003 & 0.002 & -0.000 \\ 
			$\beta_{a_1^m}$ & -0.002 & -0.000 & -0.000 & -0.006 & 0.001 & 0.003 & -0.023 & -0.000 & 0.000 & 0.002 & -0.000 & 0.000 & 0.011 & 0.000 & -0.000 & -0.000 & 0.000 \\ 
			$\alpha_{a_2^m}$ & 0.000 & 0.012 & 0.000 & -0.001 & -0.000 & 0.005 & 0.003 & 0.000 & 0.000 & -0.003 & -0.000 & 0.000 & -0.002 & 0.000 & 0.000 & -0.000 & -0.000 \\ 
			$\beta_{a_2^m}$ & -0.000 & -0.004 & -0.000 & 0.000 & -0.000 & -0.000 & -0.002 & 0.000 & -0.000 & 0.001 & 0.000 & -0.000 & 0.002 & -0.000 & -0.000 & 0.000 & 0.000 \\
			$\alpha_{a_3^m}$ & 0.002 & 0.019 & 0.011 & 0.059 & -0.009 & 0.002 & 0.295 & 0.001 & -0.000 & -0.034 & 0.002 & -0.001 & -0.094 & -0.000 & -0.001 & -0.001 & 0.000 \\ 
			$\beta_{a_3^m}$ & -0.000 & -0.001 & -0.001 & -0.003 & 0.000 & -0.000 & -0.015 & -0.000 & 0.000 & 0.002 & -0.000 & 0.000 & 0.005 & 0.000 & 0.000 & 0.000 & -0.000 \\ 
			$\alpha_{a_4}$ & 0.000 & -0.001 & 0.000 & 0.020 & -0.001 & -0.004 & 0.048 & -0.000 & 0.000 & -0.003 & 0.000 & -0.000 & -0.011 & -0.000 & -0.000 & -0.000 & 0.000 \\ 
			$\beta_{a_4}$ & 0.000 & -0.001 & 0.000 & 0.025 & -0.001 & -0.005 & 0.064 & -0.000 & 0.000 & -0.004 & 0.000 & -0.000 & -0.013 & -0.000 & -0.000 & -0.000 & 0.000 \\ 
			$\alpha_{k^m}$ & -0.030 & -0.004 & -0.027 & -0.617 & 0.168 & 0.290 & \textbf{-1.591} & -0.032 & -0.004 & 0.186 & -0.007 & 0.010 & 0.870 & 0.000 & -0.003 & 0.005 & -0.000 \\ 
			$\beta_{k^m}$ & -0.012 & 0.015 & -0.012 & -0.316 & 0.062 & 0.162 & -0.610 & -0.000 & -0.003 & 0.085 & -0.004 & 0.005 & 0.399 & 0.001 & 0.004 & 0.005 & -0.000 \\ 
			$\alpha_{\sigma^2_{a_2^m}}$ & 0.001 & -0.025 & -0.000 & 0.025 & -0.002 & -1.104 & -0.051 & 0.000 & 0.002 & -0.003 & 0.003 & -0.001 & -0.136 & -0.000 & -0.002 & 0.007 & 0.000 \\ 
			$\beta_{\sigma^2_{a^m_2}}$ & -0.000 & 0.000 & -0.000 & -0.000 & 0.000 & 0.009 & 0.000 & -0.000 & -0.000 & 0.000 & -0.000 & 0.000 & 0.001 & 0.000 & 0.000 & -0.000 & -0.000 \\ 
			$\alpha_{\sigma^2_{a_4}}$ & -0.008 & -0.026 & -0.007 & -0.260 & 0.012 & -0.066 & -3.916 & 0.001 & -0.002 & -0.014 & 0.010 & -0.004 & 0.144 & 0.000 & -0.006 & 0.006 & 0.000 \\ 
			$\beta_{\sigma^2_{a_4}}$ & 0.000 & 0.000 & 0.000 & 0.001 & -0.000 & 0.000 & 0.015 & -0.000 & 0.000 & 0.000 & -0.000 & 0.000 & -0.001 & -0.000 & 0.000 & -0.000 & -0.000 \\ 
			$\alpha_{\sigma^2_{k^m}}$ & -0.001 & -0.001 & -0.002 & 0.068 & 0.016 & -0.010 & 0.178 & -0.029 & -0.001 & 0.051 & 0.001 & -0.001 & 0.084 & -0.000 & -0.004 & 0.009 & -0.000 \\ 
			$\beta_{\sigma^2_{k_m}}$ & 0.012 & -0.063 & 0.055 & \textbf{-2.644} & -0.468 & -0.067 & \textbf{-10.187} & \textbf{1.004} & 0.058 & \textbf{-1.964} & -0.051 & 0.048 & \textbf{-3.475} & 0.004 & 0.123 & -0.378 & 0.000 \\ 
			$\alpha_{a_1^f}$ & -0.000 & 0.001 & -0.000 & 0.012 & -0.001 & -0.057 & 0.024 & 0.000 & 0.013 & -0.051 & 0.004 & -0.002 & -0.065 & -0.000 & 0.002 & 0.001 & -0.000 \\ 
			$\beta_{a_1^f}$ & 0.000 & -0.000 & 0.000 & -0.001 & 0.000 & 0.006 & -0.003 & -0.000 & -0.001 & 0.005 & -0.000 & 0.000 & 0.007 & 0.000 & -0.000 & -0.000 & 0.000 \\ 
			$\alpha_{\triangle_{a_2}}$ & -0.000 & -0.005 & -0.000 & -0.006 & 0.001 & 0.001 & 0.008 & -0.000 & -0.001 & 0.040 & -0.000 & 0.001 & 0.069 & 0.000 & -0.001 & -0.000 & 0.000 \\ 
			$\beta_{\triangle_{a_2}}$ & -0.001 & -0.013 & -0.001 & -0.020 & 0.002 & 0.002 & 0.030 & -0.001 & -0.002 & 0.123 & -0.001 & 0.002 & 0.222 & 0.000 & -0.002 & 0.000 & -0.000 \\ 
			$\alpha_{a_3^f}$ & 0.000 & 0.002 & 0.001 & 0.003 & -0.002 & -0.074 & -0.286 & -0.000 & 0.004 & -0.040 & 0.069 & -0.006 & -0.263 & -0.002 & -0.002 & 0.004 & 0.000 \\ 
			$\beta_{a_3^f}$ & -0.000 & 0.001 & -0.000 & -0.000 & 0.000 & 0.019 & 0.070 & 0.000 & -0.001 & 0.008 & -0.018 & 0.001 & 0.059 & 0.000 & 0.001 & -0.002 & -0.000 \\ 
			$\alpha_{k^f}$ & -0.005 & 0.052 & -0.002 & -0.071 & 0.008 & 0.134 & 0.500 & 0.002 & -0.007 & 0.249 & -0.022 & 0.110 & 0.909 & -0.022 & 0.020 & 0.010 & -0.000 \\ 
			$\beta_{k^f}$ & 0.012 & -0.119 & 0.004 & 0.151 & -0.019 & -0.289 & \textbf{-1.155} & -0.005 & 0.015 & -0.551 & 0.049 & -0.249 & \textbf{-1.948} & 0.049 & -0.048 & -0.027 & 0.000 \\ 
			$\alpha_{\sigma^2_{\triangle_{a_2}}}$ & 0.003 & 0.013 & 0.003 & 0.063 & -0.006 & -0.129 & 0.217 & 0.001 & 0.002 & -0.184 & 0.008 & -0.007 &\textbf{ -2.611} & -0.002 & -0.004 & 0.001 & 0.000 \\ 
			$\beta_{\sigma^2_{\triangle_{a_2}}}$  & -0.000 & -0.000 & -0.000 & -0.000 & 0.000 & 0.001 & -0.001 & -0.000 & -0.000 & 0.001 & -0.000 & 0.000 & 0.012 & 0.000 & 0.000 & -0.000 & -0.000 \\ 
			$\alpha_{\sigma^2_{k^f}}$ & 0.000 & -0.001 & 0.000 & 0.052 & 0.000 & -0.021 & 0.044 & -0.000 & 0.001 & -0.018 & 0.007 & 0.021 & -0.214 & -0.016 & 0.002 & 0.008 & 0.000 \\ 
			$\beta_{\sigma^2_{k_f}}$ & -0.025 & -0.139 & -0.021 & \textbf{-3.454} & -0.022 & \textbf{1.940} &\textbf{ -2.054} & 0.029 & -0.042 & \textbf{1.225 }& -0.486 & \textbf{-1.595 }& \textbf{14.056} & \textbf{1.093} & -0.124 & -0.655 & -0.000 \\ 
			$\alpha_{\nu}$ & 0.001 & 0.013 & -0.000 & -0.012 & -0.000 & 0.018 & 0.034 & 0.001 & 0.001 & -0.014 & -0.001 & 0.002 & 0.019 & -0.000 & 0.079 & -0.002 & -0.000 \\ 
			$\beta_{\nu}$ & -0.001 & -0.012 & 0.000 & 0.014 & 0.000 & -0.029 & -0.013 & -0.001 & -0.001 & 0.019 & 0.001 & -0.001 & -0.016 & 0.000 & -0.098 & 0.005 & 0.000 \\ 
			$\alpha_{\rho^2}$ & -0.000 & 0.003 & 0.000 & 0.010 & -0.000 & 0.051 & 0.039 & 0.000 & -0.000 & 0.001 & -0.002 & -0.001 & 0.017 & 0.000 & 0.002 & -0.165 & -0.000 \\ 
			$\beta_{\rho^2}$ & 0.000 & -0.001 & -0.000 & -0.003 & -0.000 & -0.014 & -0.013 & -0.000 & 0.000 & -0.000 & 0.000 & 0.000 & -0.005 & -0.000 & -0.001 & 0.046 & 0.000 \\ 
			$\alpha_{\sigma^2_{h}}$ & 0.000 & 0.001 & -0.000 & -0.000 & 0.001 & 0.016 & 0.043 & -0.000 & 0.001 & -0.000 & -0.000 & 0.000 & 0.021 & 0.000 & 0.003 & -0.007 & -0.000 \\ 
			$\beta_{\sigma^2_{h}}$ & \textbf{-37.688} & \textbf{-116.830} & \textbf{16.011} & \textbf{17.100 }& \textbf{-82.440} & \textbf{-2098.717 }& \textbf{-5650.674 }& \textbf{27.425} & \textbf{-71.201} & \textbf{10.789 }& \textbf{37.786} & \textbf{-66.317 }& \textbf{-2863.039} & \textbf{-1.673} & \textbf{-434.118 }& \textbf{958.762 }& \textbf{4.054 }\\ 
			\hline
	\end{tabular}}
\end{sidewaystable}

\begin{table}[tbp]
	\caption{Out-of-sample validation results of ASAF for both male and female with $\beta_{\sigma^2_{k_m}}$ changed. ``Bayes (mod)" is the BHM with changed hyperparameter. }\label{tb:sigma2km}
	\centering
	\begin{tabular}{c|c|c|c|c|c|ccc}
		\multirow{2}{*}{Gender} &	\multirow{2}{*}{Train} & \multirow{2}{*}{Test} &\multirow{2}{*}{num}	&\multirow{2}{*}{Method} &\multirow{2}{*}{MAE} & \multicolumn{3}{c}{Coverage} \\
		& & & & & & $80\%$ & $90\%$ & $95\%$  \\
		\hline
		\multirow{2}{*}{Male} &	\multirow{2}{*}{1950-2000} & \multirow{2}{*}{2000-2015} &\multirow{2}{*}{63} 
		&  Bayes(mod) & 0.016 &0.63 &0.76 & 0.85\\
		&&	& & Bayes &0.016  &0.65  &0.76 & 0.84 \\
		\hline
		\multirow{2}{*}{Female} &	\multirow{2}{*}{1950-2000} & \multirow{2}{*}{2000-2015} &\multirow{2}{*}{63} 
		&  Bayes(mod) & 0.011 &0.80 & 0.89  & 0.94 \\
		&	& &  & Bayes & 0.011& 0.81 & 0.90 & 0.95\\
		\hline
	\end{tabular}
\end{table}

\begin{table}[tbp]
	\caption{Out-of-sample validation results of ASAF for both male and female with $\beta_{\sigma^2_{k_f}}$ changed. ``Bayes (mod)" is the BHM with changed hyperparameter.}\label{tb:sigma2kf}
	\centering
	\begin{tabular}{c|c|c|c|c|c|ccc}
		\multirow{2}{*}{Gender} &	\multirow{2}{*}{Train} & \multirow{2}{*}{Test} &\multirow{2}{*}{num}	&\multirow{2}{*}{Method} &\multirow{2}{*}{MAE} & \multicolumn{3}{c}{Coverage} \\
		& & & & & & $80\%$ & $90\%$ & $95\%$  \\
		\hline
		\multirow{2}{*}{Male} &	\multirow{2}{*}{1950-2000} & \multirow{2}{*}{2000-2015} &\multirow{2}{*}{63} 
		&  Bayes(mod) & 0.016 &0.64 &0.77 & 0.85\\
		&&	& & Bayes &0.016  &0.65  &0.76 & 0.84 \\
		\hline
		\multirow{2}{*}{Female} &	\multirow{2}{*}{1950-2000} & \multirow{2}{*}{2000-2015} &\multirow{2}{*}{63} 
		&  Bayes(mod) & 0.011 &0.82 & 0.90  & 0.95 \\
		&	& &  & Bayes & 0.011& 0.81 & 0.90 & 0.95\\
		\hline
	\end{tabular}
\end{table}

\begin{table}[tbp]
	\caption{Out-of-sample validation results of ASAF for both male and female with $\beta_{\sigma^2_h}$ changed. ``Bayes (mod)" is the BHM with changed hyperparameter.}\label{tb:sigmah2}
	\centering
	\begin{tabular}{c|c|c|c|c|c|ccc}
		\multirow{2}{*}{Gender} &	\multirow{2}{*}{Train} & \multirow{2}{*}{Test} &\multirow{2}{*}{num}	&\multirow{2}{*}{Method} &\multirow{2}{*}{MAE} & \multicolumn{3}{c}{Coverage} \\
		& & & & & & $80\%$ & $90\%$ & $95\%$  \\
		\hline
		\multirow{2}{*}{Male} &	\multirow{2}{*}{1950-2000} & \multirow{2}{*}{2000-2015} &\multirow{2}{*}{63} 
		&  Bayes(mod) & 0.016 &0.68 &0.80 & 0.88\\
		&&	& & Bayes &0.016  &0.65  &0.76 & 0.84 \\
		\hline
		\multirow{2}{*}{Female} &	\multirow{2}{*}{1950-2000} & \multirow{2}{*}{2000-2015} &\multirow{2}{*}{63} 
		&  Bayes(mod) & 0.011 &0.82 & 0.90  & 0.95 \\
		&	& &  & Bayes & 0.011& 0.81 & 0.90 & 0.95\\
		\hline
	\end{tabular}
\end{table}

\begin{table}[tbp]
	\caption{Out-of-sample validation results of ASAF for both male and female with $\alpha_{a_4}$ changed. ``Bayes (mod)" is the BHM with changed hyperparameter. }\label{tb:mua4}
	\centering
	\begin{tabular}{c|c|c|c|c|c|ccc}
		\multirow{2}{*}{Gender} &	\multirow{2}{*}{Train} & \multirow{2}{*}{Test} &\multirow{2}{*}{num}	&\multirow{2}{*}{Method} &\multirow{2}{*}{MAE} & \multicolumn{3}{c}{Coverage} \\
		& & & & & & $80\%$ & $90\%$ & $95\%$  \\
		\hline
		\multirow{2}{*}{Male} &	\multirow{2}{*}{1950-2000} & \multirow{2}{*}{2000-2015} &\multirow{2}{*}{63} 
		&  Bayes(mod) & 0.016 &0.65 &0.76 & 0.85\\
		&&	& & Bayes &0.016  &0.65  &0.76 & 0.84 \\
		\hline
		\multirow{2}{*}{Female} &	\multirow{2}{*}{1950-2000} & \multirow{2}{*}{2000-2015} &\multirow{2}{*}{63} 
		&  Bayes(mod) & 0.011 &0.81 & 0.89& 0.95 \\
		&	& &  & Bayes & 0.011& 0.81 & 0.90 & 0.95\\
		\hline
	\end{tabular}
\end{table}

\section{All-age Smoking Attribuable Fraction Projection for Both Genders in Over  60 Countries} \label{app4}

\begin{figure}[H]
	\begin{center}
		\includegraphics[scale=0.45]{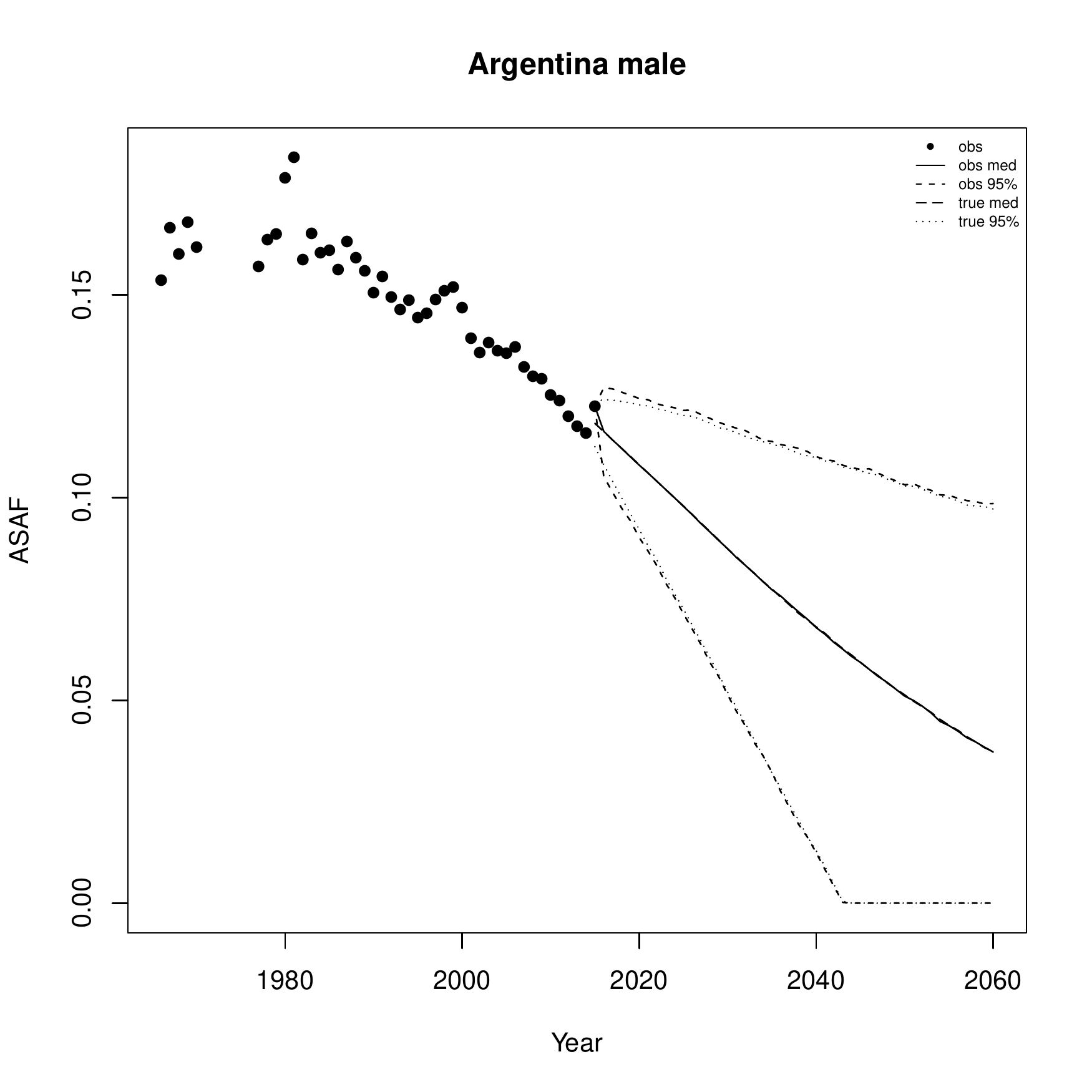}
		\includegraphics[scale=0.45]{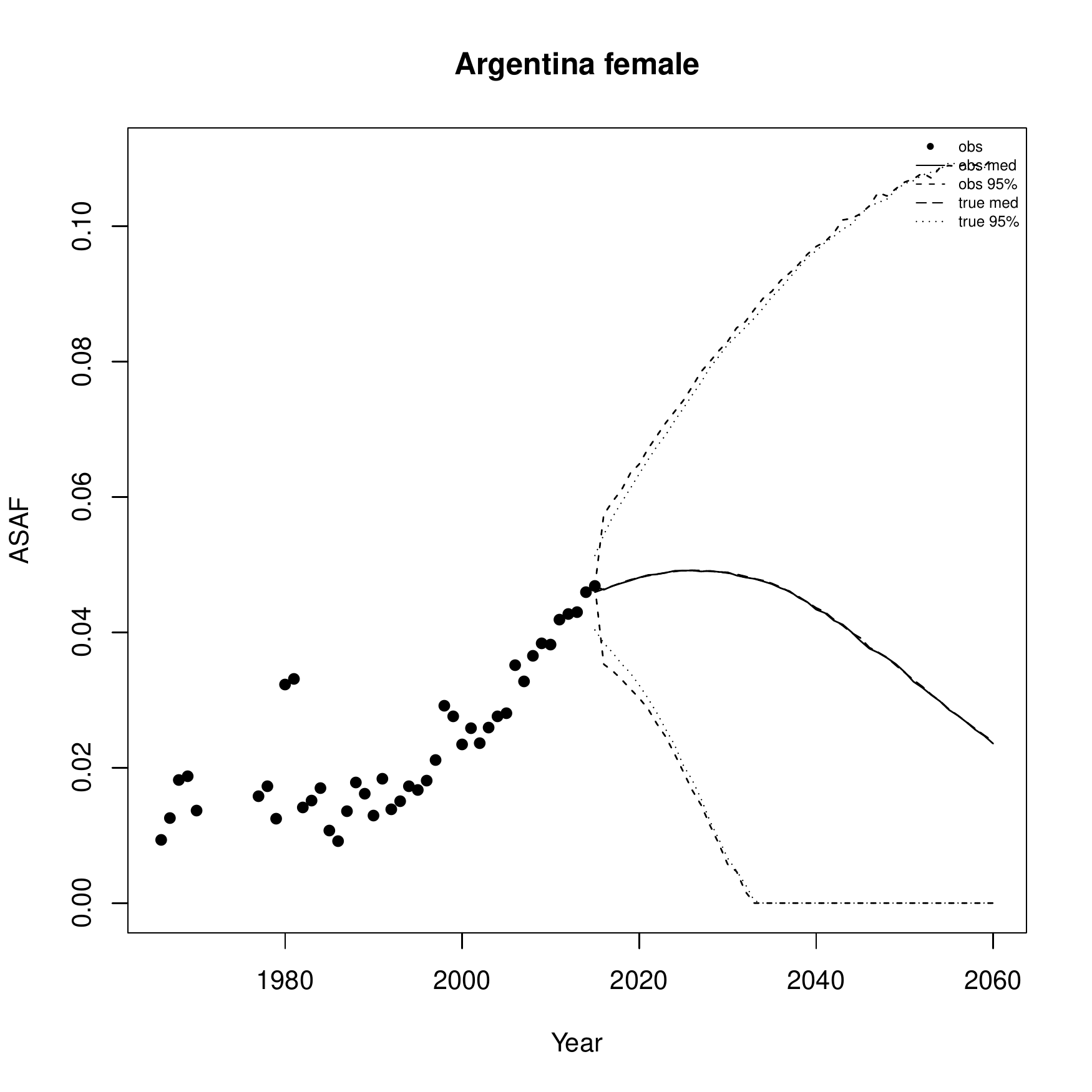}
		\includegraphics[scale=0.45]{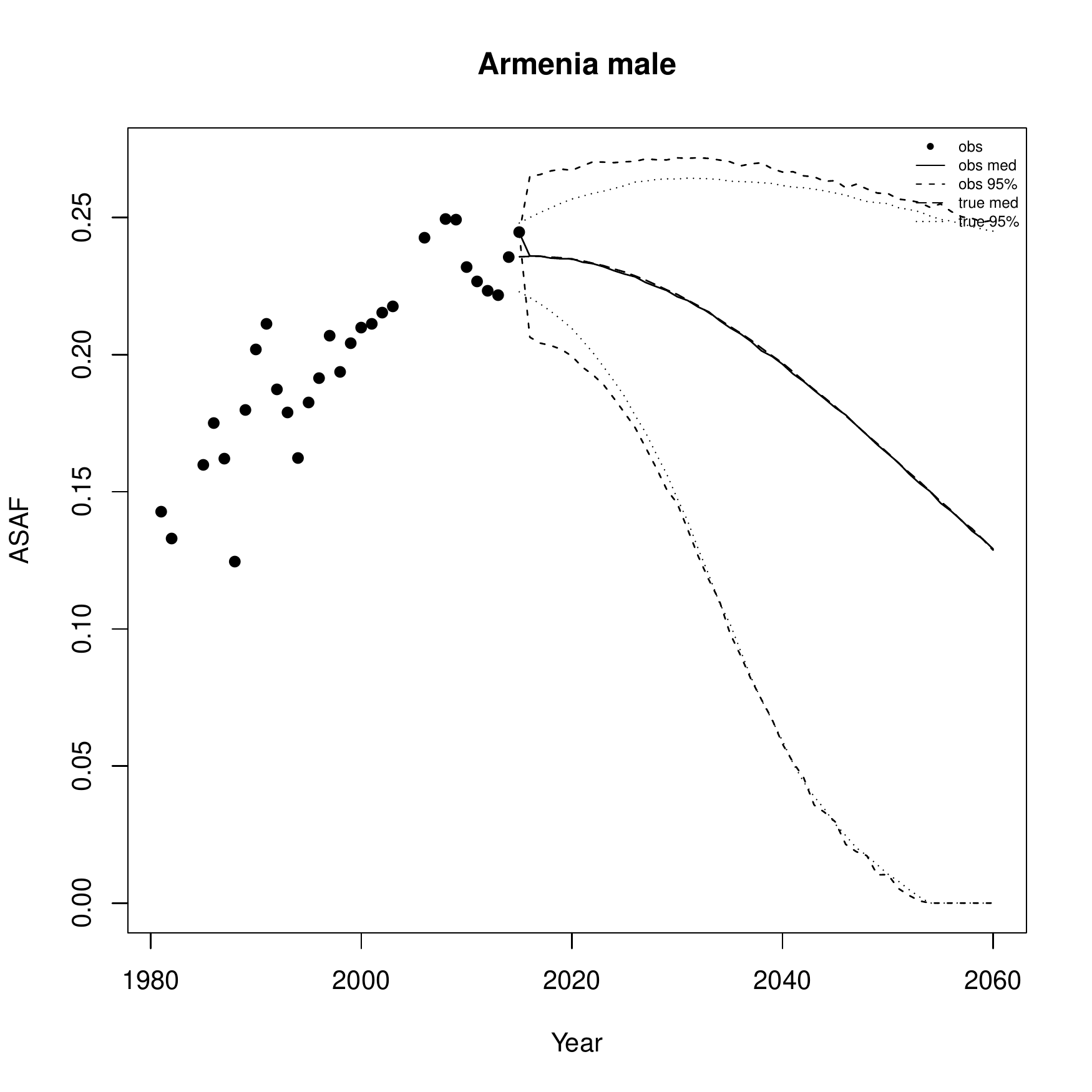}
		\includegraphics[scale=0.45]{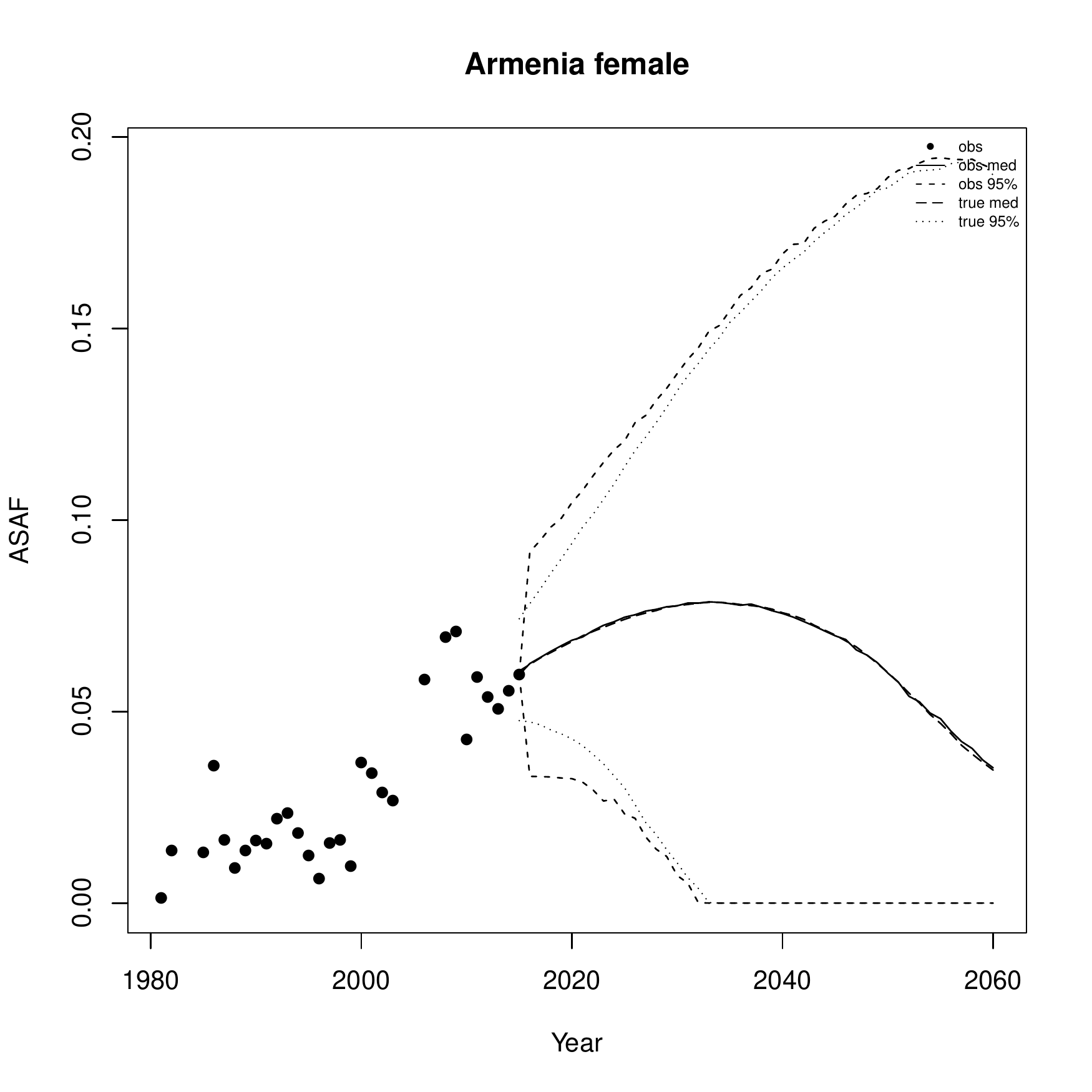}
	\end{center}
\end{figure}

\begin{figure}[H]
	\begin{center}
		\includegraphics[scale=0.43]{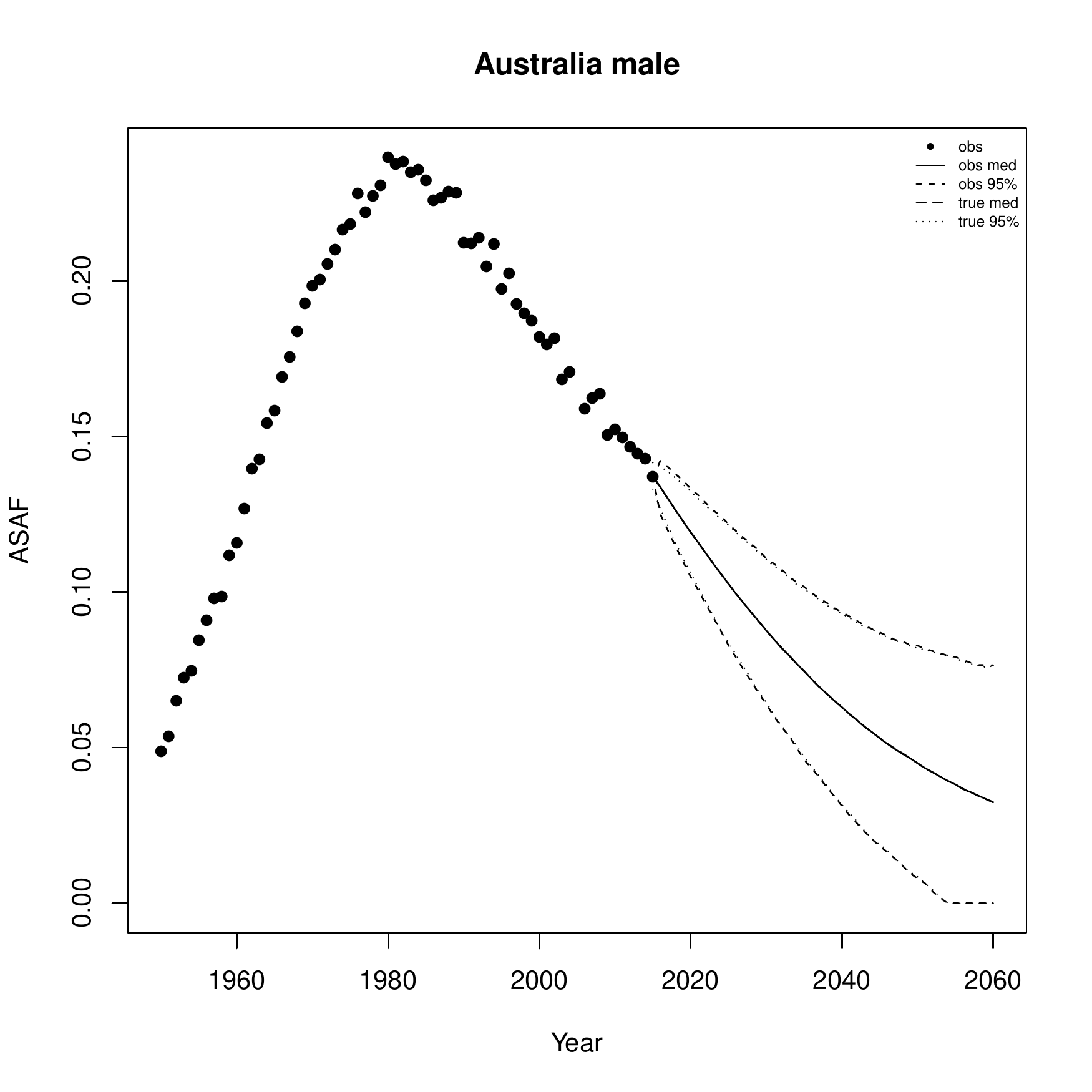}
		\includegraphics[scale=0.43]{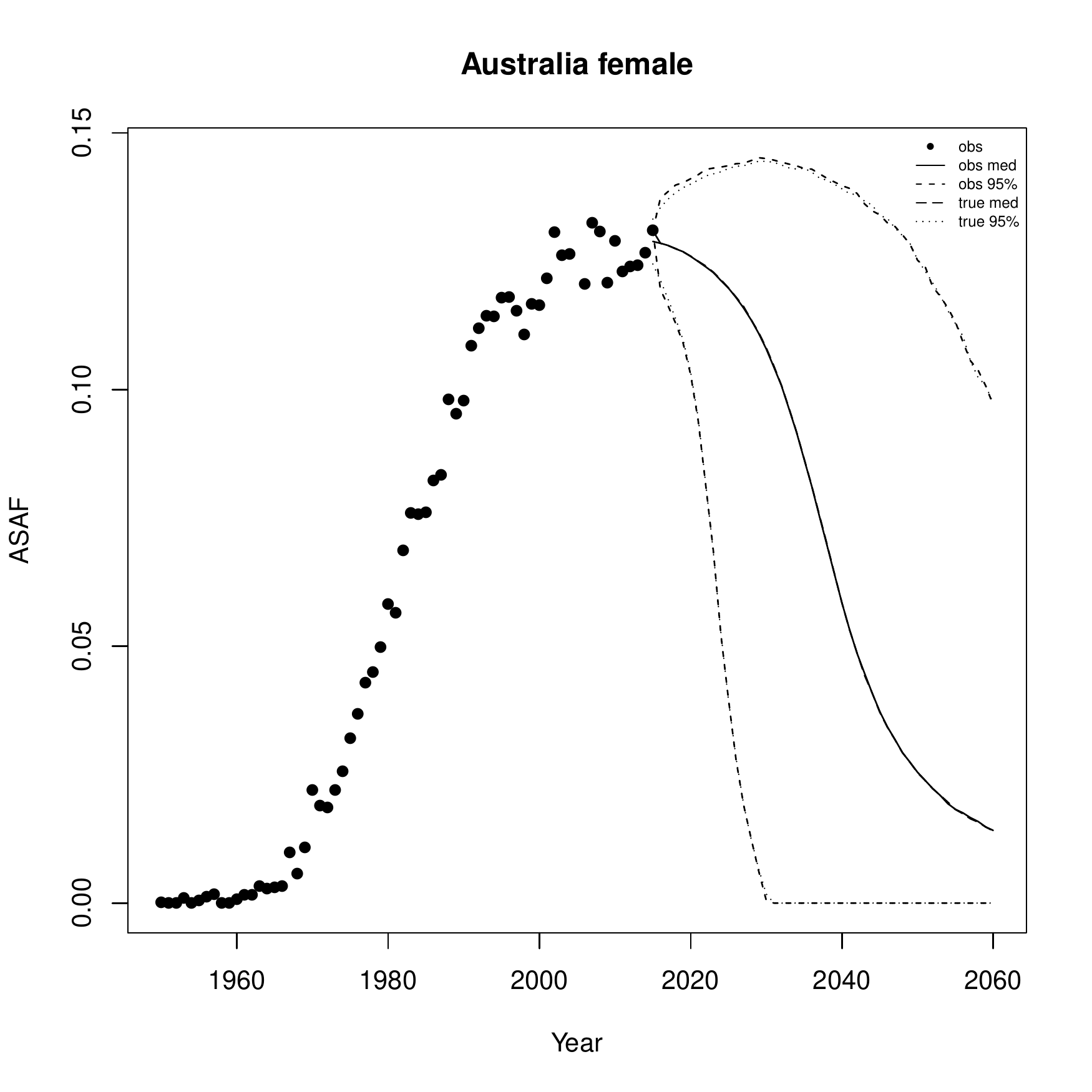}
		\includegraphics[scale=0.43]{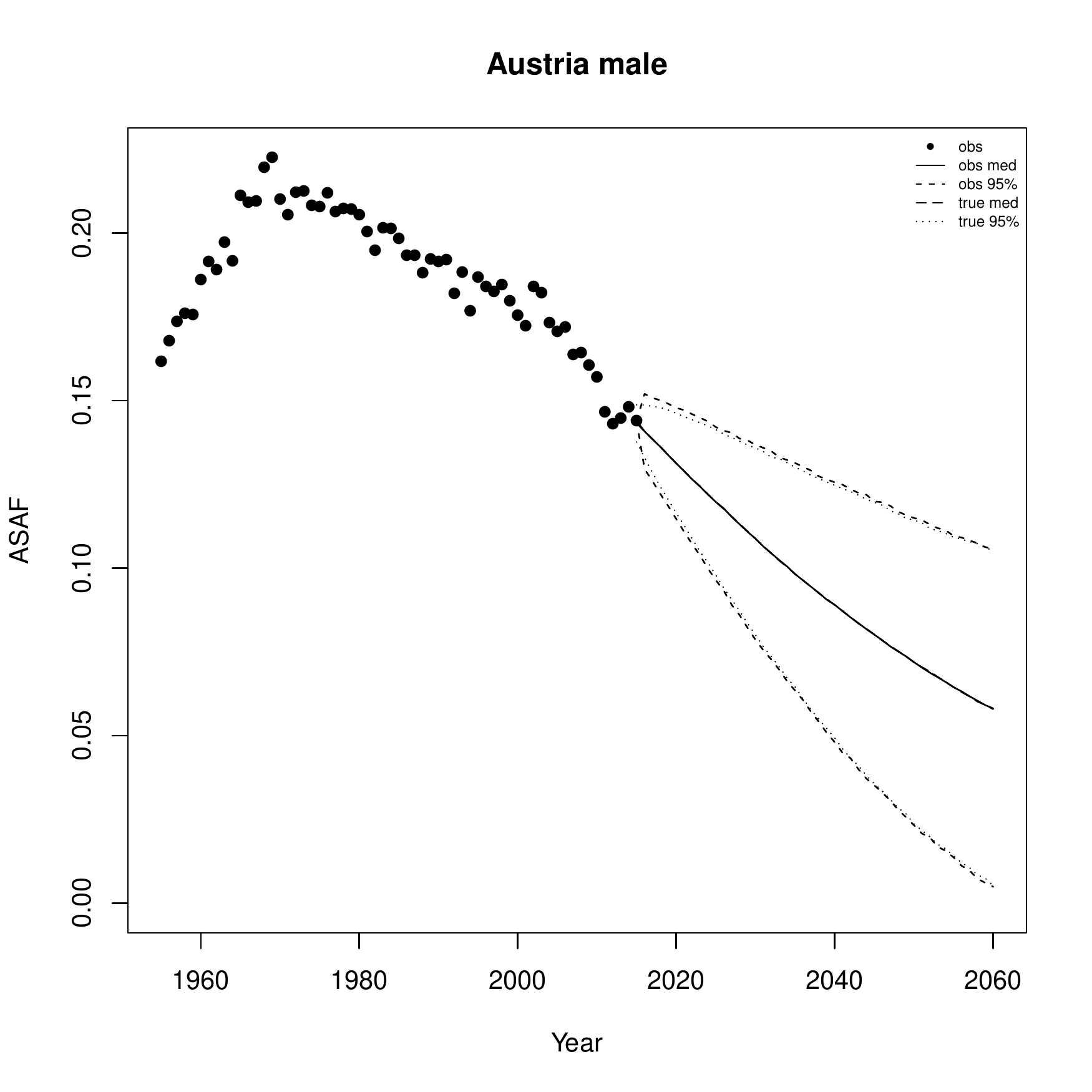}
		\includegraphics[scale=0.43]{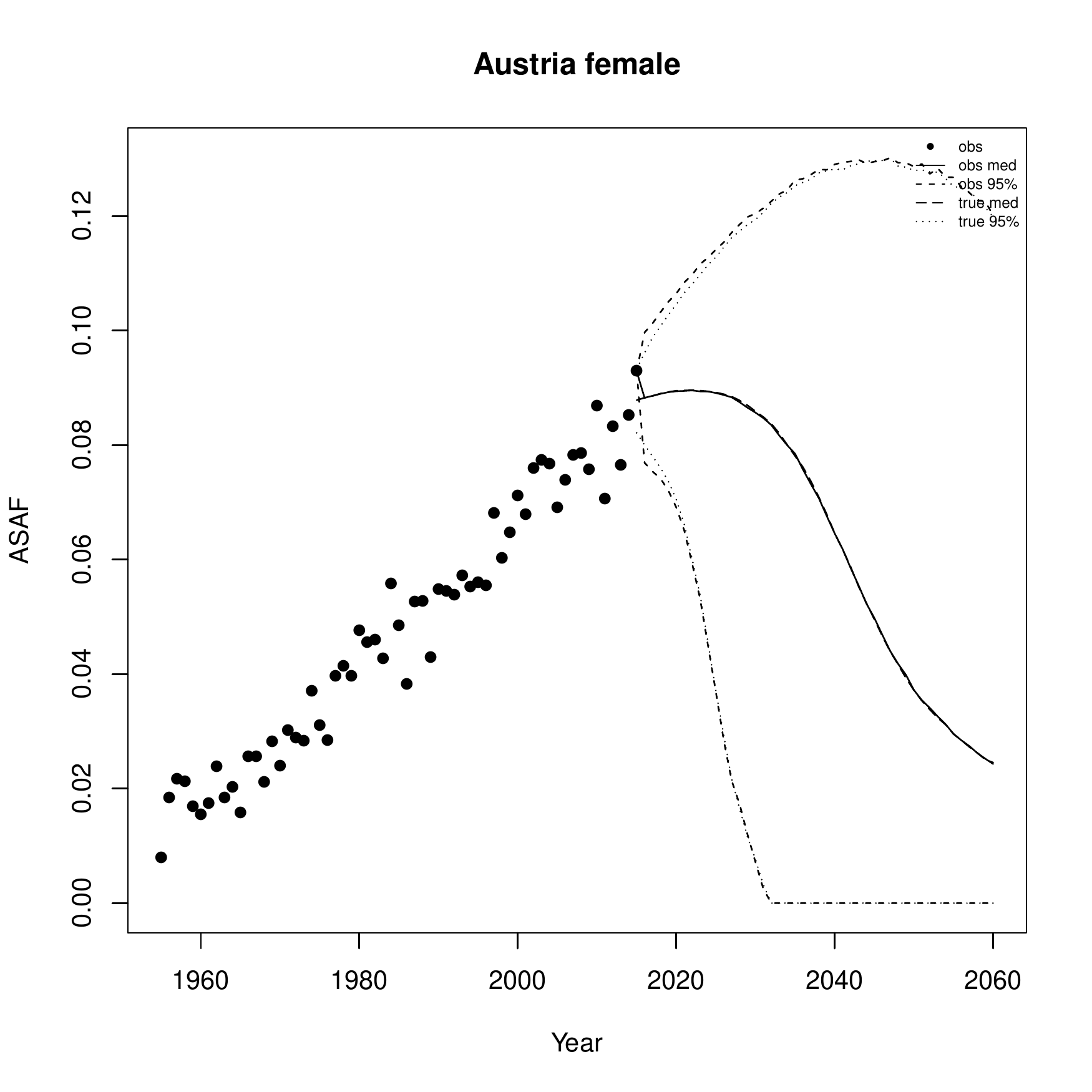}
		\includegraphics[scale=0.43]{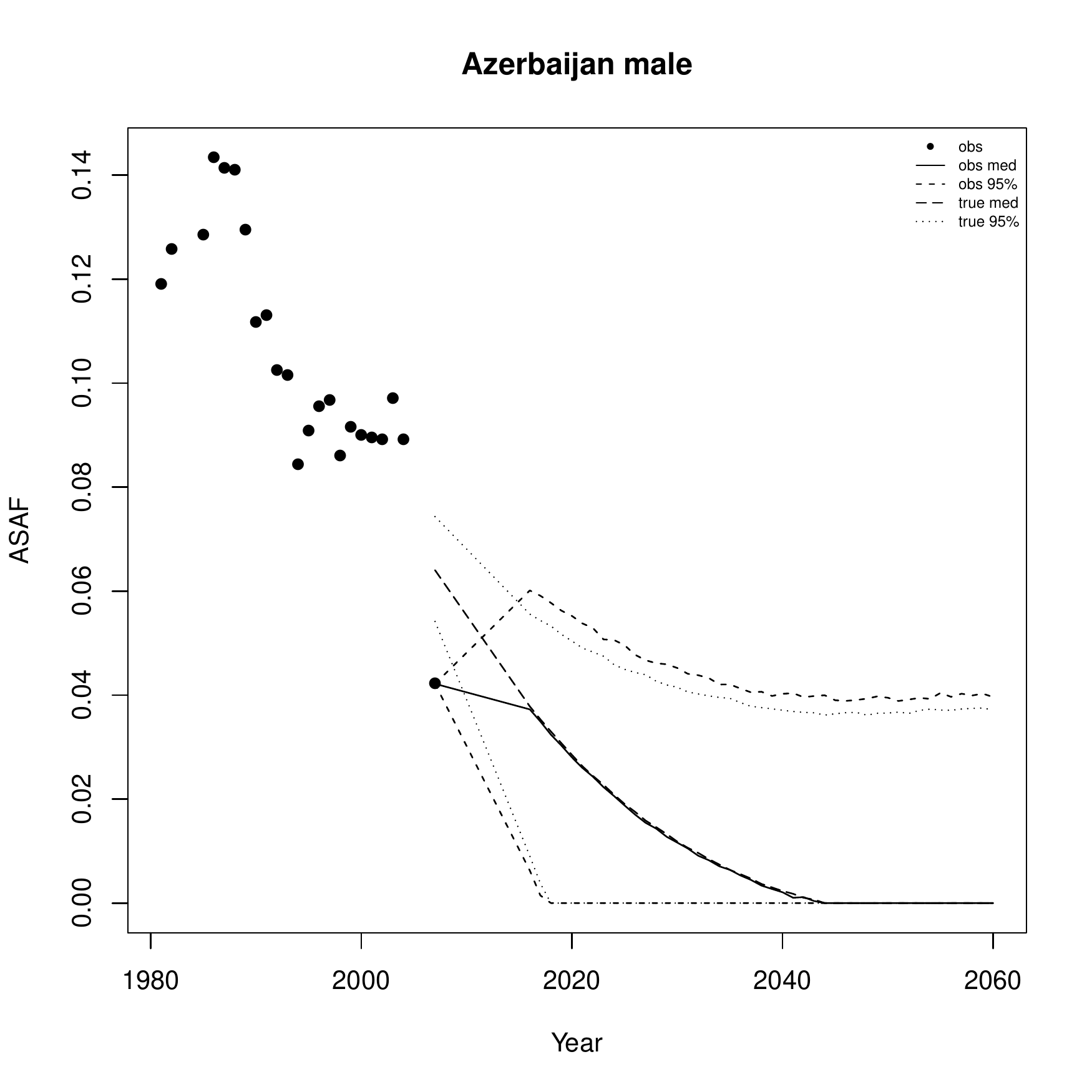}
		\includegraphics[scale=0.43]{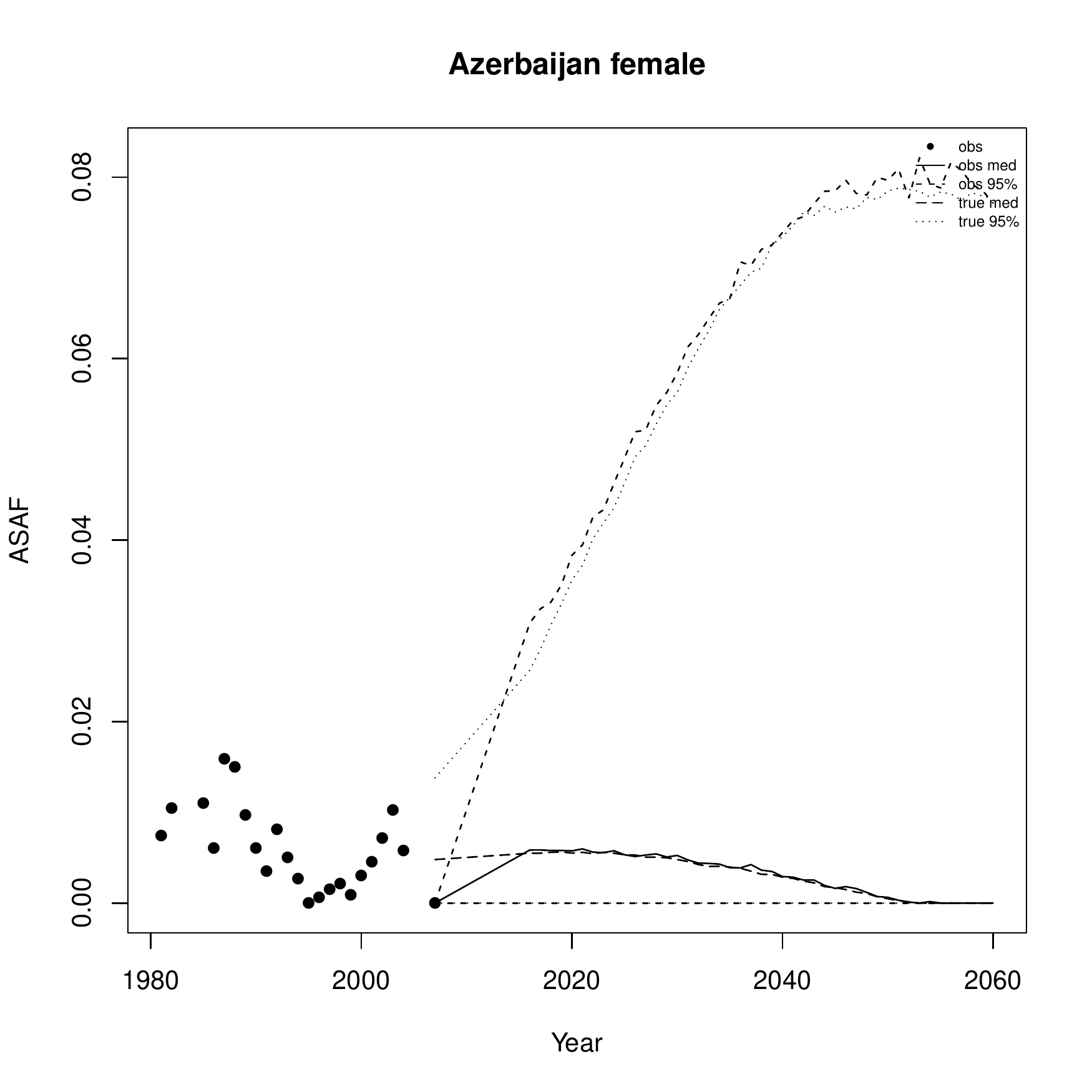}
	\end{center}
\end{figure}

\begin{figure}[H]
	\begin{center}
		\includegraphics[scale=0.43]{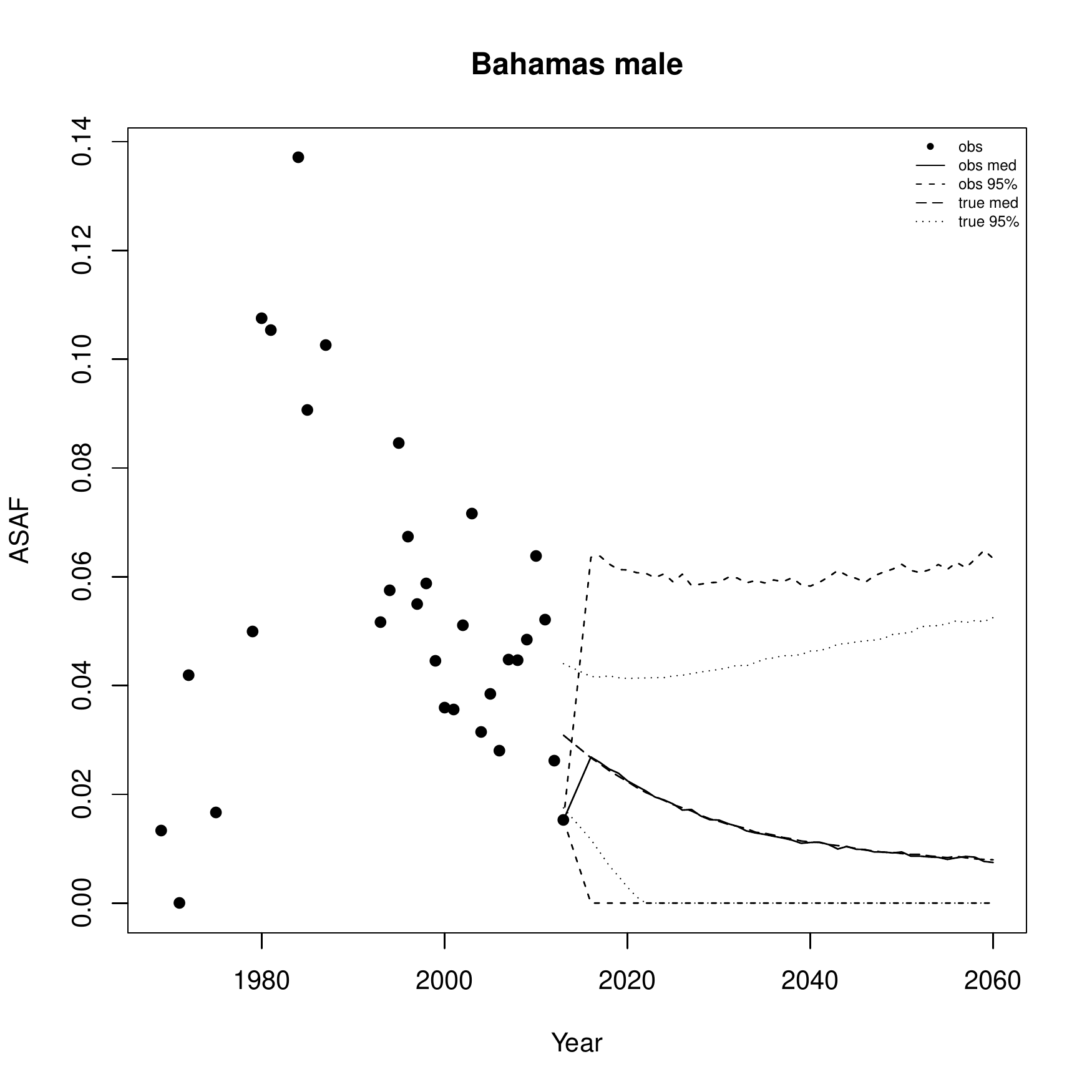}
		\includegraphics[scale=0.43]{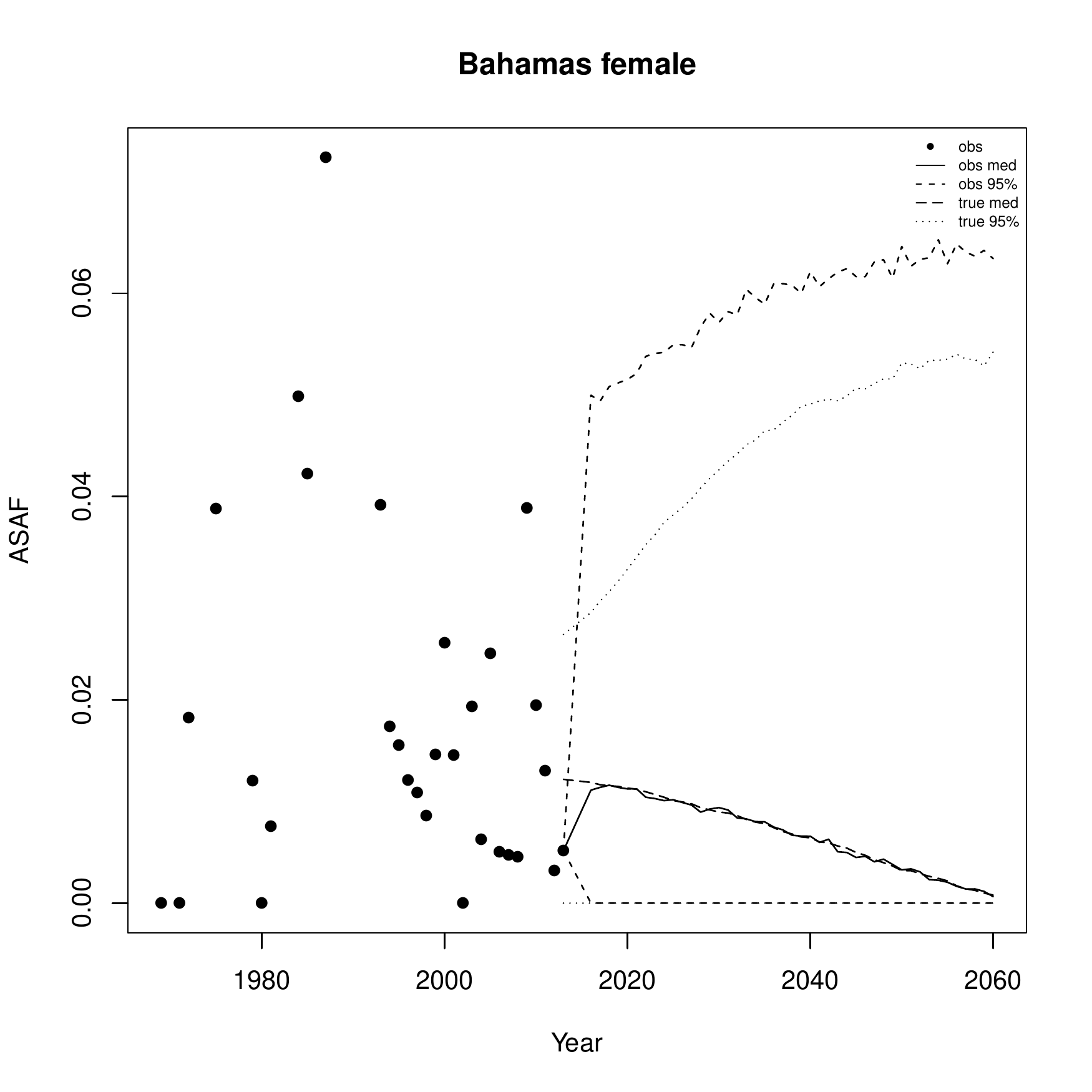}
		\includegraphics[scale=0.43]{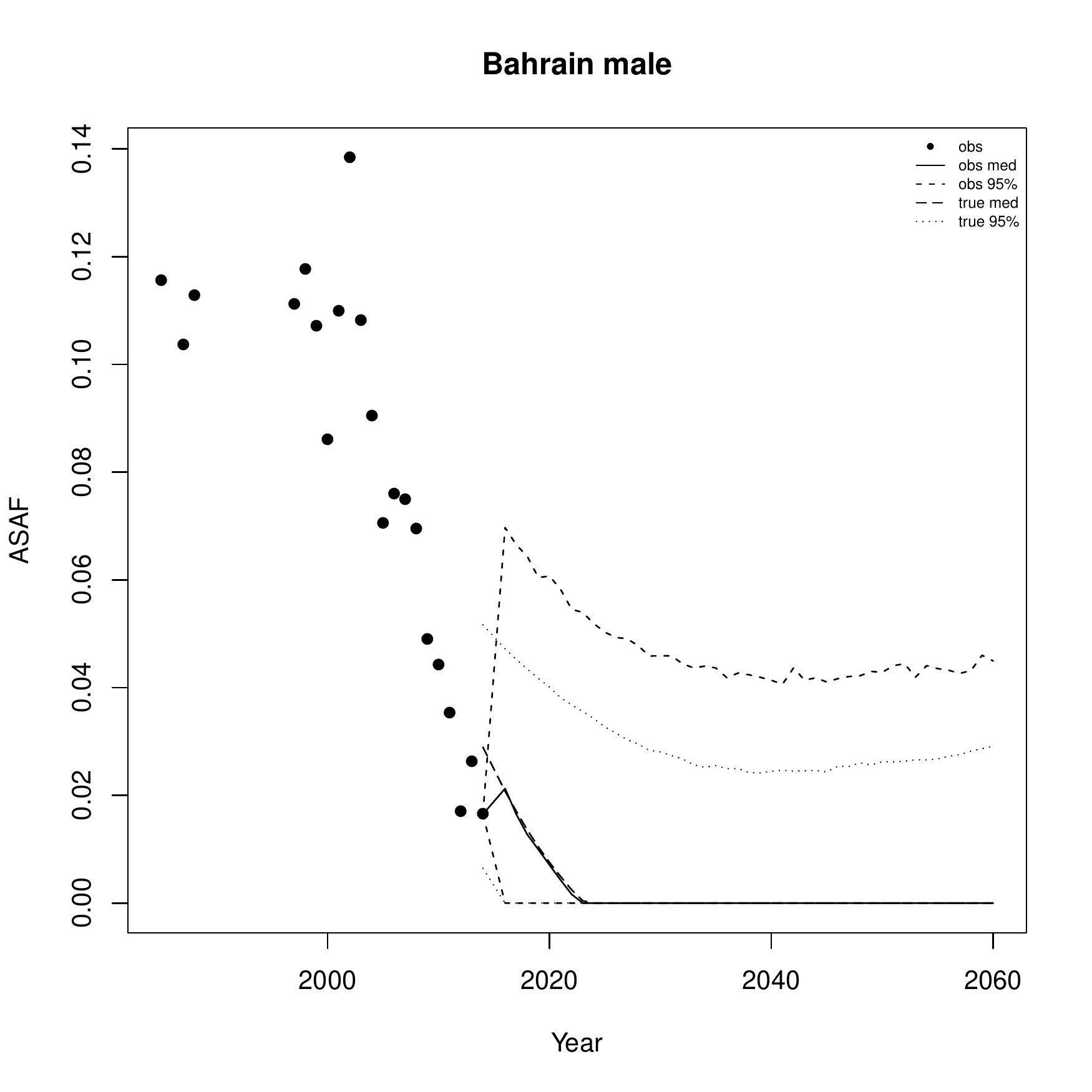}
		\includegraphics[scale=0.43]{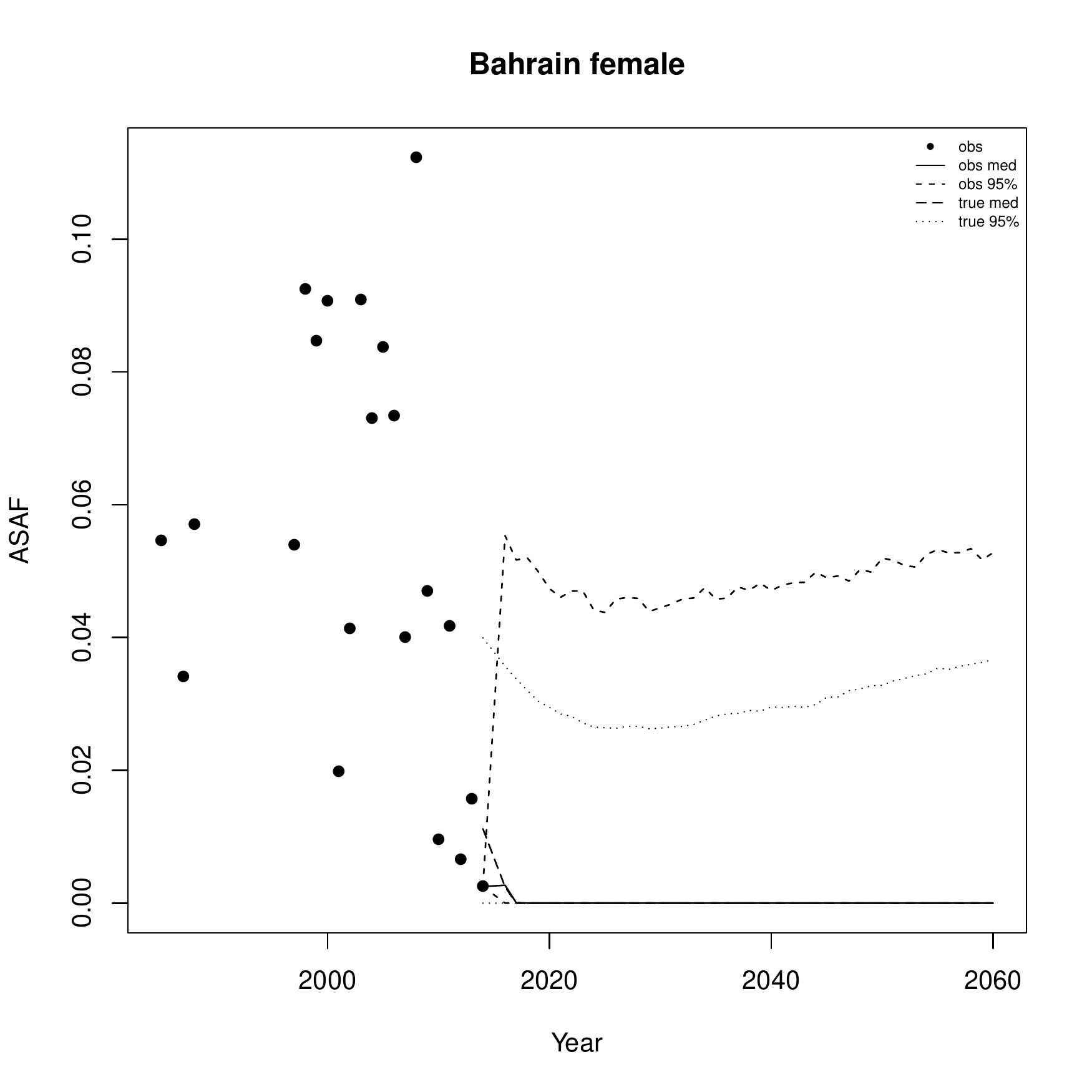}
		\includegraphics[scale=0.43]{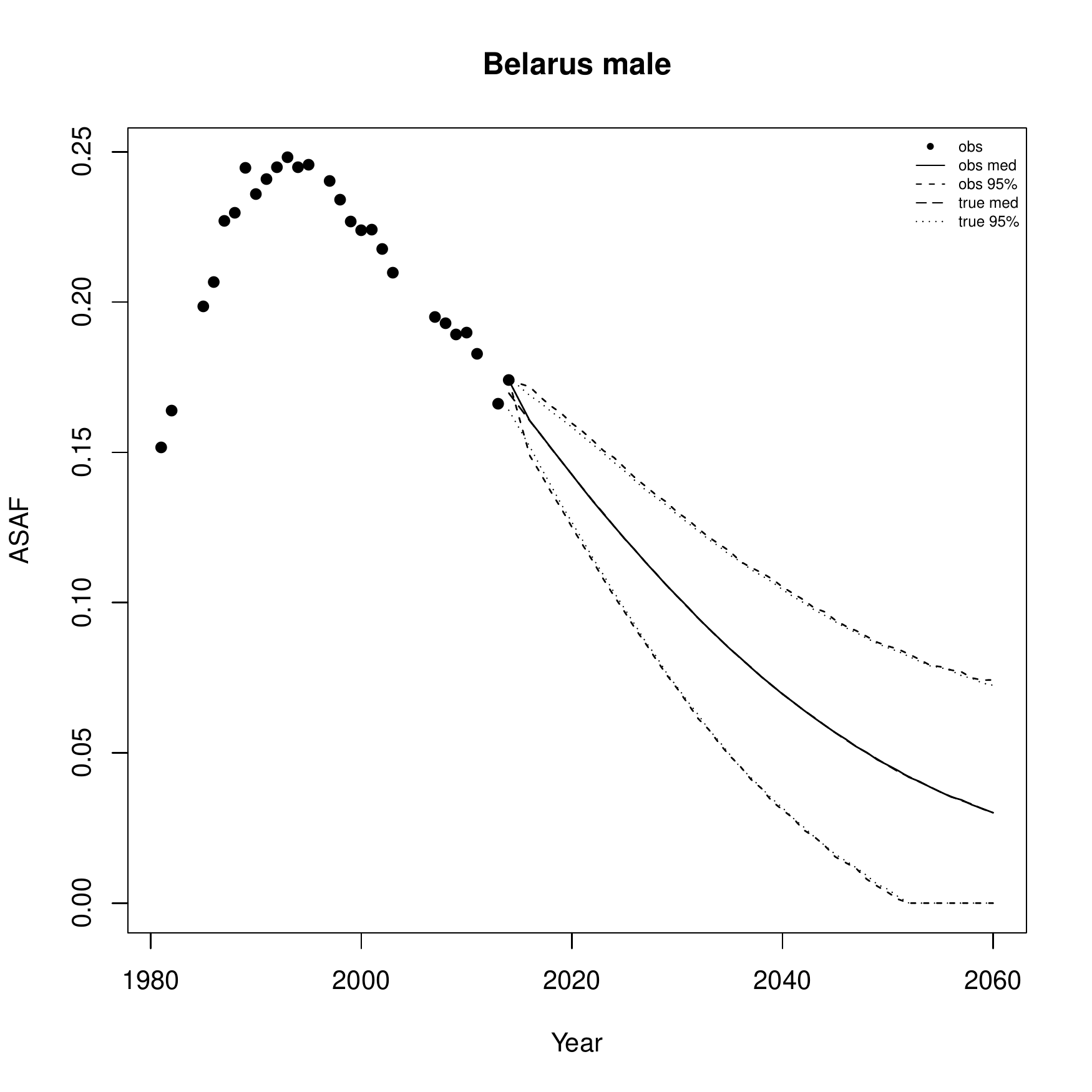}
		\includegraphics[scale=0.43]{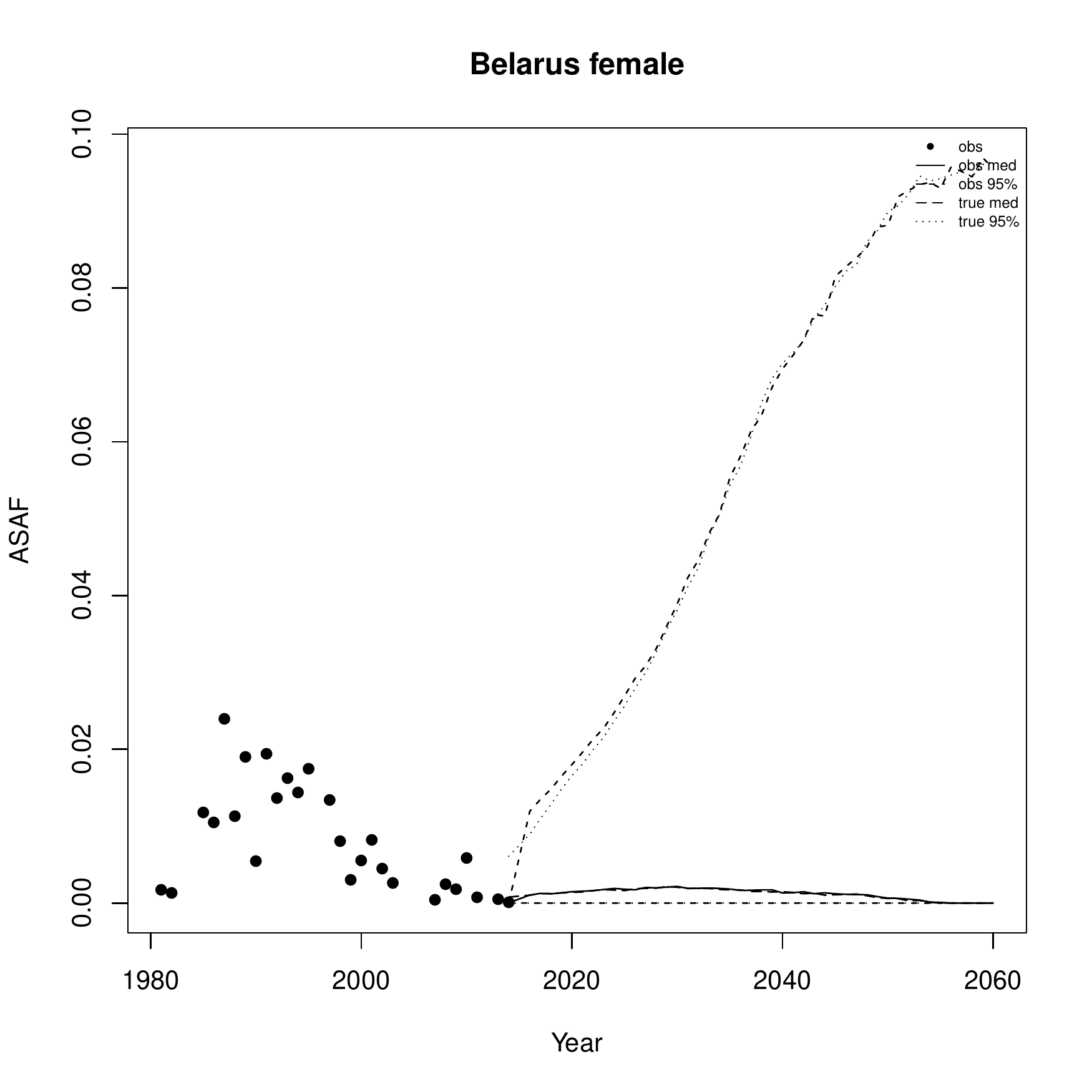}
	\end{center}
\end{figure}

\begin{figure}[H]
	\begin{center}
		\includegraphics[scale=0.43]{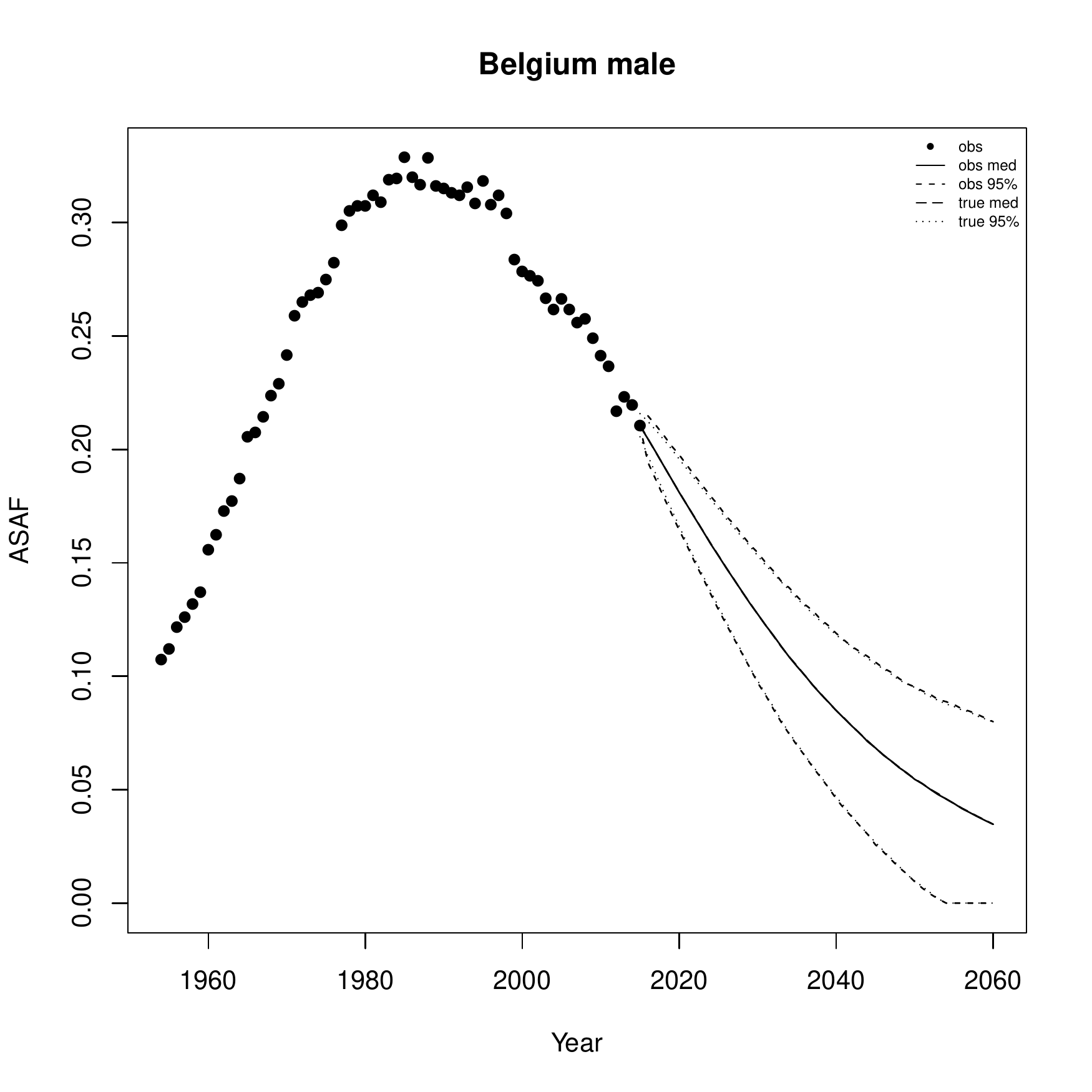}
		\includegraphics[scale=0.43]{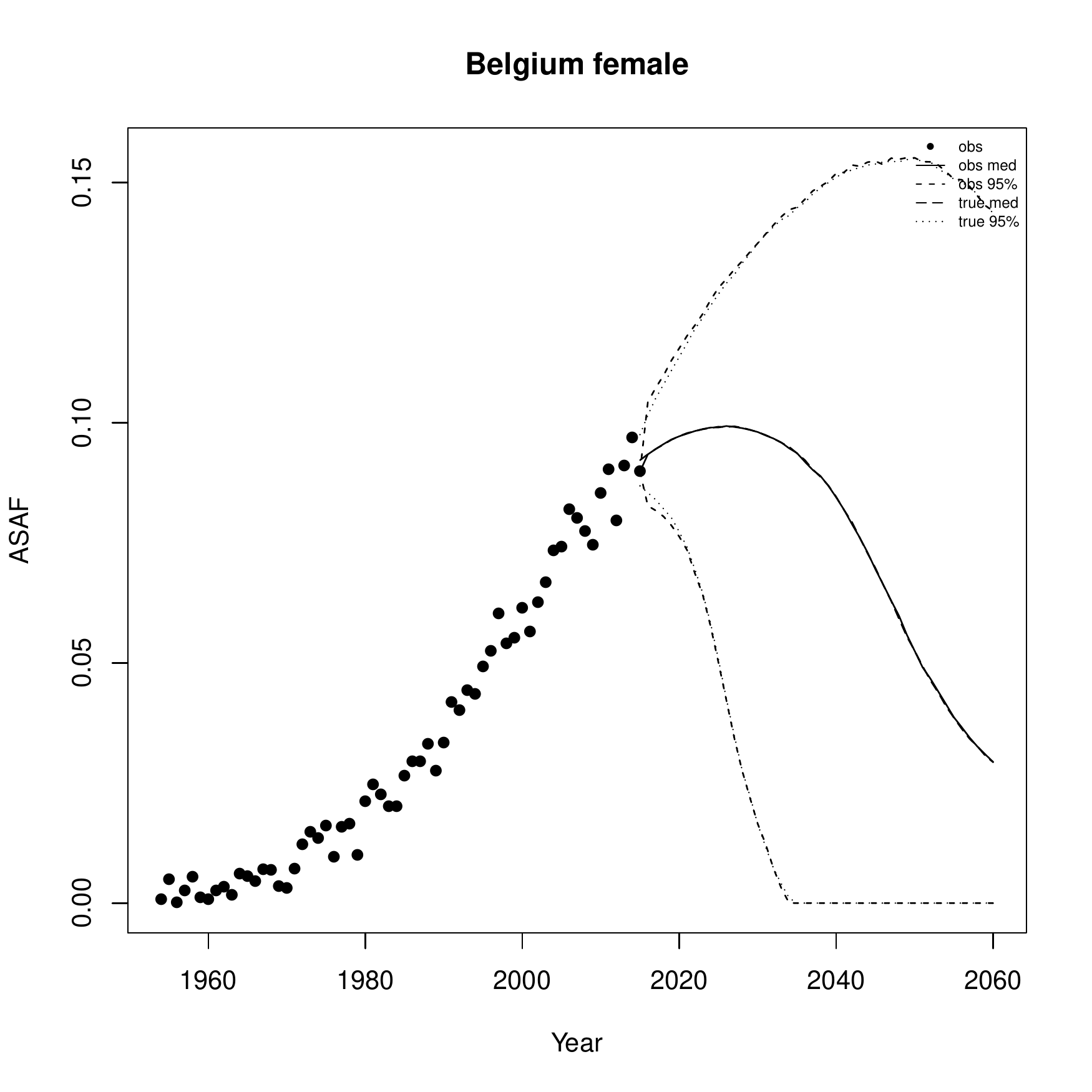}
		\includegraphics[scale=0.43]{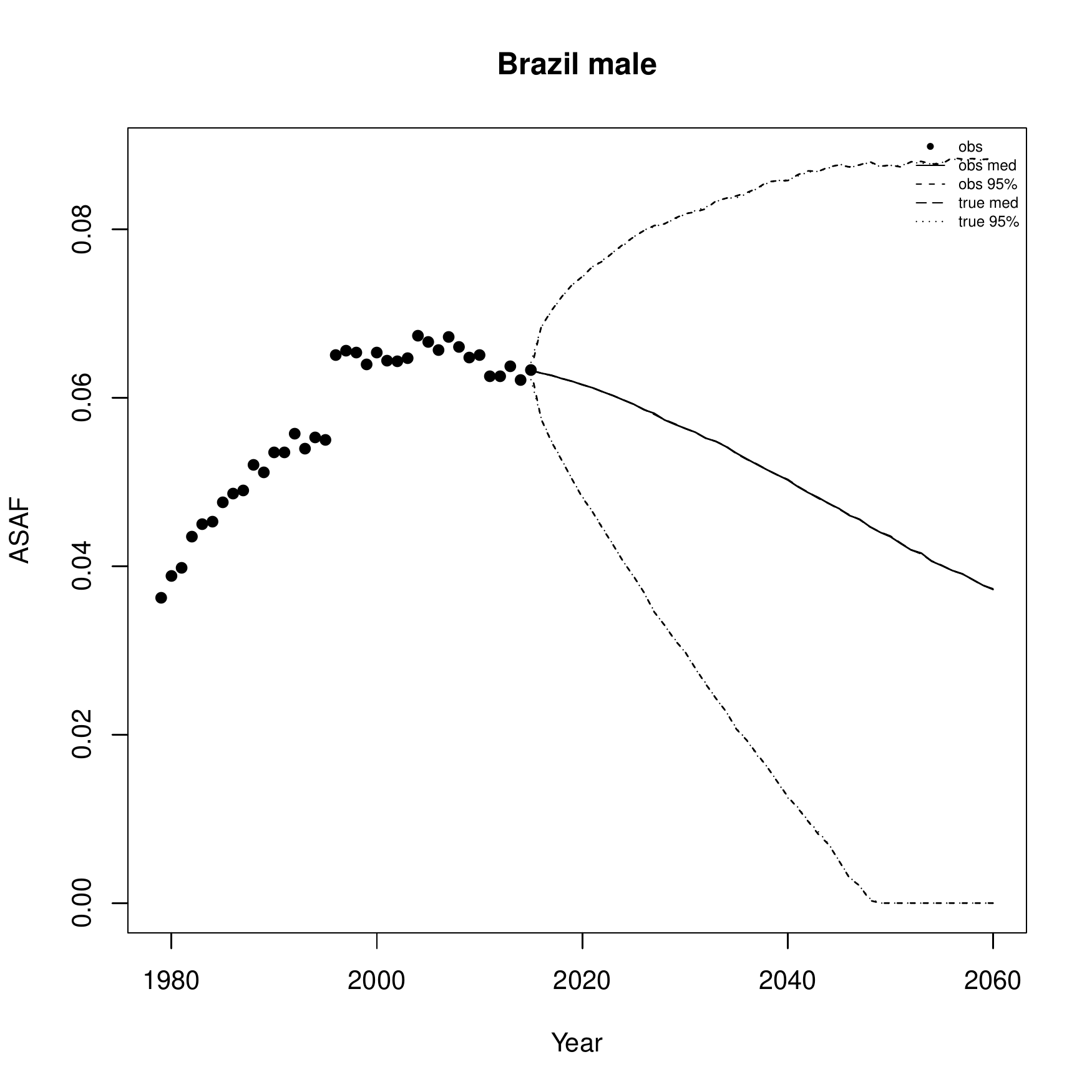}
		\includegraphics[scale=0.43]{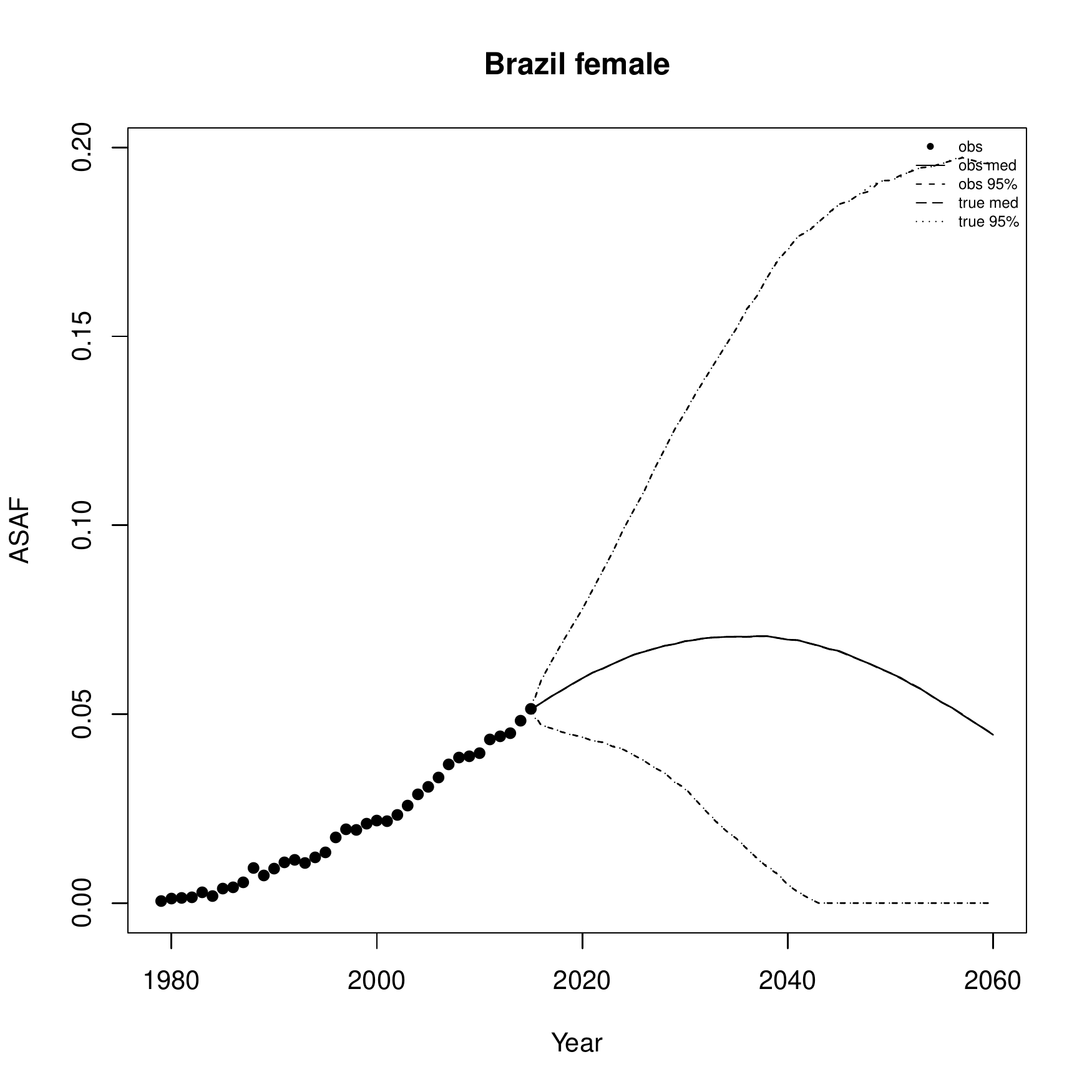}
		\includegraphics[scale=0.43]{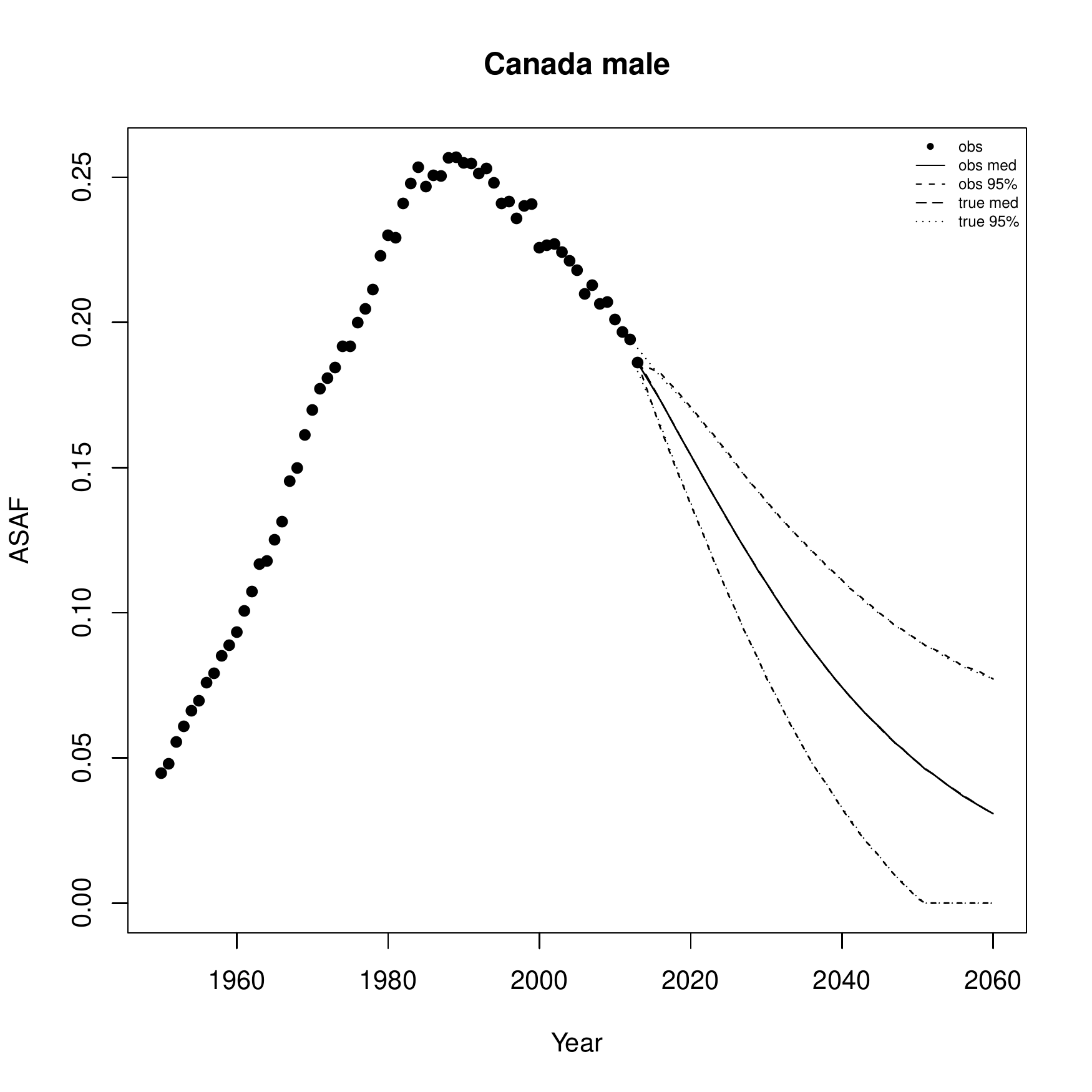}
		\includegraphics[scale=0.43]{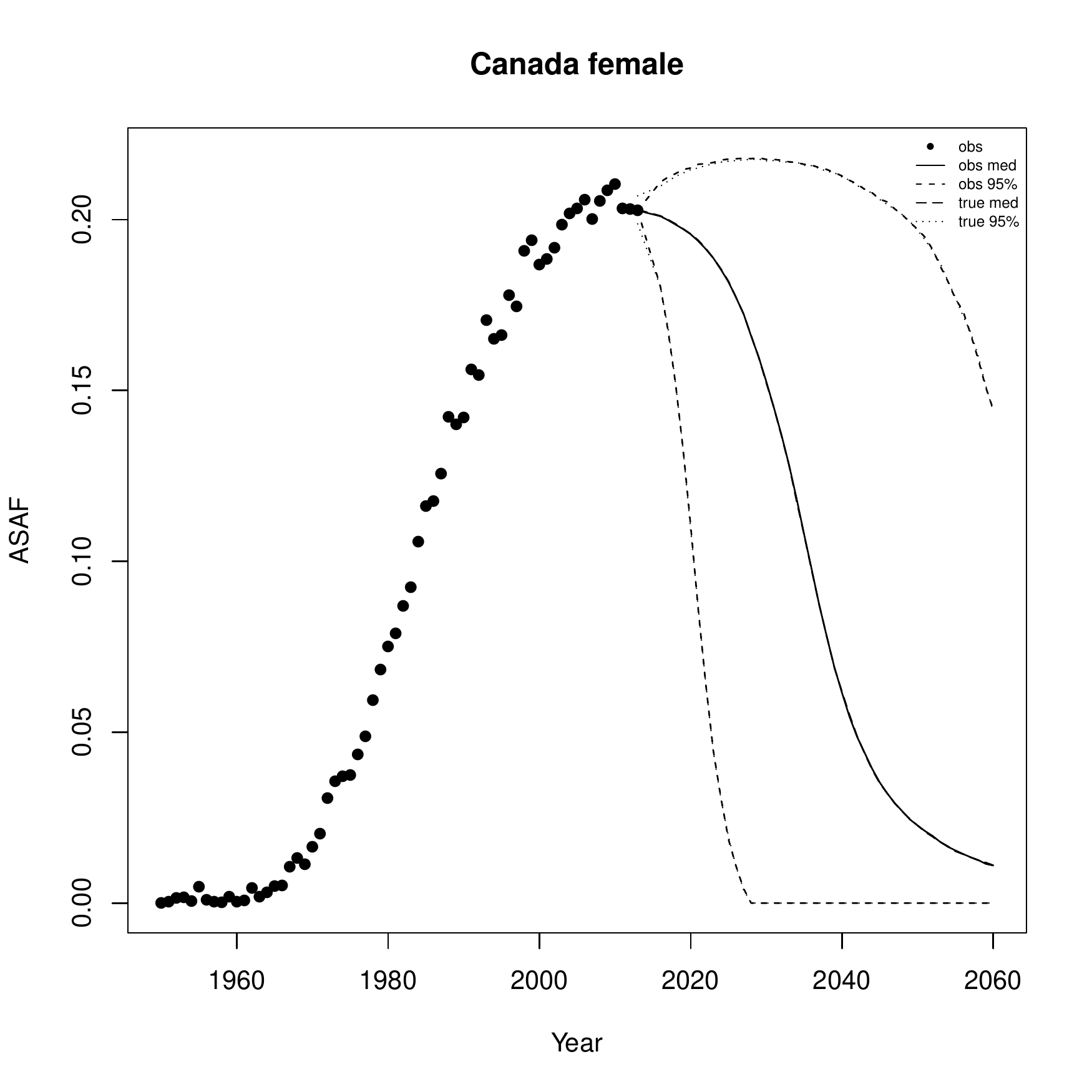}
	\end{center}
\end{figure}

\begin{figure}[H]
	\begin{center}
		\includegraphics[scale=0.43]{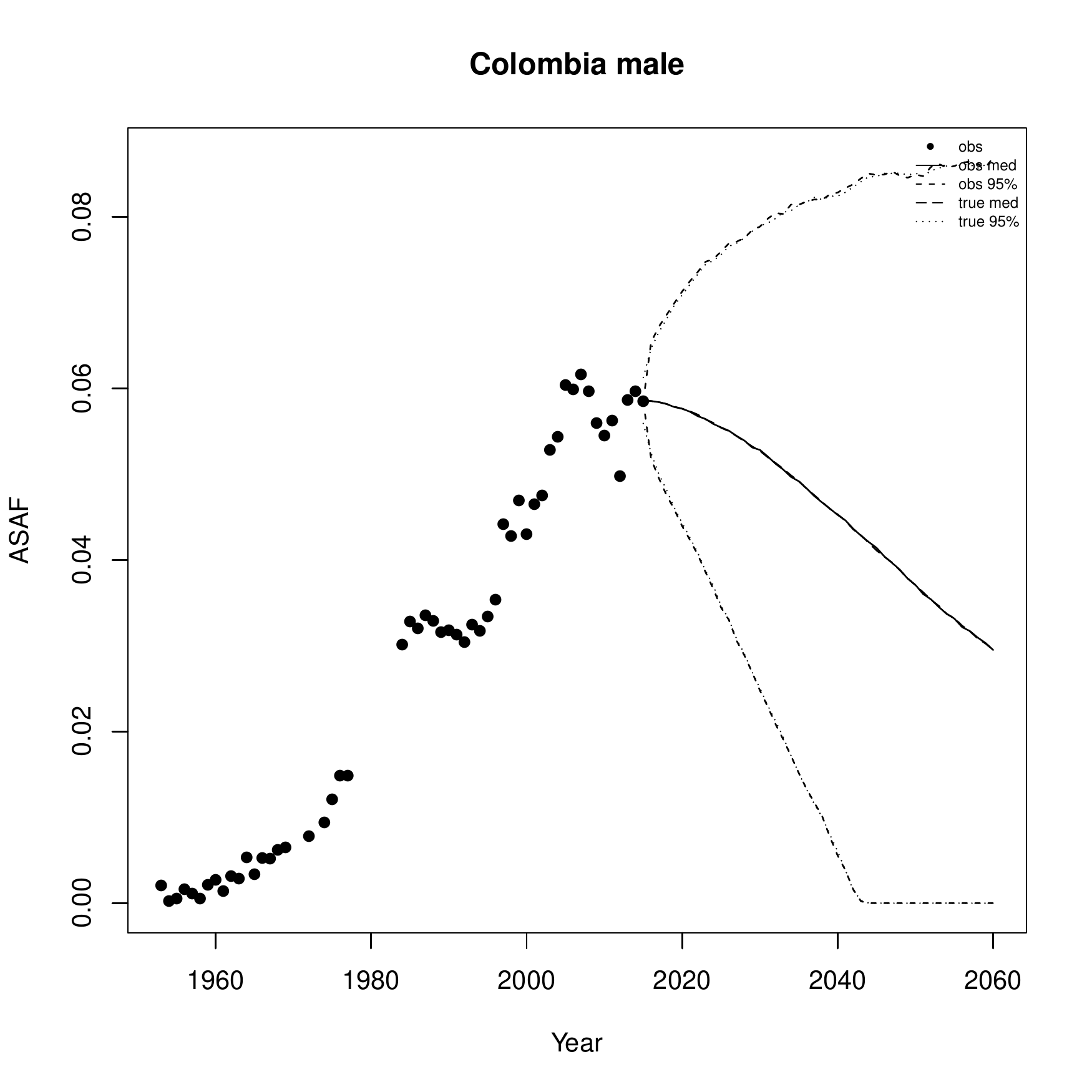}
		\includegraphics[scale=0.43]{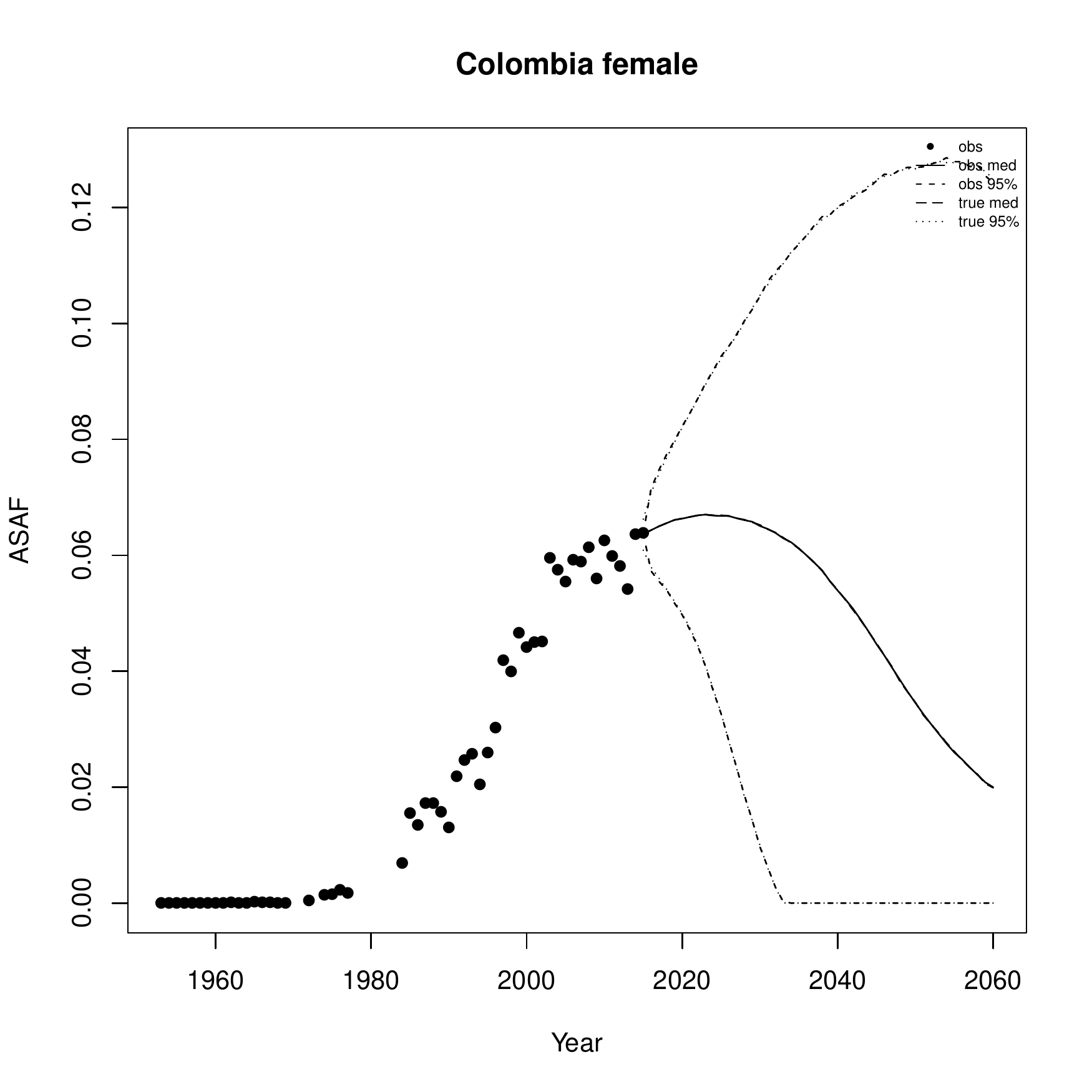}
		\includegraphics[scale=0.43]{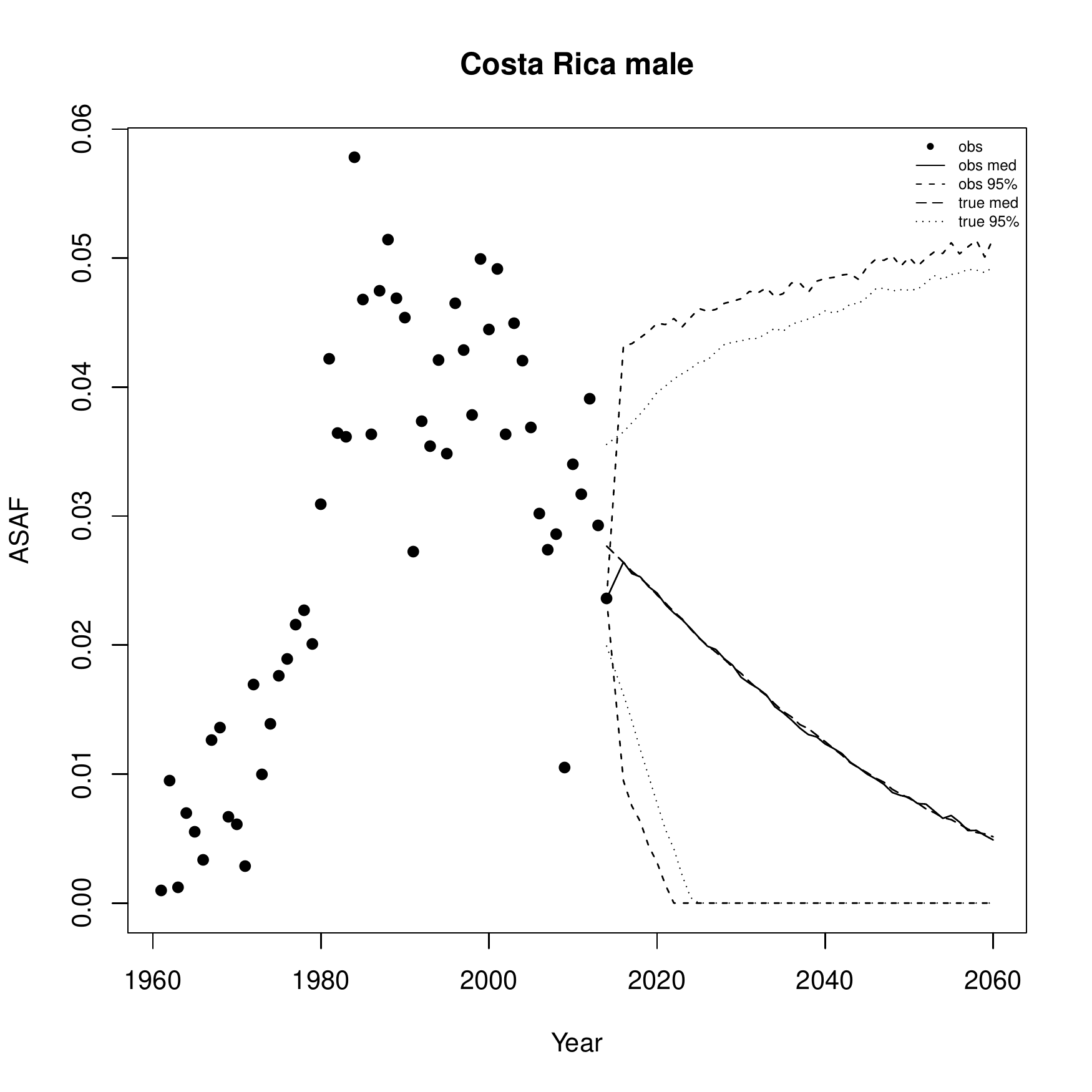}
		\includegraphics[scale=0.43]{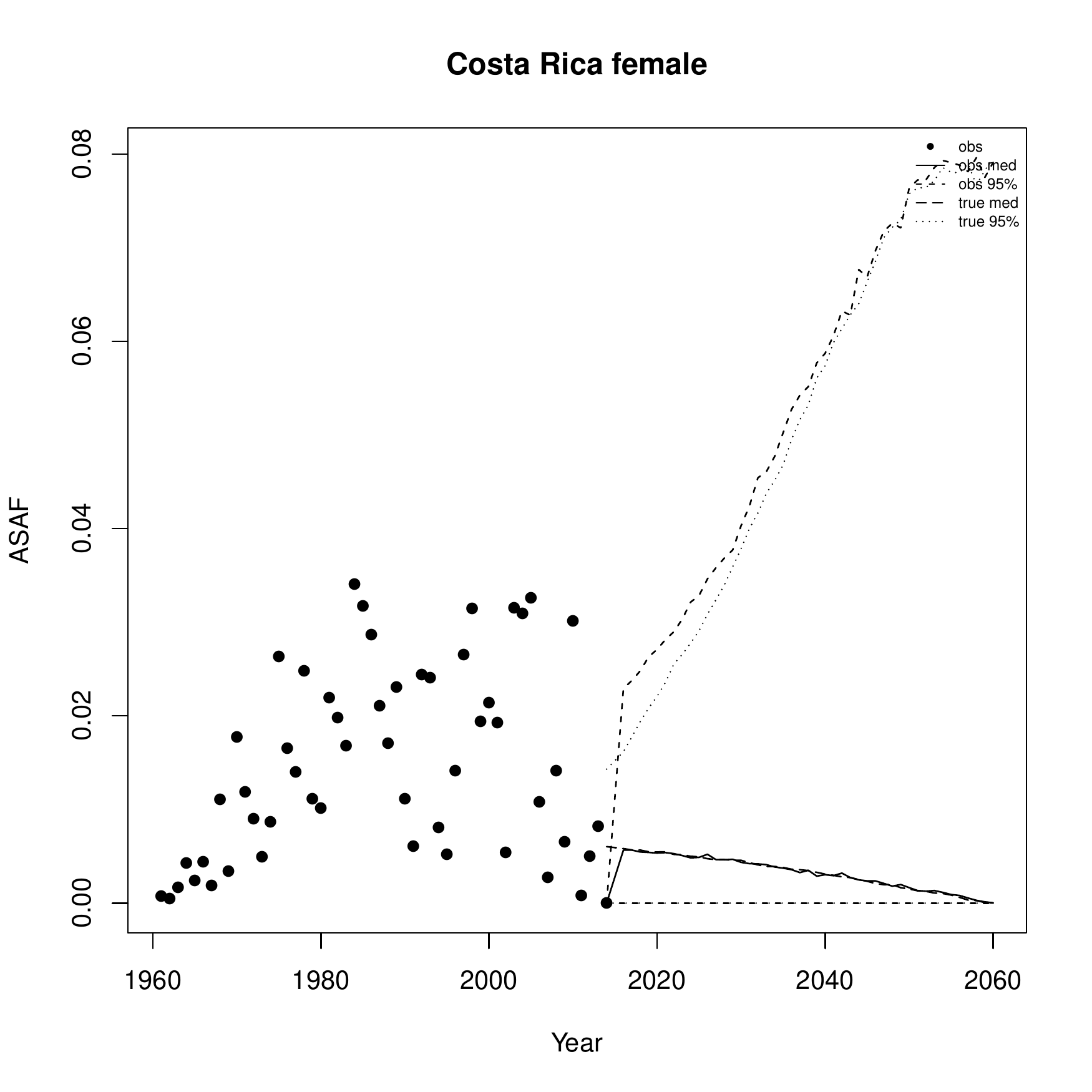}
		\includegraphics[scale=0.43]{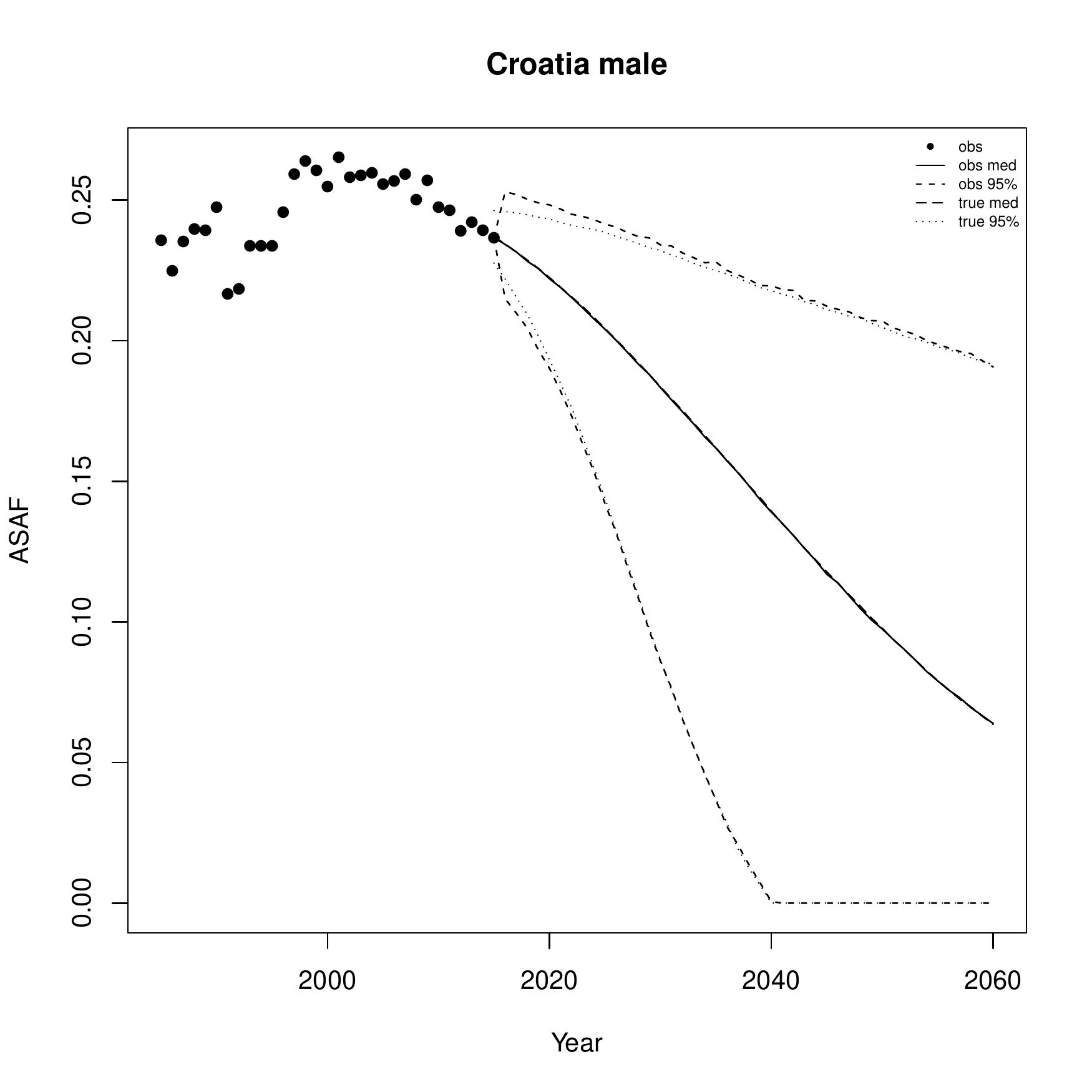}
		\includegraphics[scale=0.43]{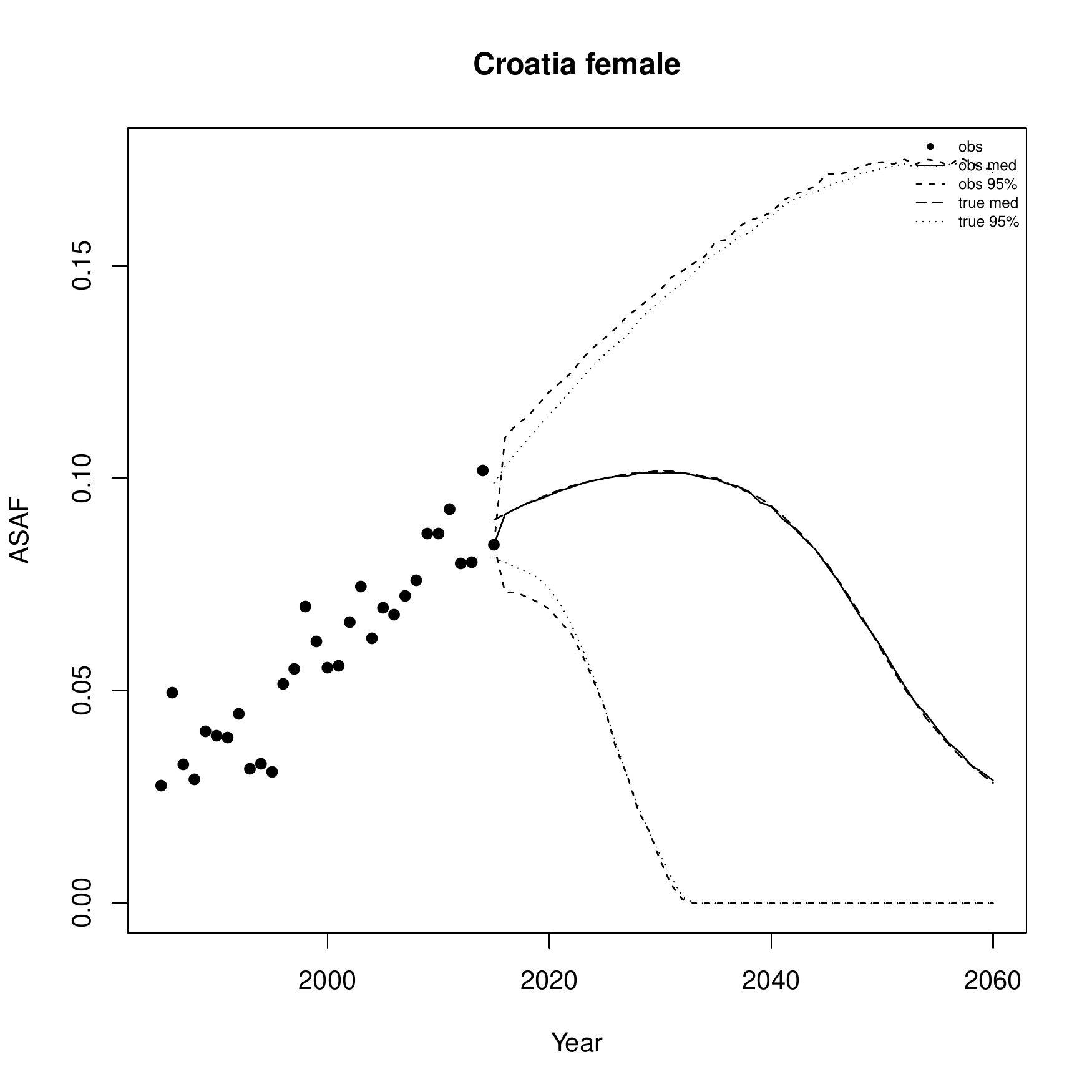}
	\end{center}
\end{figure}

\begin{figure}[H]
	\begin{center}
		\includegraphics[scale=0.43]{plot/Croatiamaleboth}
		\includegraphics[scale=0.43]{plot/Croatiafemaleboth}
		\includegraphics[scale=0.43]{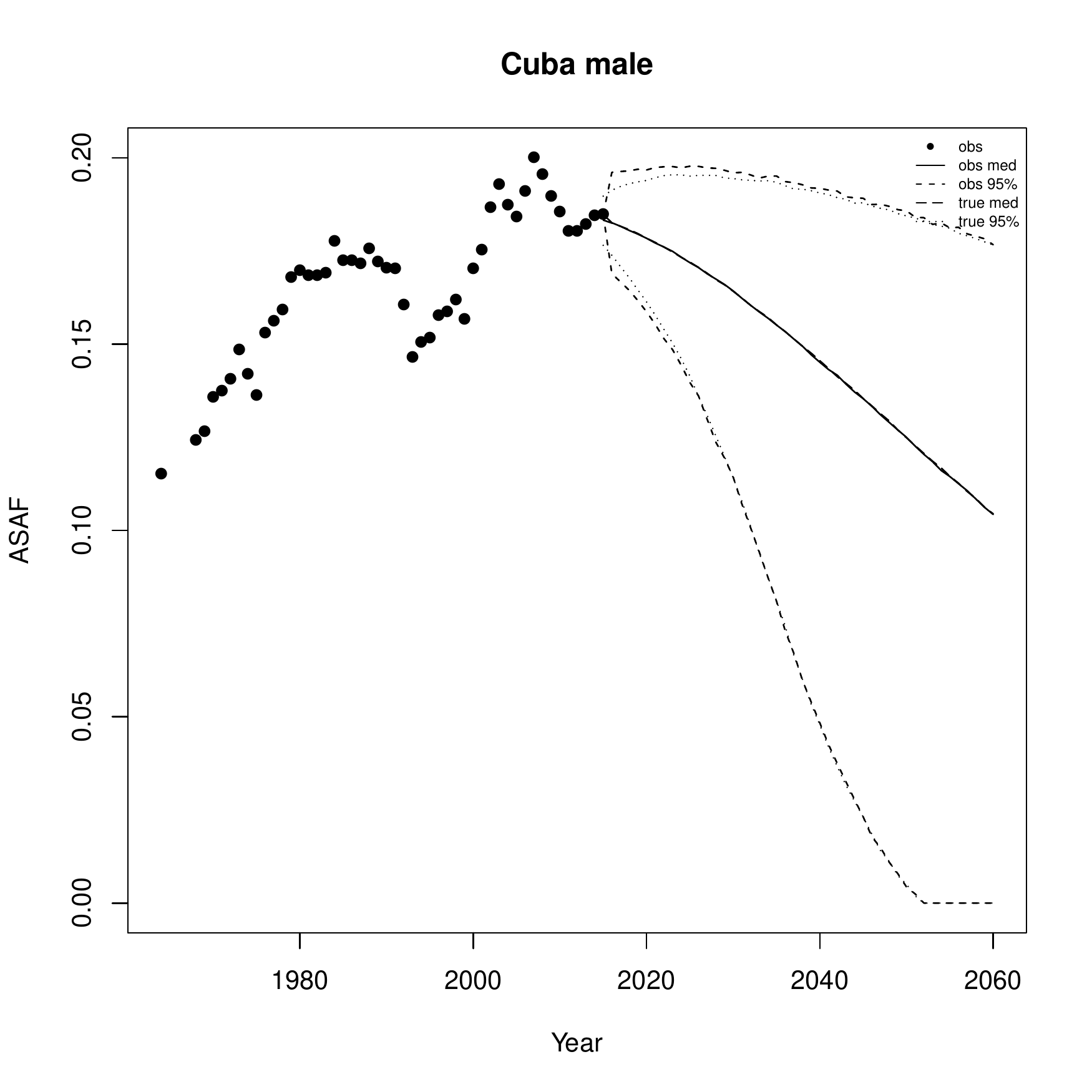}
		\includegraphics[scale=0.43]{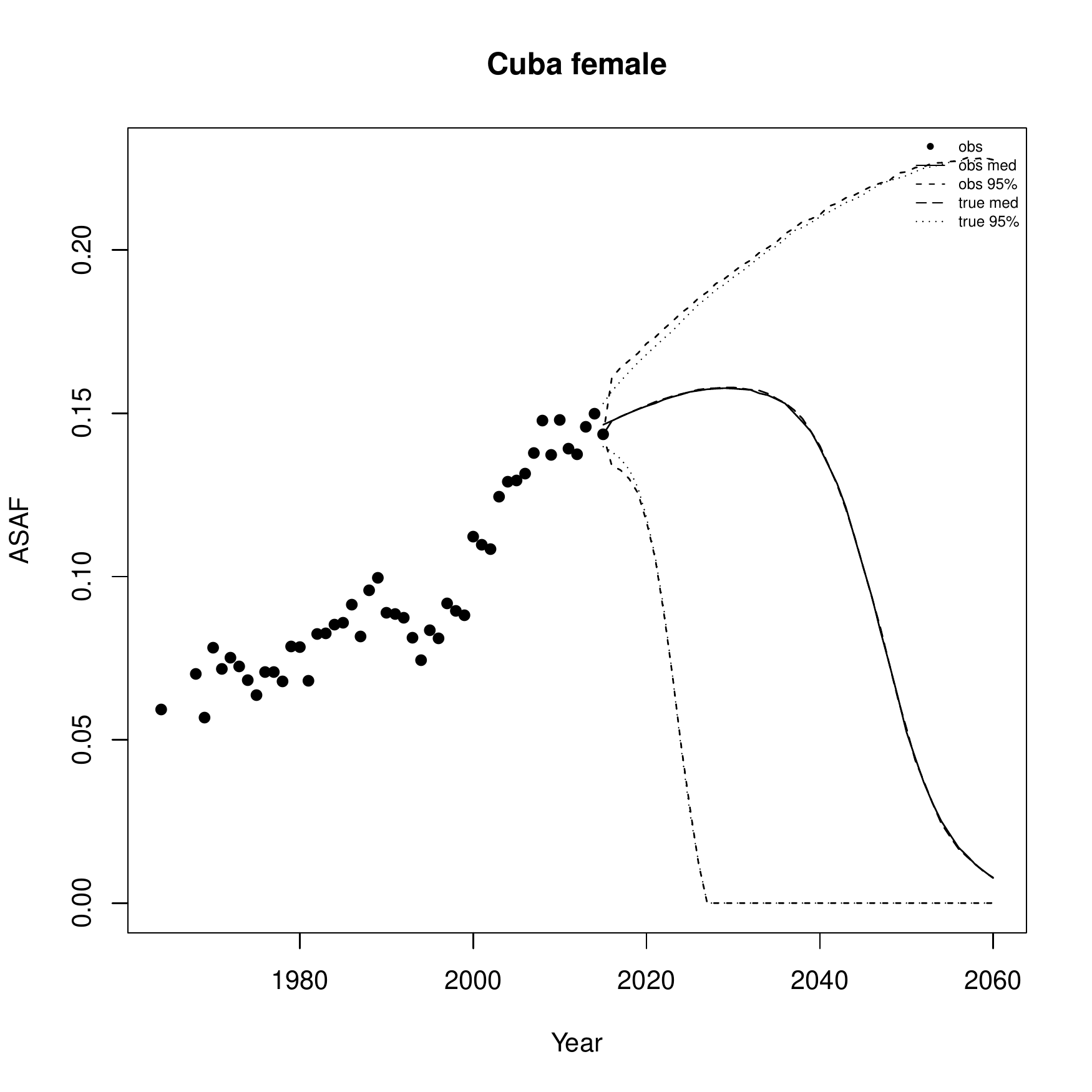}
		\includegraphics[scale=0.43]{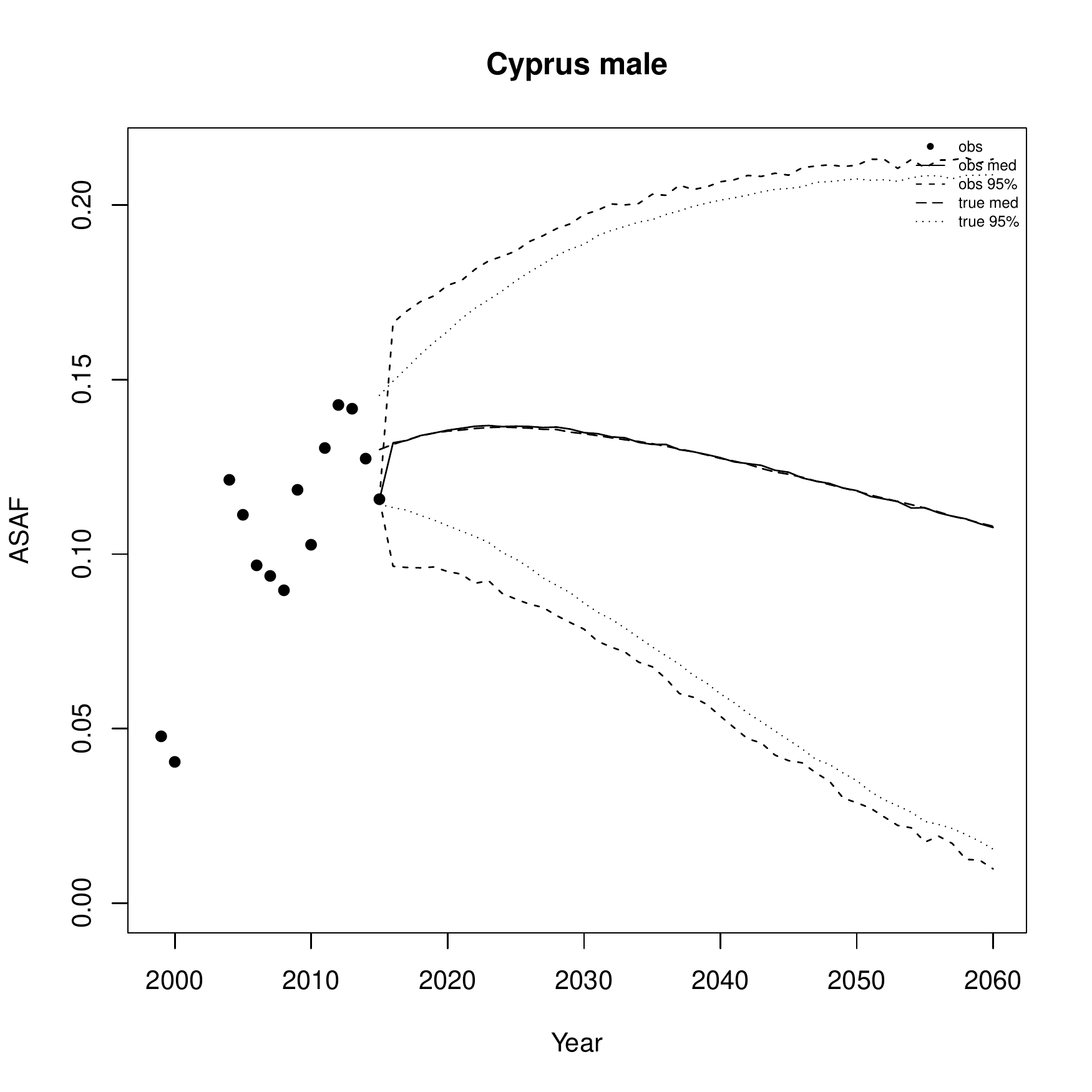}
		\includegraphics[scale=0.43]{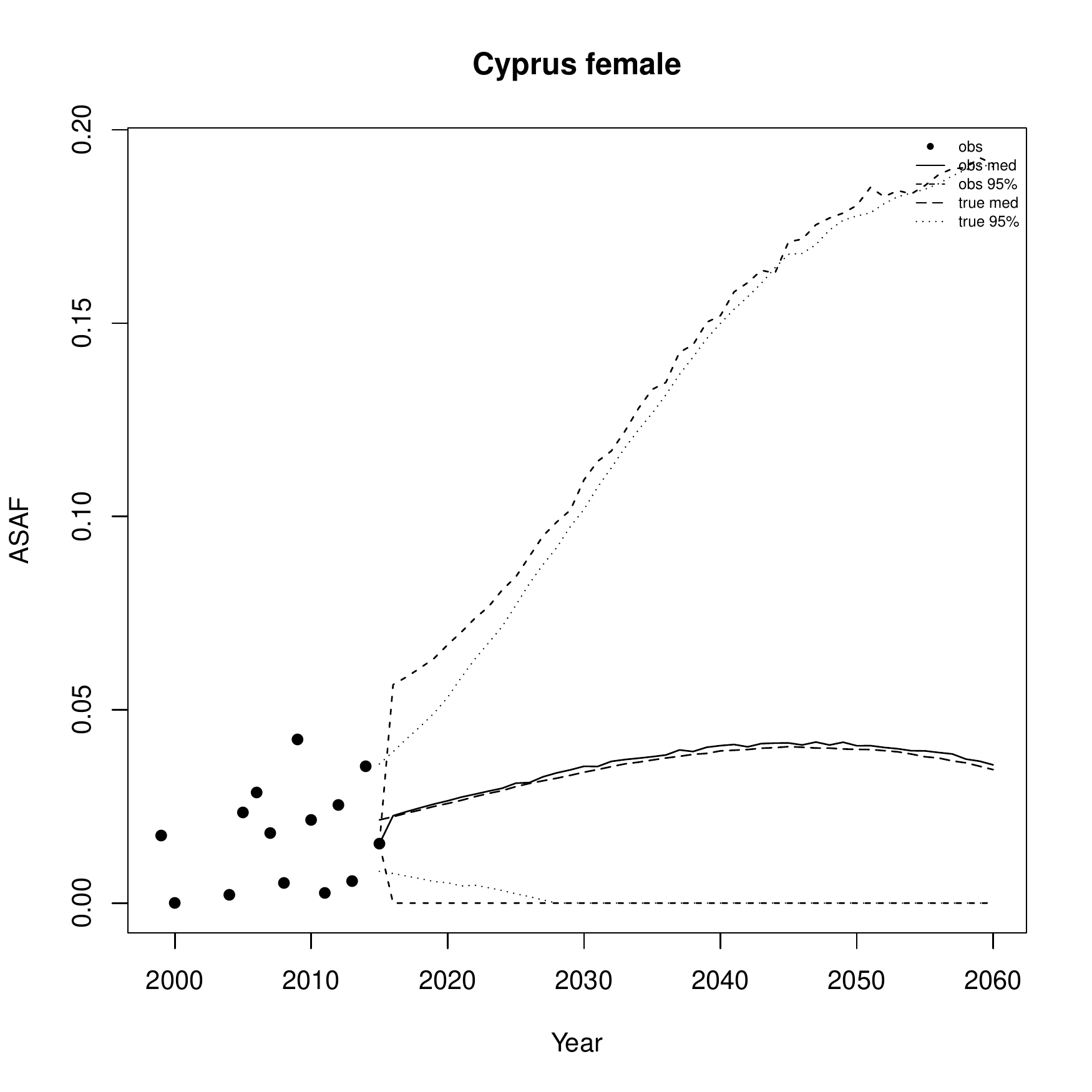}
	\end{center}
\end{figure}

\begin{figure}[H]
	\begin{center}
		\includegraphics[scale=0.43]{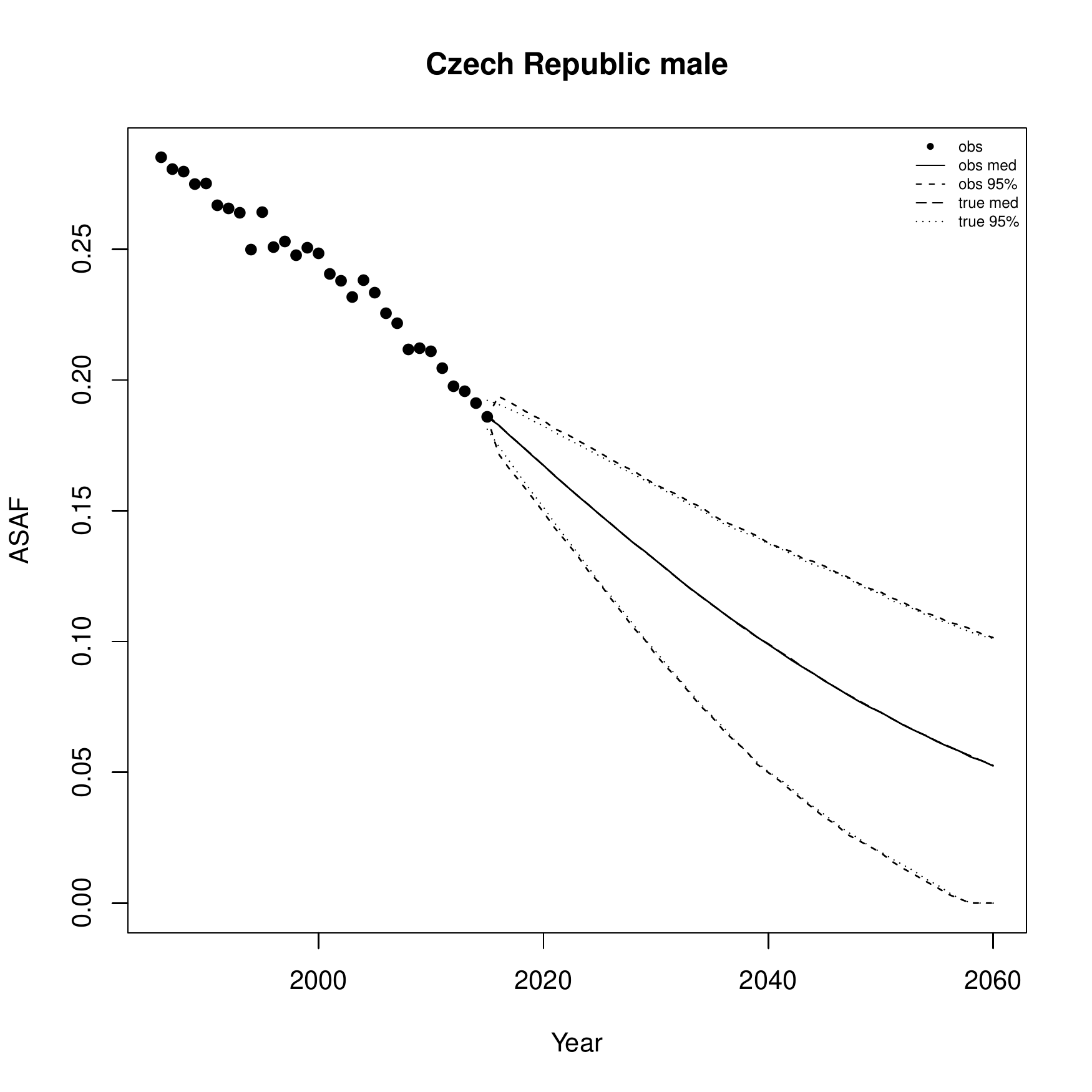}
		\includegraphics[scale=0.43]{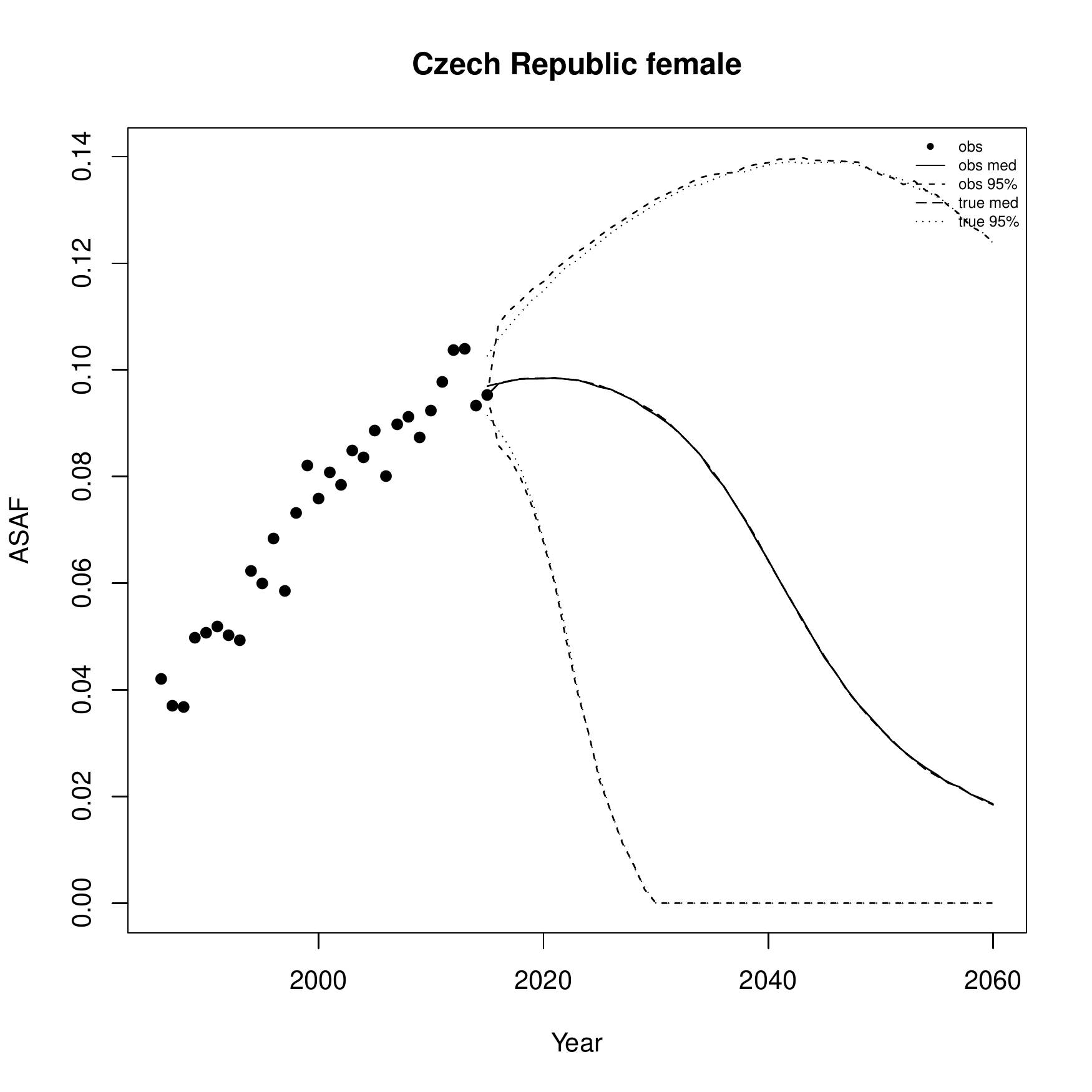}
		\includegraphics[scale=0.43]{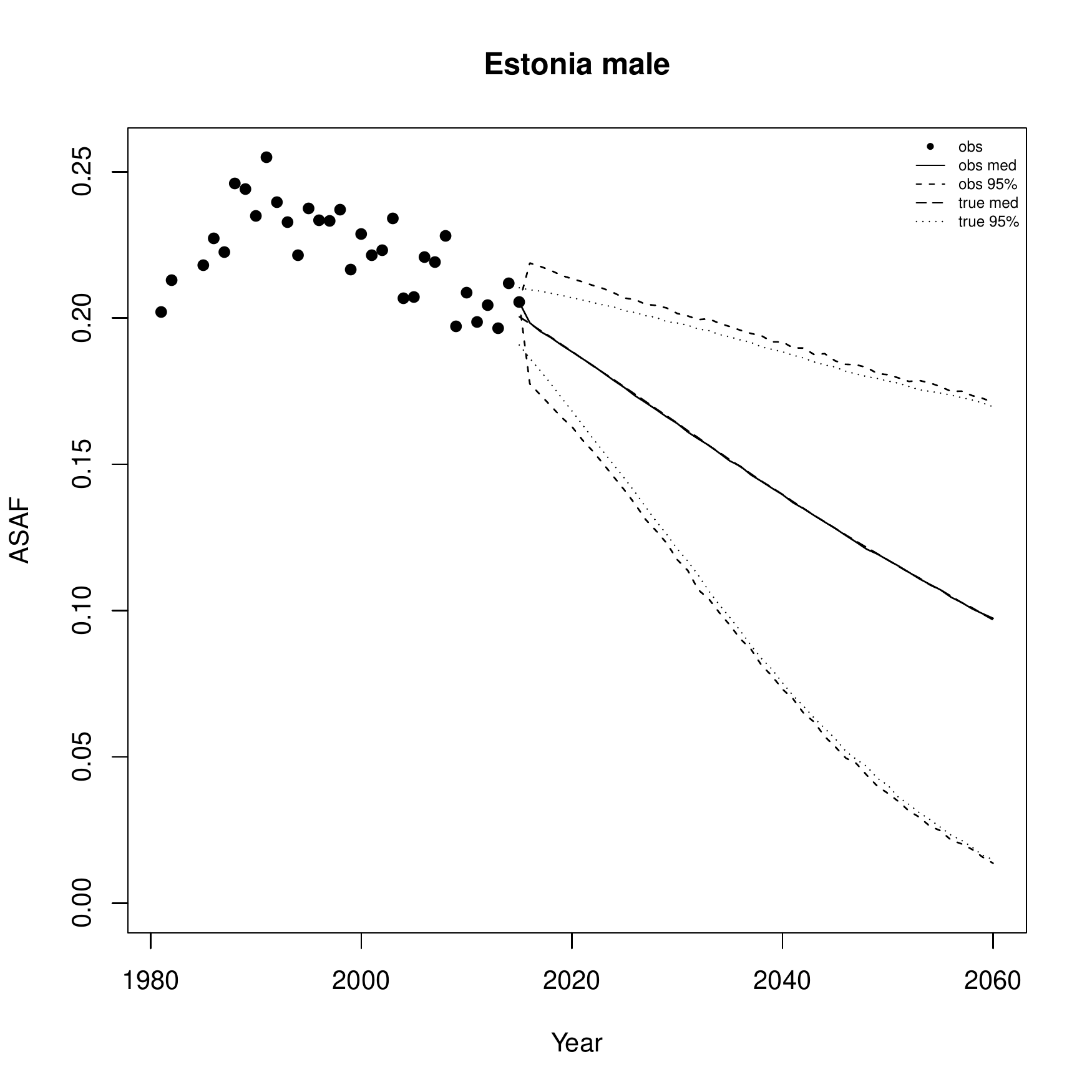}
		\includegraphics[scale=0.43]{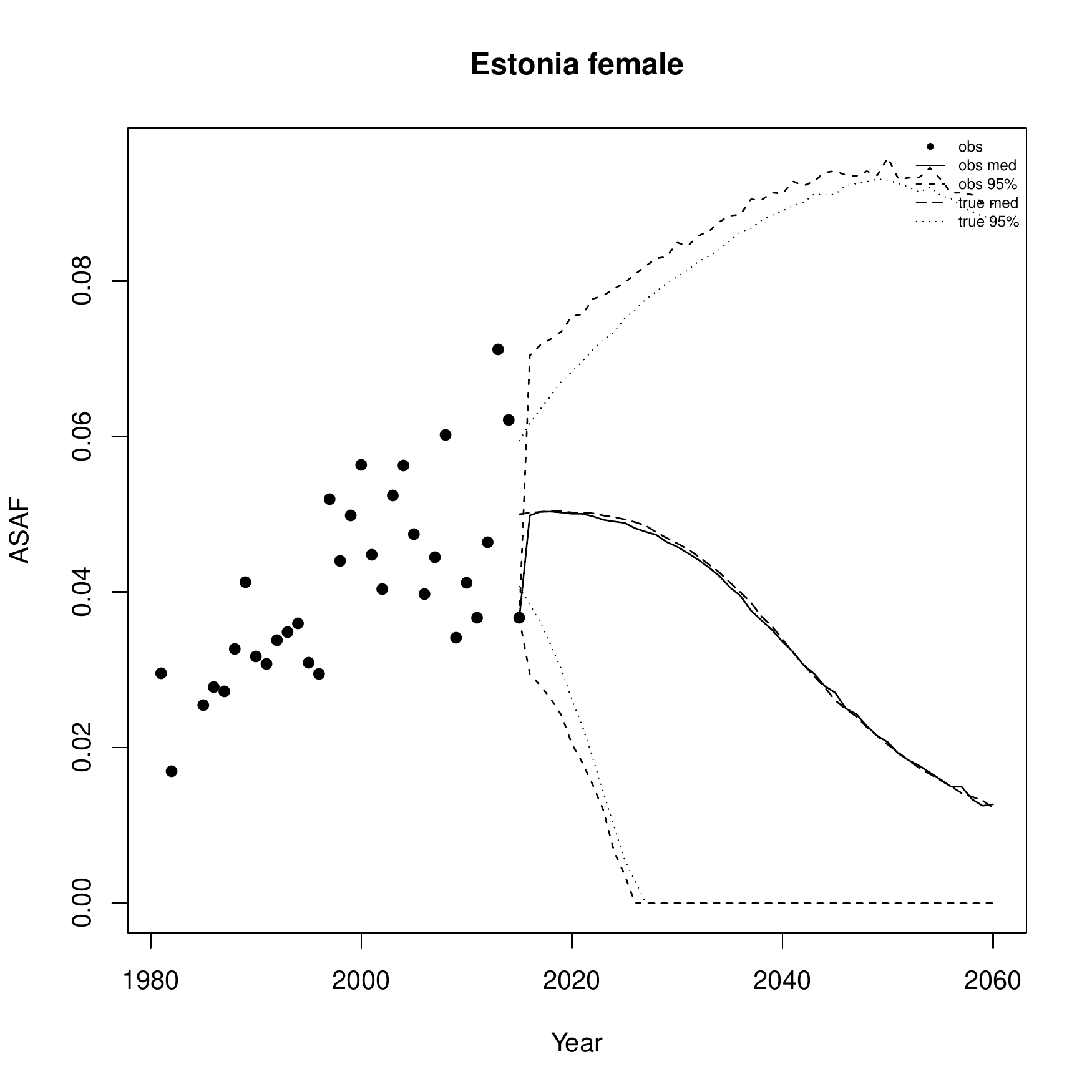}
		\includegraphics[scale=0.43]{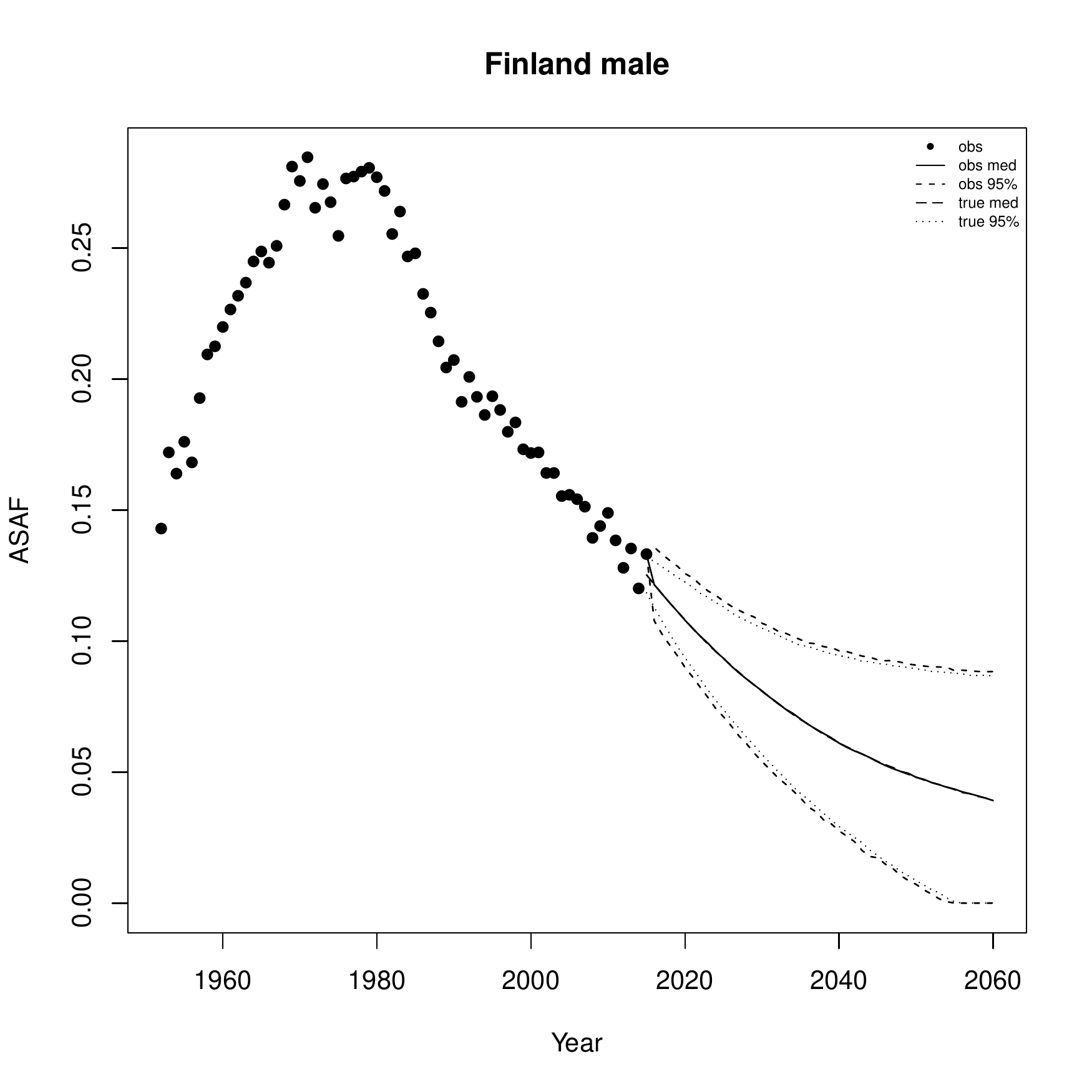}
		\includegraphics[scale=0.43]{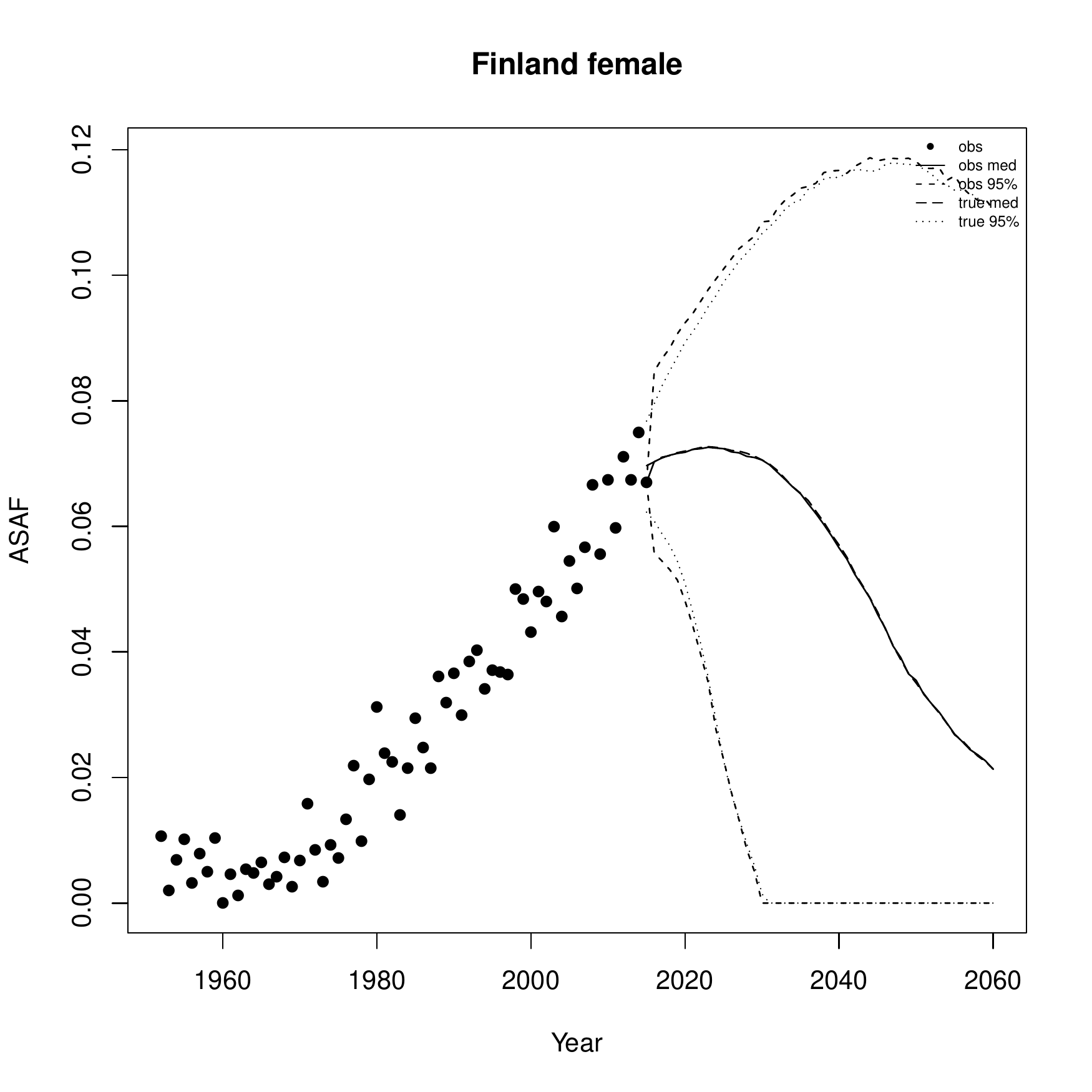}
	\end{center}
\end{figure}

\begin{figure}[H]
	\begin{center}
		\includegraphics[scale=0.43]{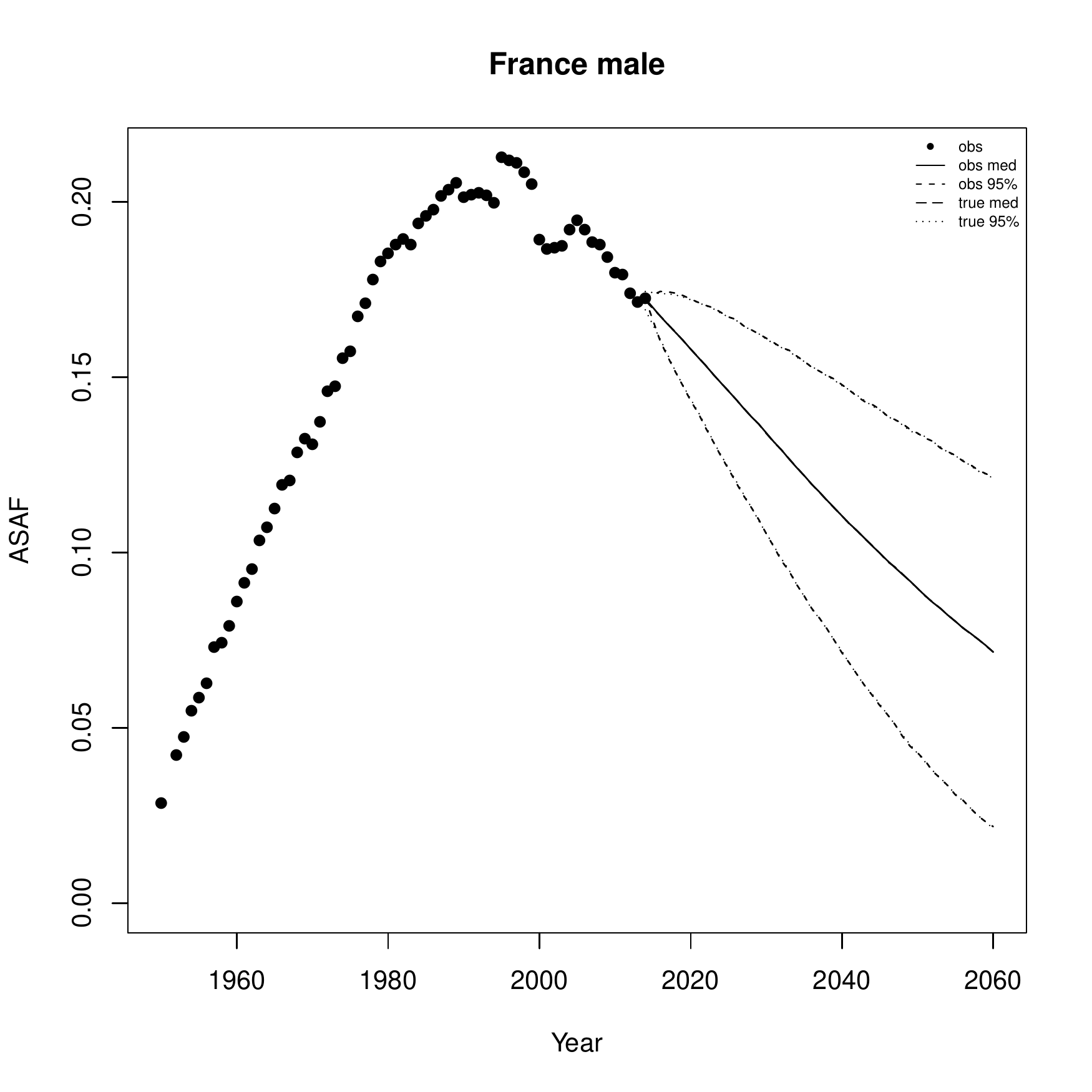}
		\includegraphics[scale=0.43]{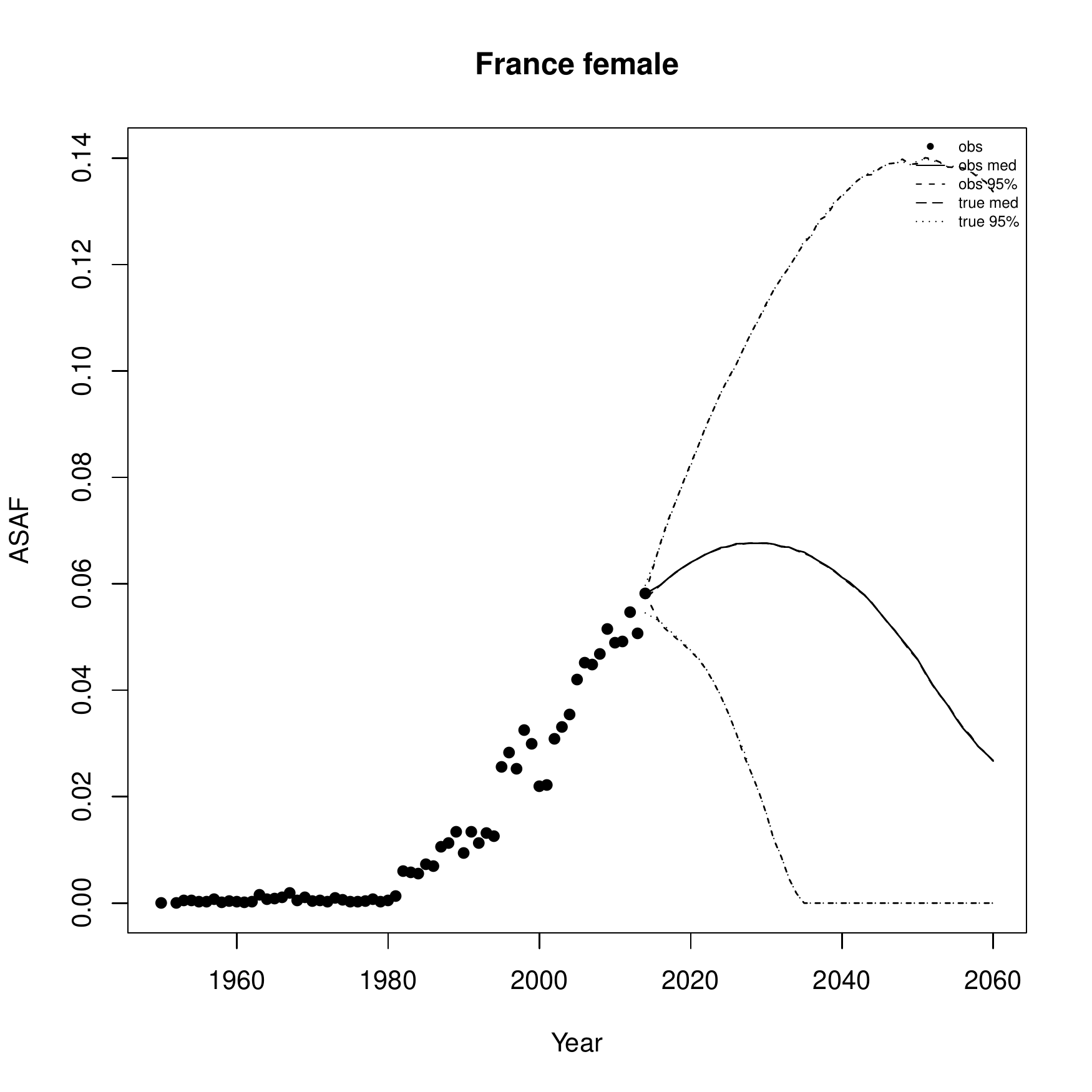}
		\includegraphics[scale=0.43]{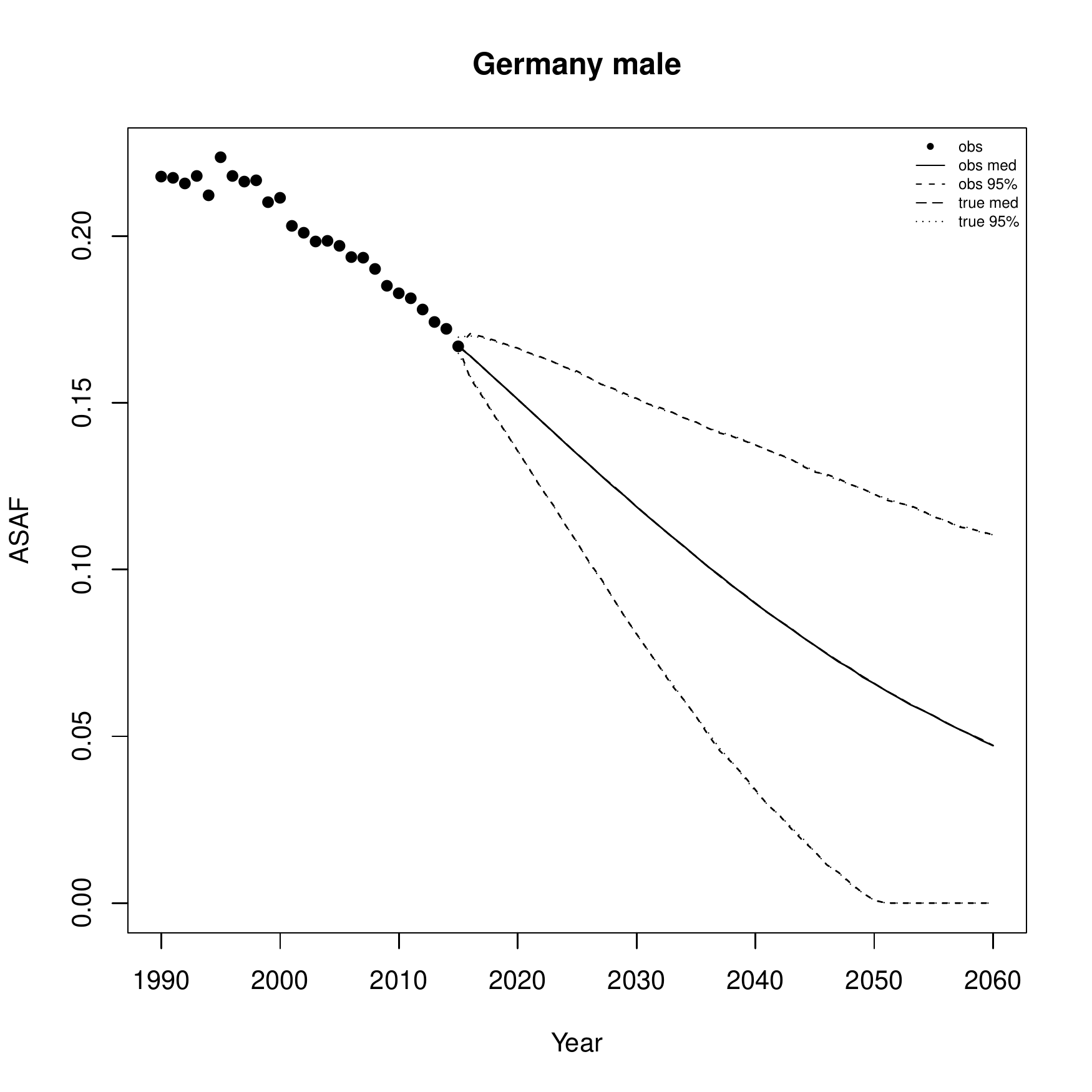}
		\includegraphics[scale=0.43]{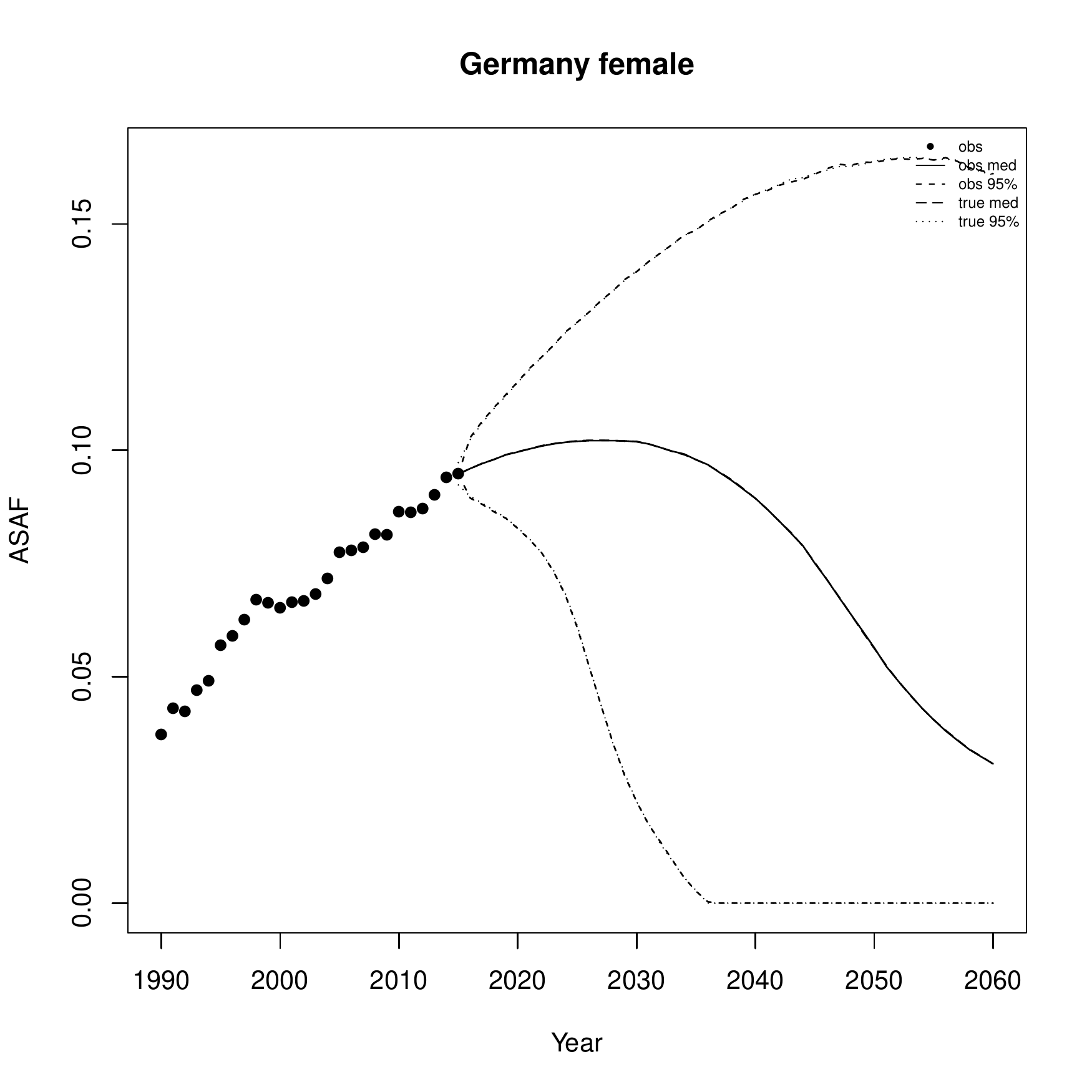}
		\includegraphics[scale=0.43]{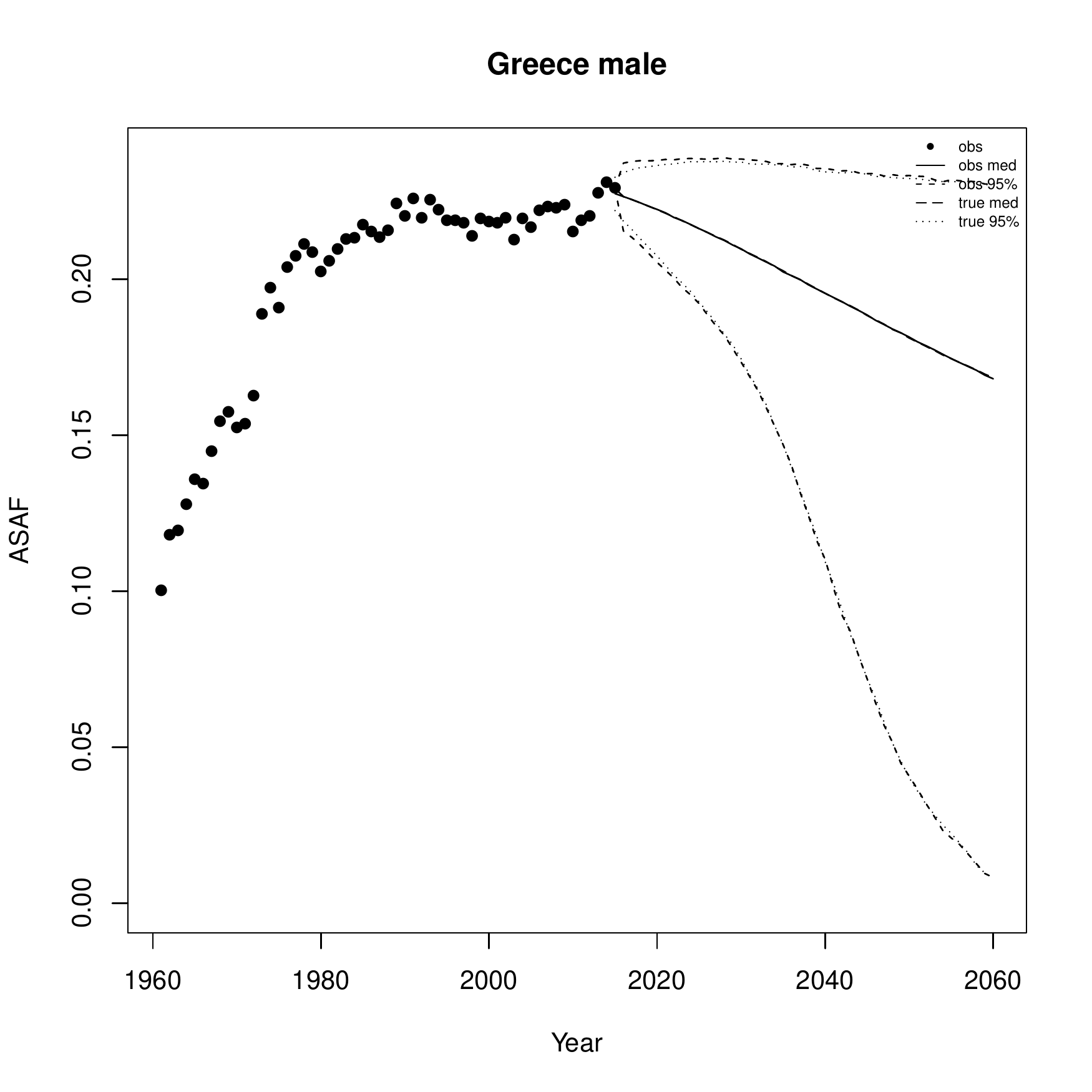}
		\includegraphics[scale=0.43]{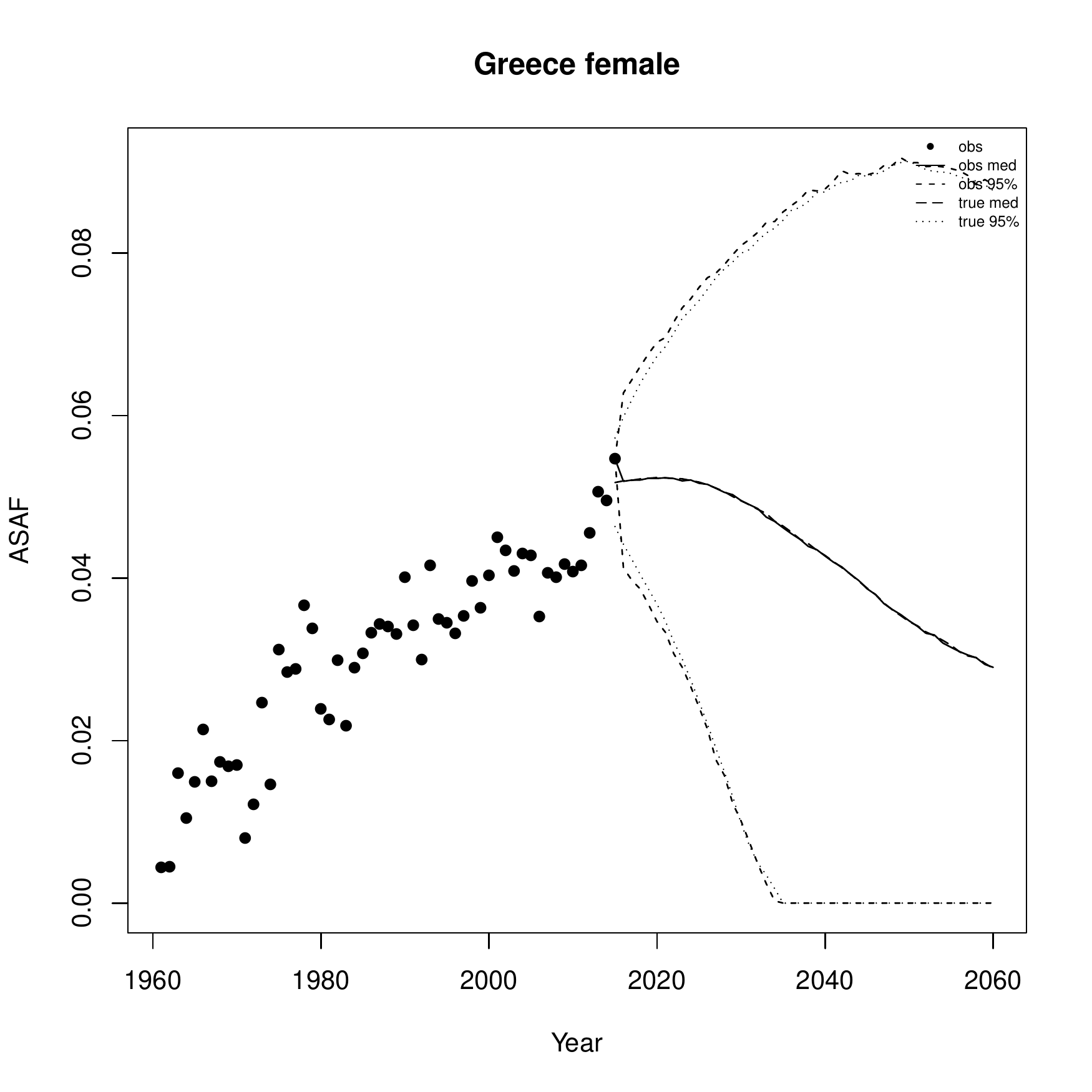}
	\end{center}
\end{figure}

\begin{figure}[H]
	\begin{center}
		\includegraphics[scale=0.43]{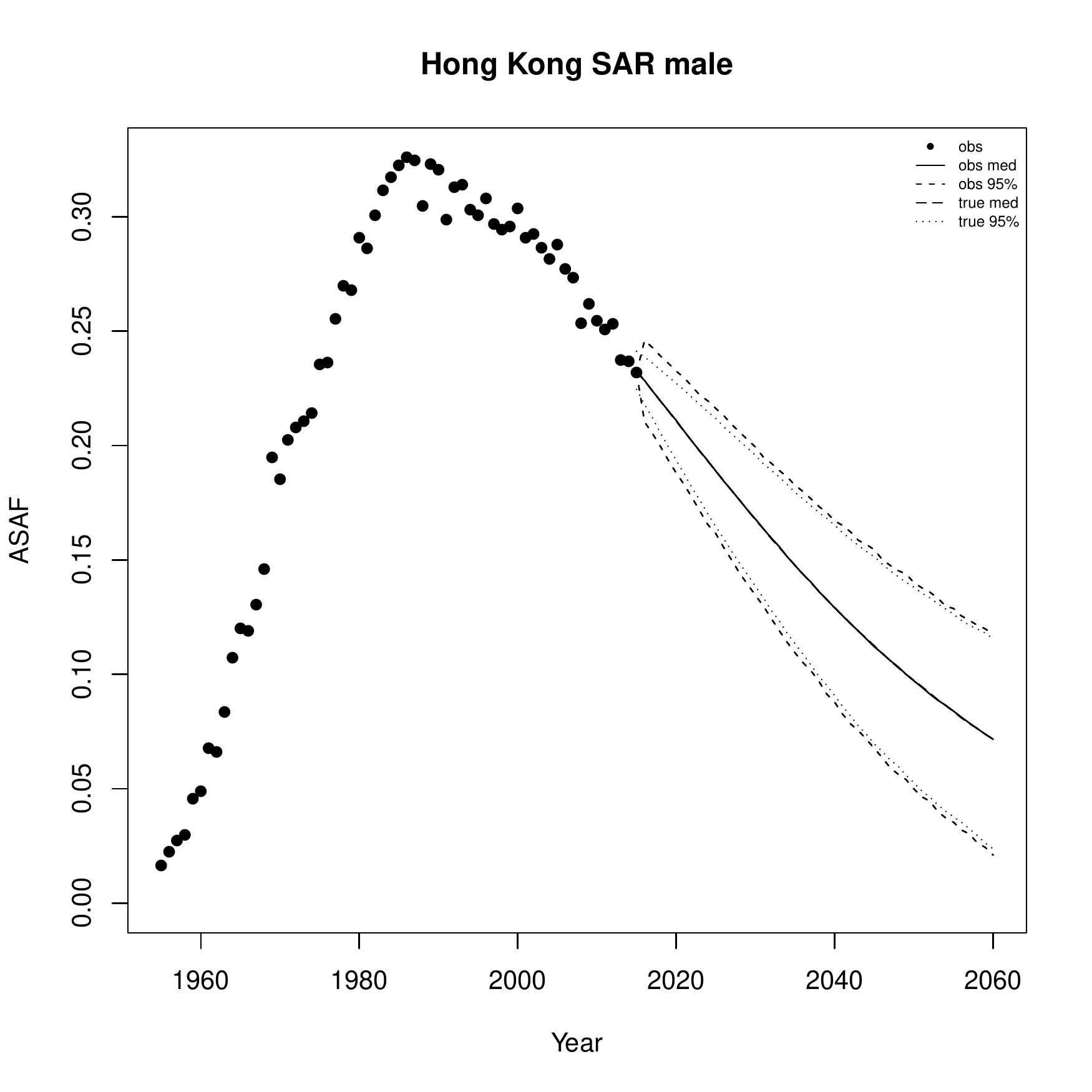}
		\includegraphics[scale=0.43]{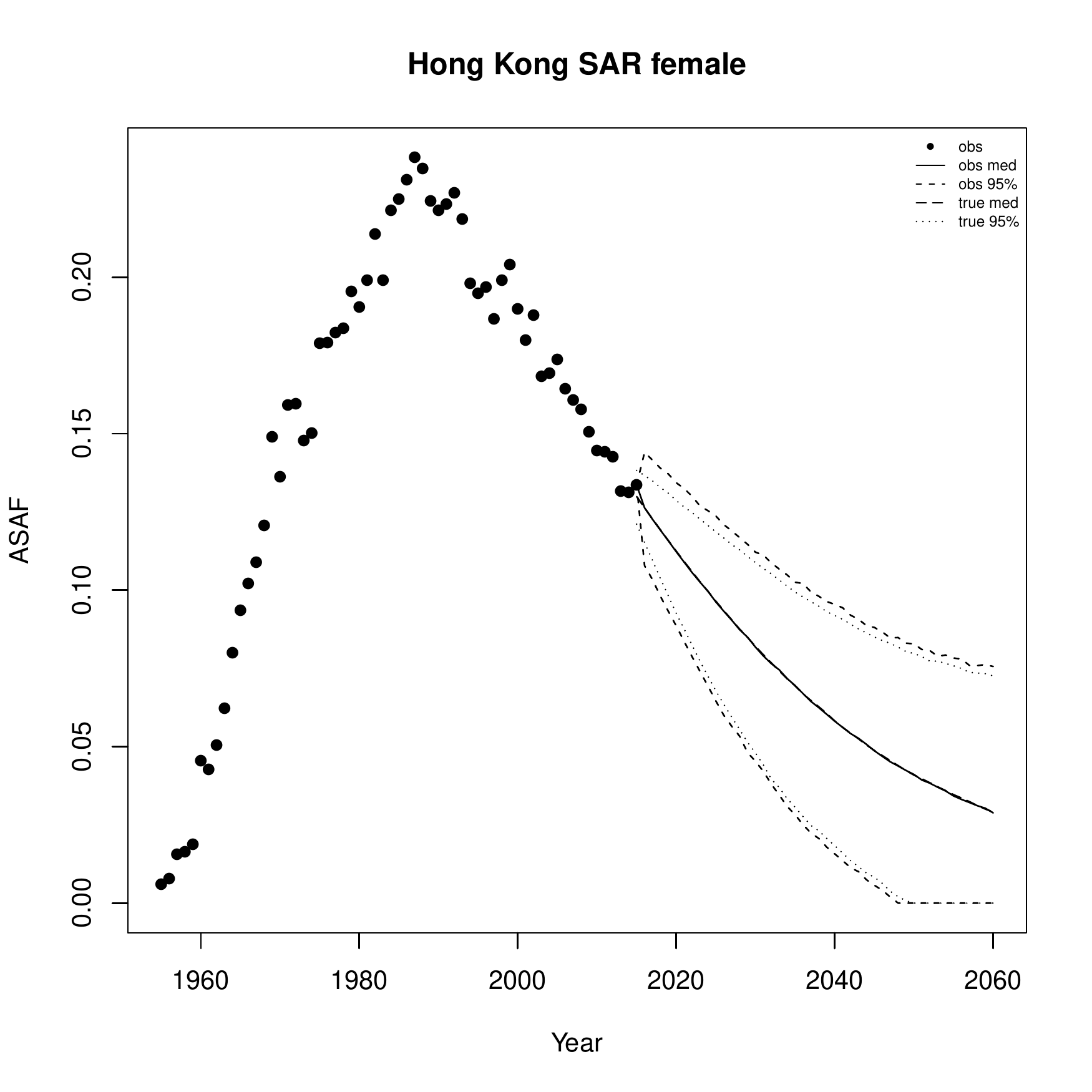}
		\includegraphics[scale=0.43]{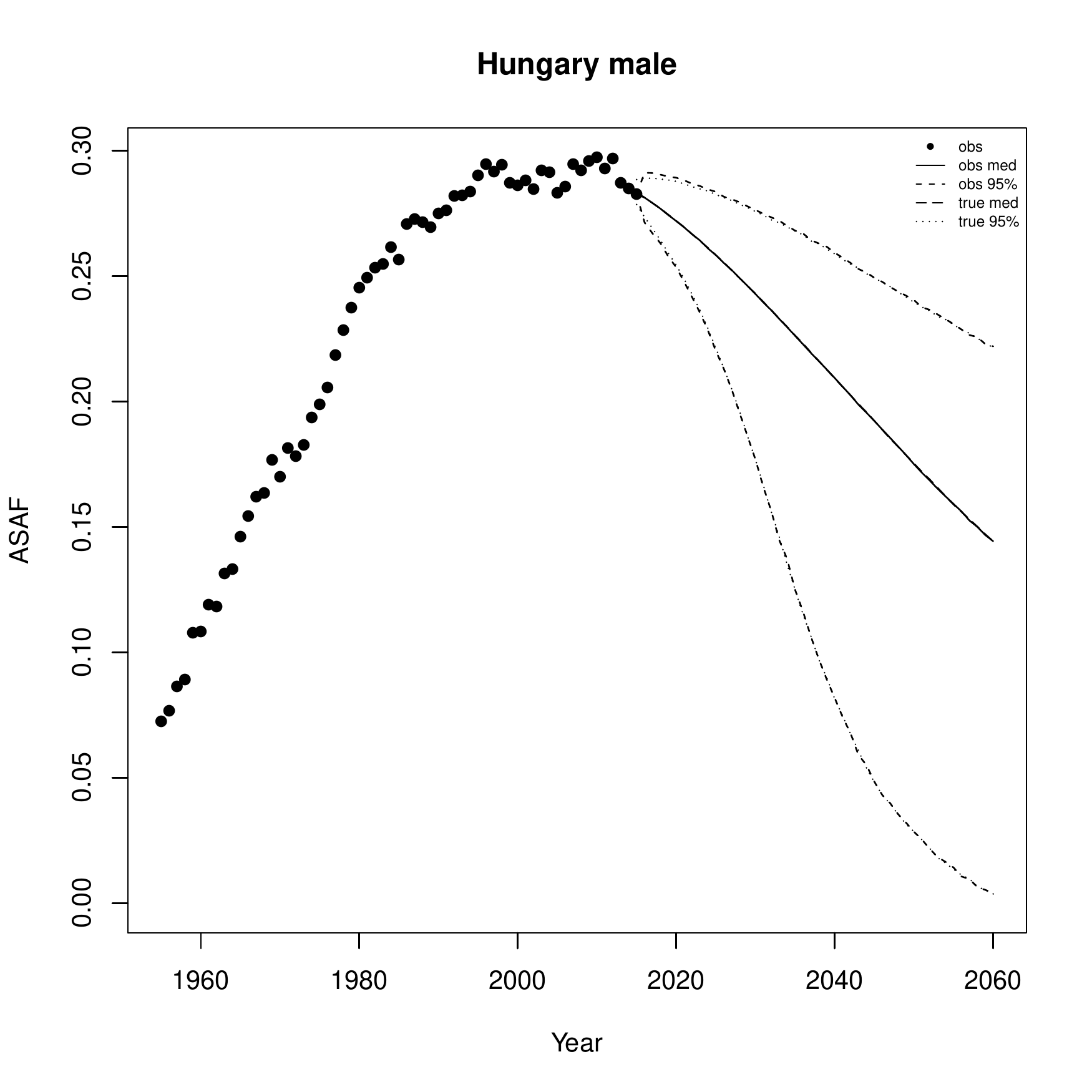}
		\includegraphics[scale=0.43]{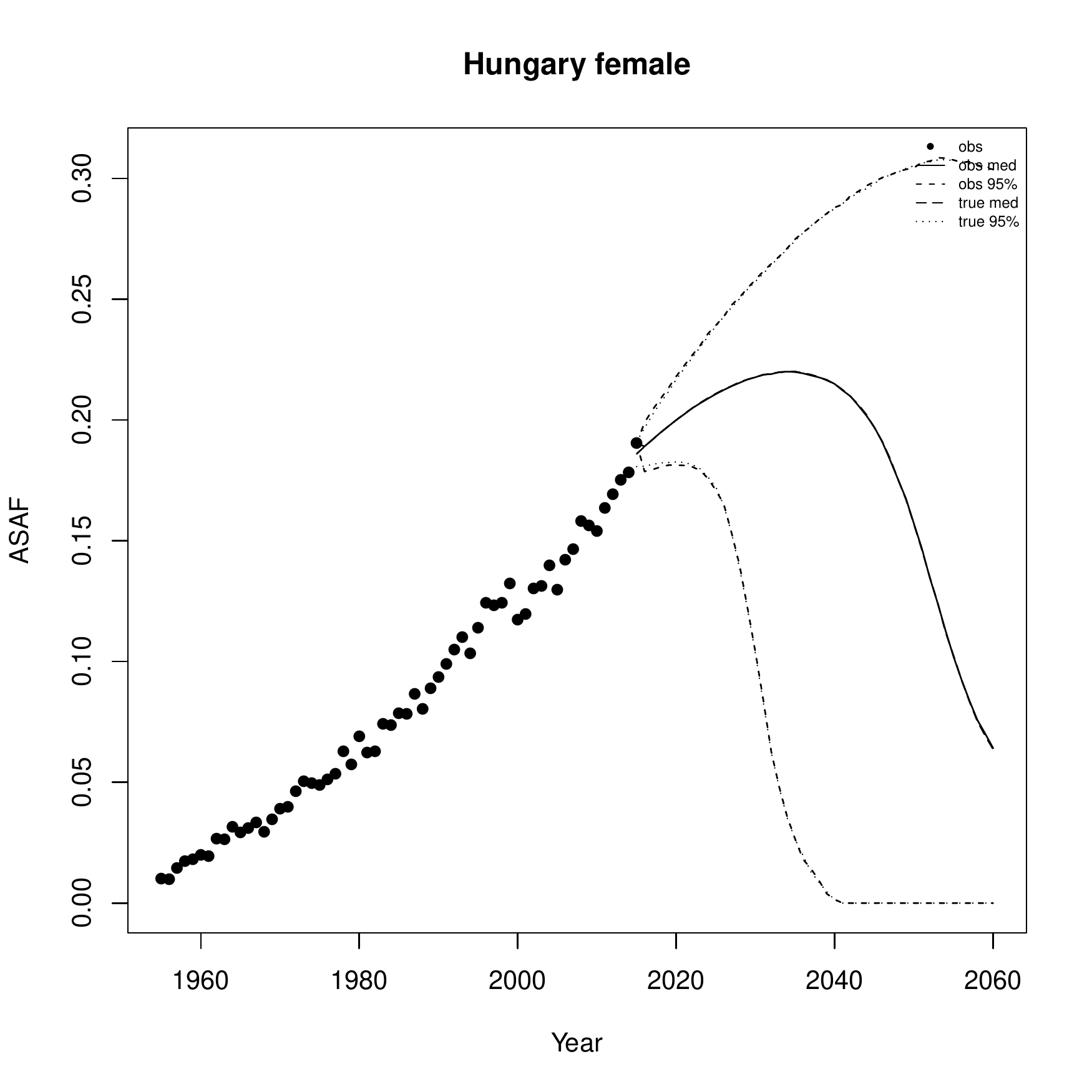}
		\includegraphics[scale=0.43]{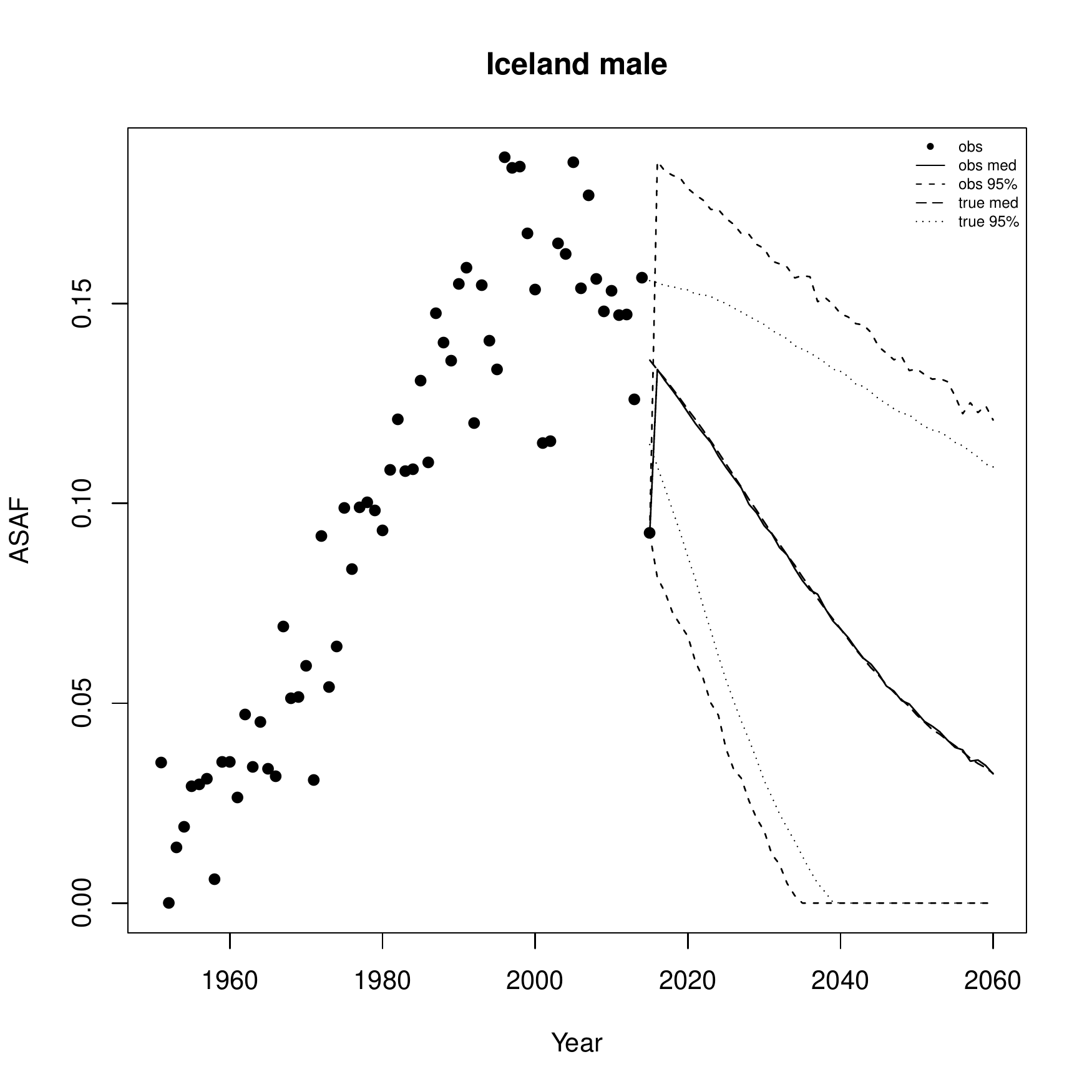}
		\includegraphics[scale=0.43]{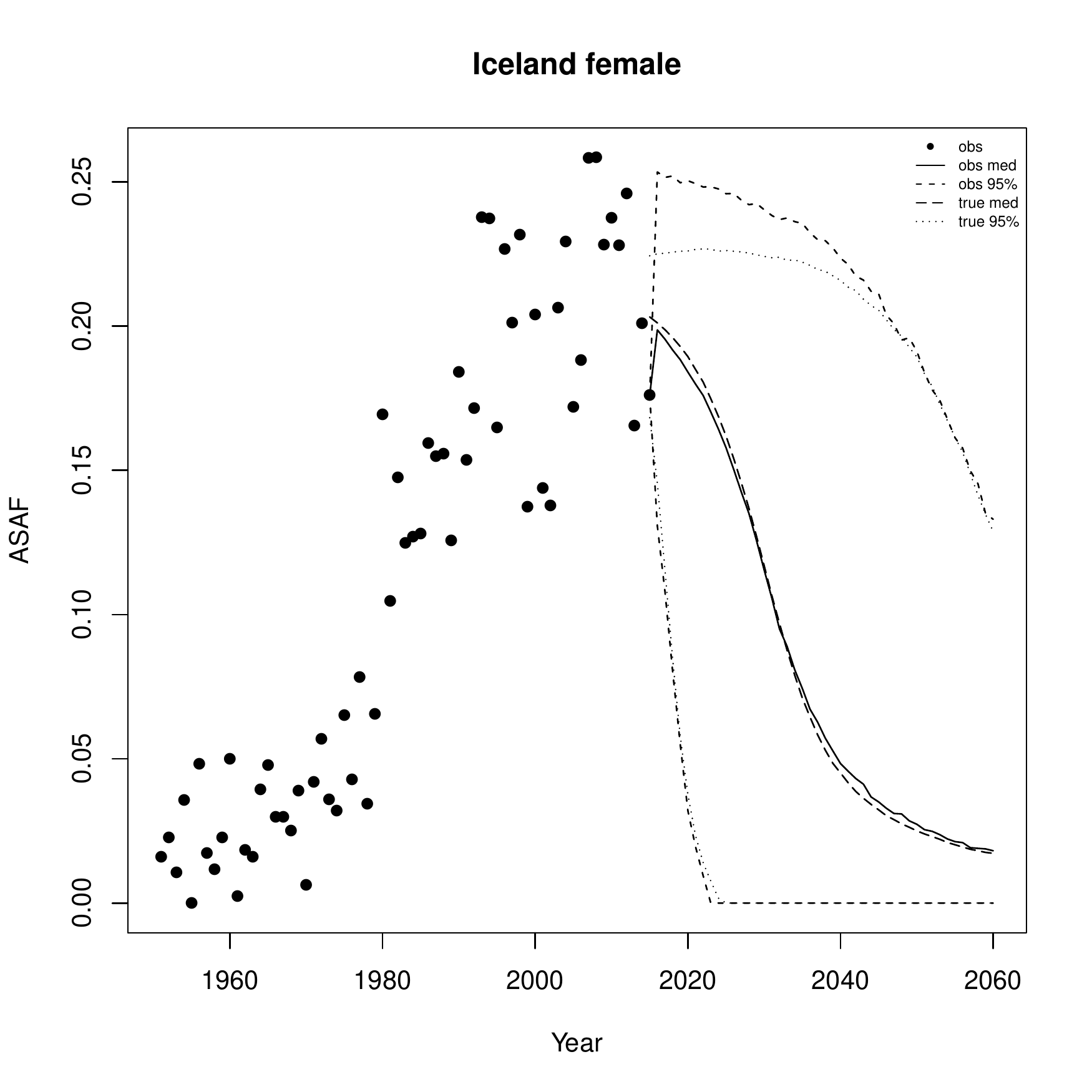}
	\end{center}
\end{figure}

\begin{figure}[H]
	\begin{center}
		\includegraphics[scale=0.43]{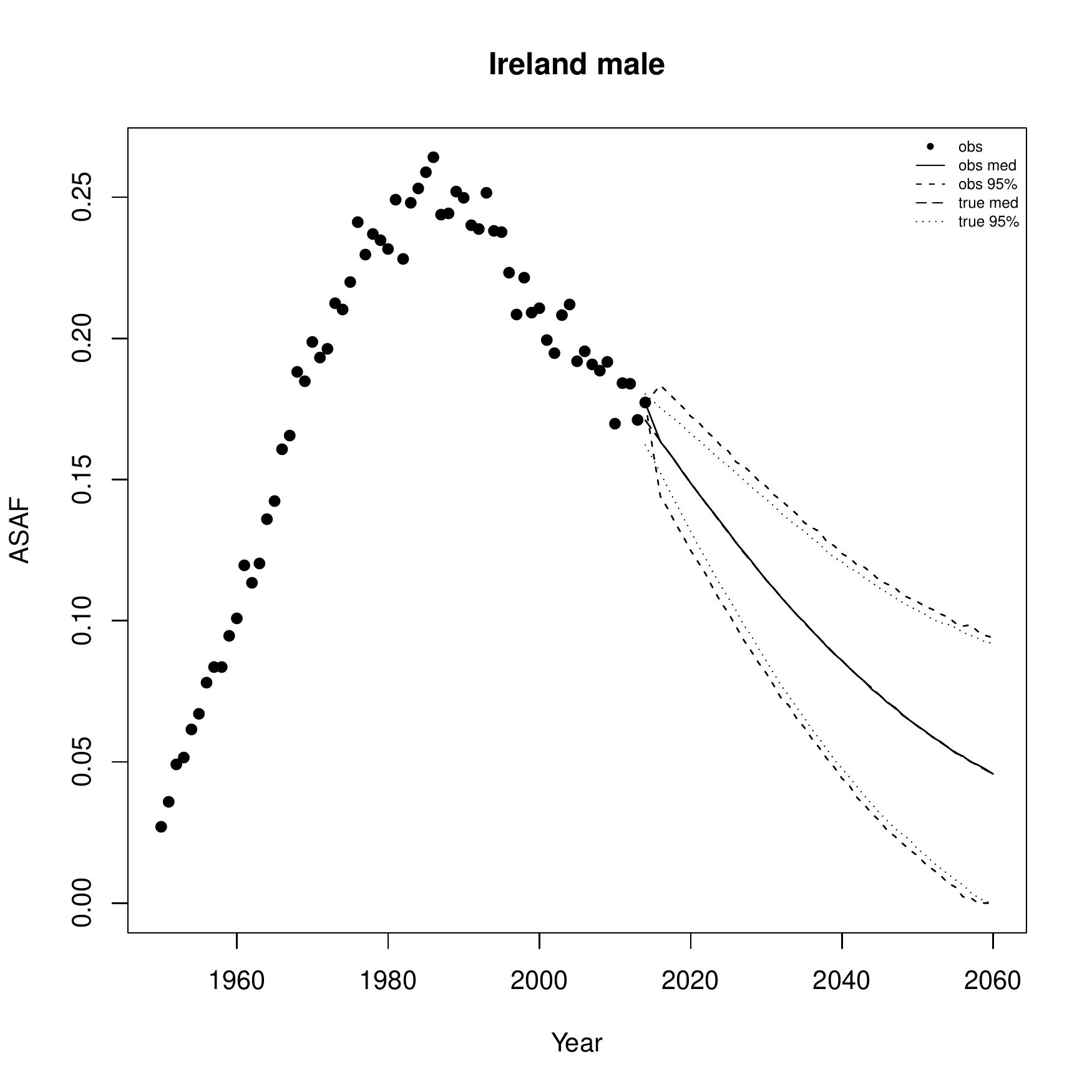}
		\includegraphics[scale=0.43]{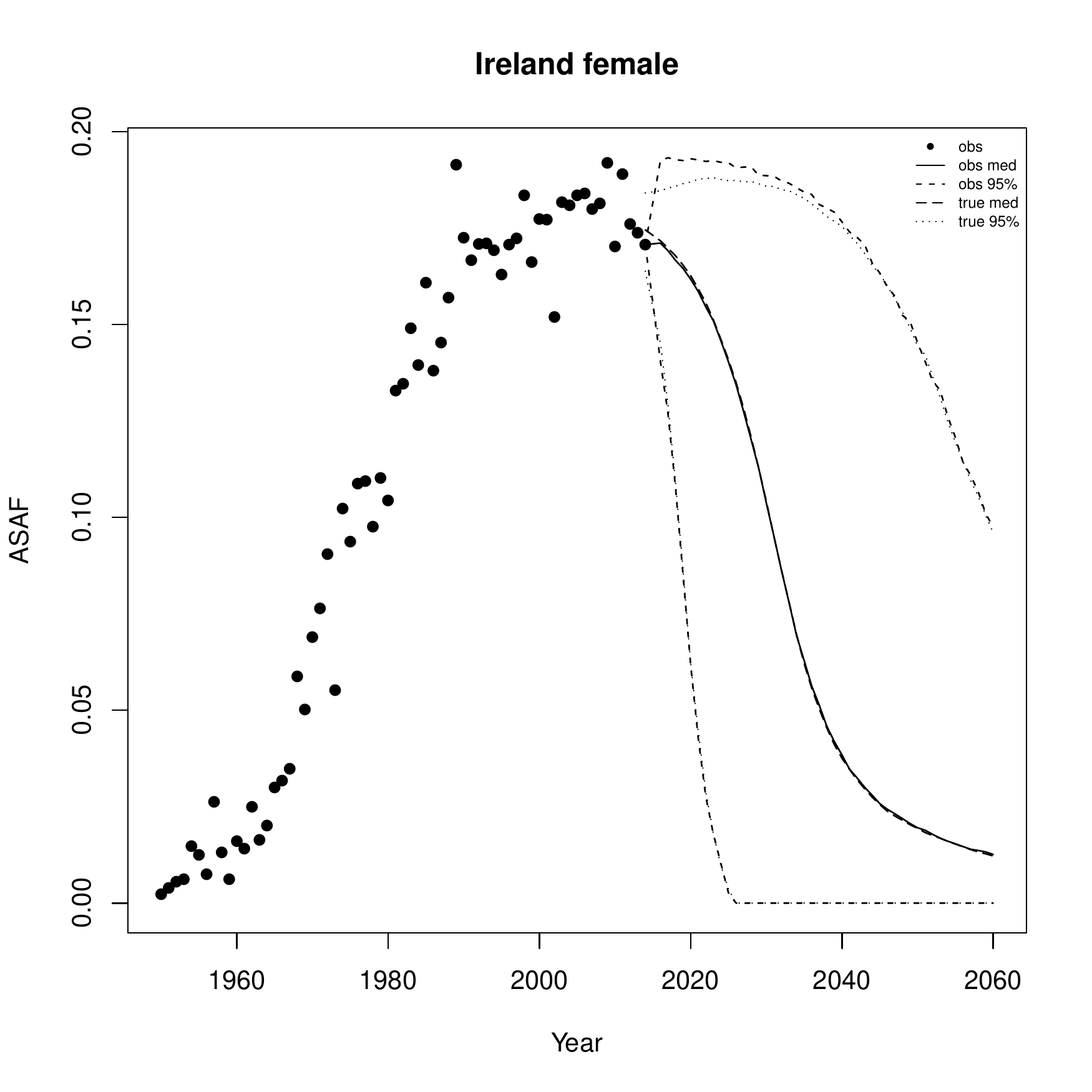}
		\includegraphics[scale=0.43]{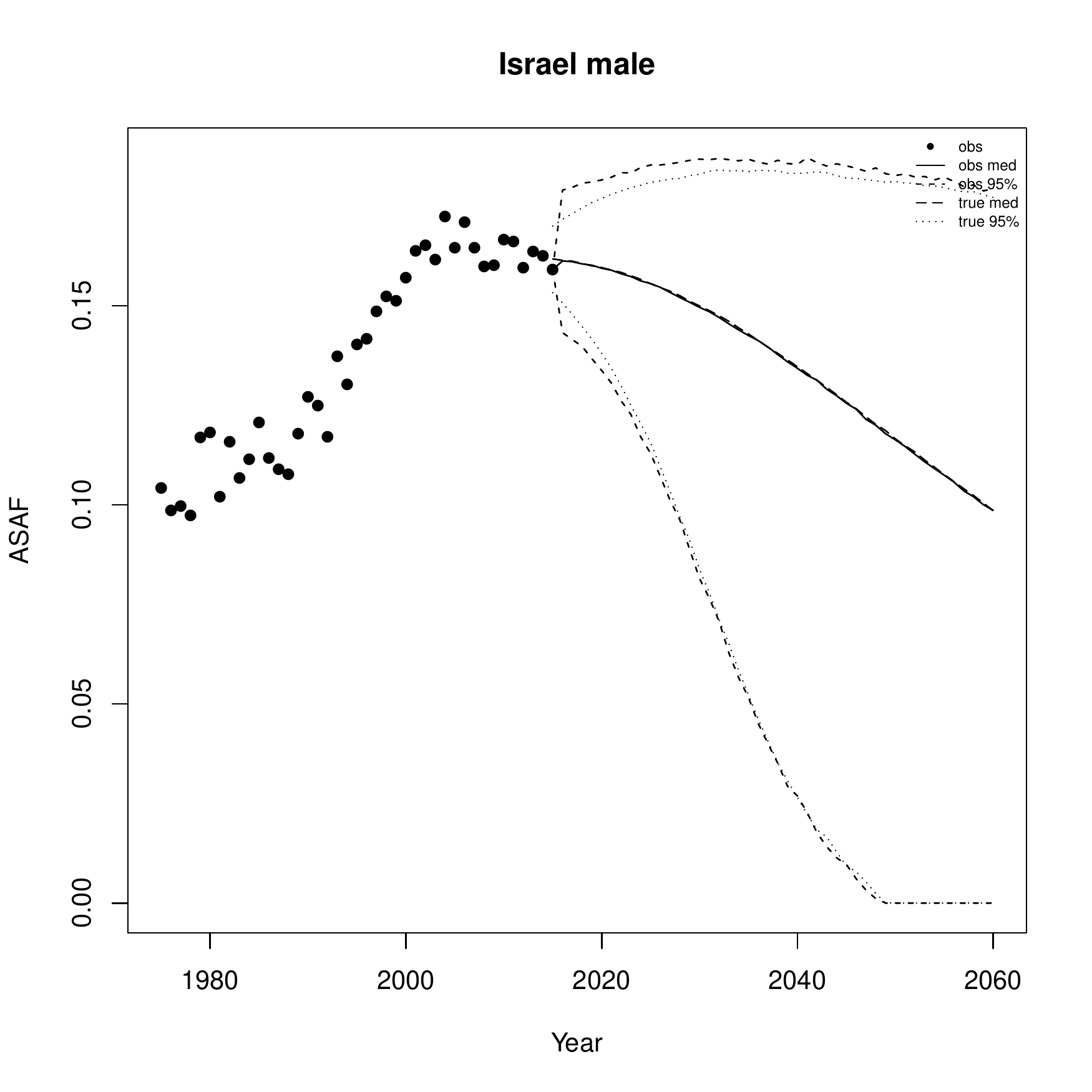}
		\includegraphics[scale=0.43]{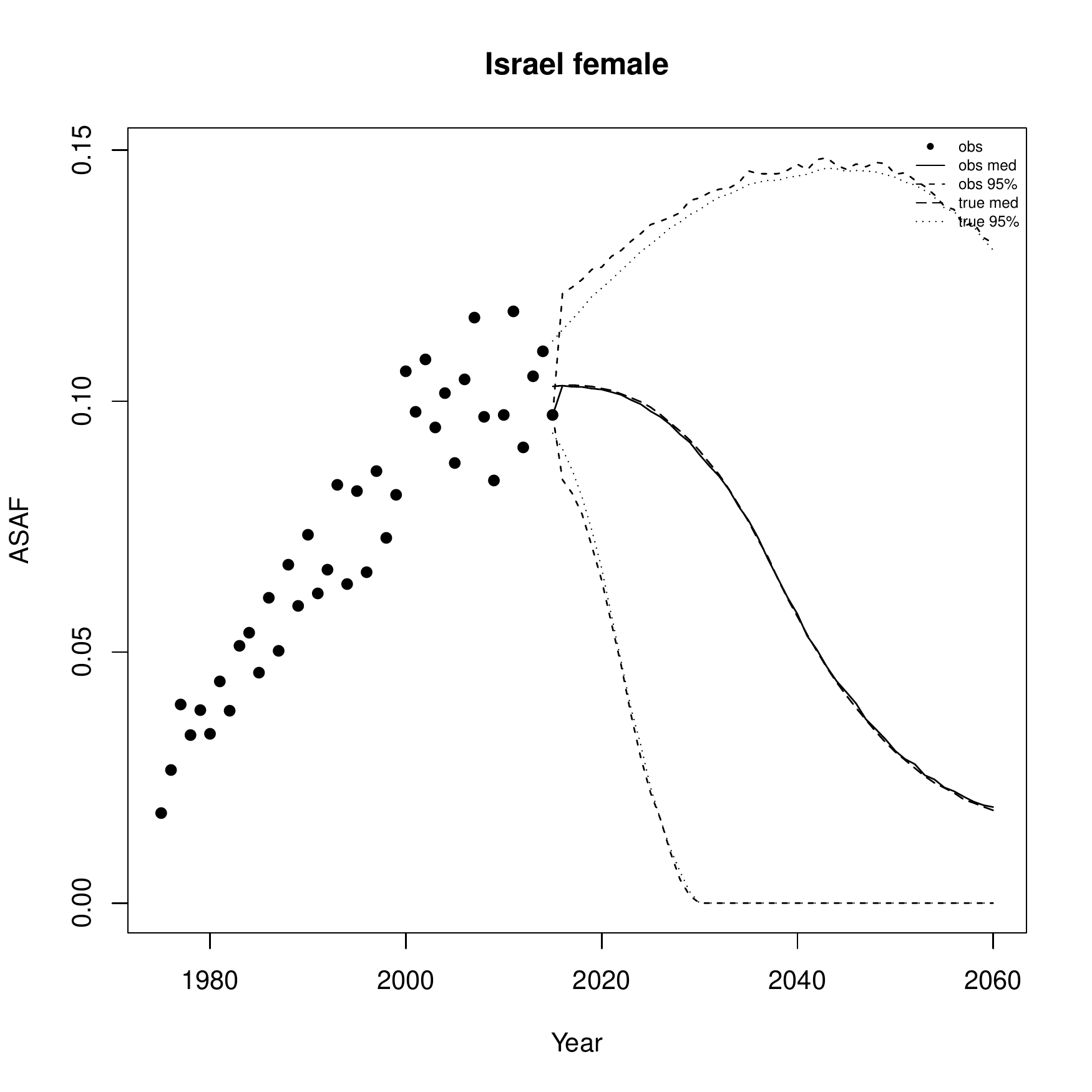}
		\includegraphics[scale=0.43]{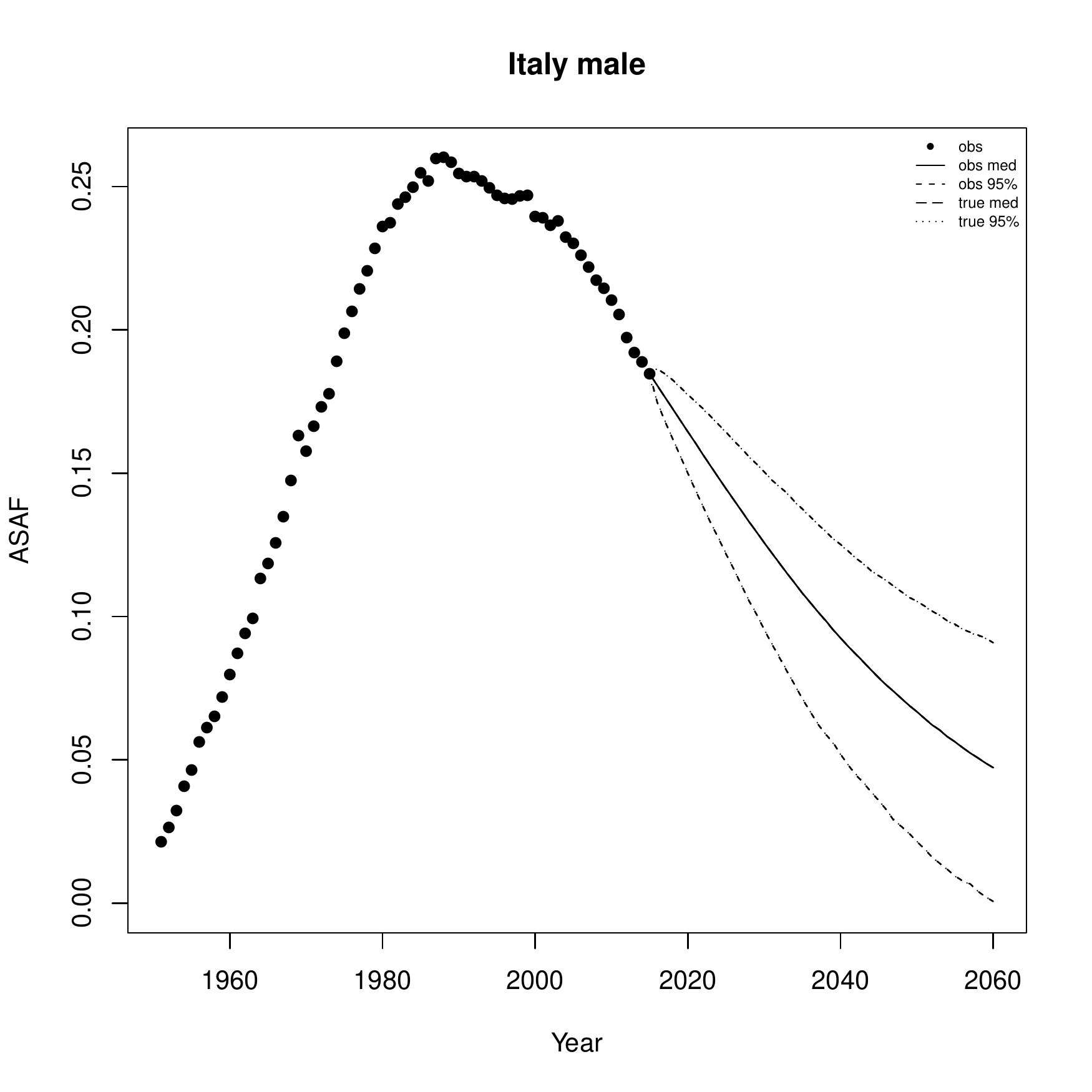}
		\includegraphics[scale=0.43]{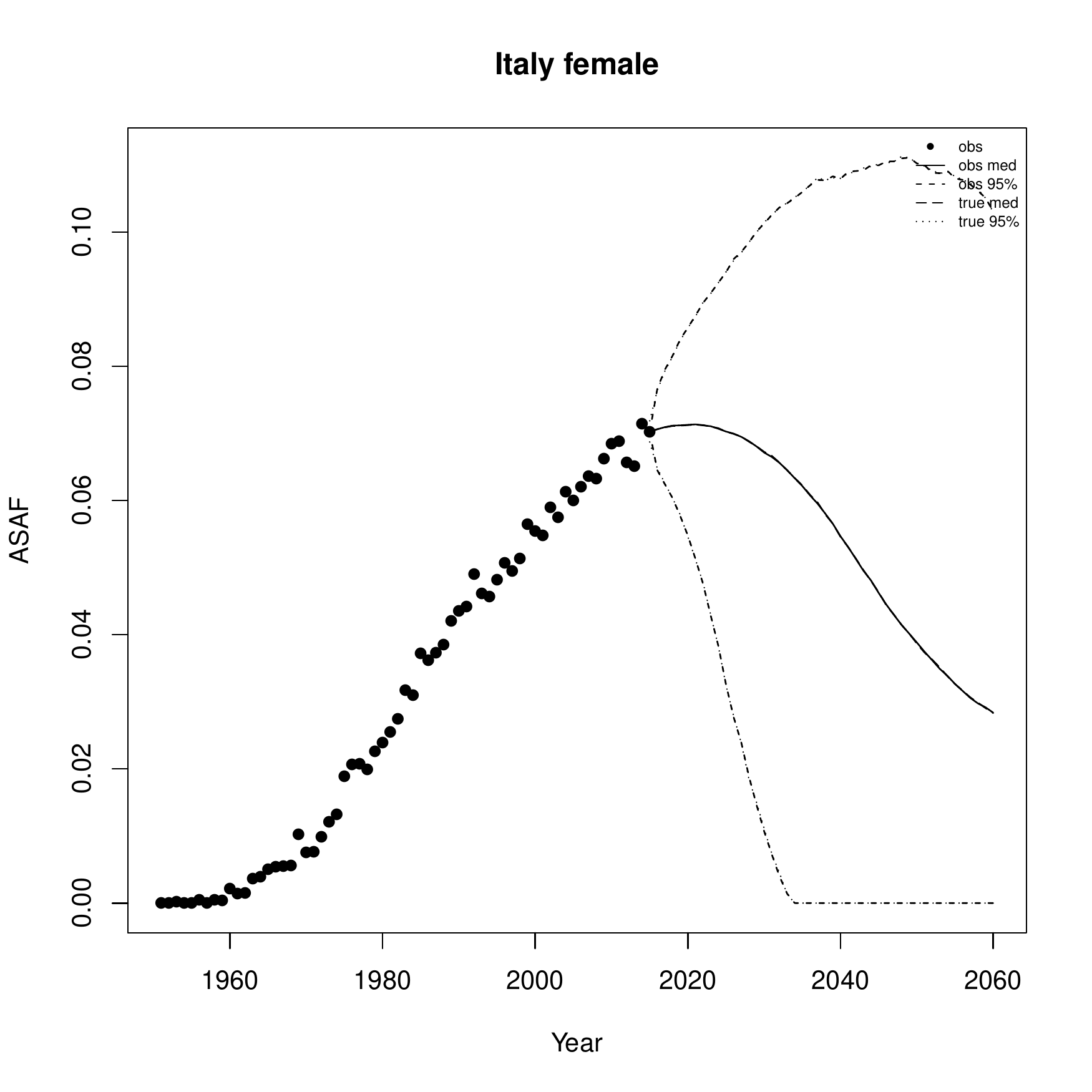}
	\end{center}
\end{figure}

\begin{figure}[H]
	\begin{center}
		\includegraphics[scale=0.43]{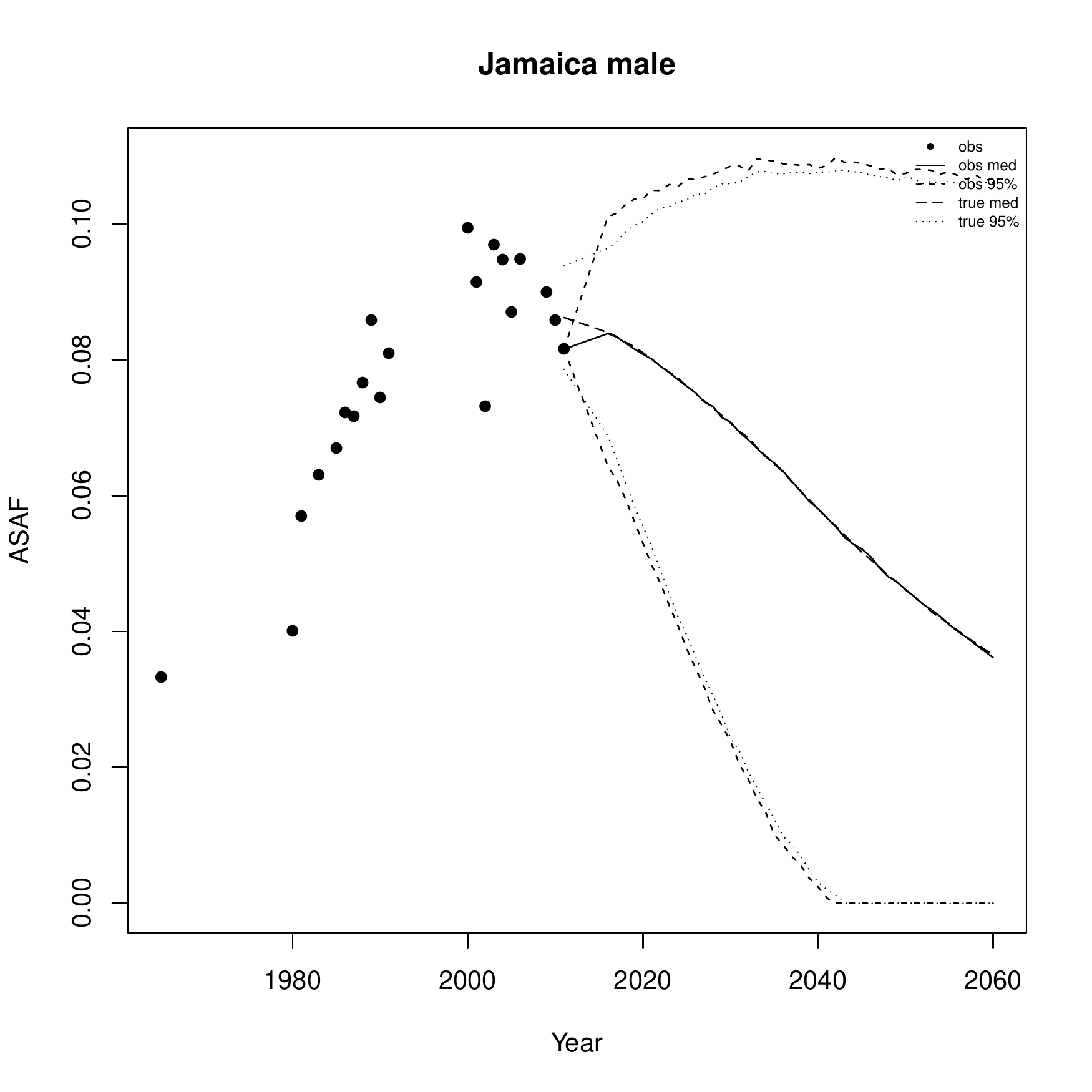}
		\includegraphics[scale=0.43]{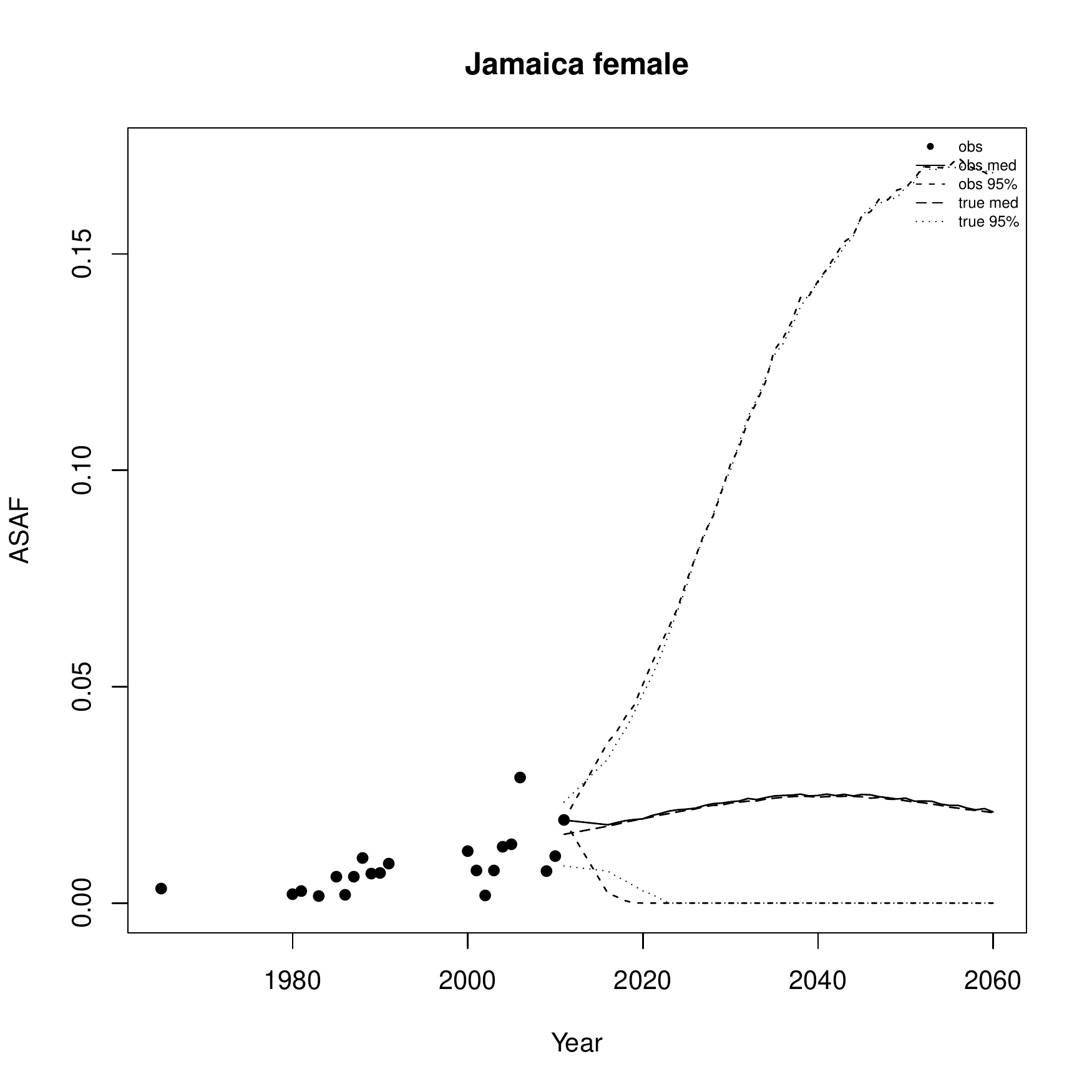}
		\includegraphics[scale=0.43]{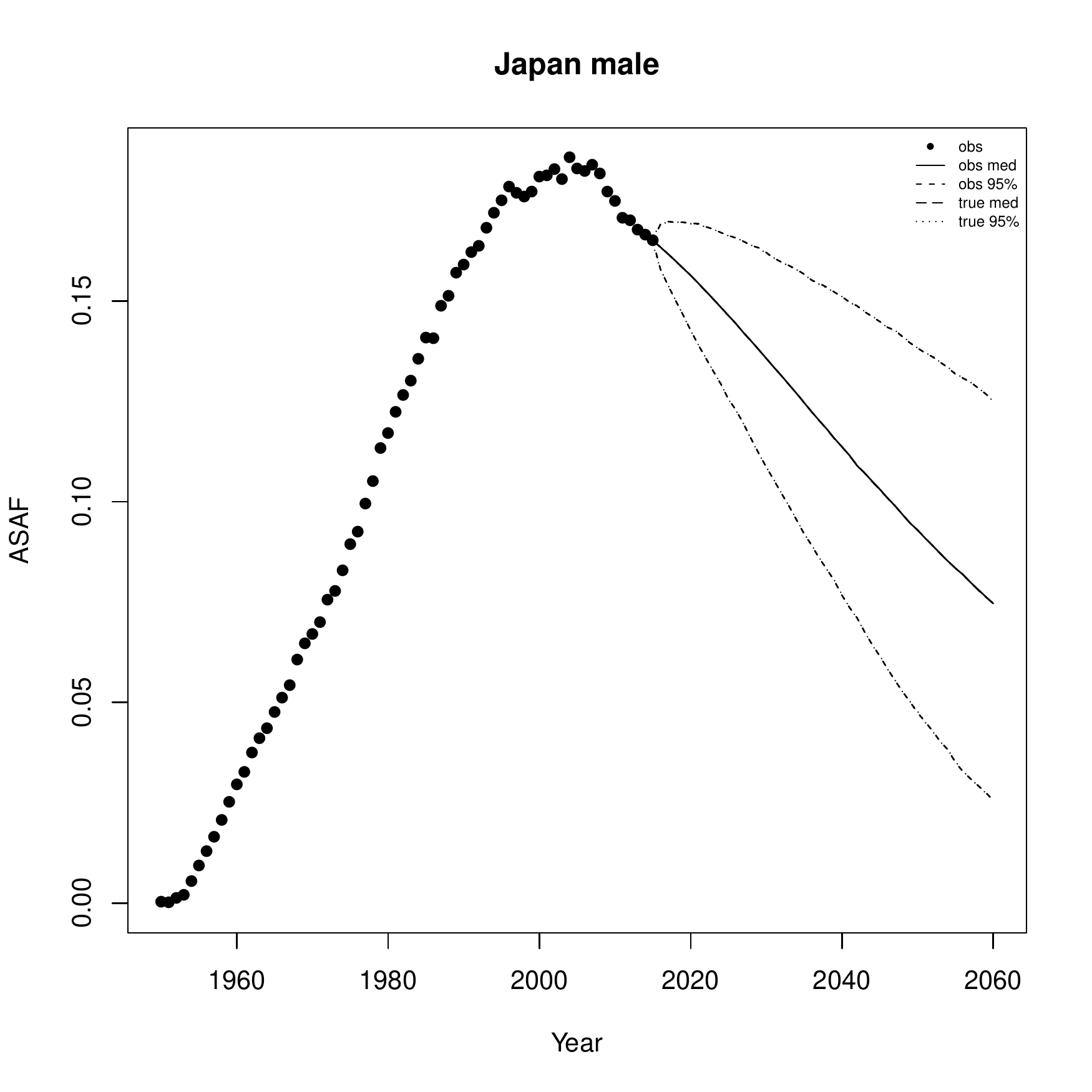}
		\includegraphics[scale=0.43]{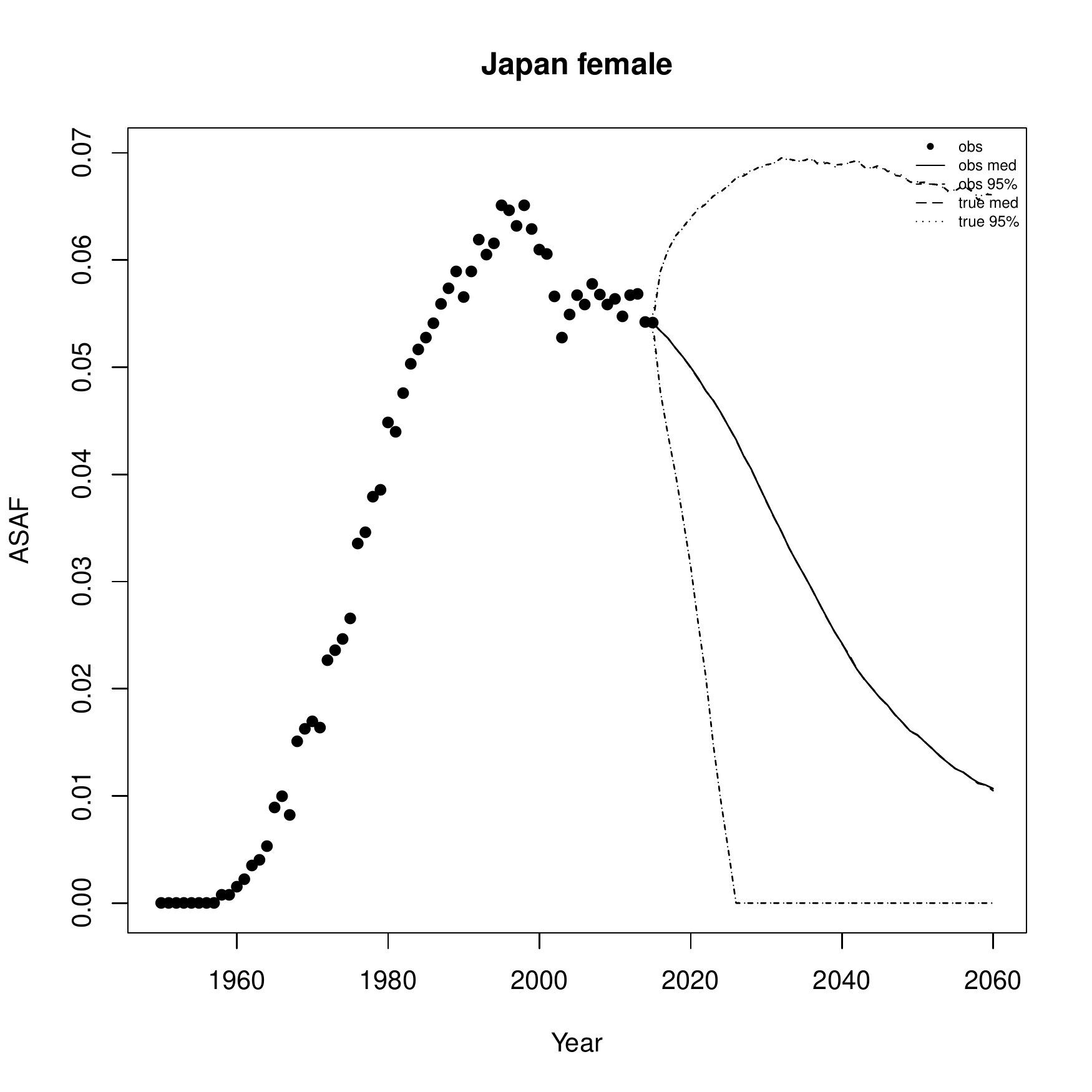}
		\includegraphics[scale=0.43]{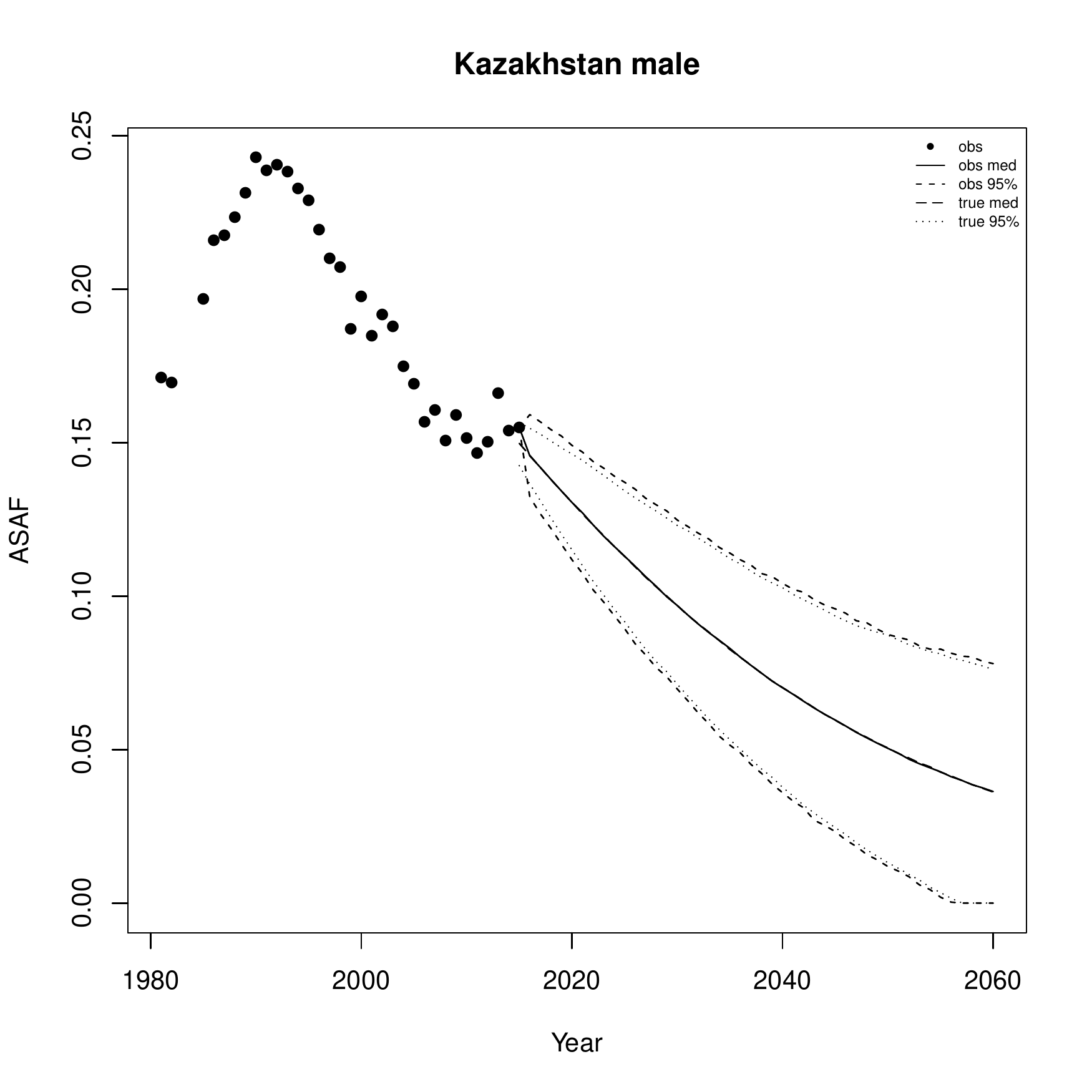}
		\includegraphics[scale=0.43]{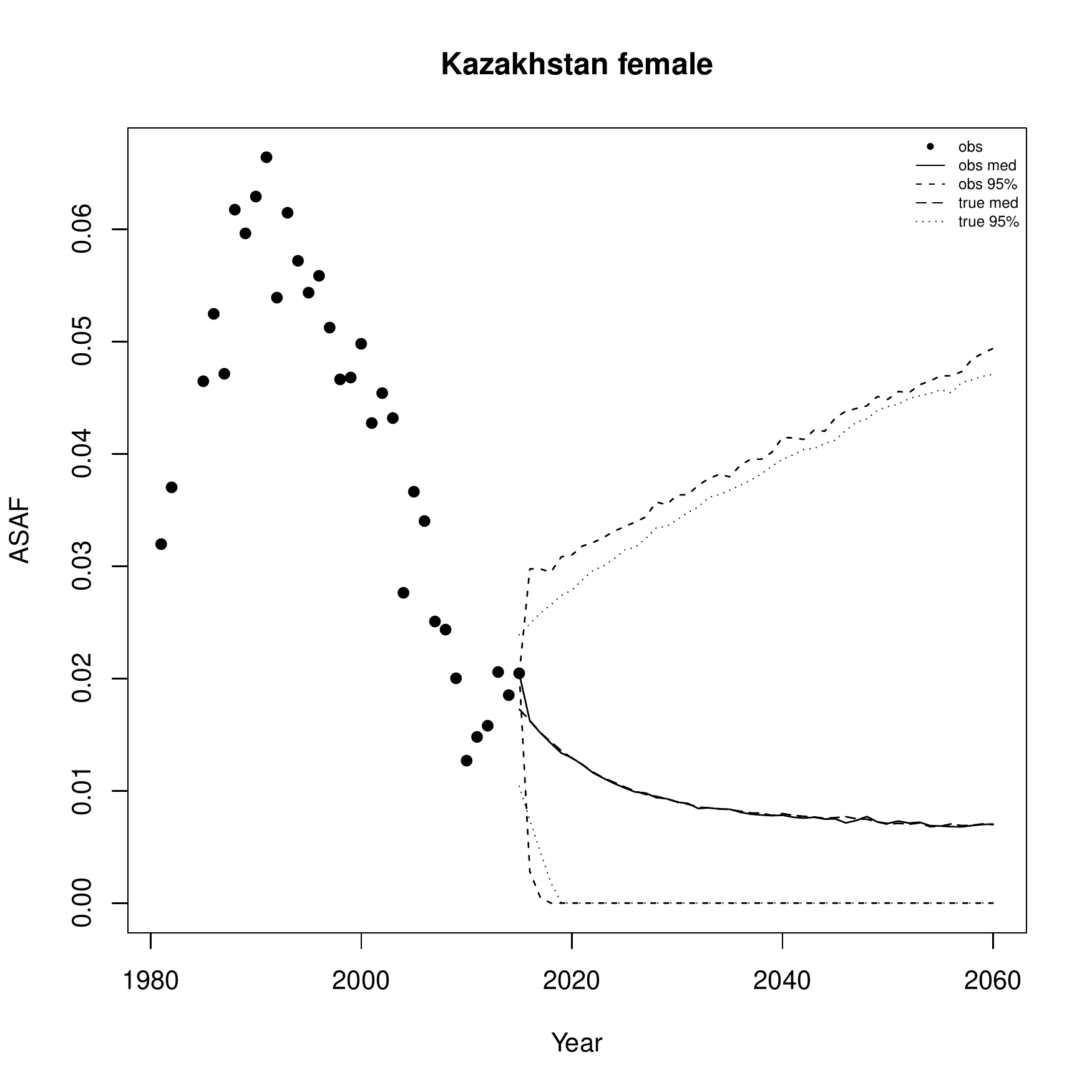}
	\end{center}
\end{figure}

\begin{figure}[H]
	\begin{center}
		\includegraphics[scale=0.43]{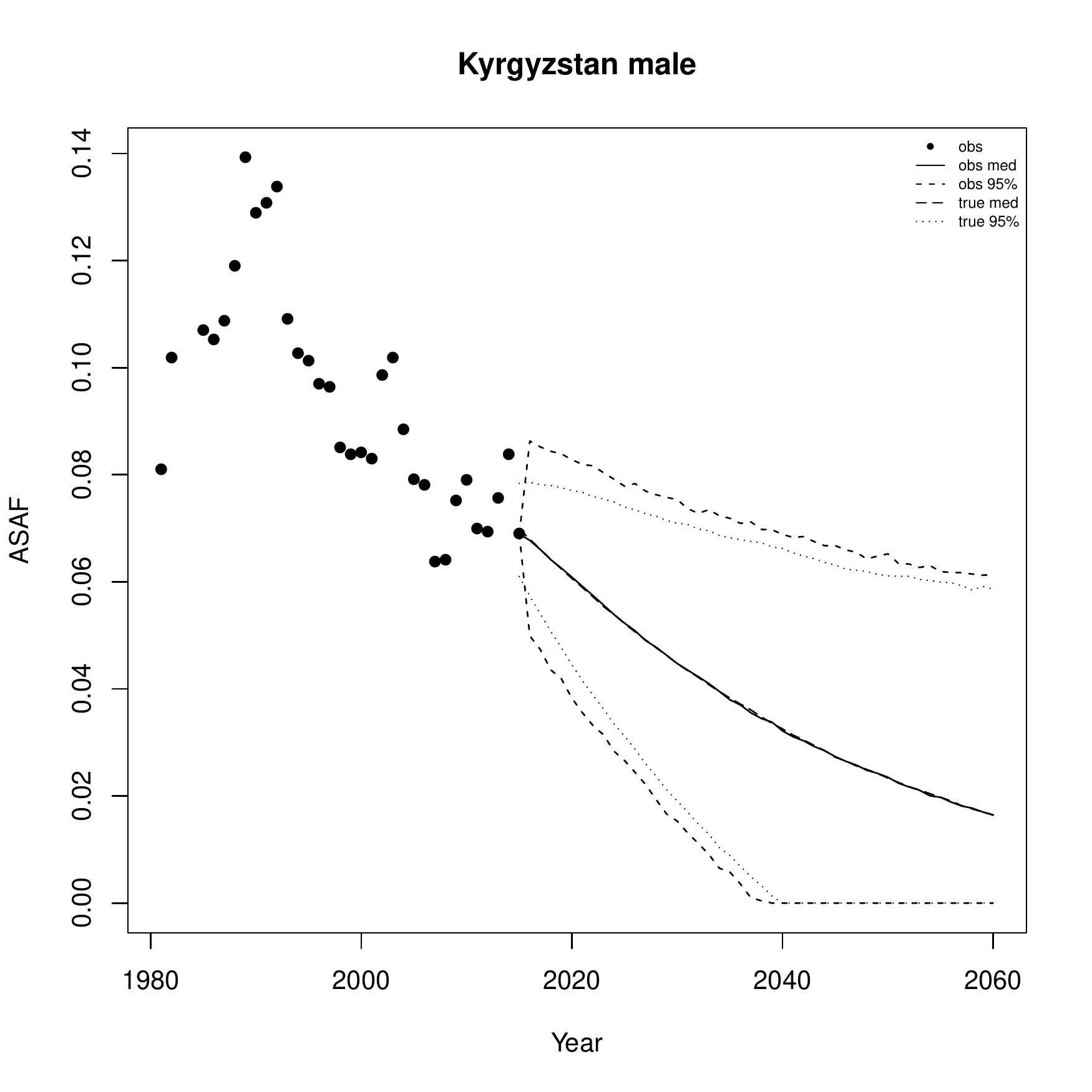}
		\includegraphics[scale=0.43]{plot/Kyrgyzstanmaleboth}
		\includegraphics[scale=0.43]{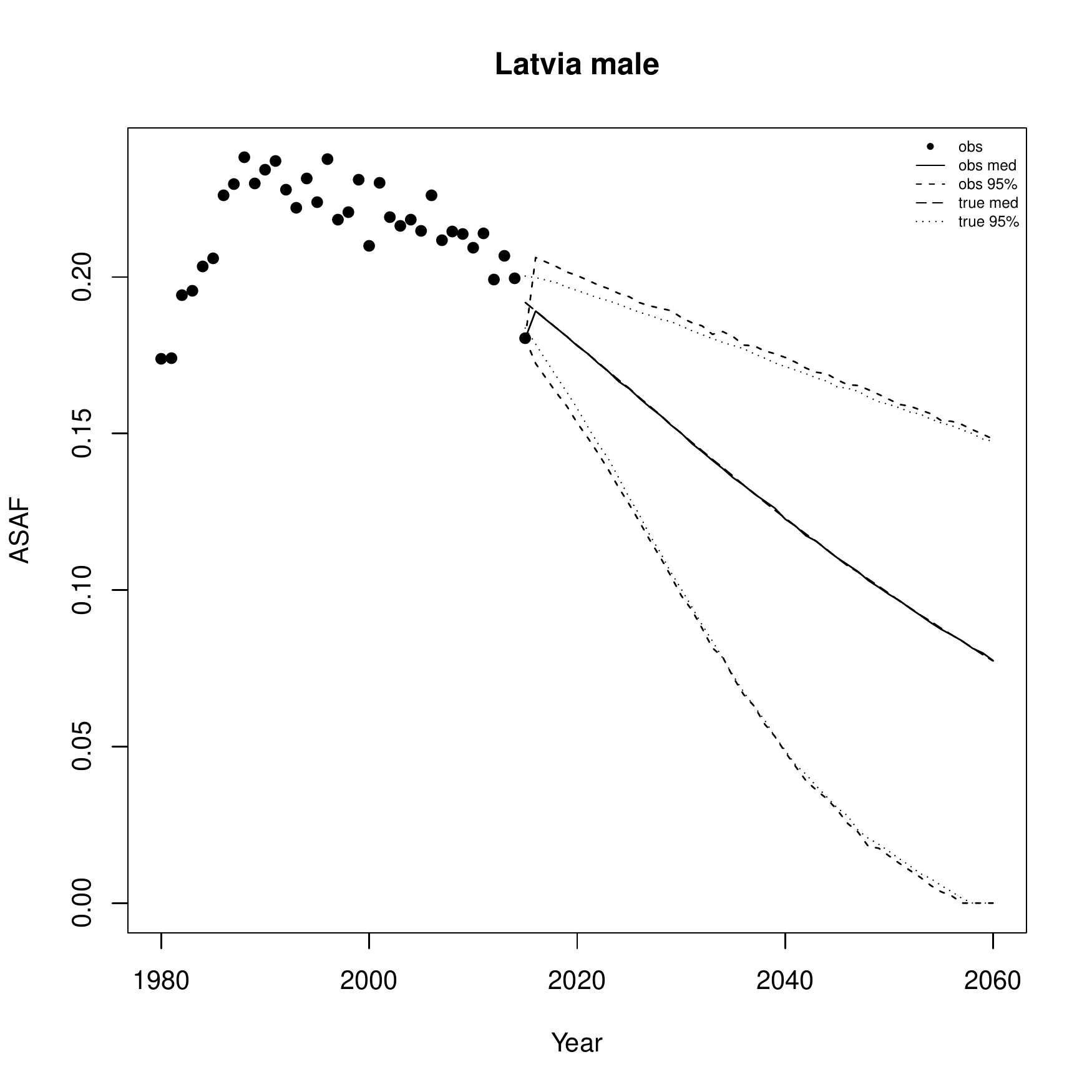}
		\includegraphics[scale=0.43]{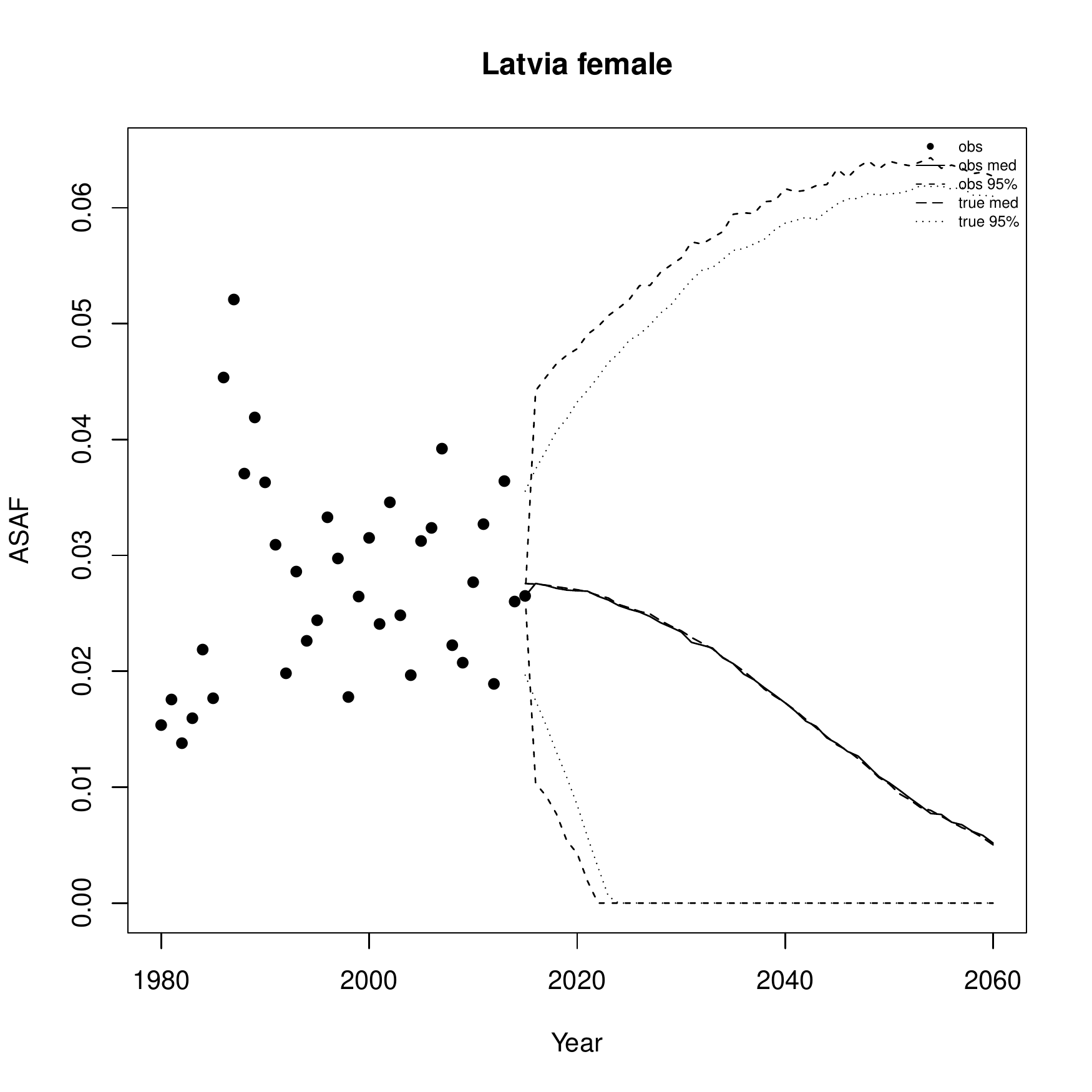}
		\includegraphics[scale=0.43]{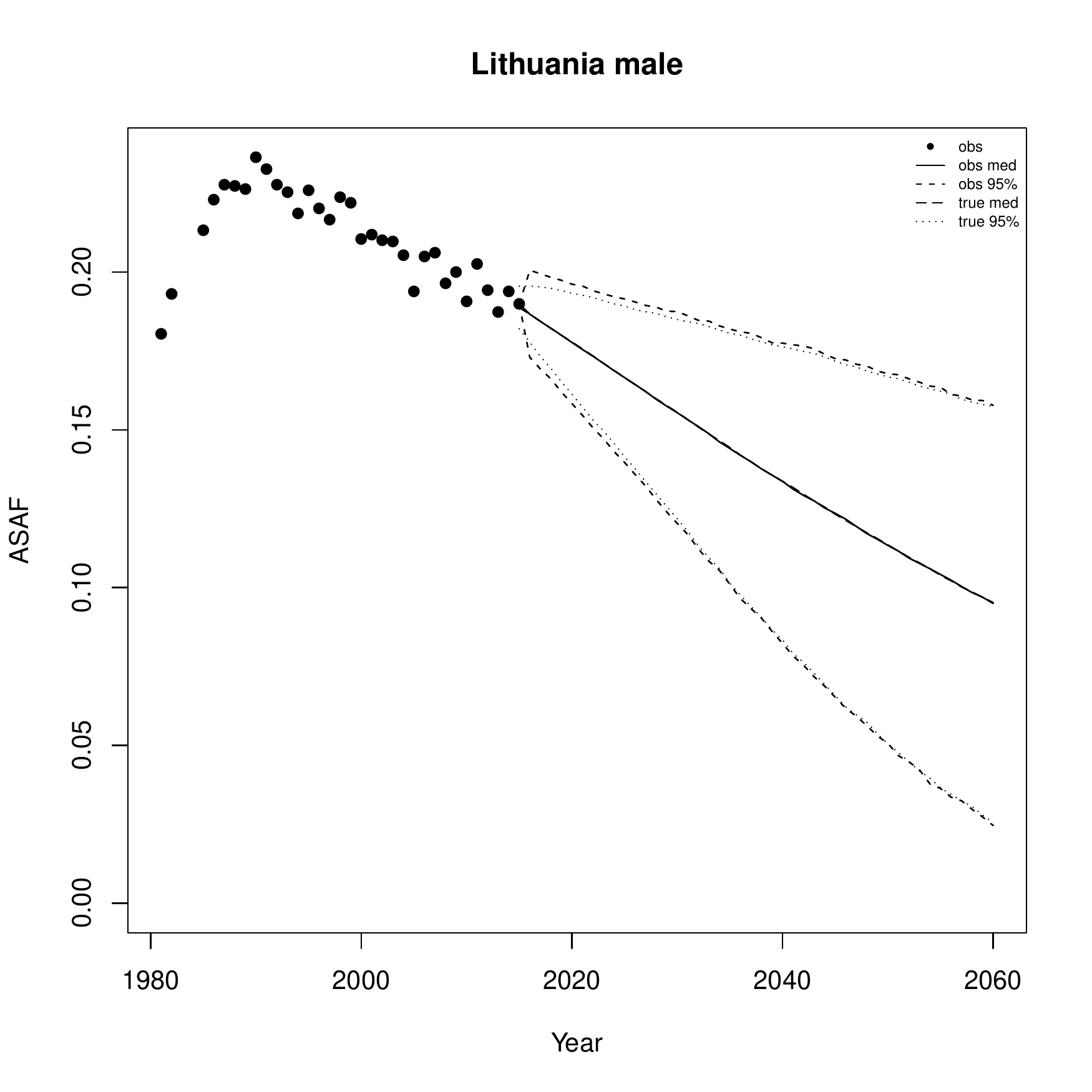}
		\includegraphics[scale=0.43]{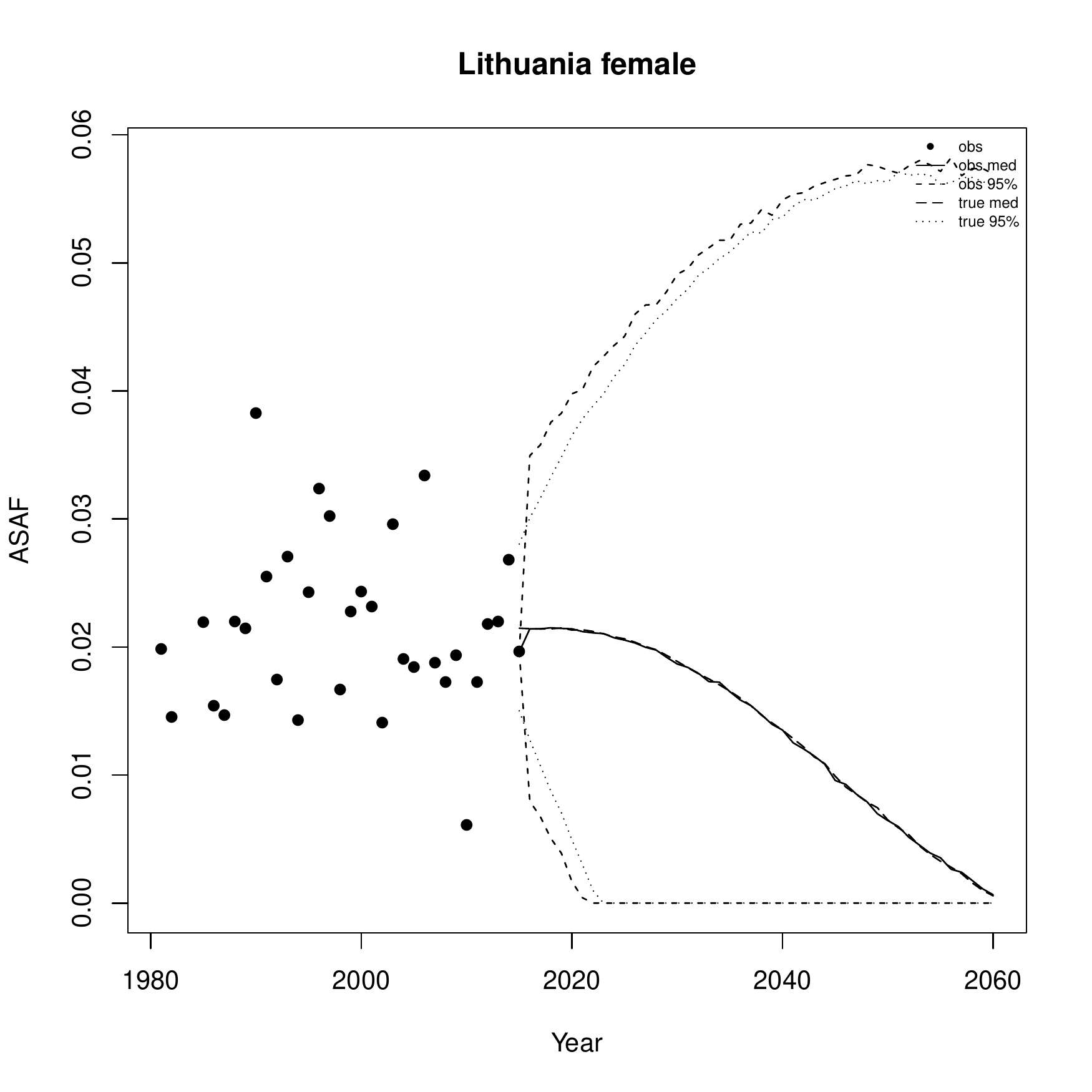}
	\end{center}
\end{figure}

\begin{figure}[H]
	\begin{center}
		\includegraphics[scale=0.43]{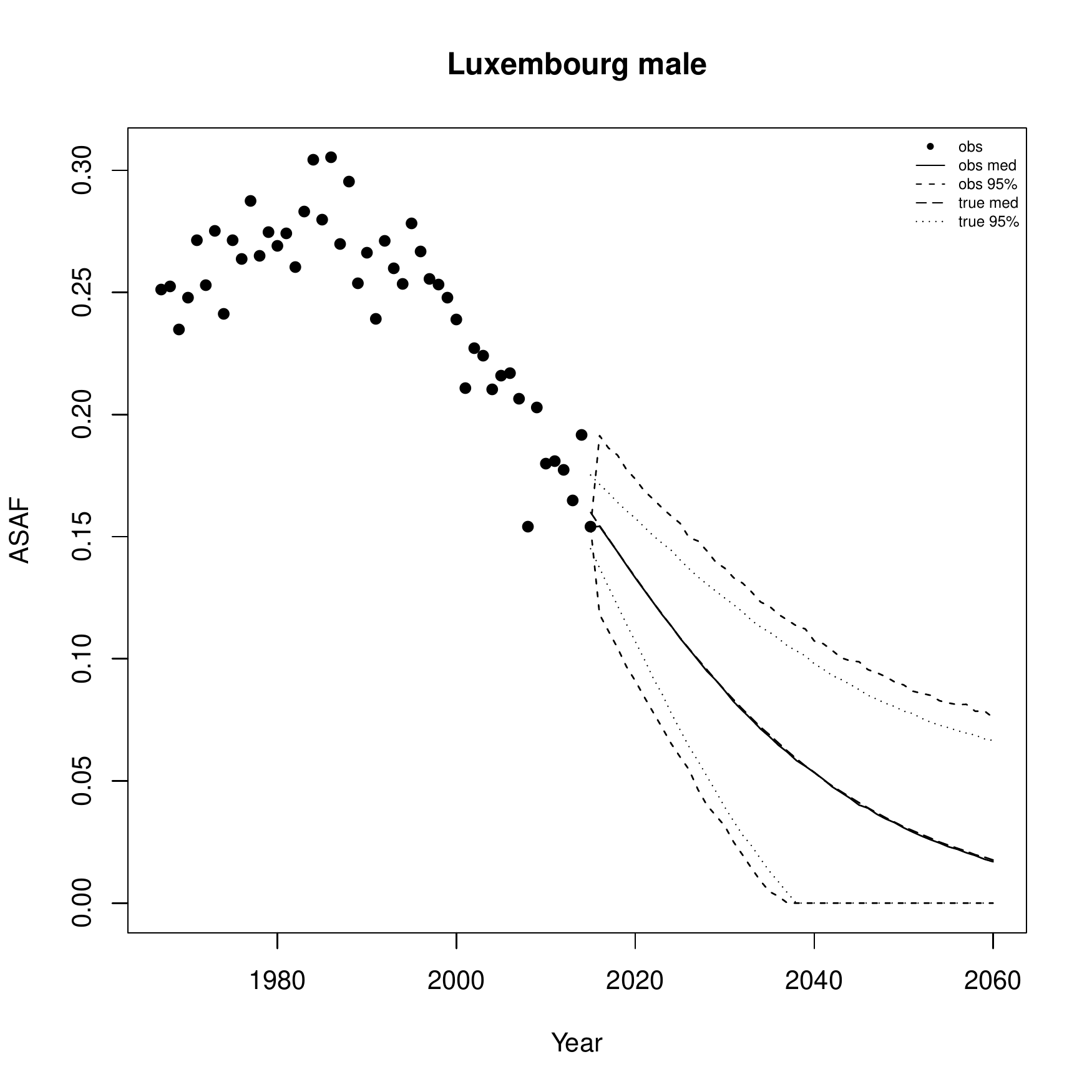}
		\includegraphics[scale=0.43]{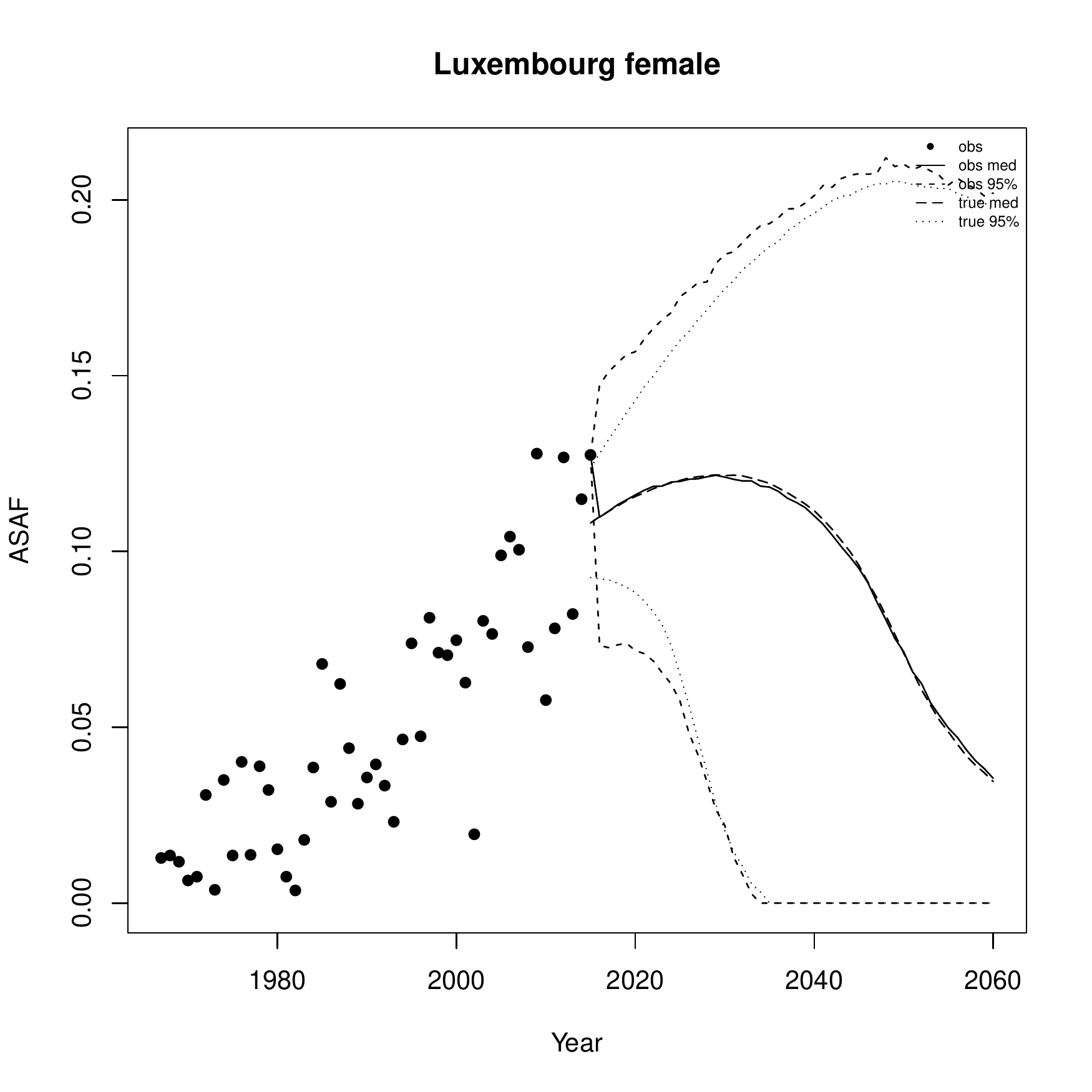}
		\includegraphics[scale=0.43]{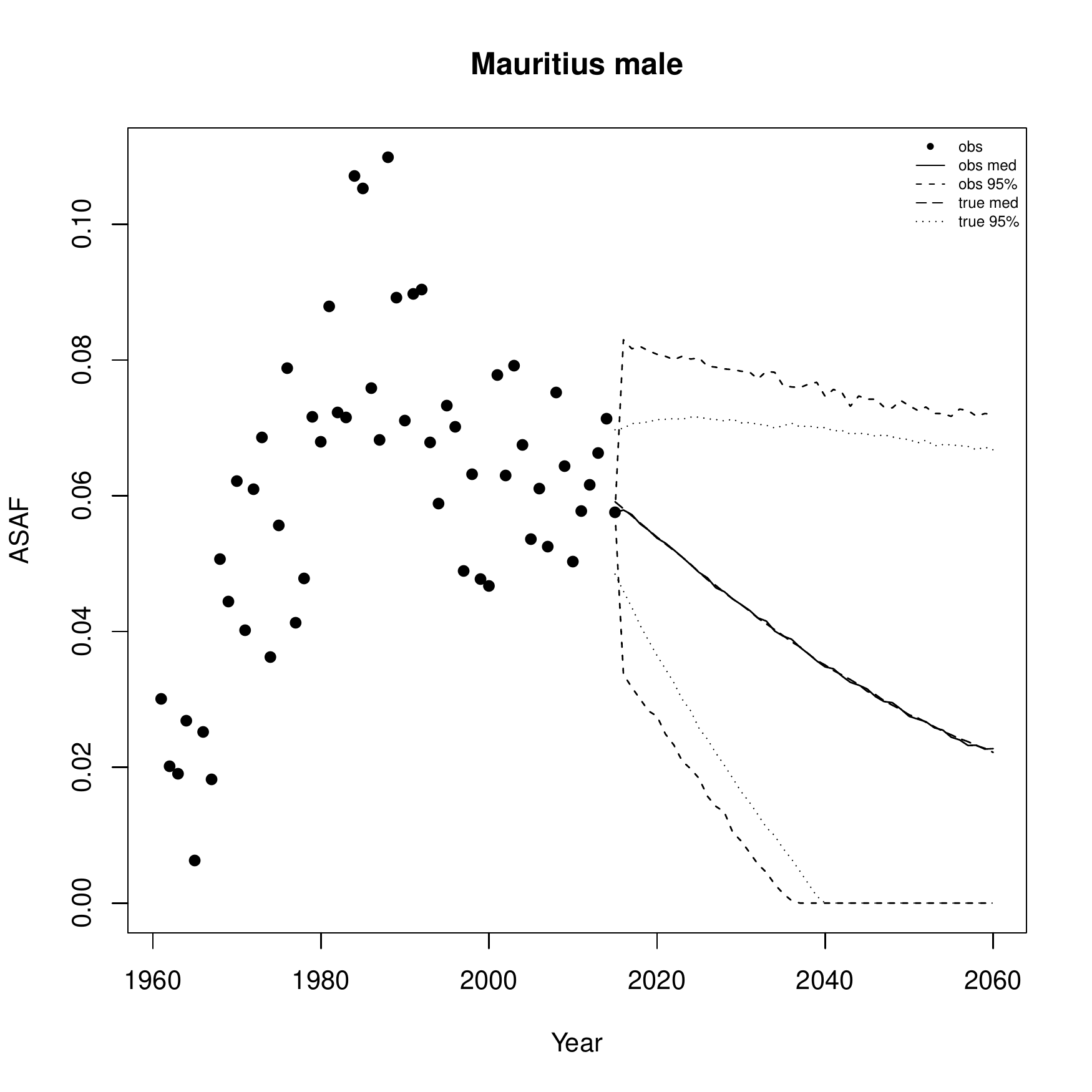}
		\includegraphics[scale=0.43]{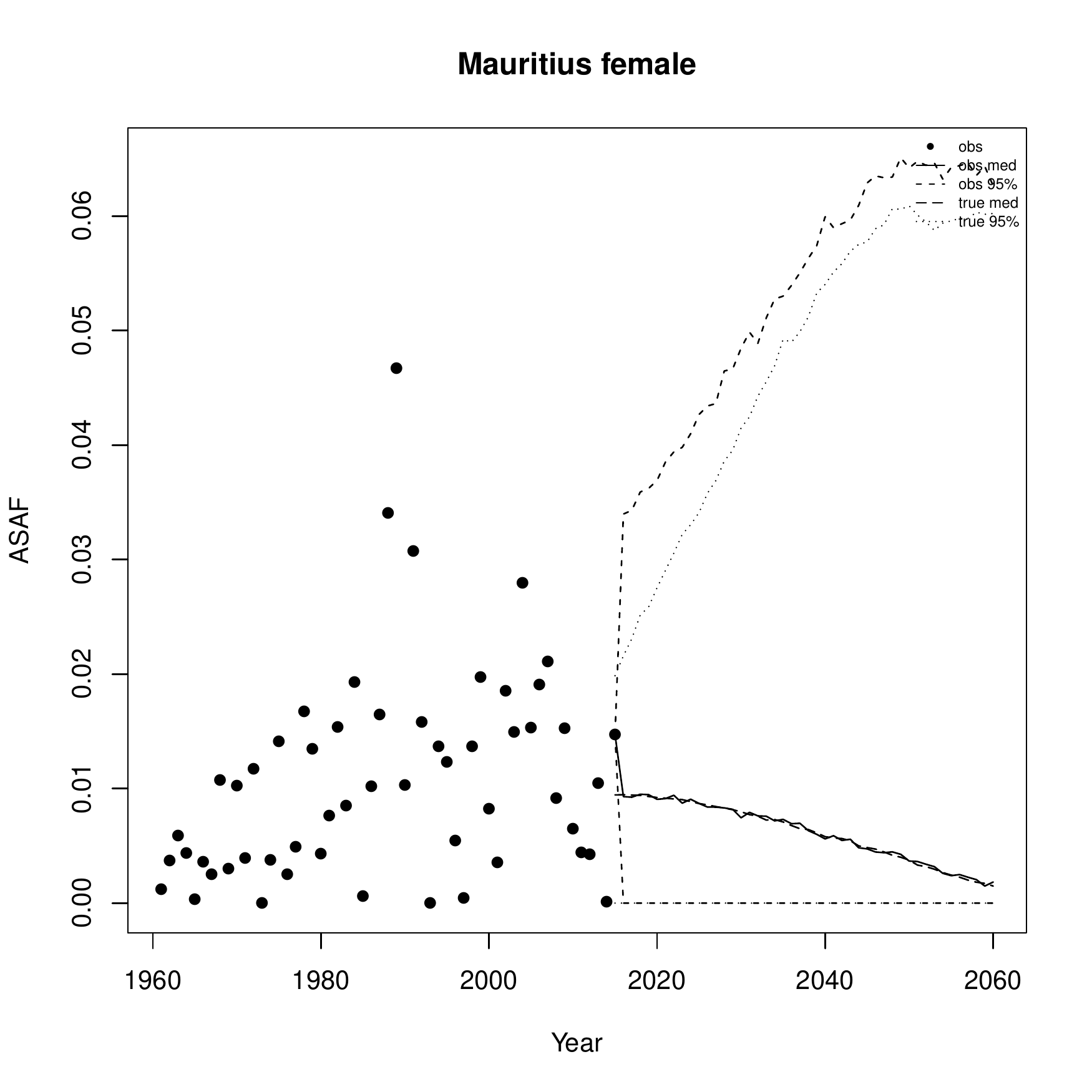}
		\includegraphics[scale=0.43]{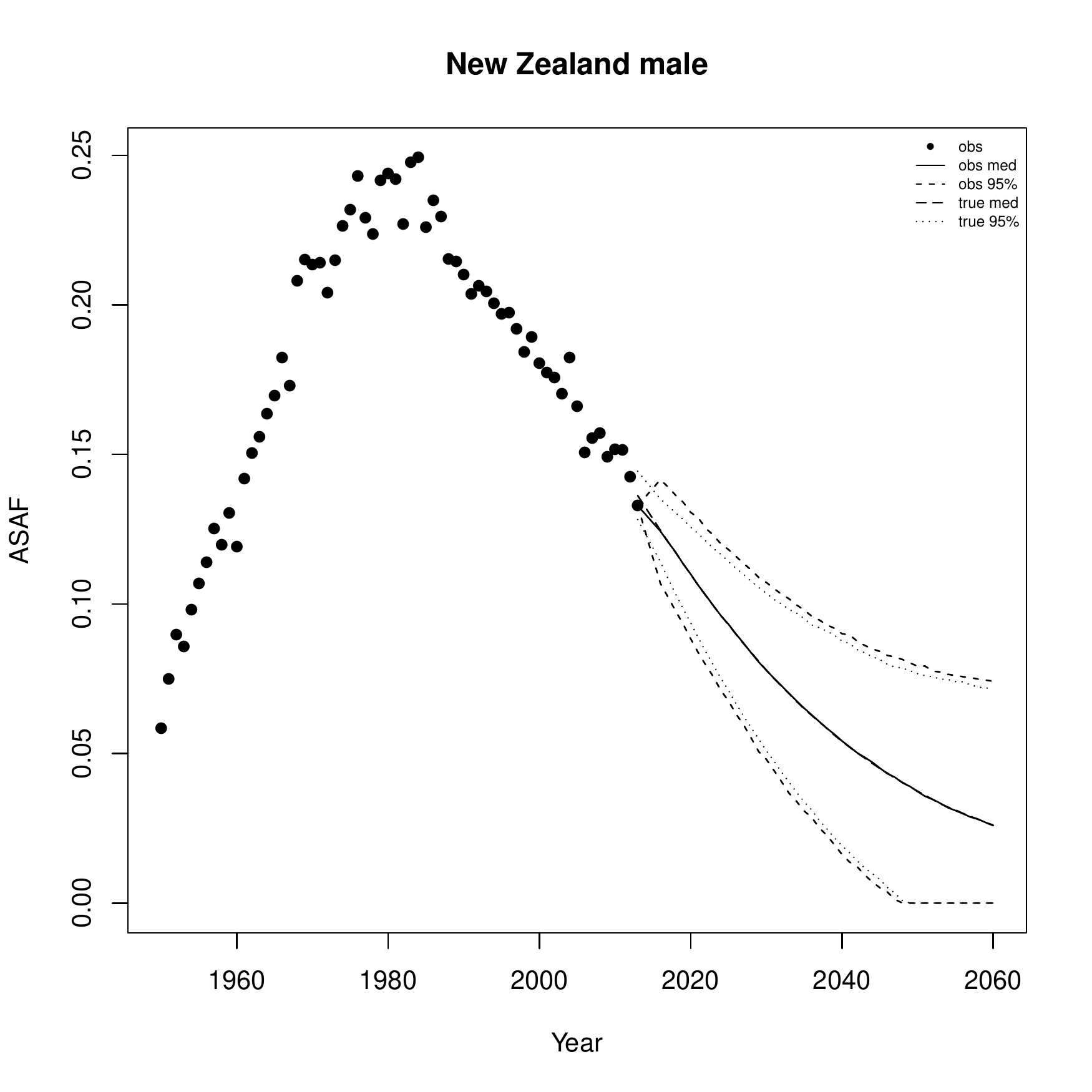}
		\includegraphics[scale=0.43]{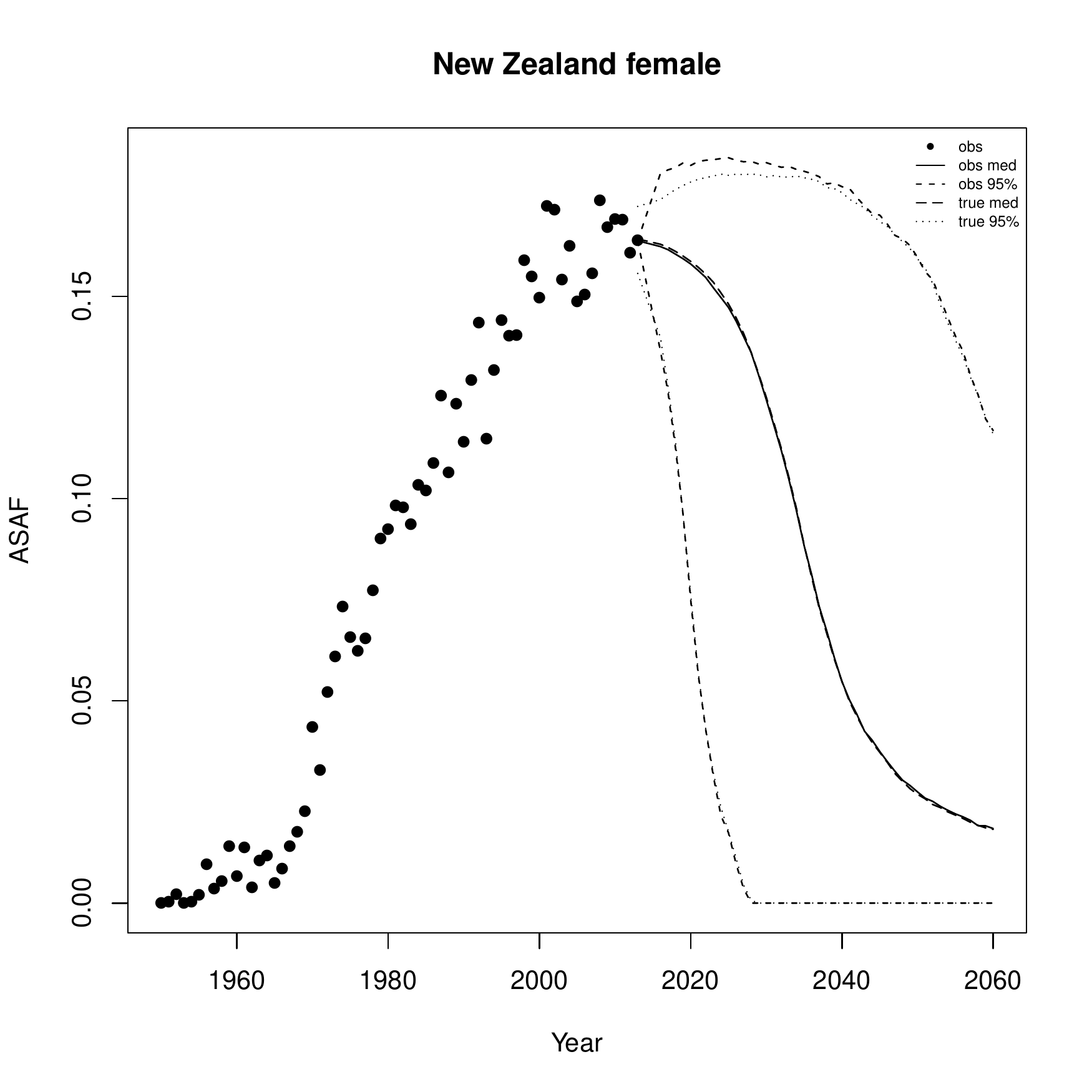}
	\end{center}
\end{figure}

\begin{figure}[H]
	\begin{center}
		\includegraphics[scale=0.43]{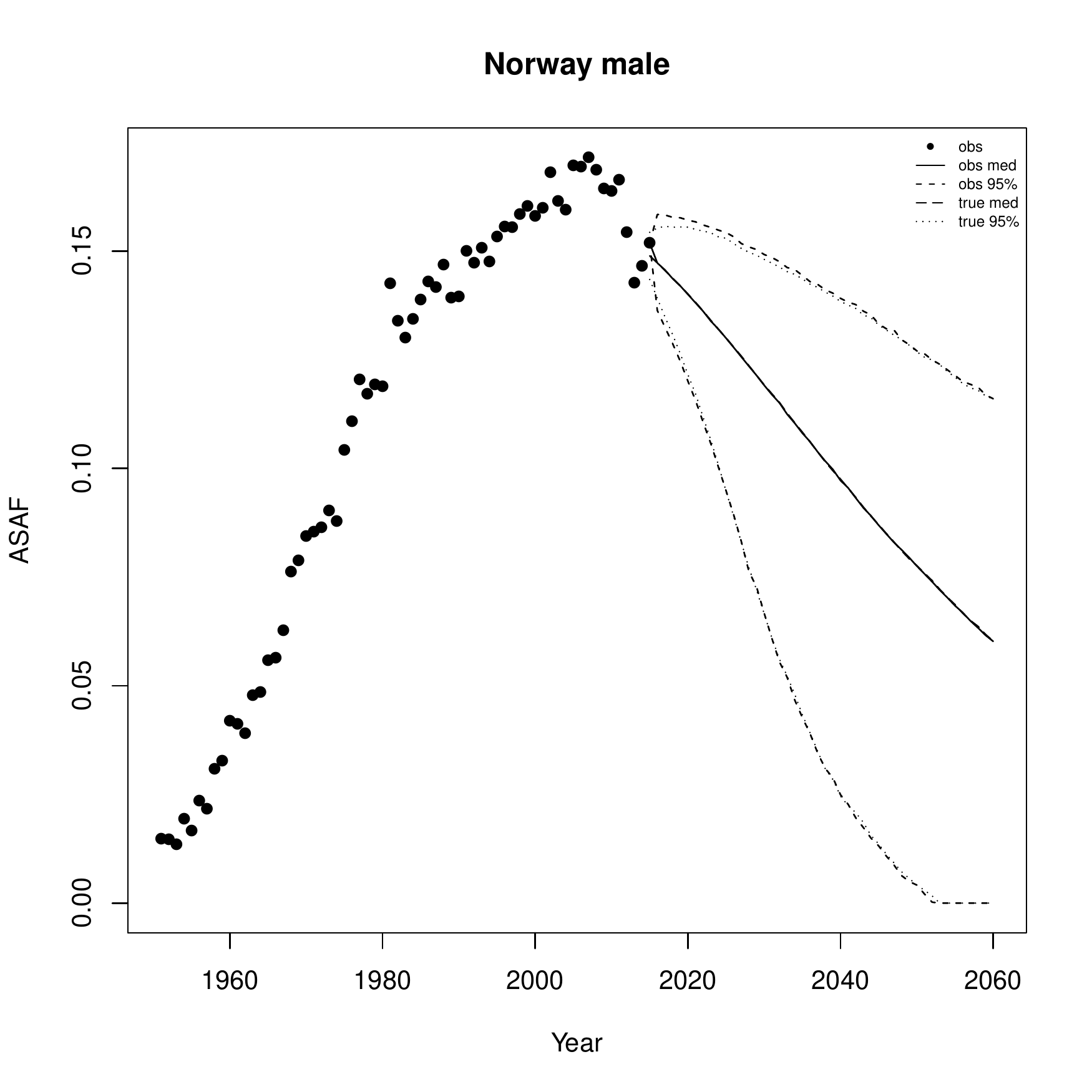}
		\includegraphics[scale=0.43]{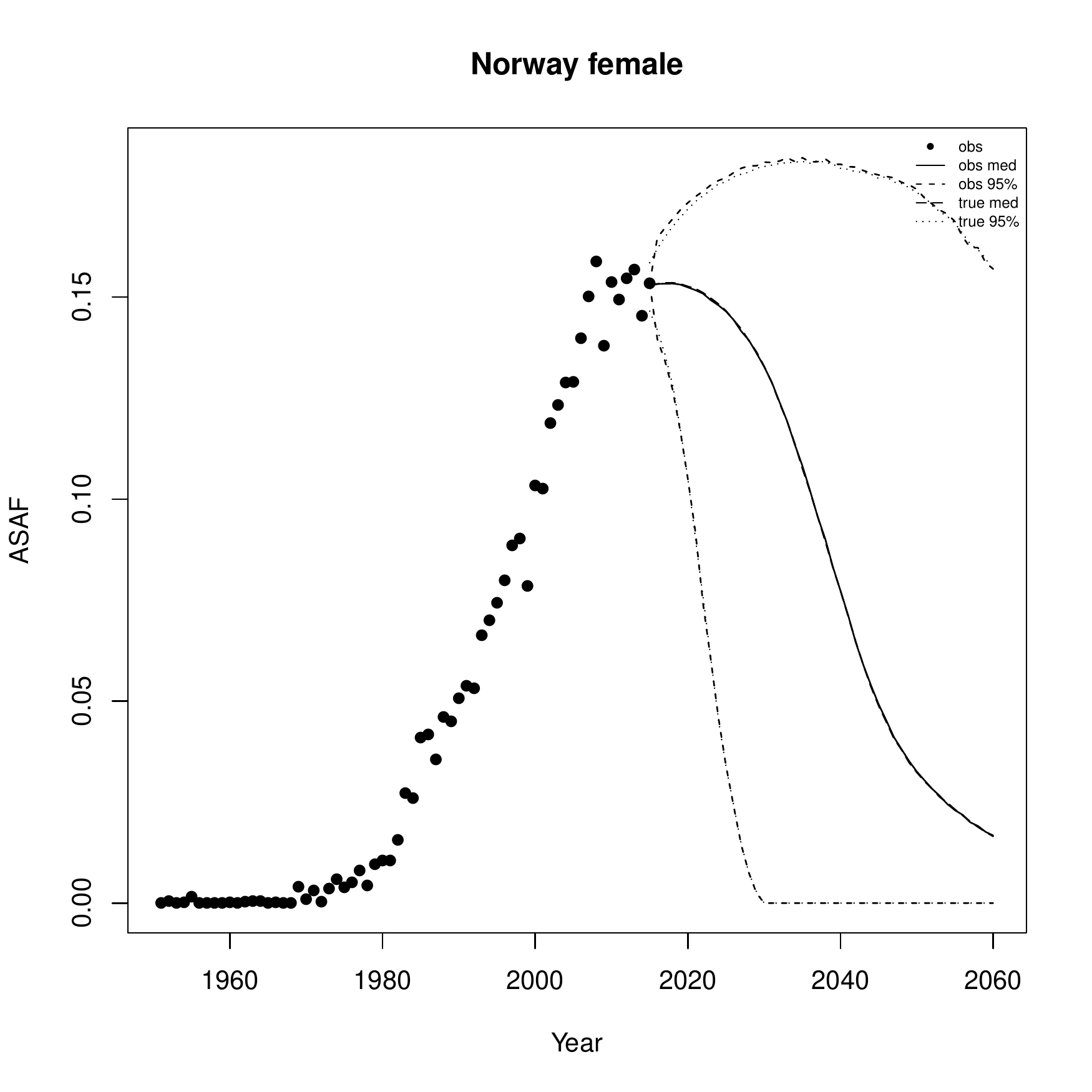}
		\includegraphics[scale=0.43]{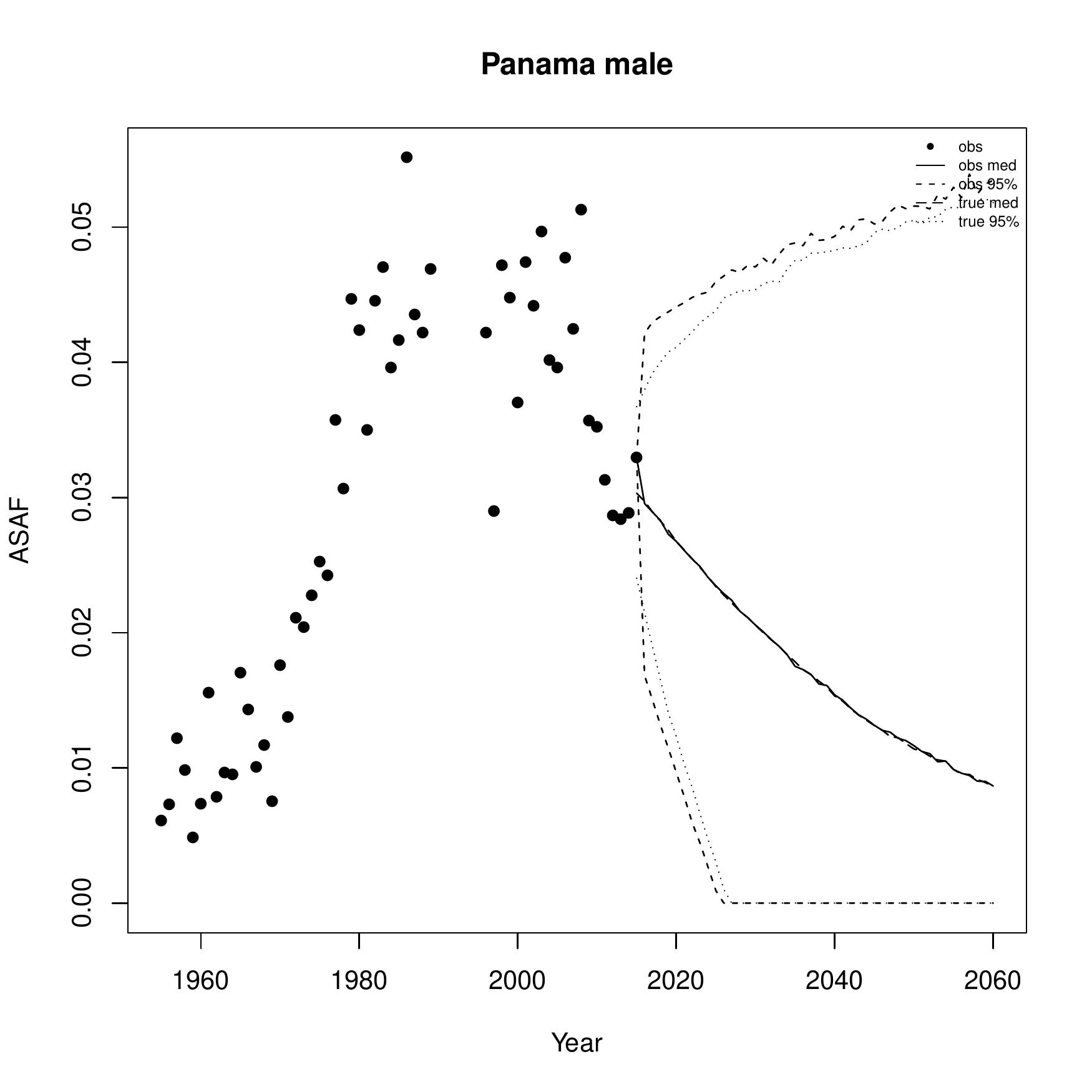}
		\includegraphics[scale=0.43]{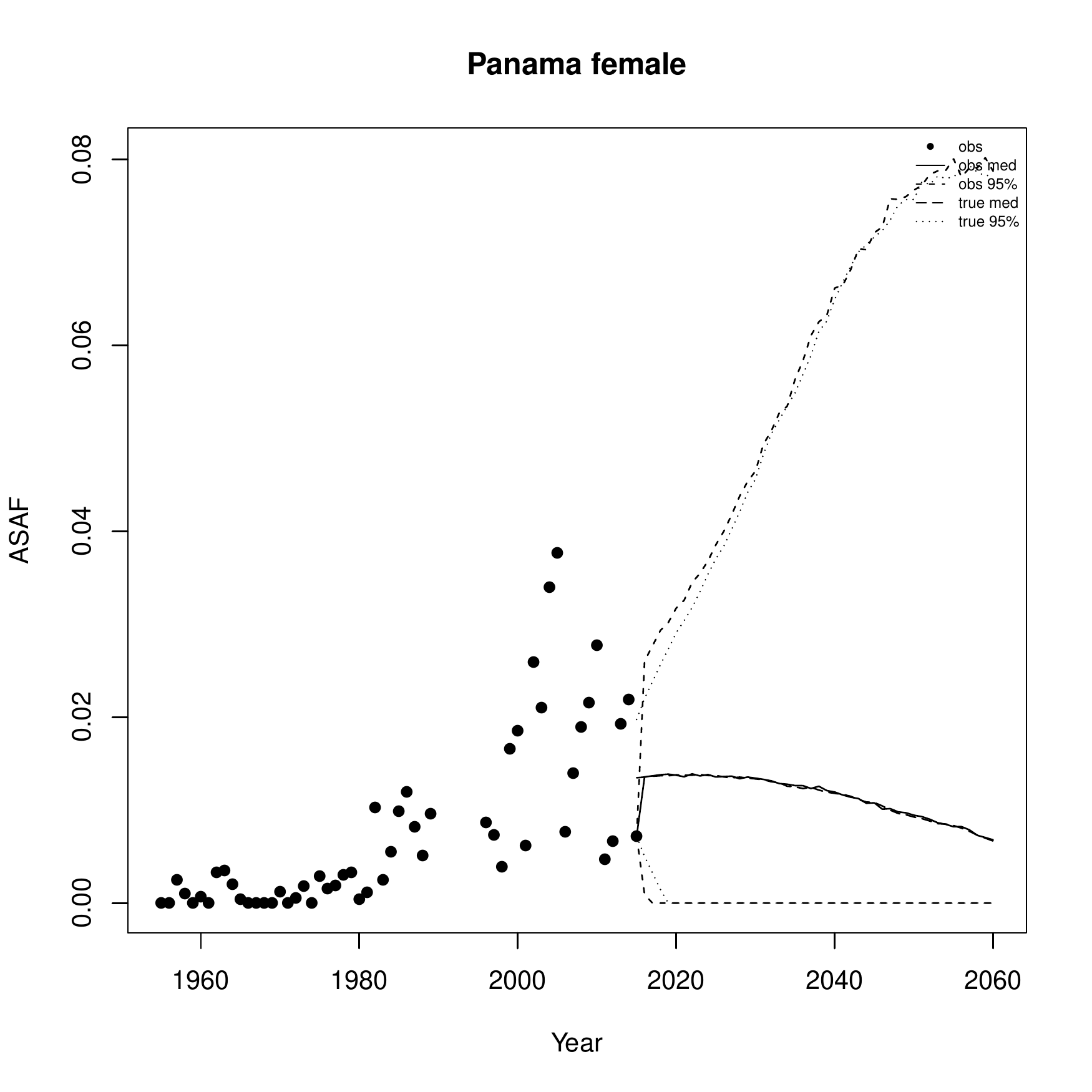}
		\includegraphics[scale=0.43]{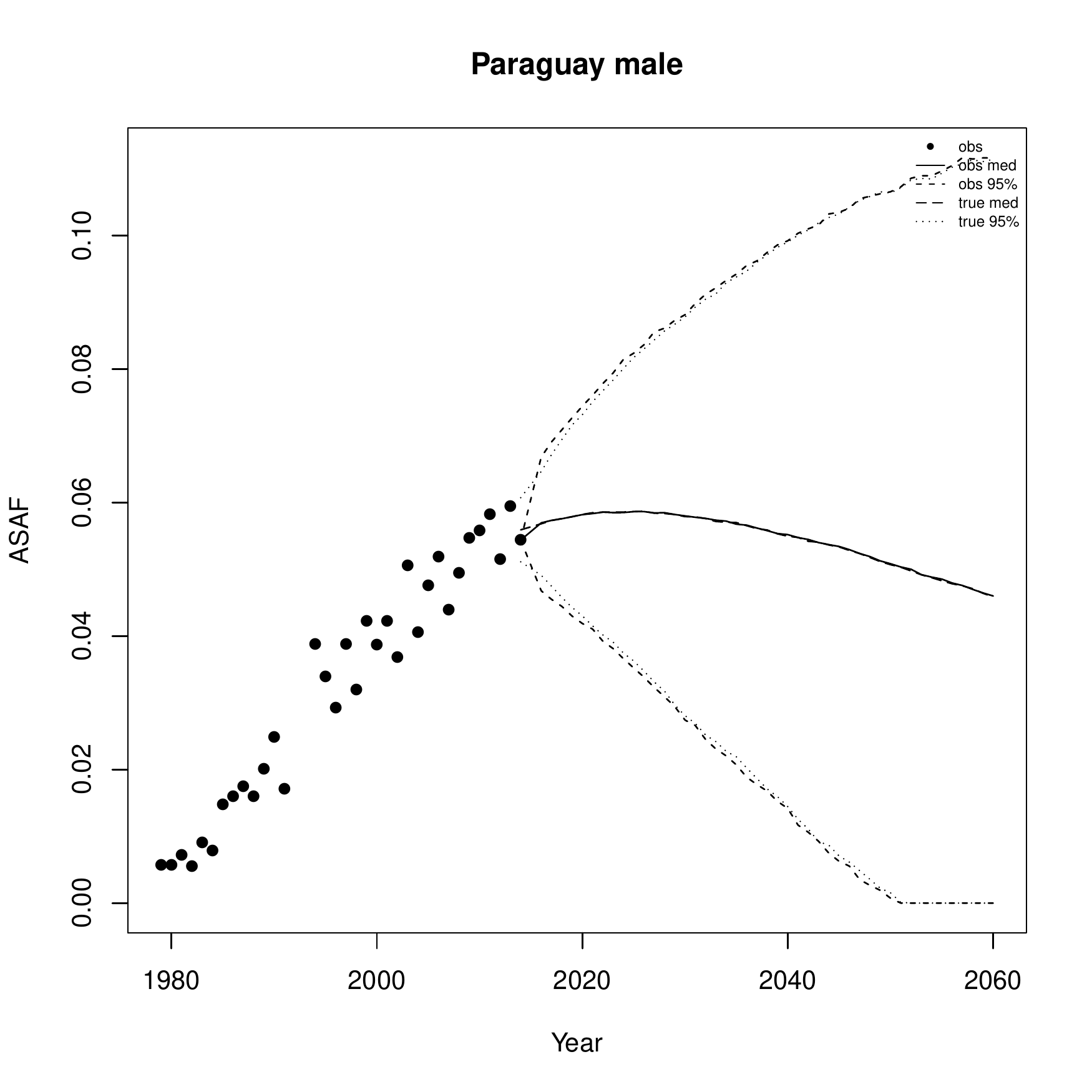}
		\includegraphics[scale=0.43]{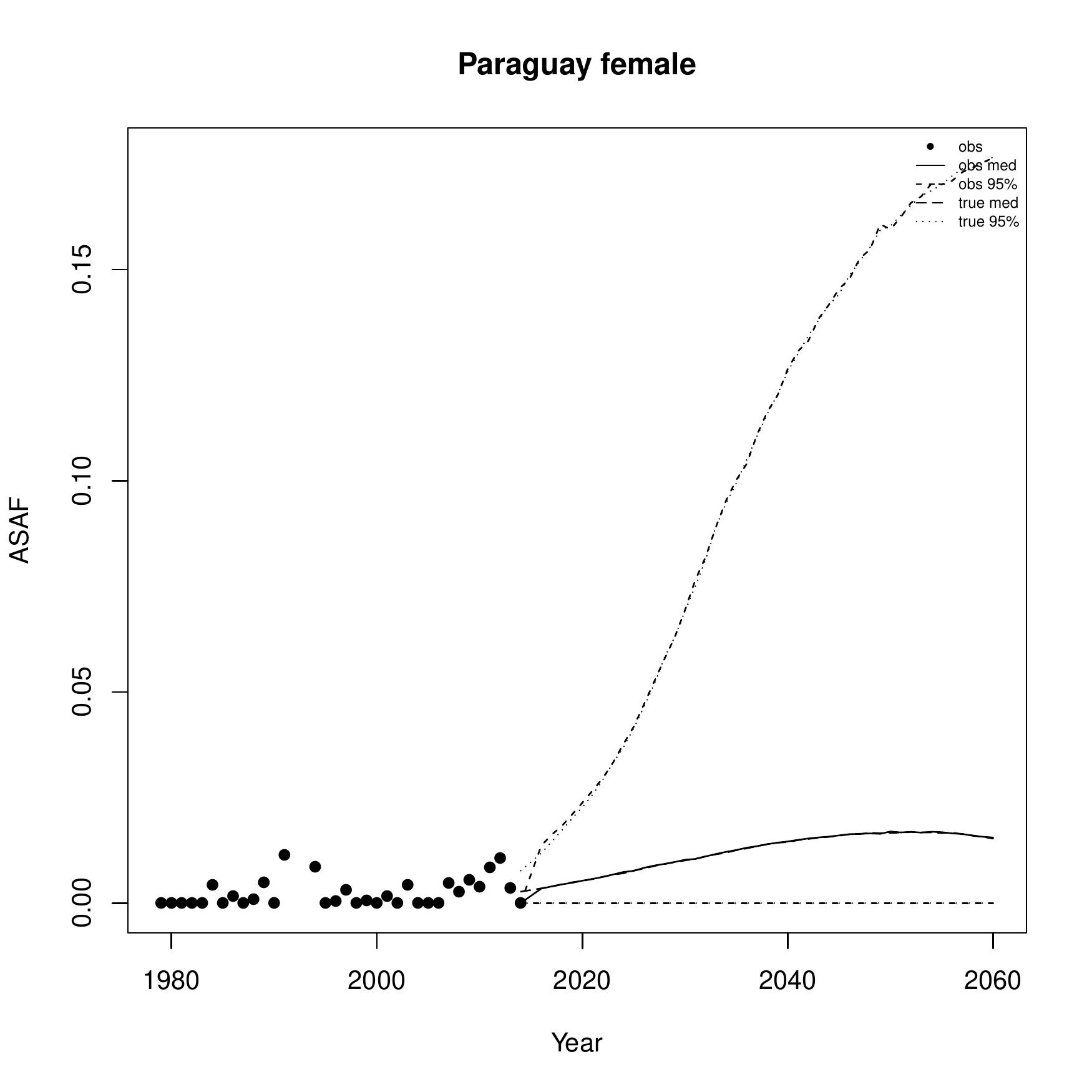}
	\end{center}
\end{figure}

\begin{figure}[H]
	\begin{center}
		\includegraphics[scale=0.43]{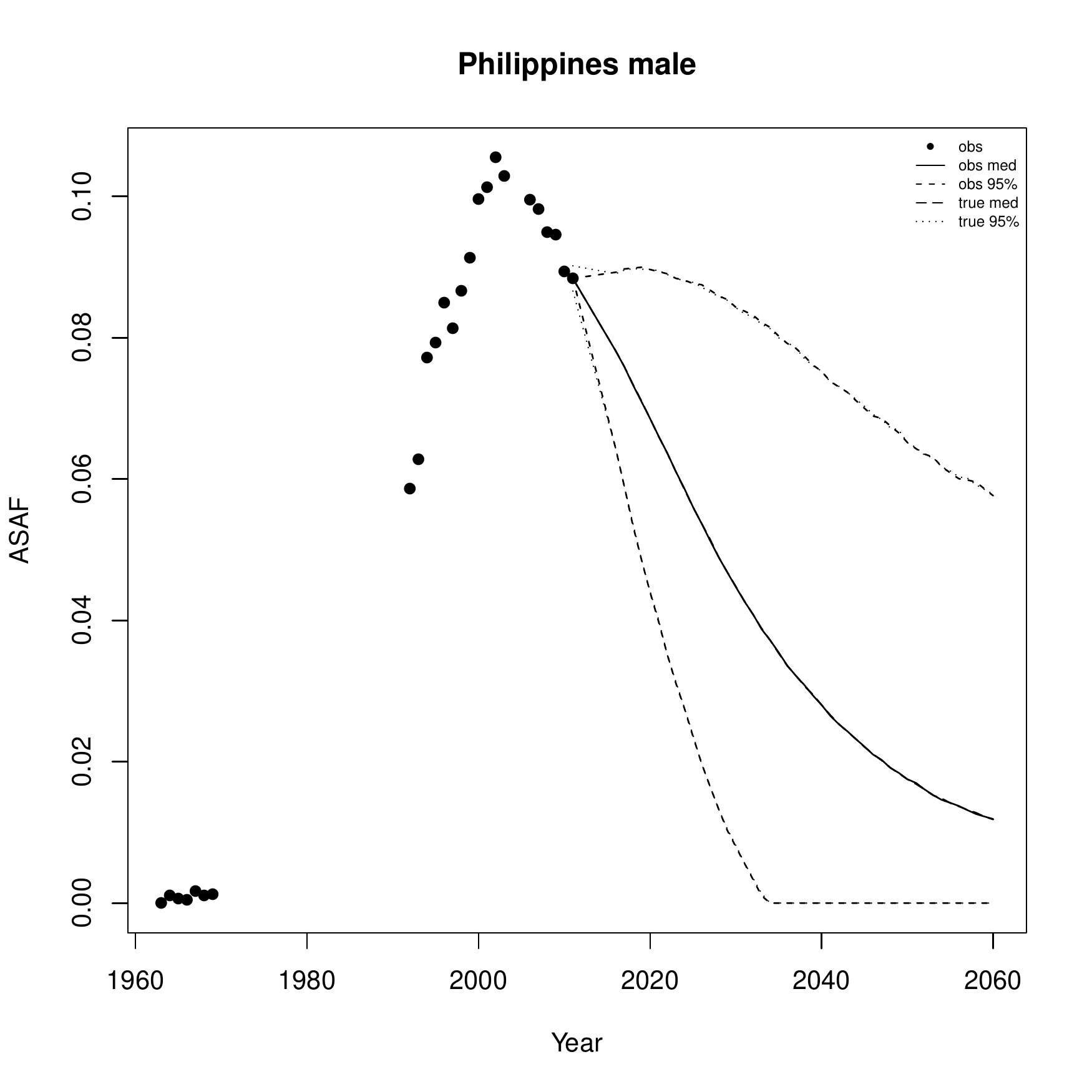}
		\includegraphics[scale=0.43]{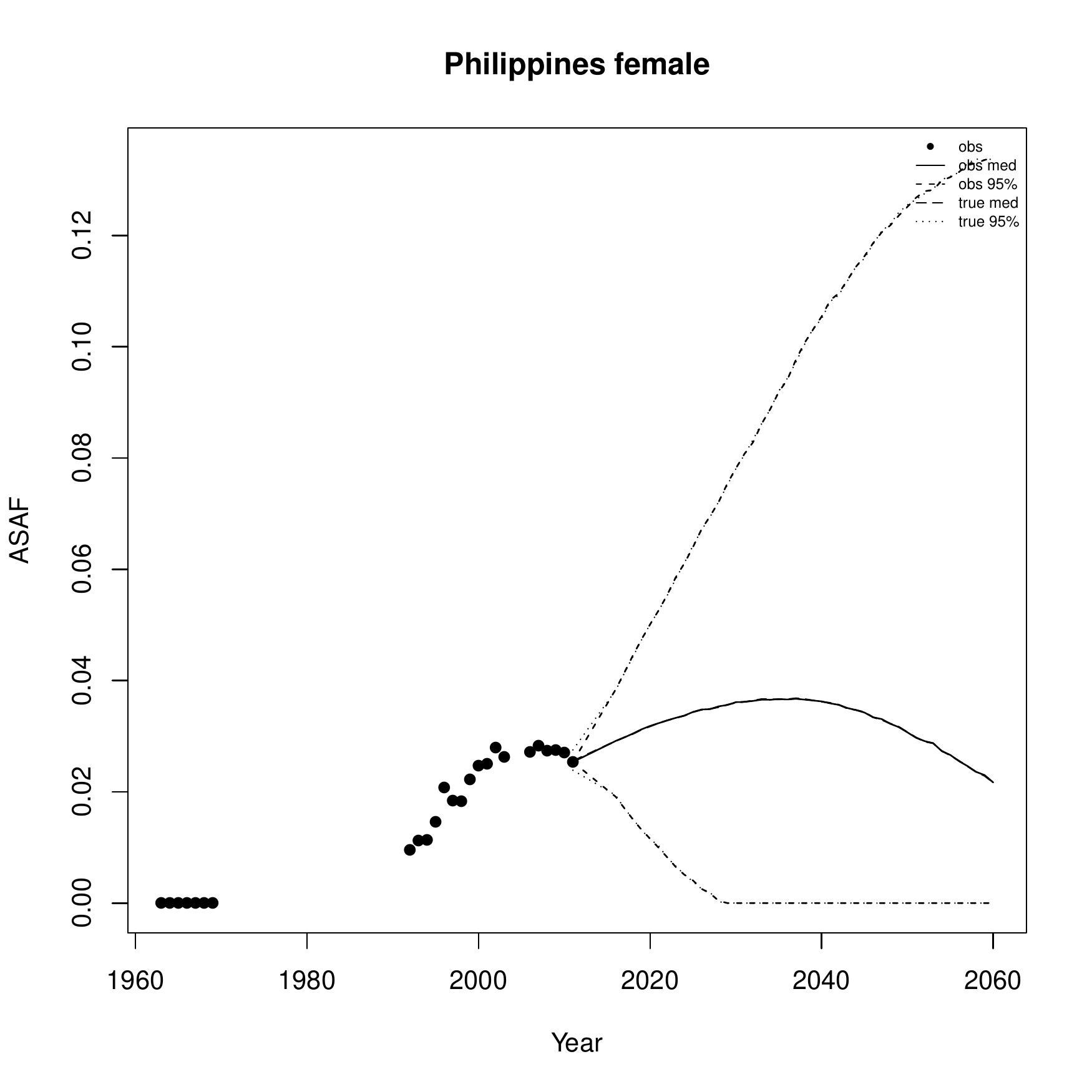}
		\includegraphics[scale=0.43]{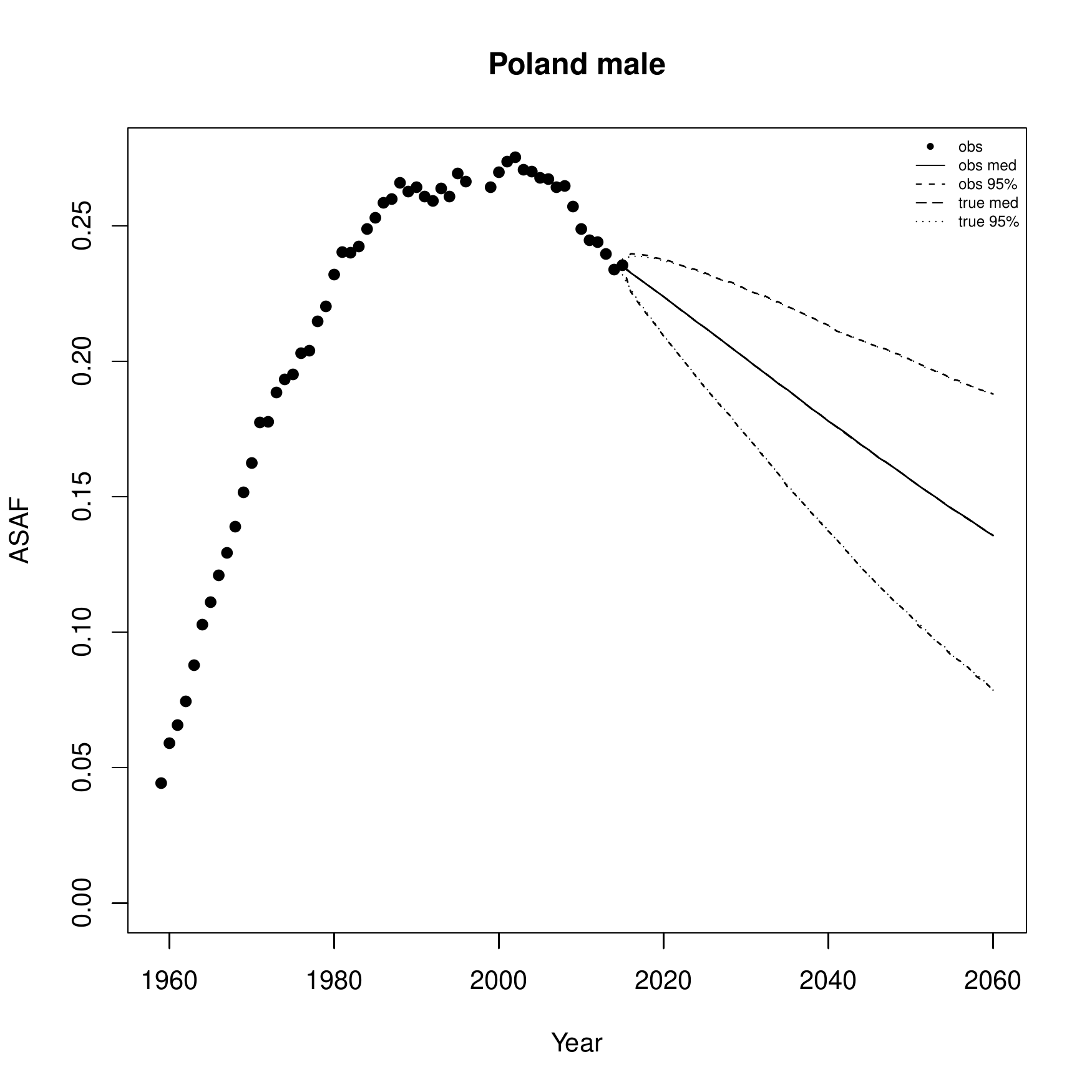}
		\includegraphics[scale=0.43]{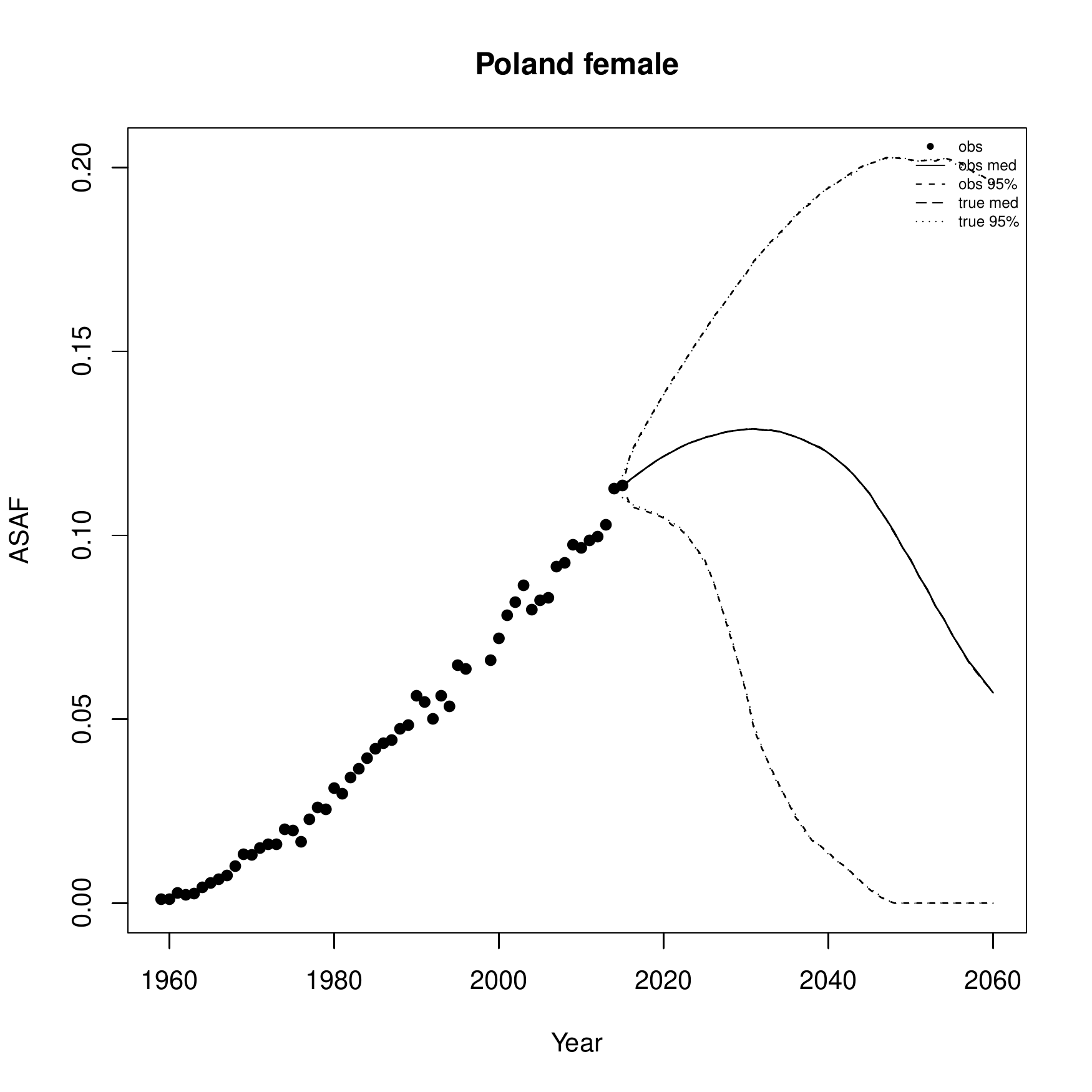}
		\includegraphics[scale=0.43]{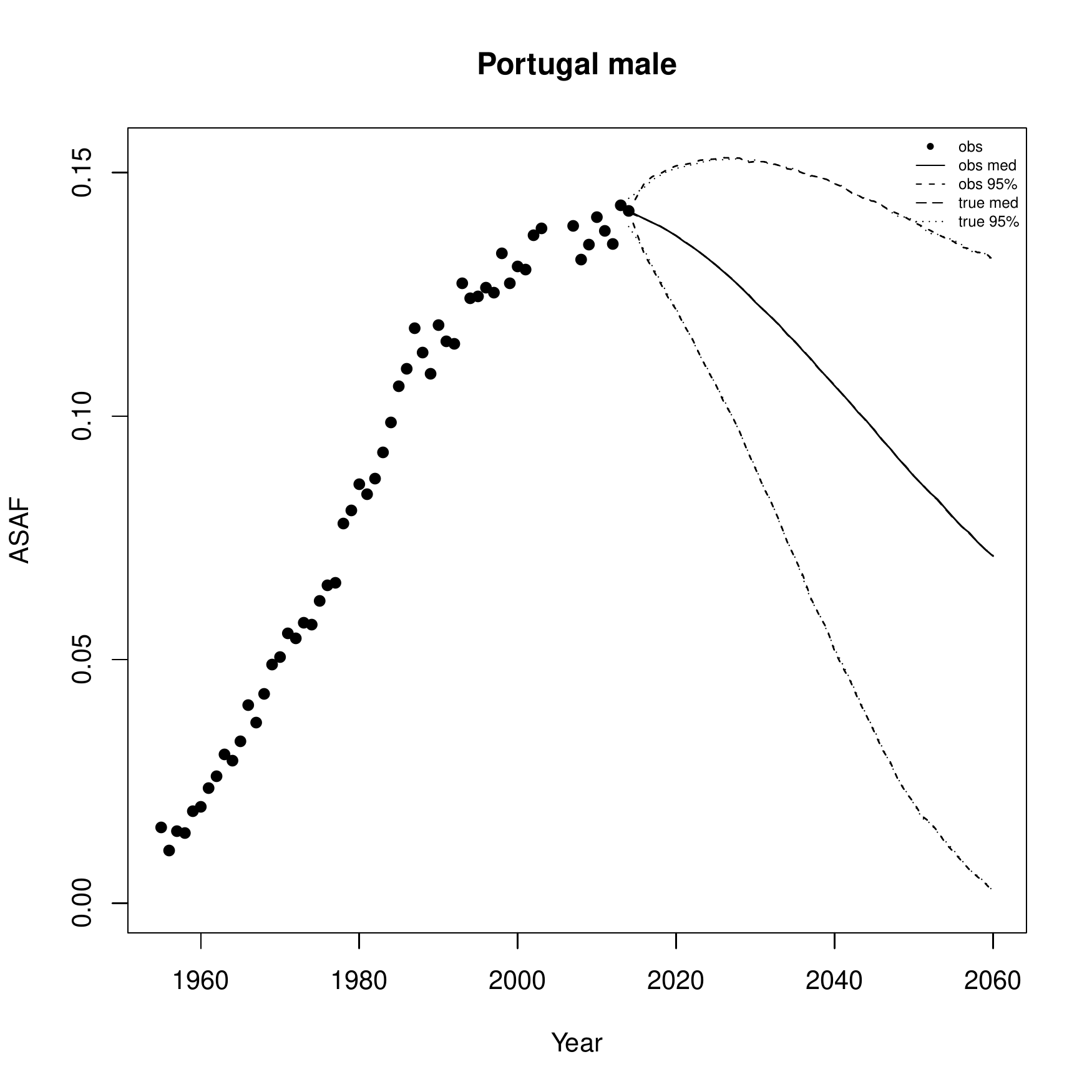}
		\includegraphics[scale=0.43]{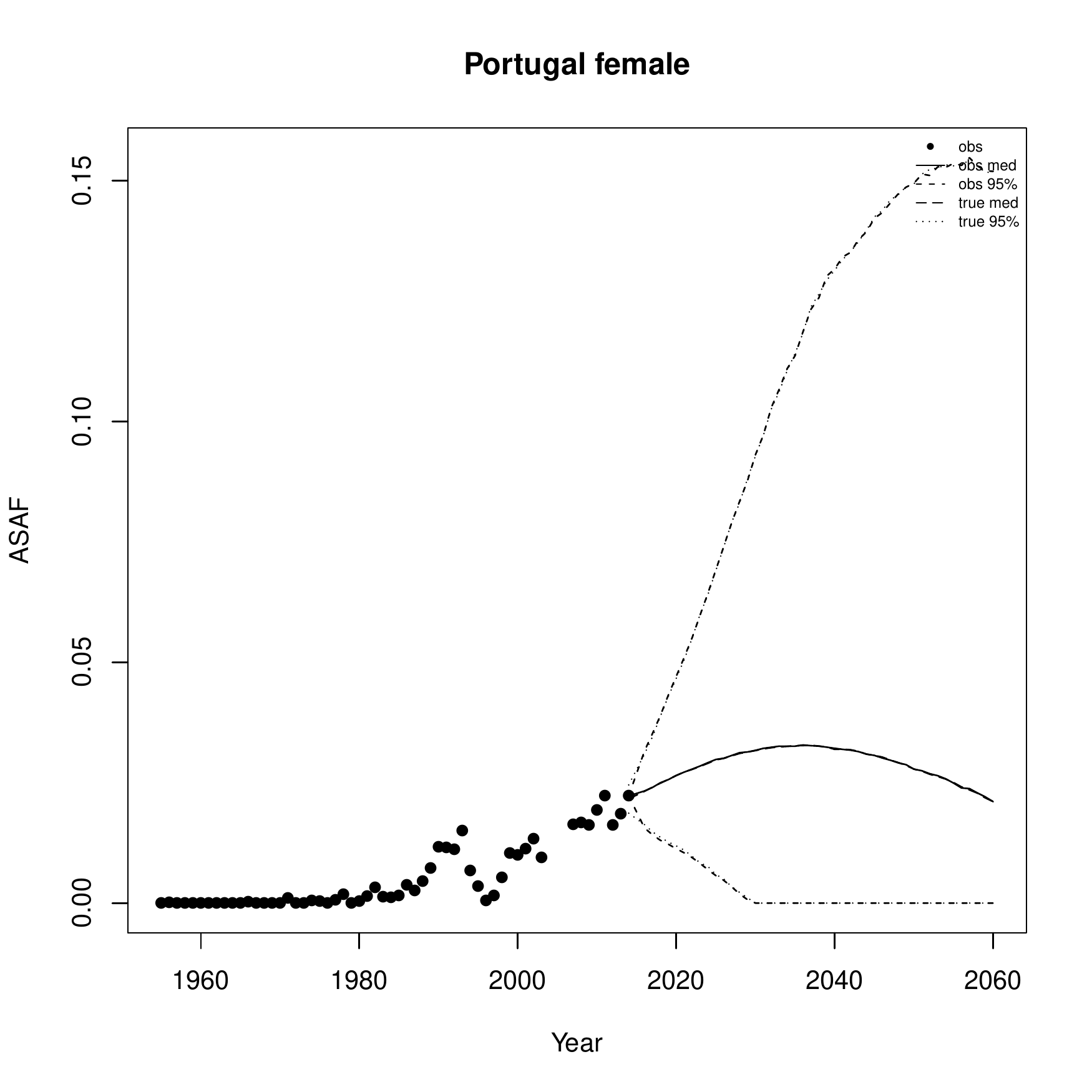}
	\end{center}
\end{figure}

\begin{figure}[H]
	\begin{center}
		\includegraphics[scale=0.43]{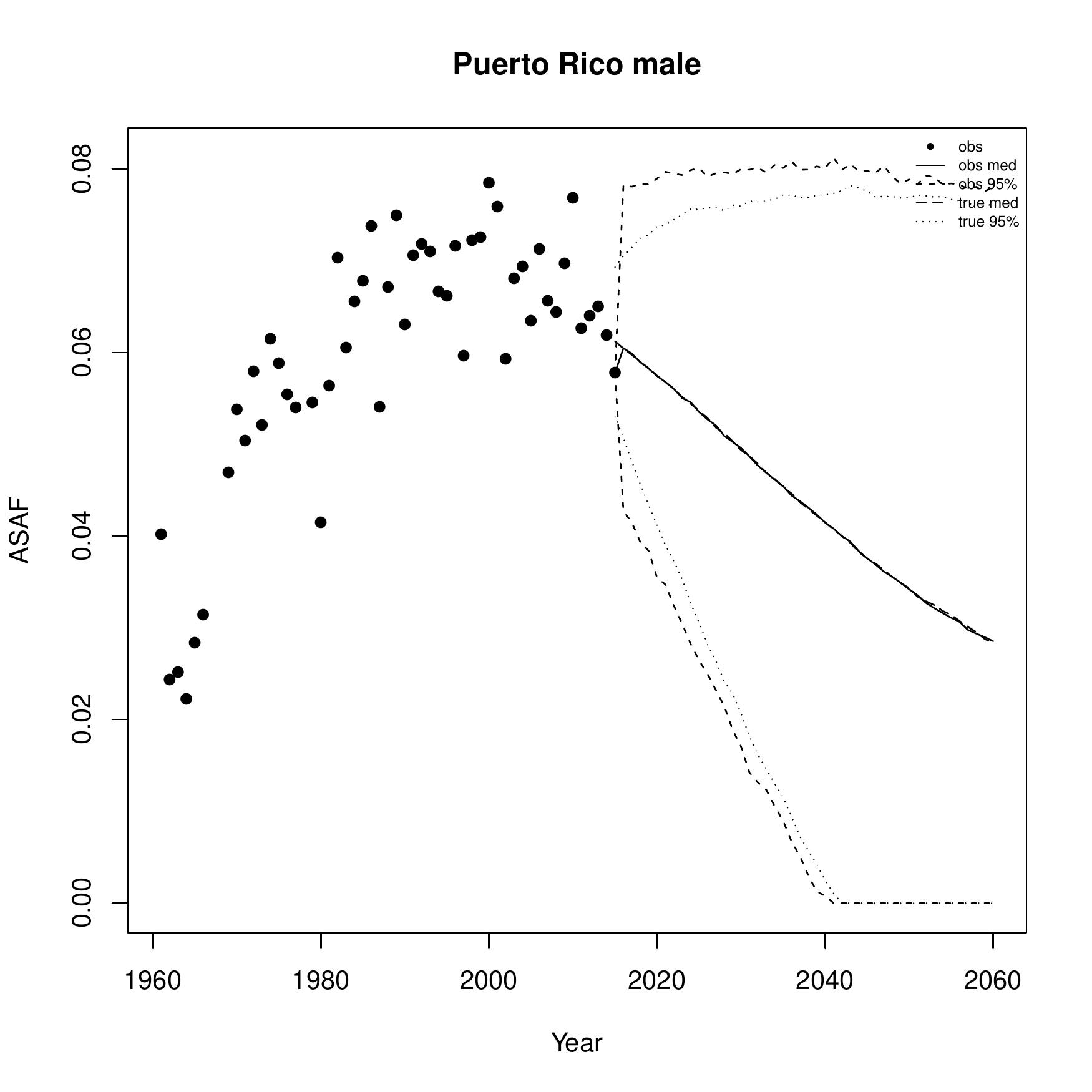}
		\includegraphics[scale=0.43]{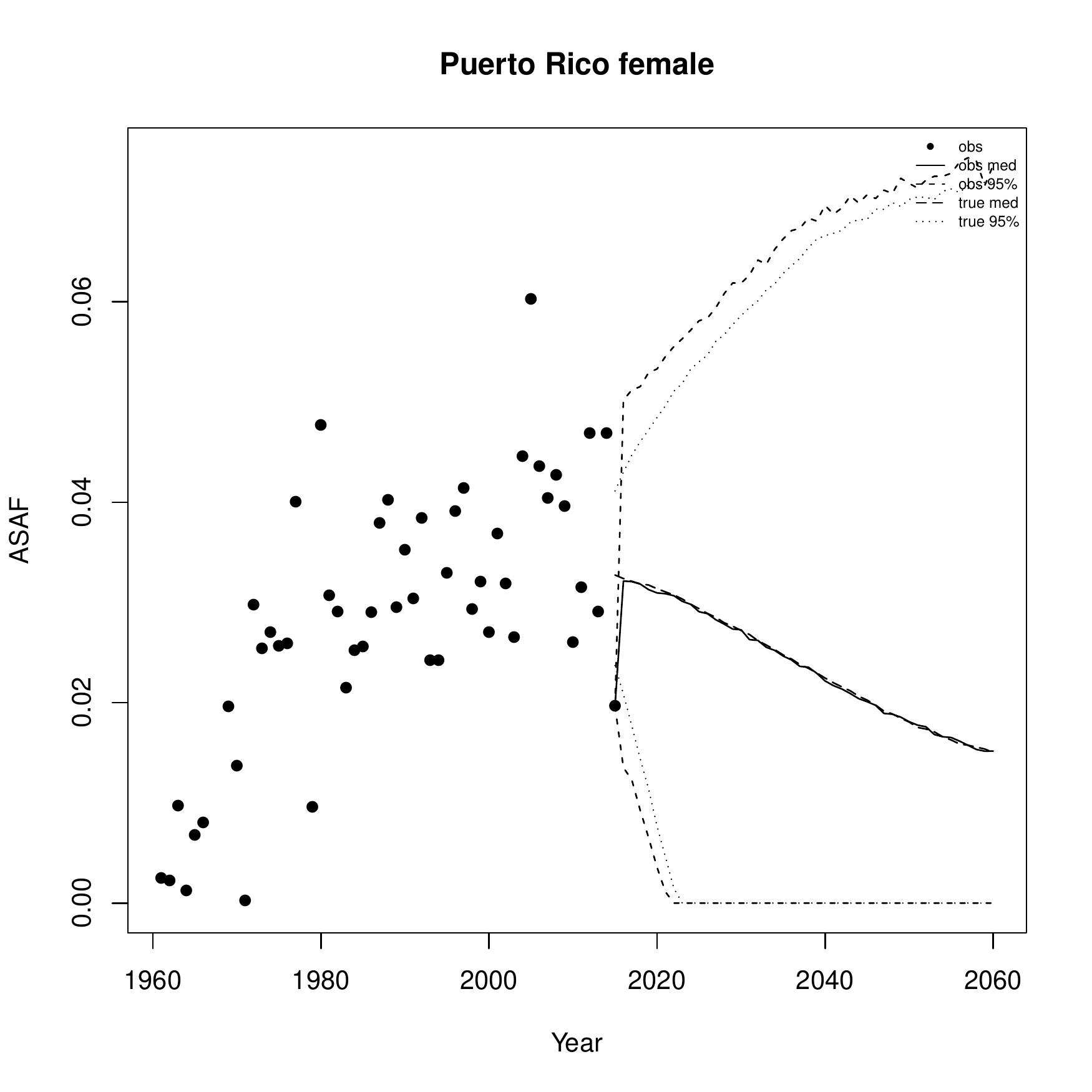}
		\includegraphics[scale=0.43]{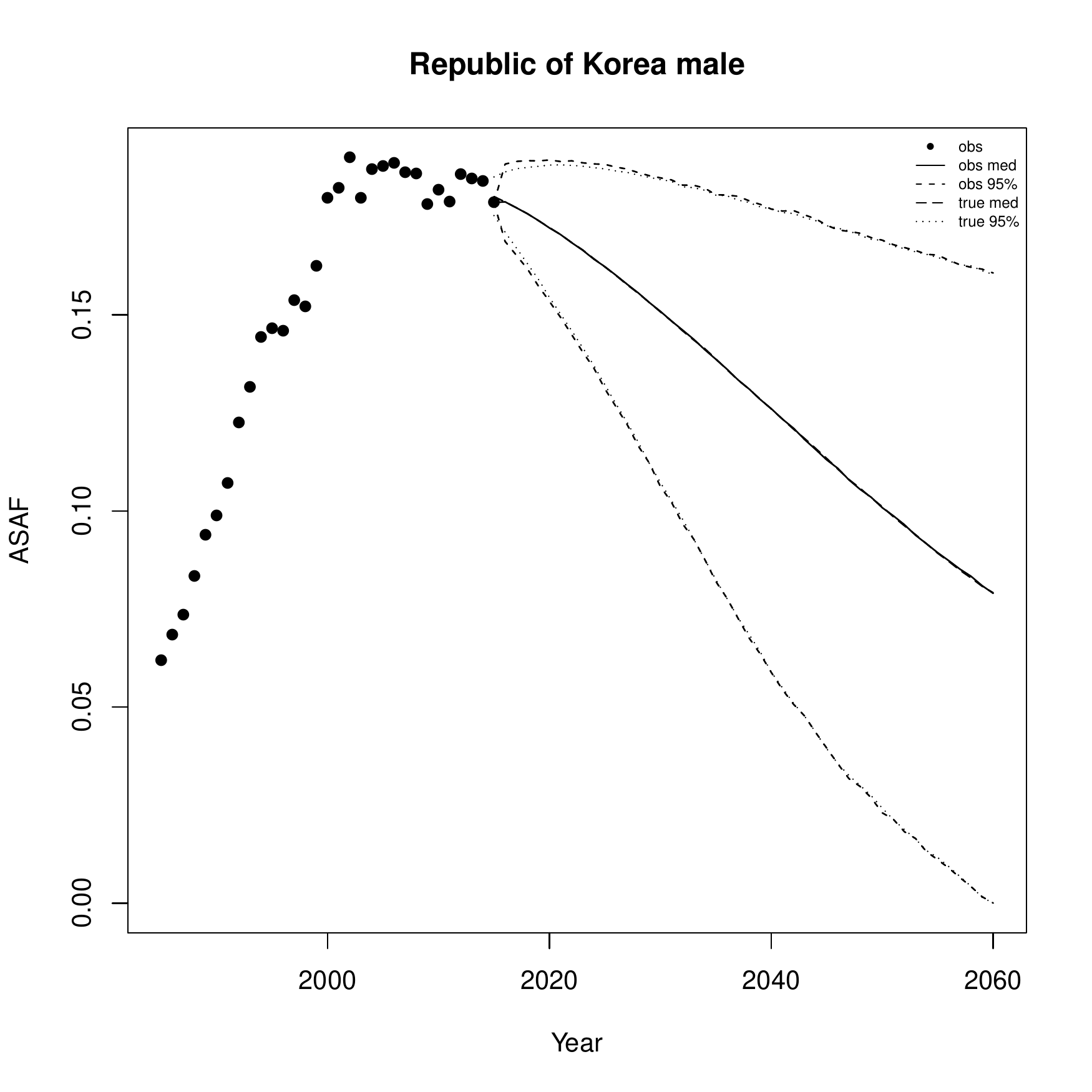}
		\includegraphics[scale=0.43]{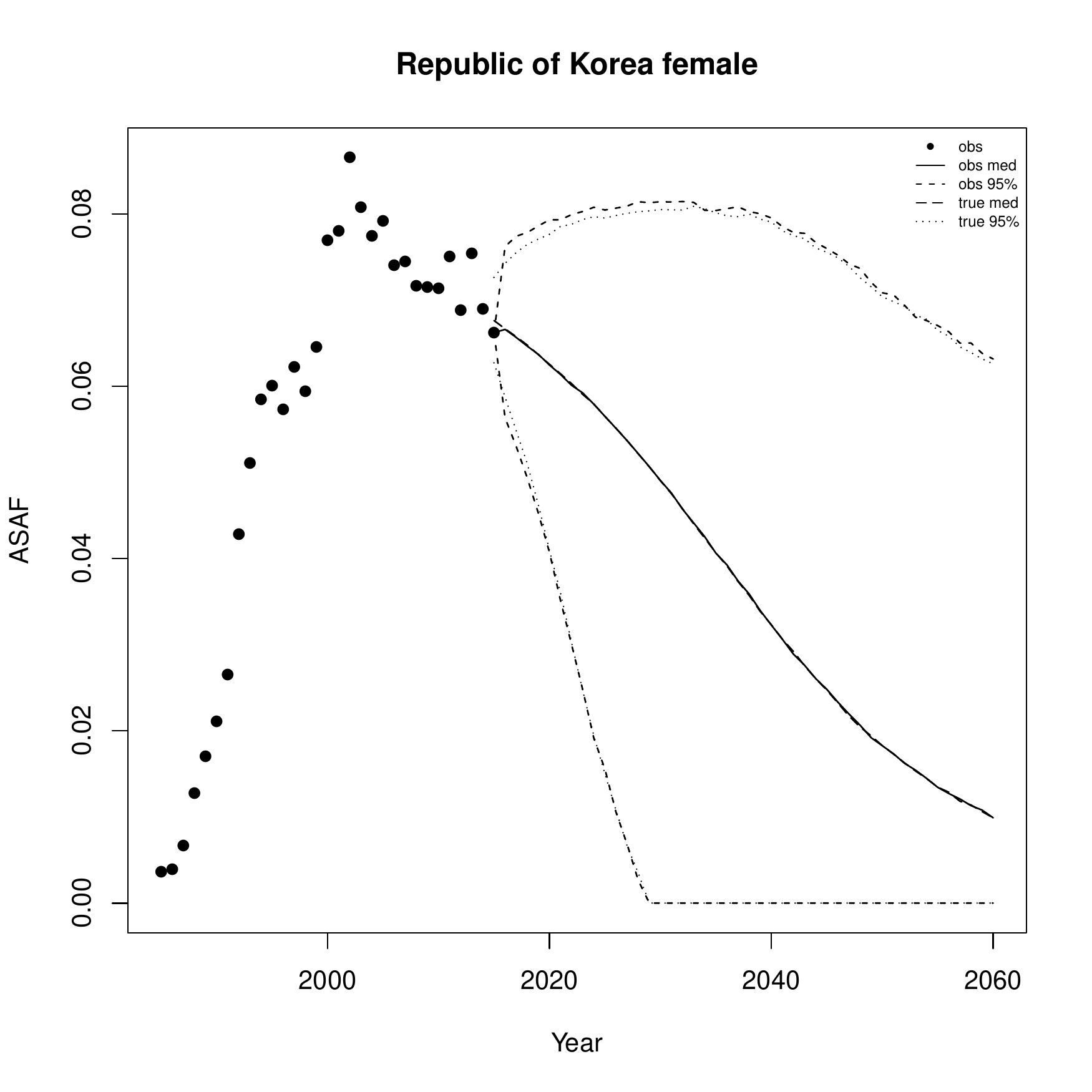}
		\includegraphics[scale=0.43]{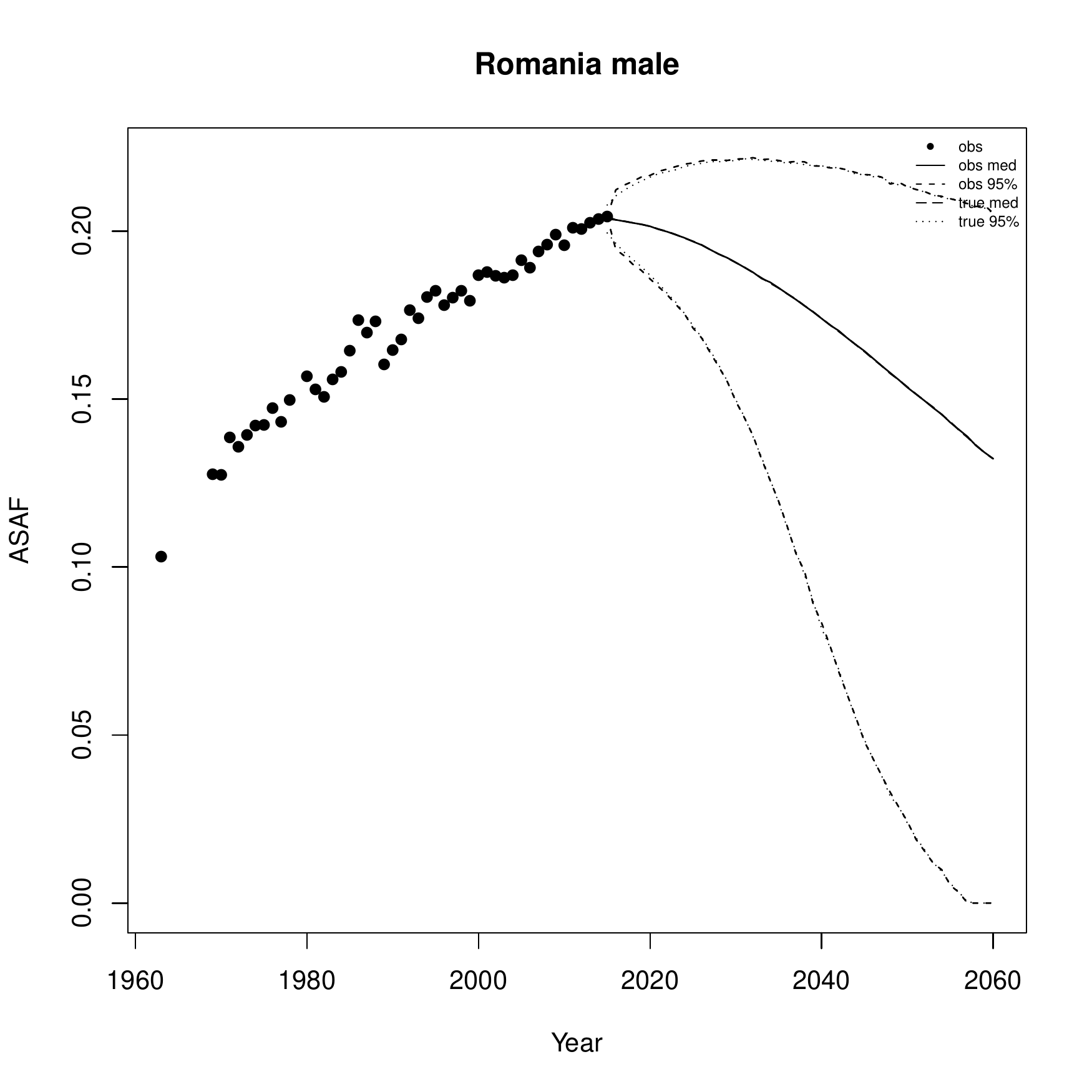}
		\includegraphics[scale=0.43]{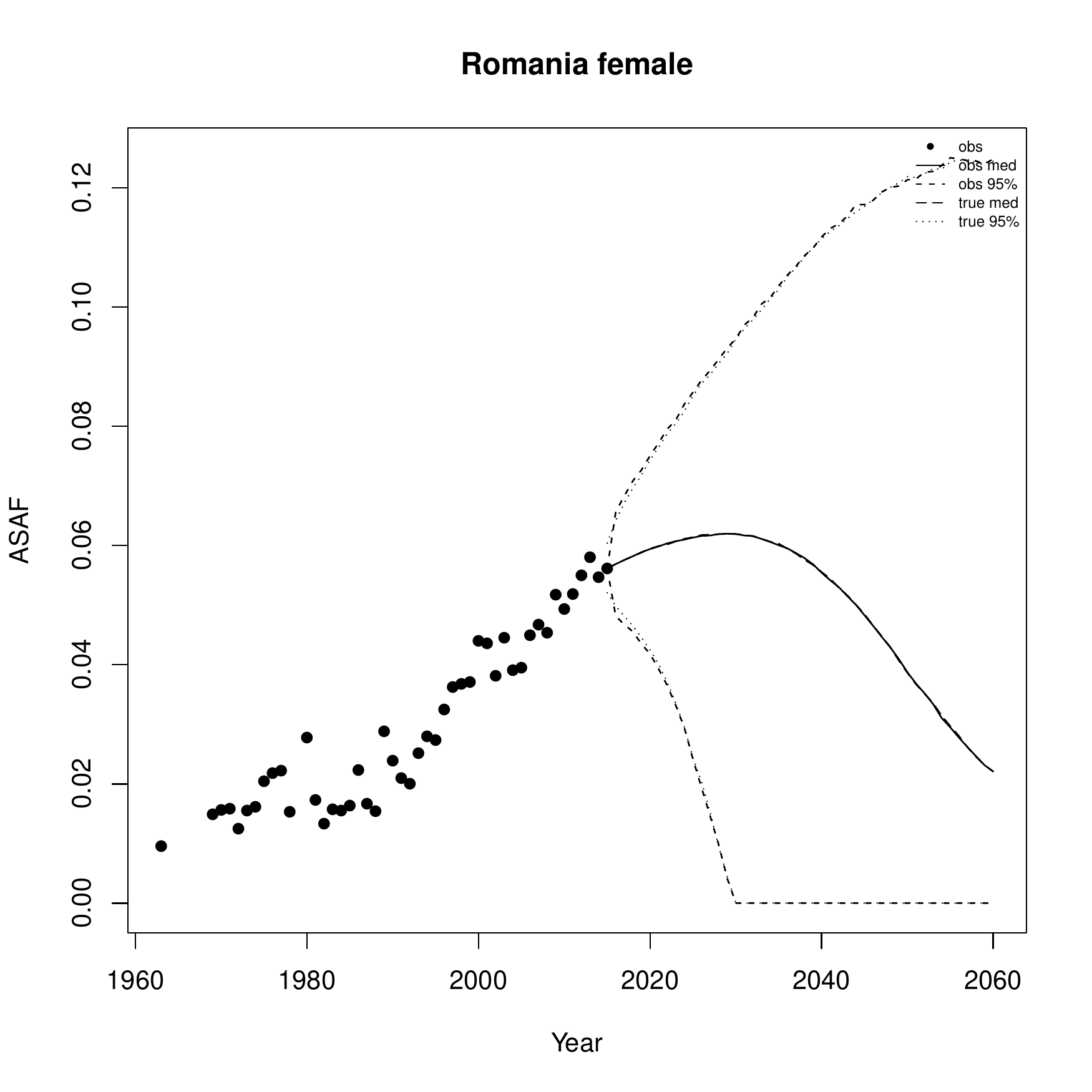}
	\end{center}
\end{figure}

\begin{figure}[H]
	\begin{center}
		\includegraphics[scale=0.43]{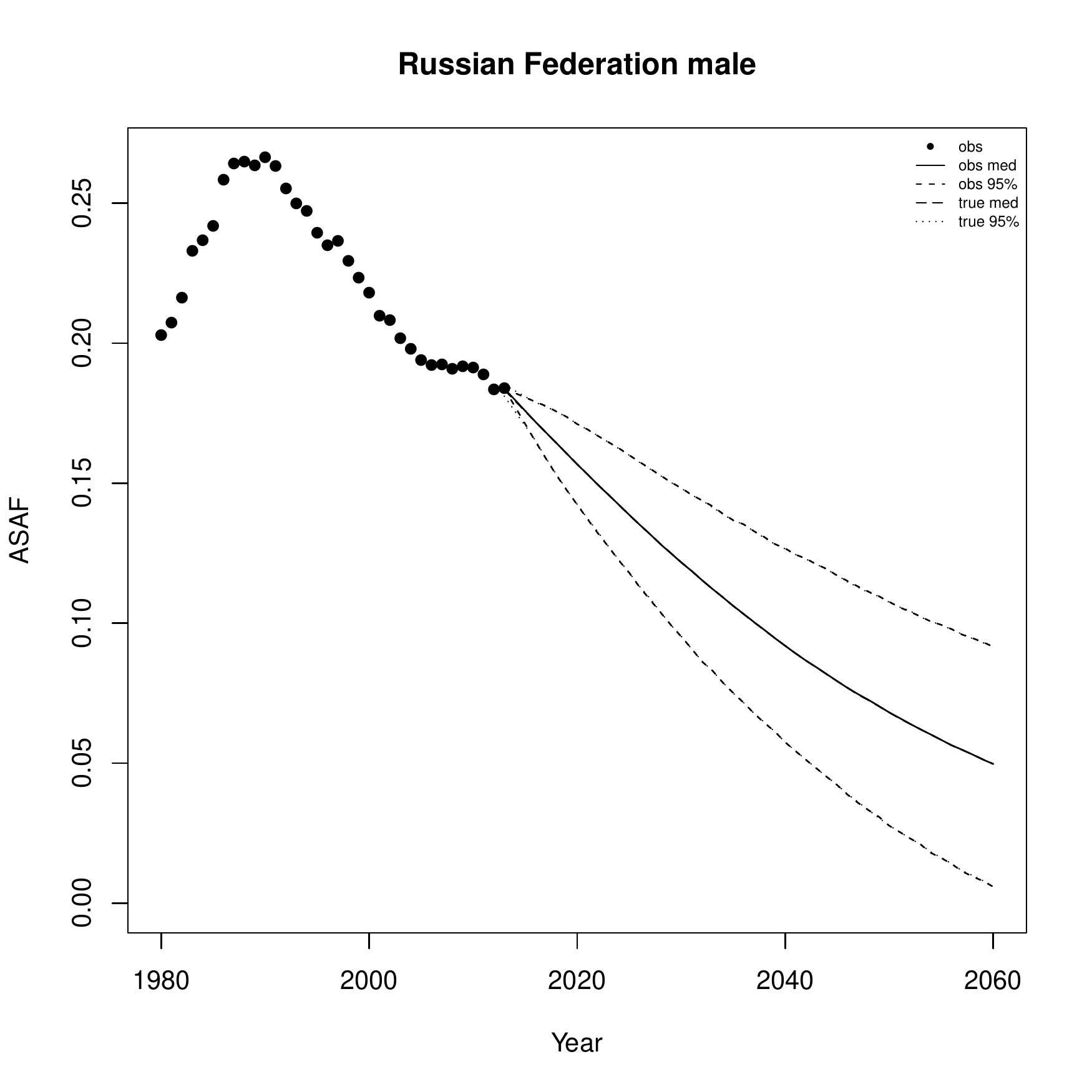}
		\includegraphics[scale=0.43]{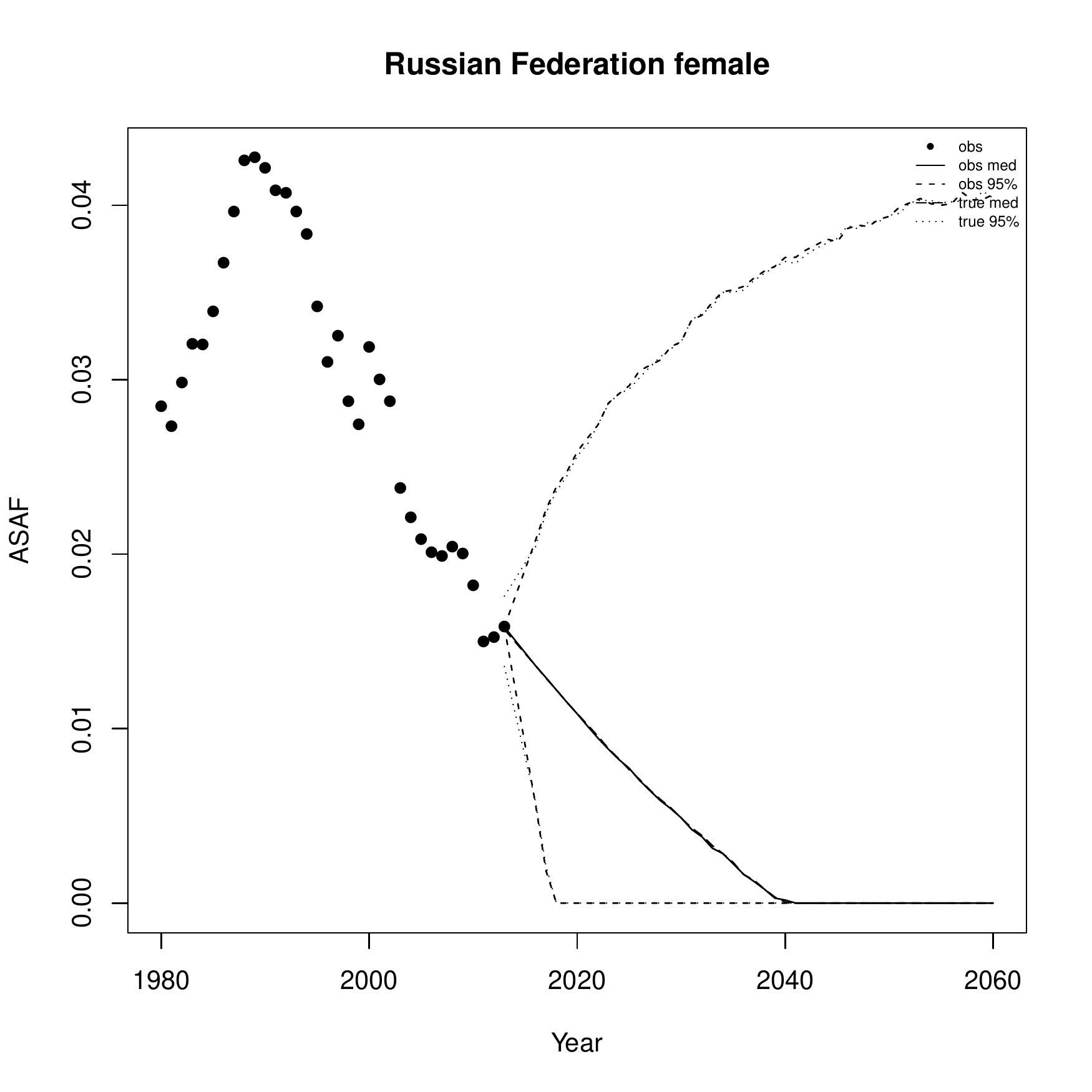}
		\includegraphics[scale=0.43]{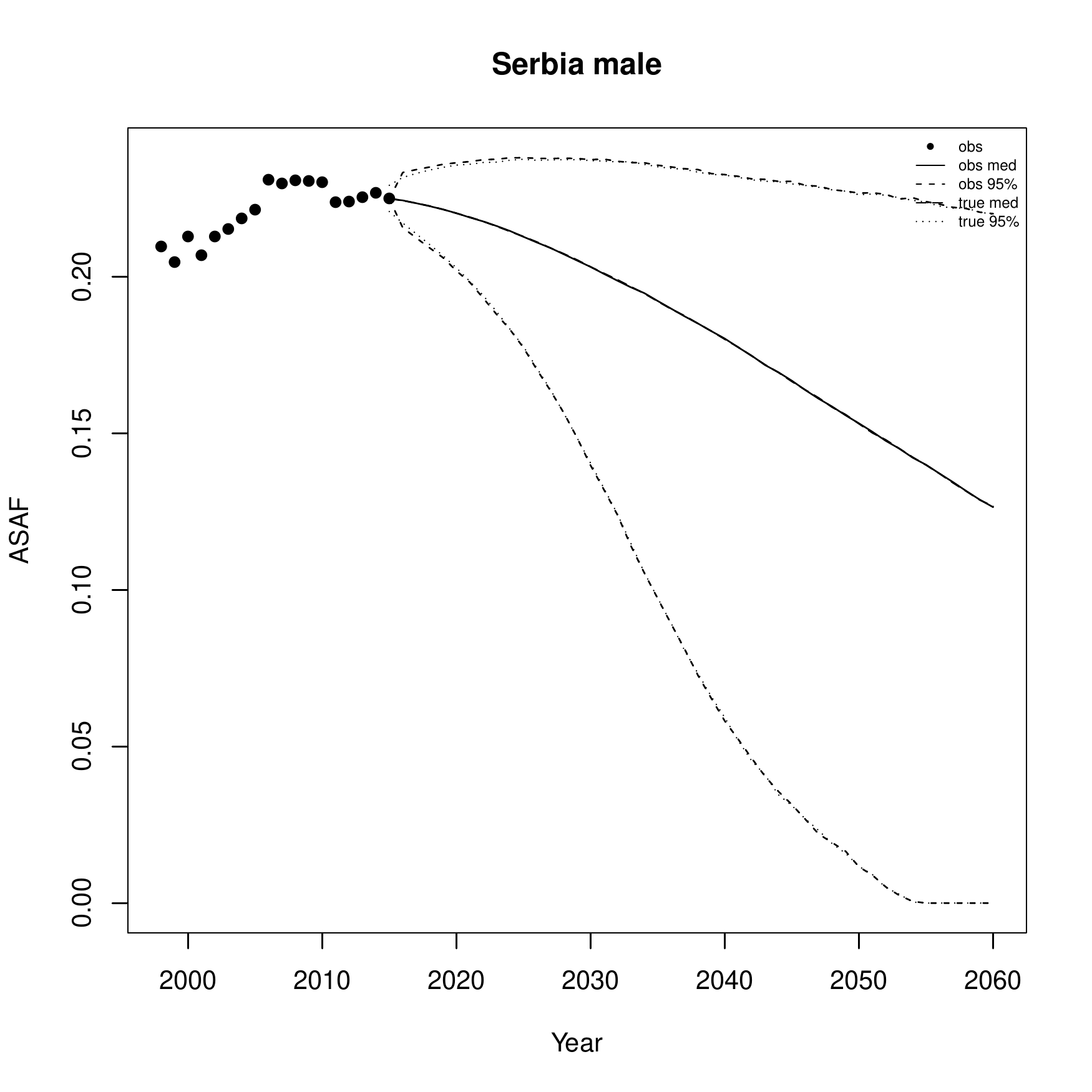}
		\includegraphics[scale=0.43]{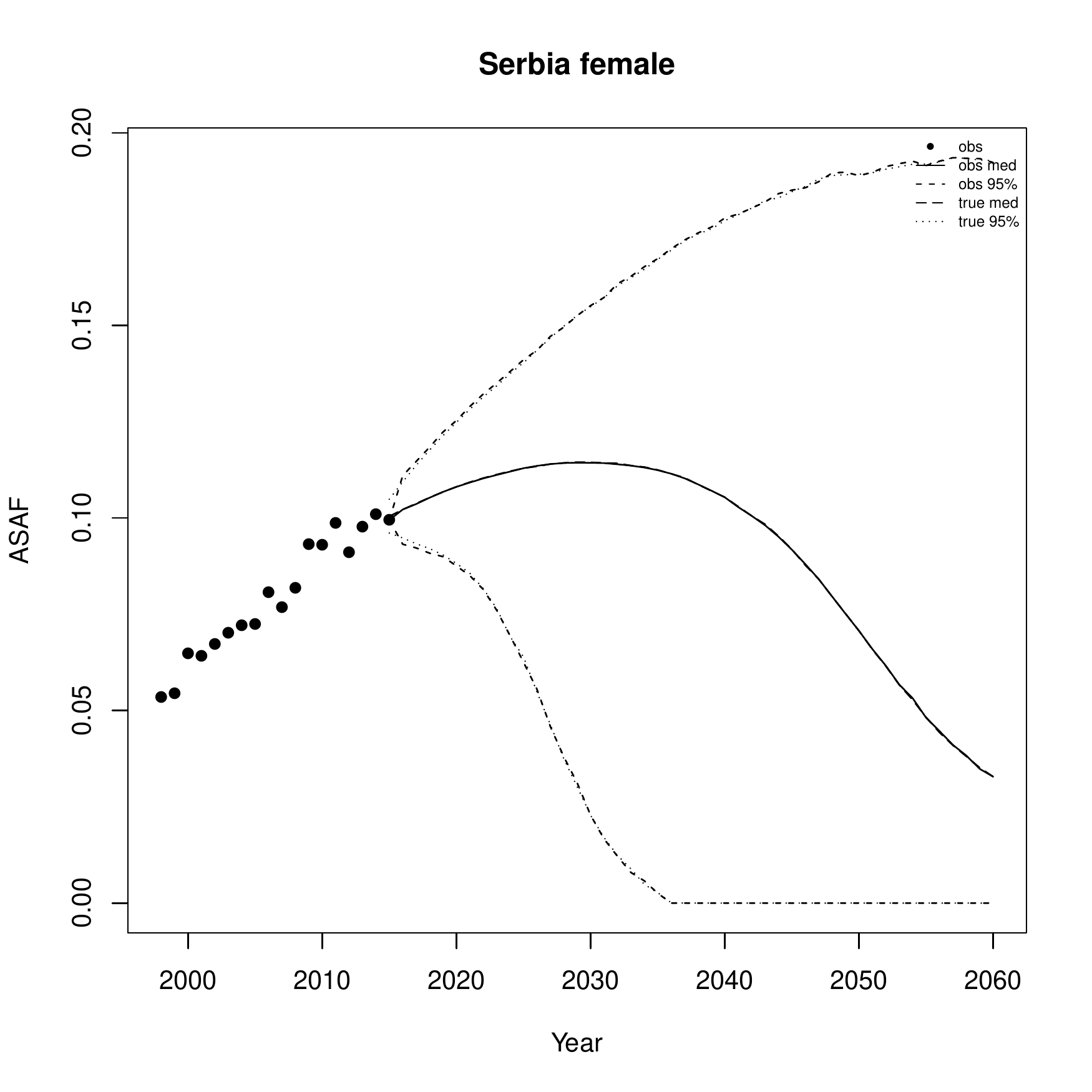}
		\includegraphics[scale=0.43]{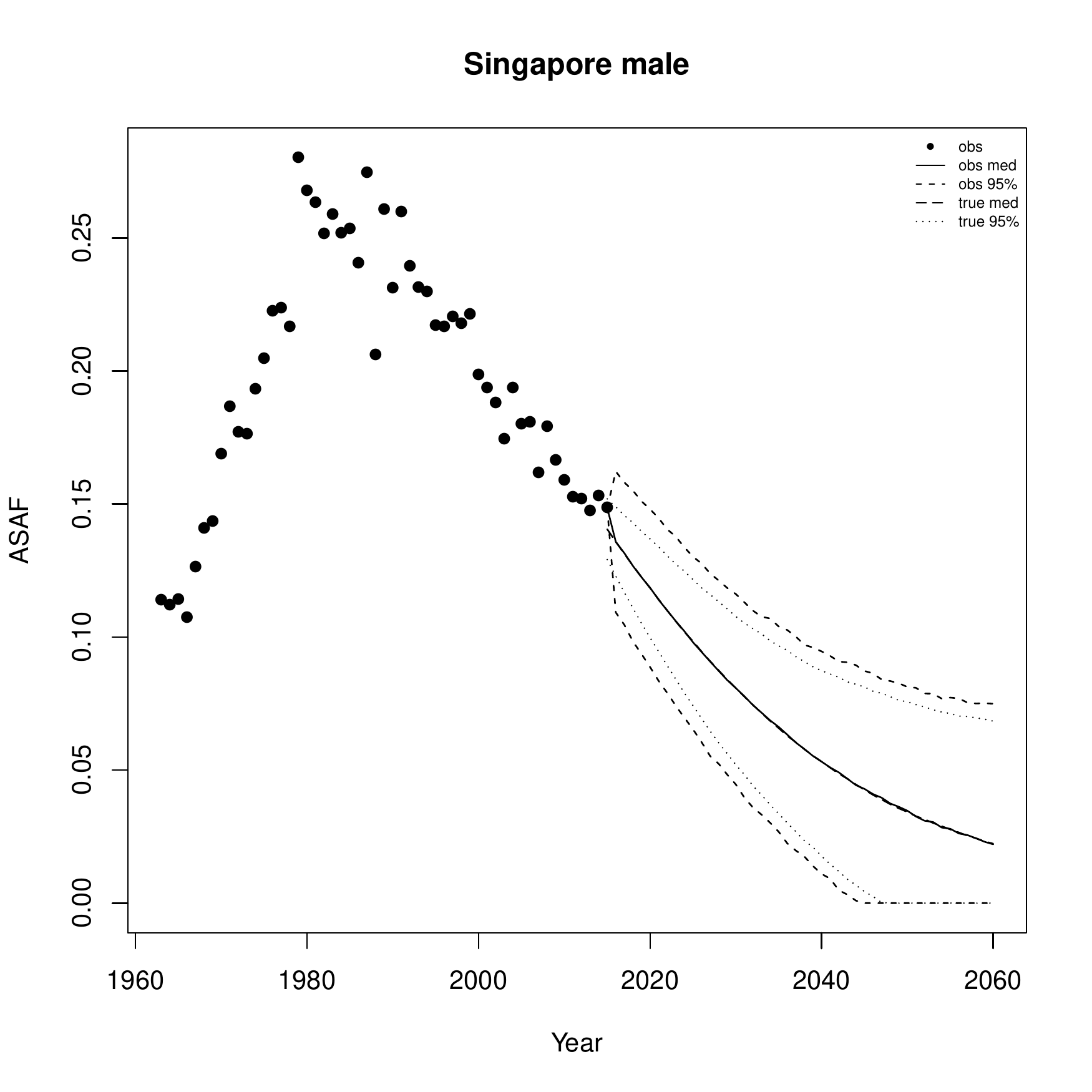}
		\includegraphics[scale=0.43]{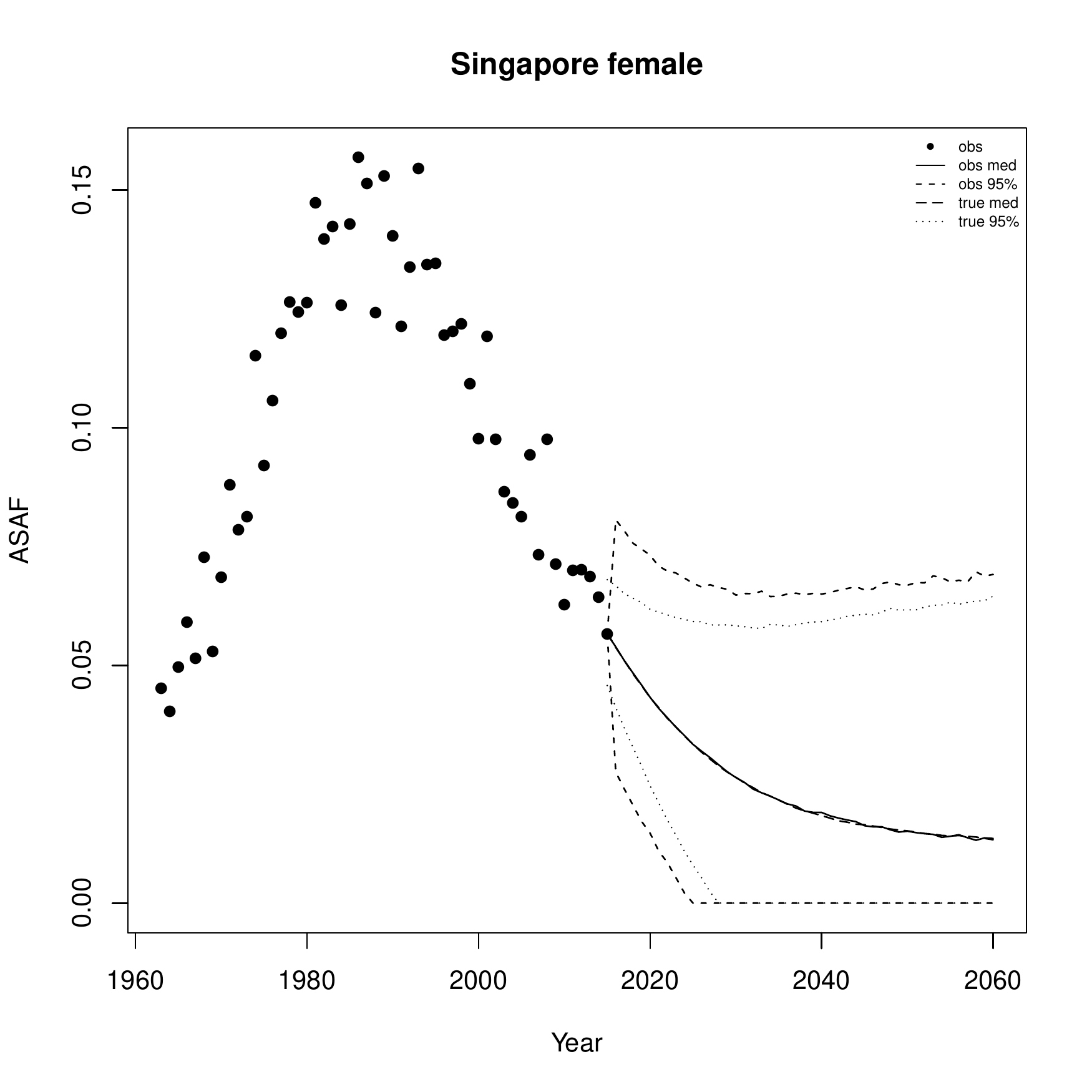}
	\end{center}
\end{figure}

\begin{figure}[H]
	\begin{center}
		\includegraphics[scale=0.43]{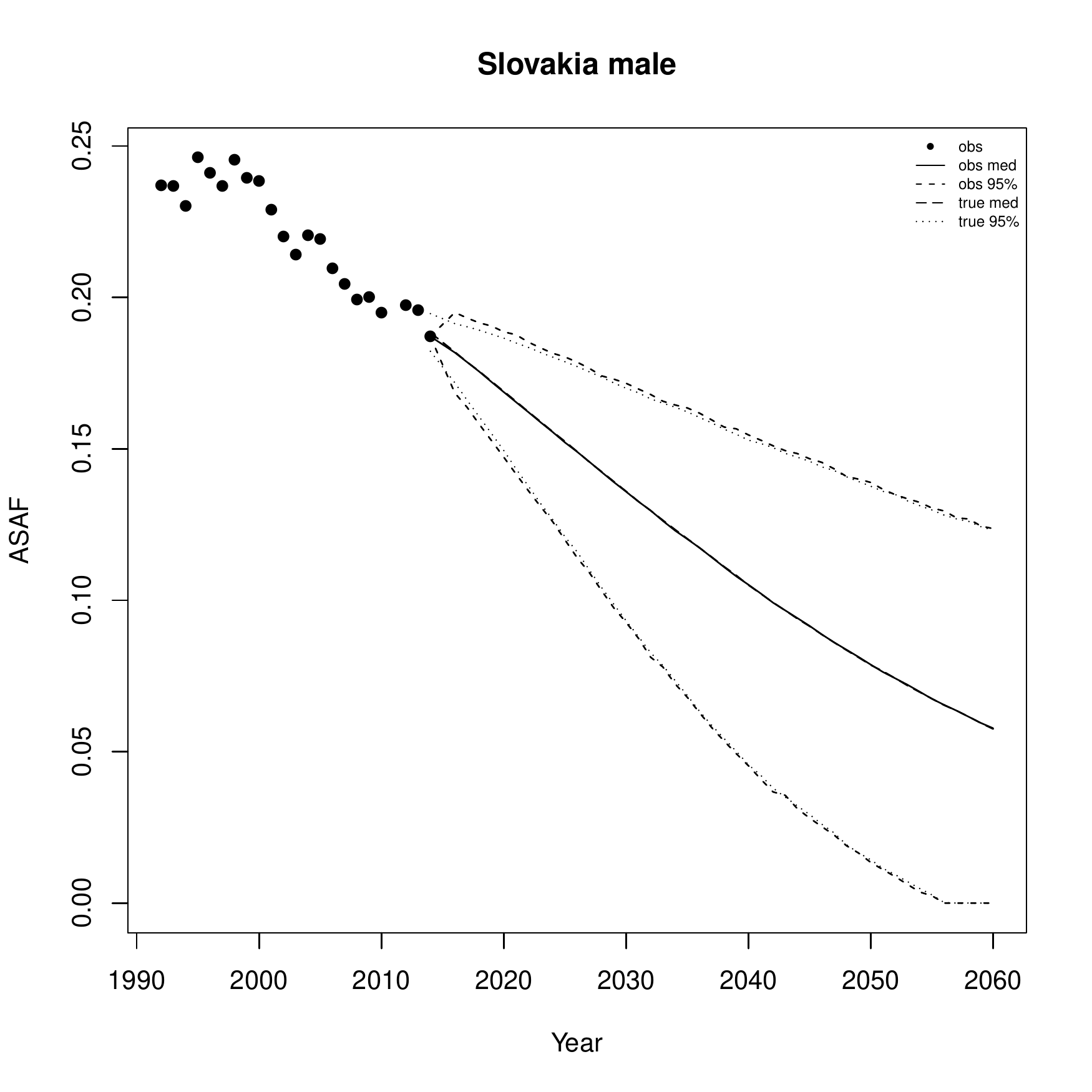}
		\includegraphics[scale=0.43]{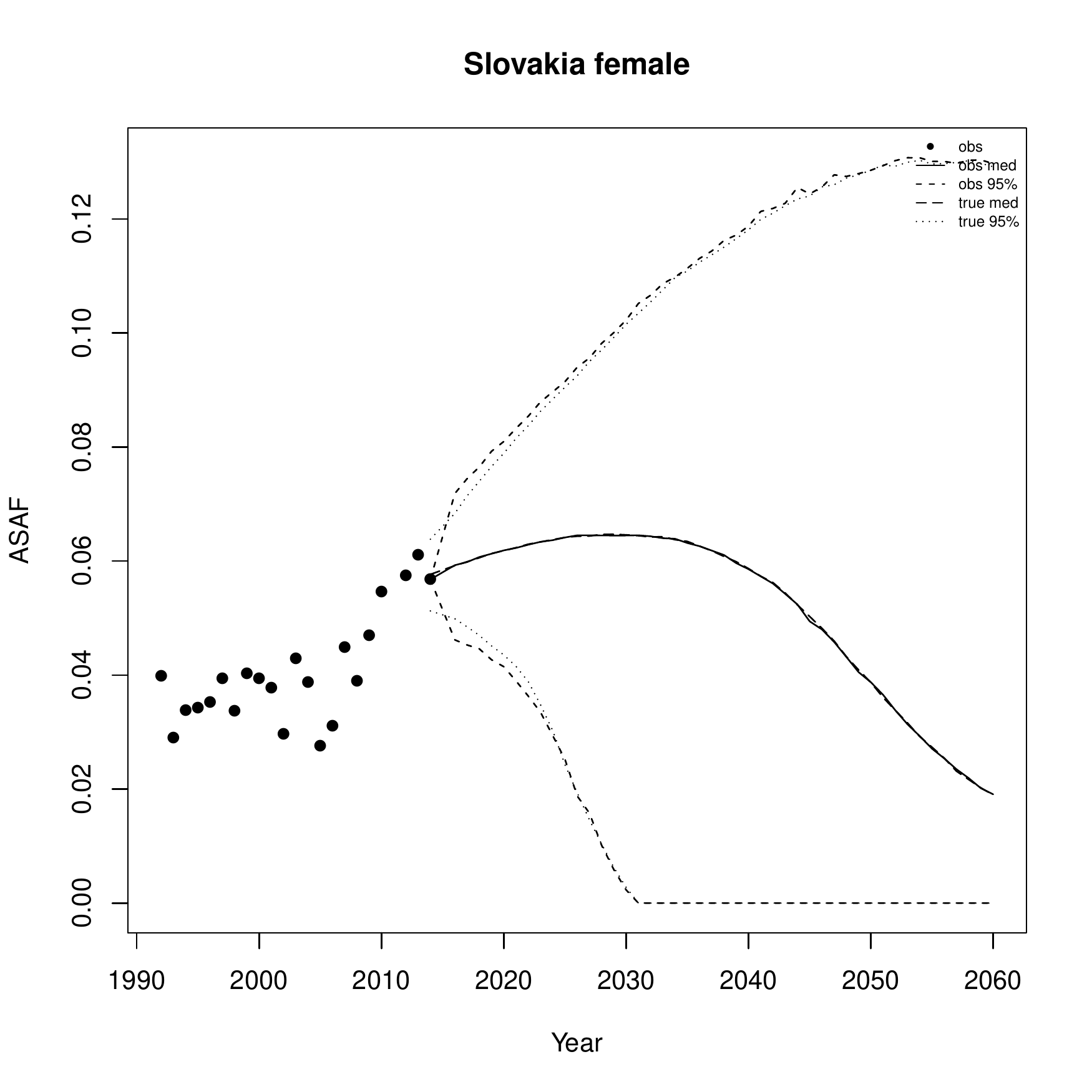}
		\includegraphics[scale=0.43]{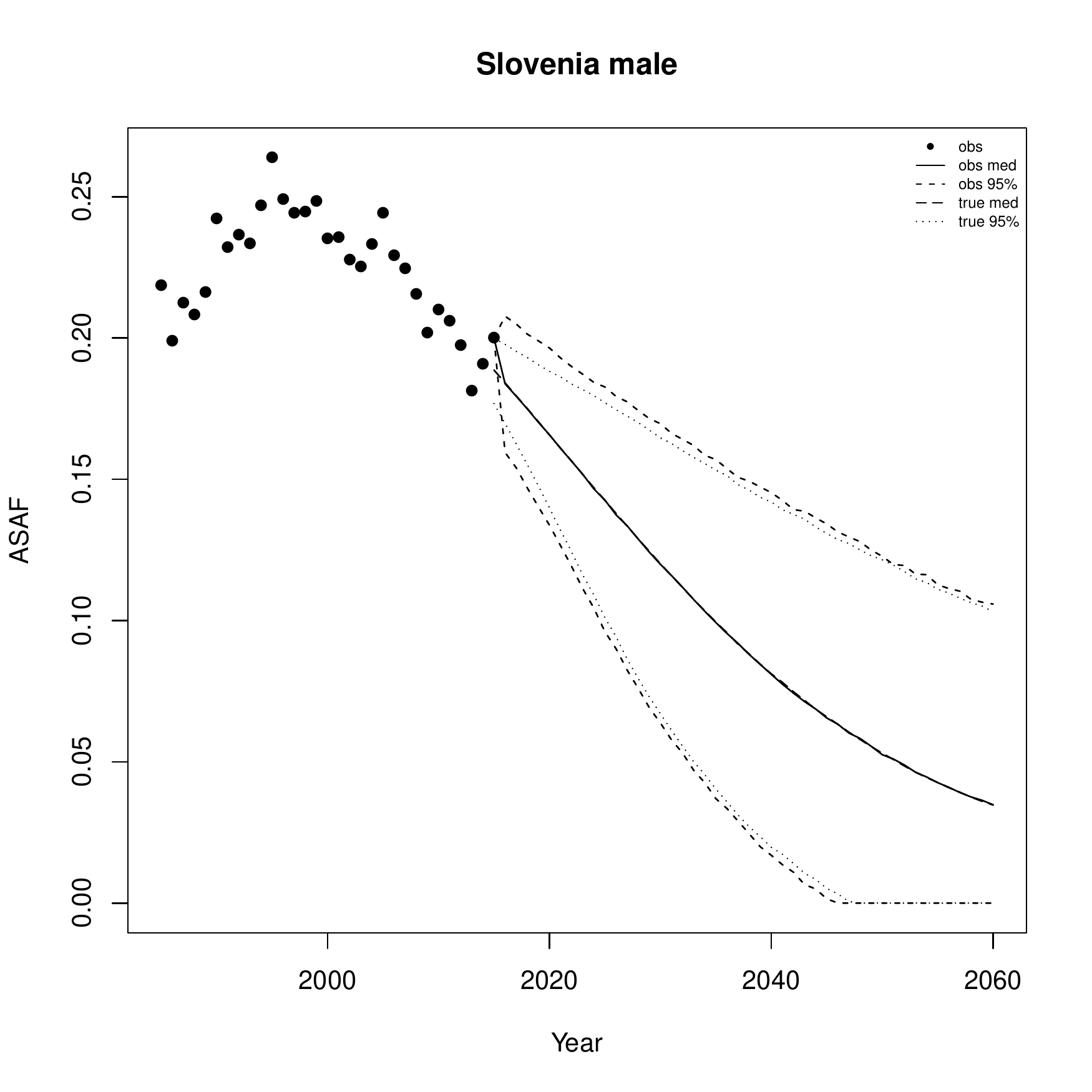}
		\includegraphics[scale=0.43]{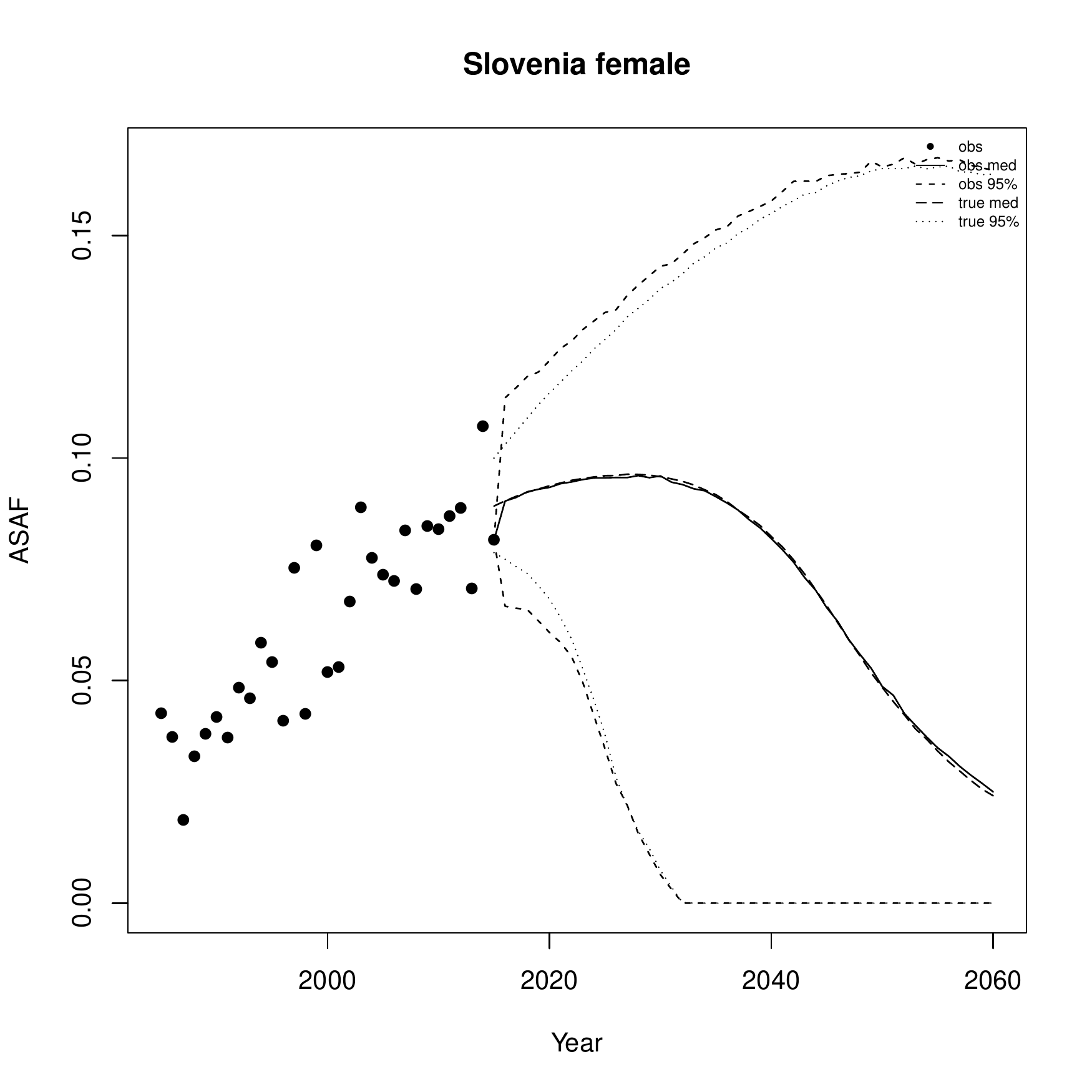}
		\includegraphics[scale=0.43]{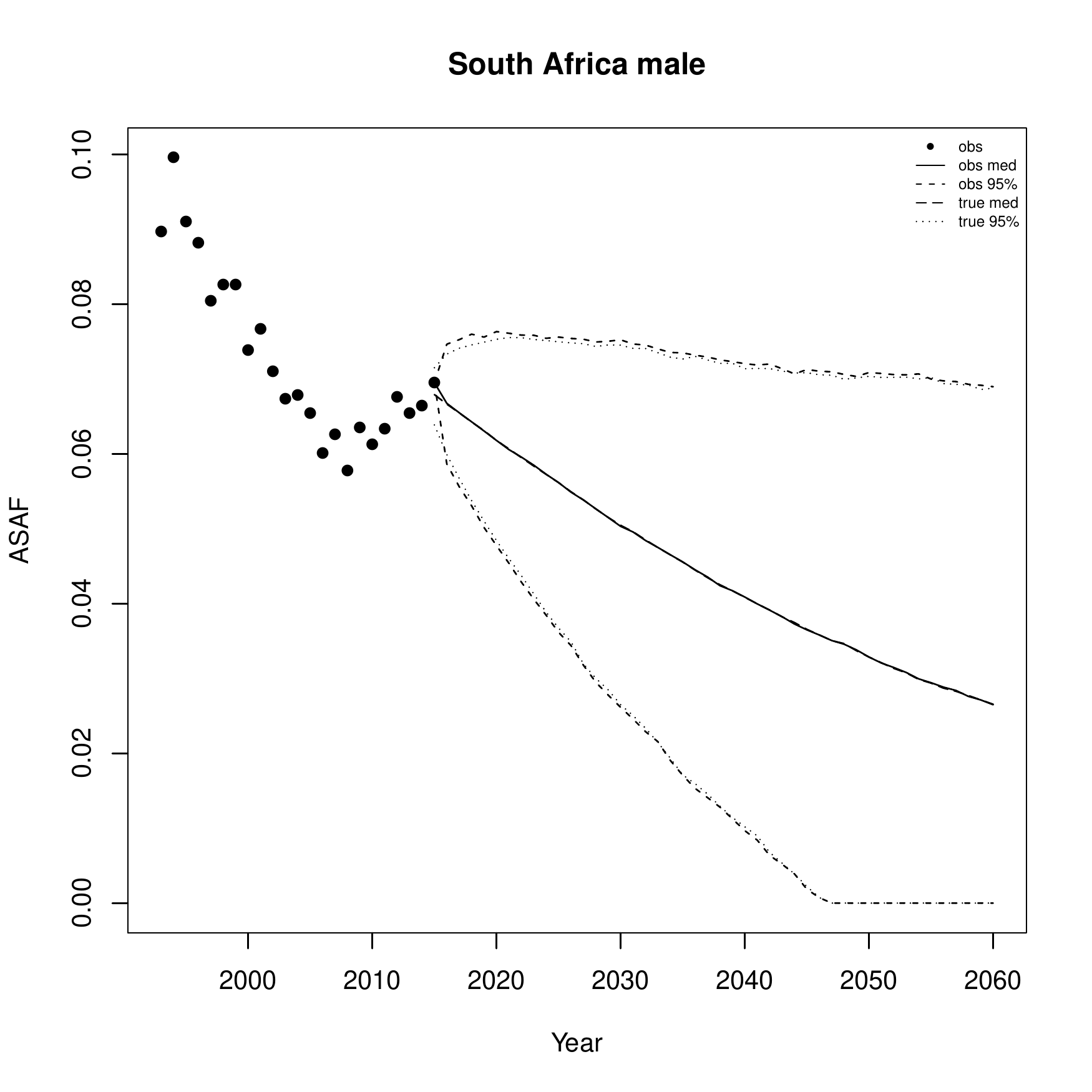}
		\includegraphics[scale=0.43]{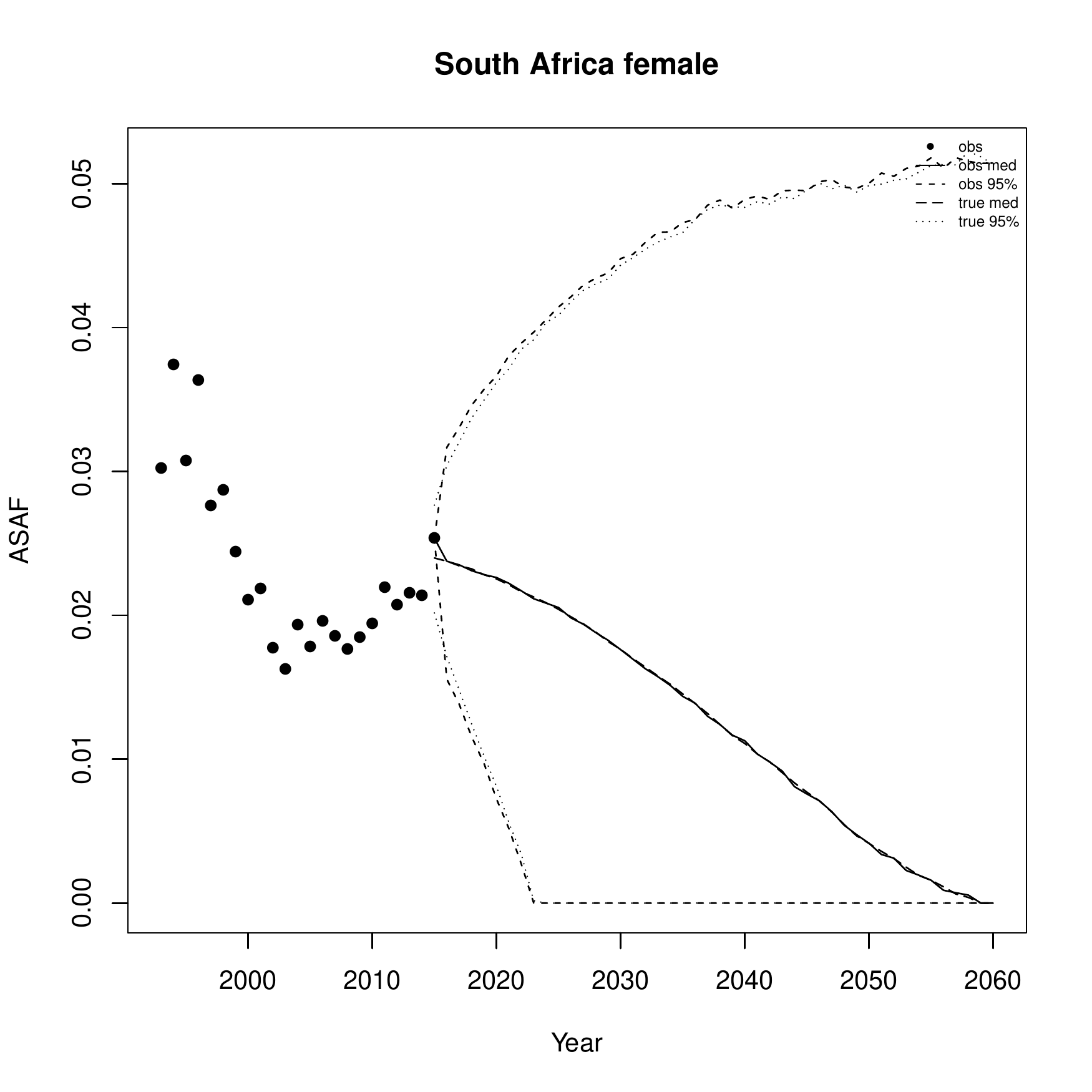}
	\end{center}
\end{figure}

\begin{figure}[H]
	\begin{center}
		\includegraphics[scale=0.43]{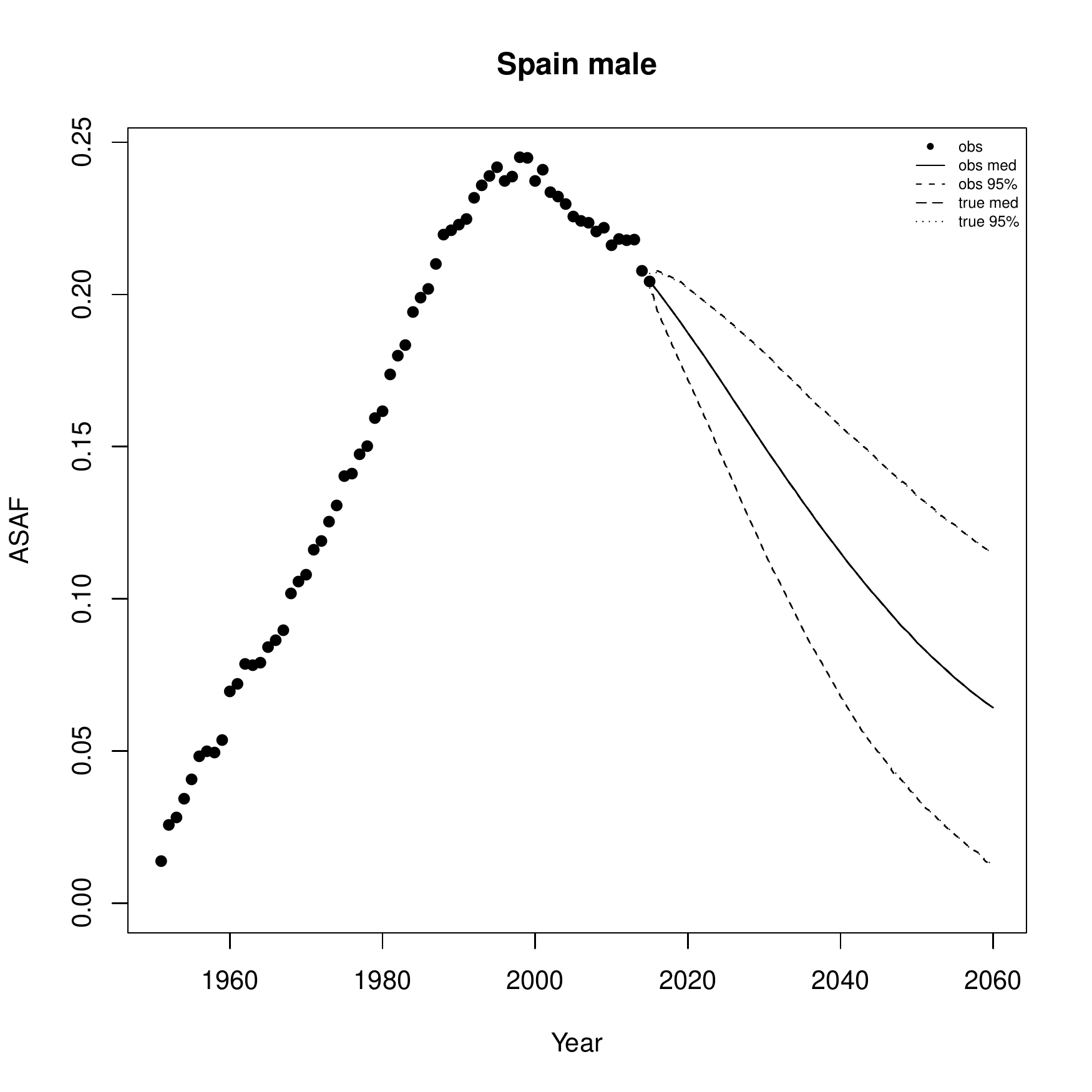}
		\includegraphics[scale=0.43]{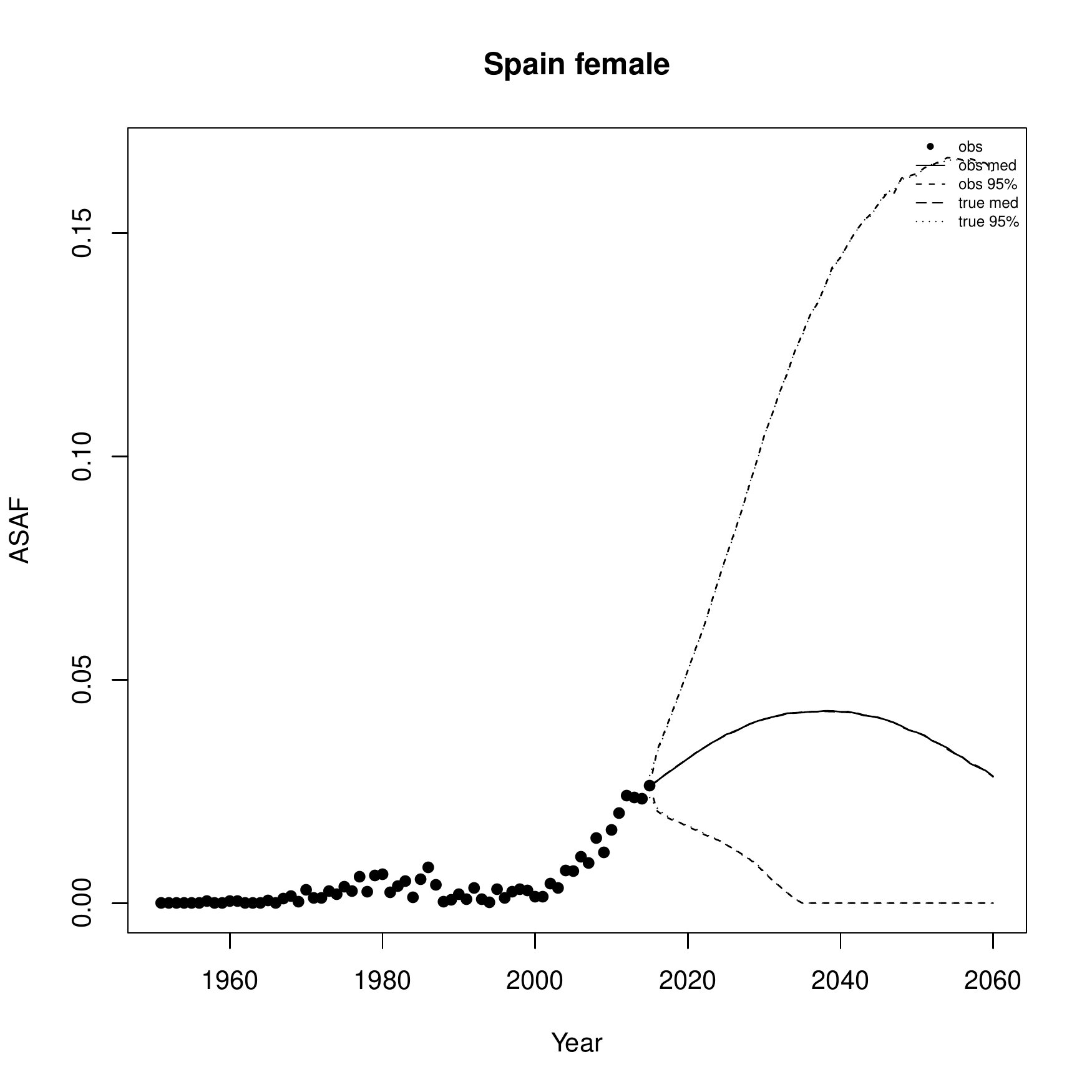}
		\includegraphics[scale=0.43]{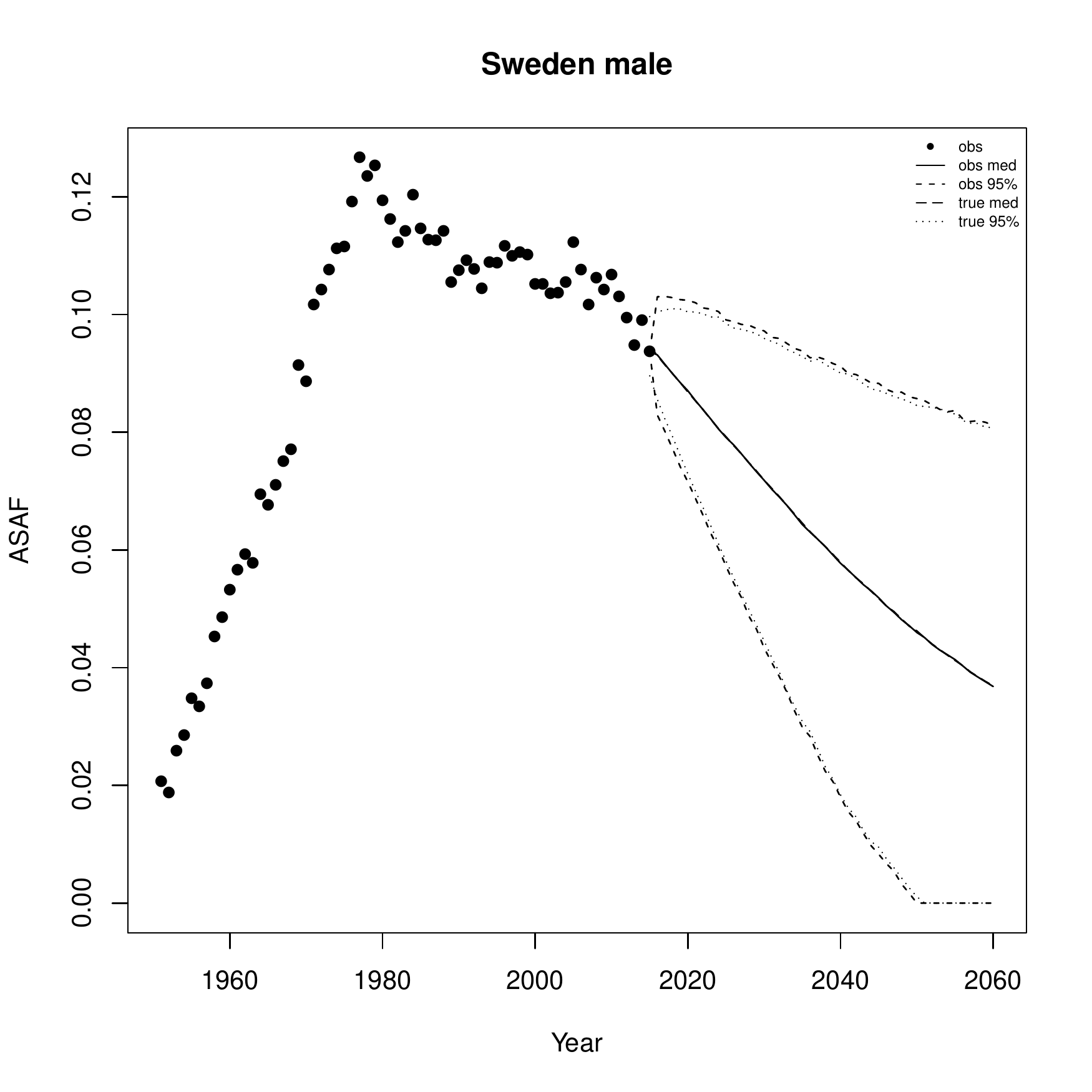}
		\includegraphics[scale=0.43]{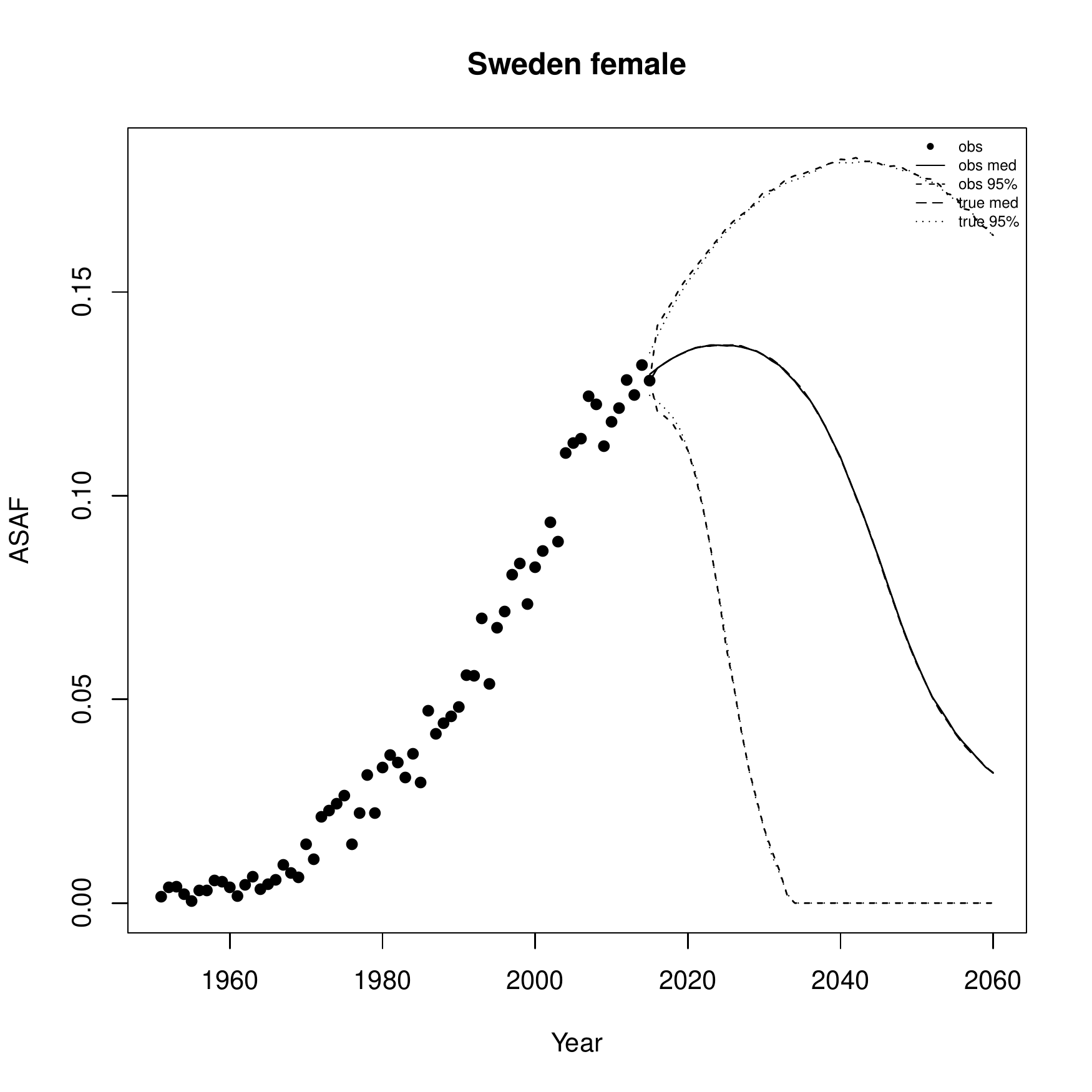}
		\includegraphics[scale=0.43]{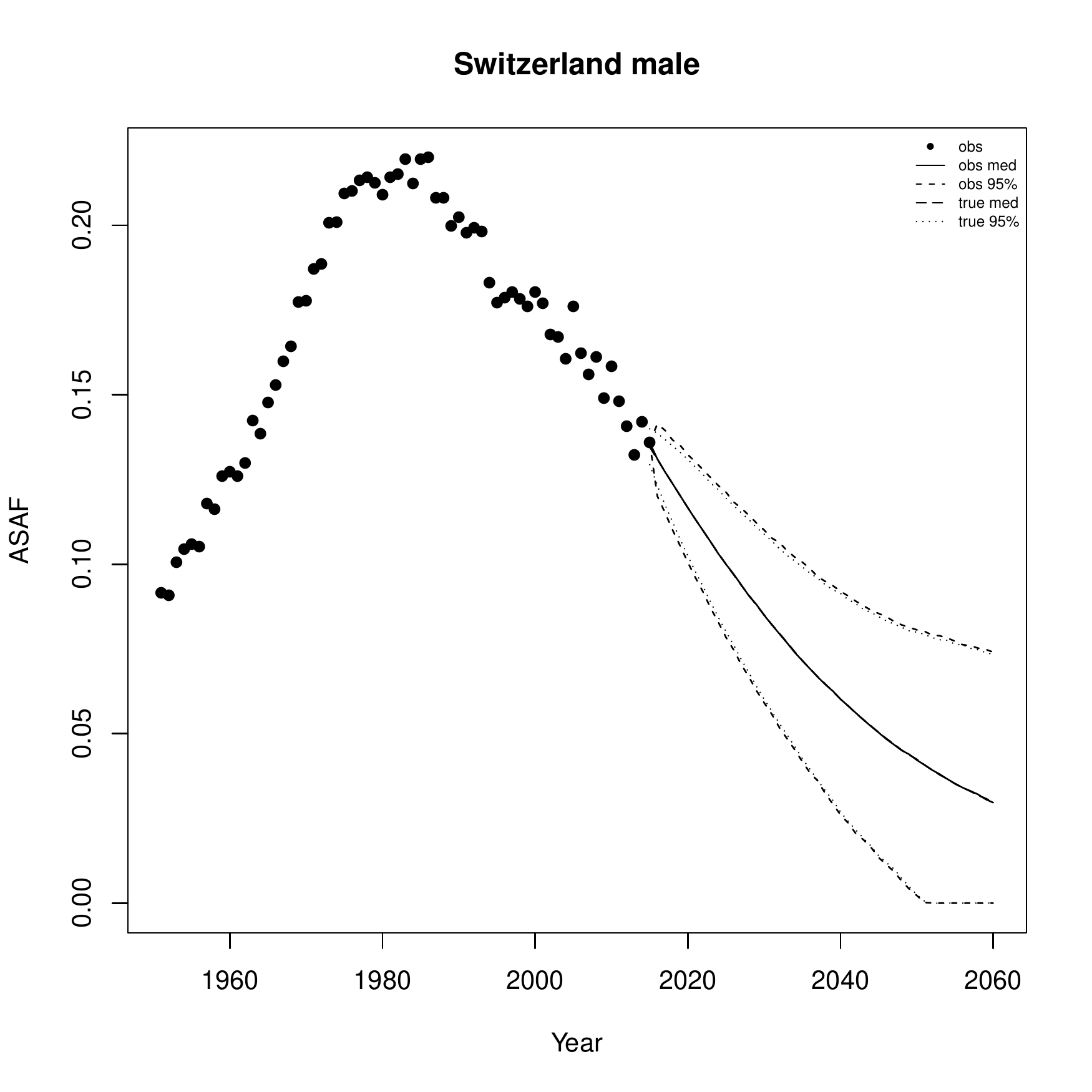}
		\includegraphics[scale=0.43]{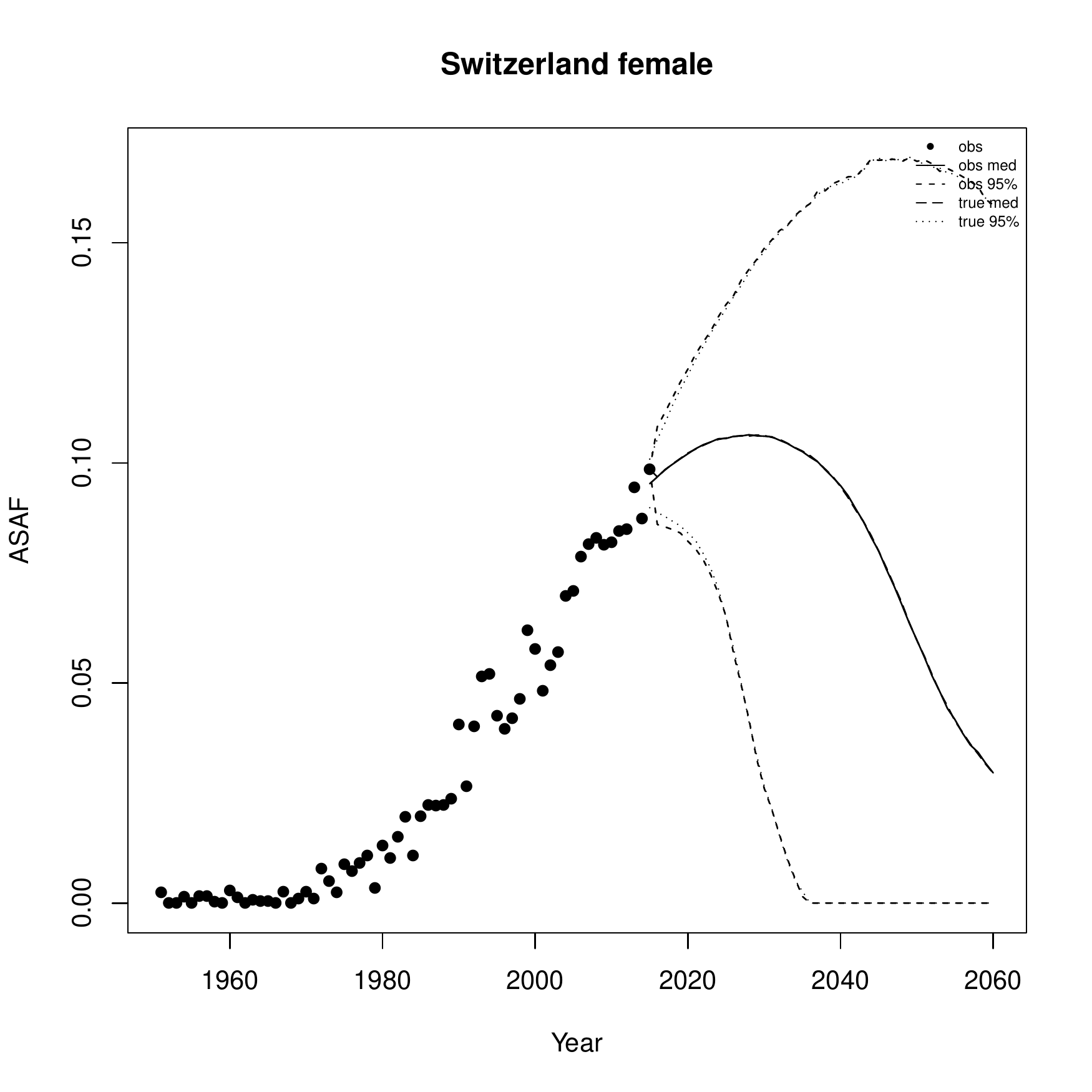}
	\end{center}
\end{figure}

\begin{figure}[H]
	\begin{center}
		\includegraphics[scale=0.43]{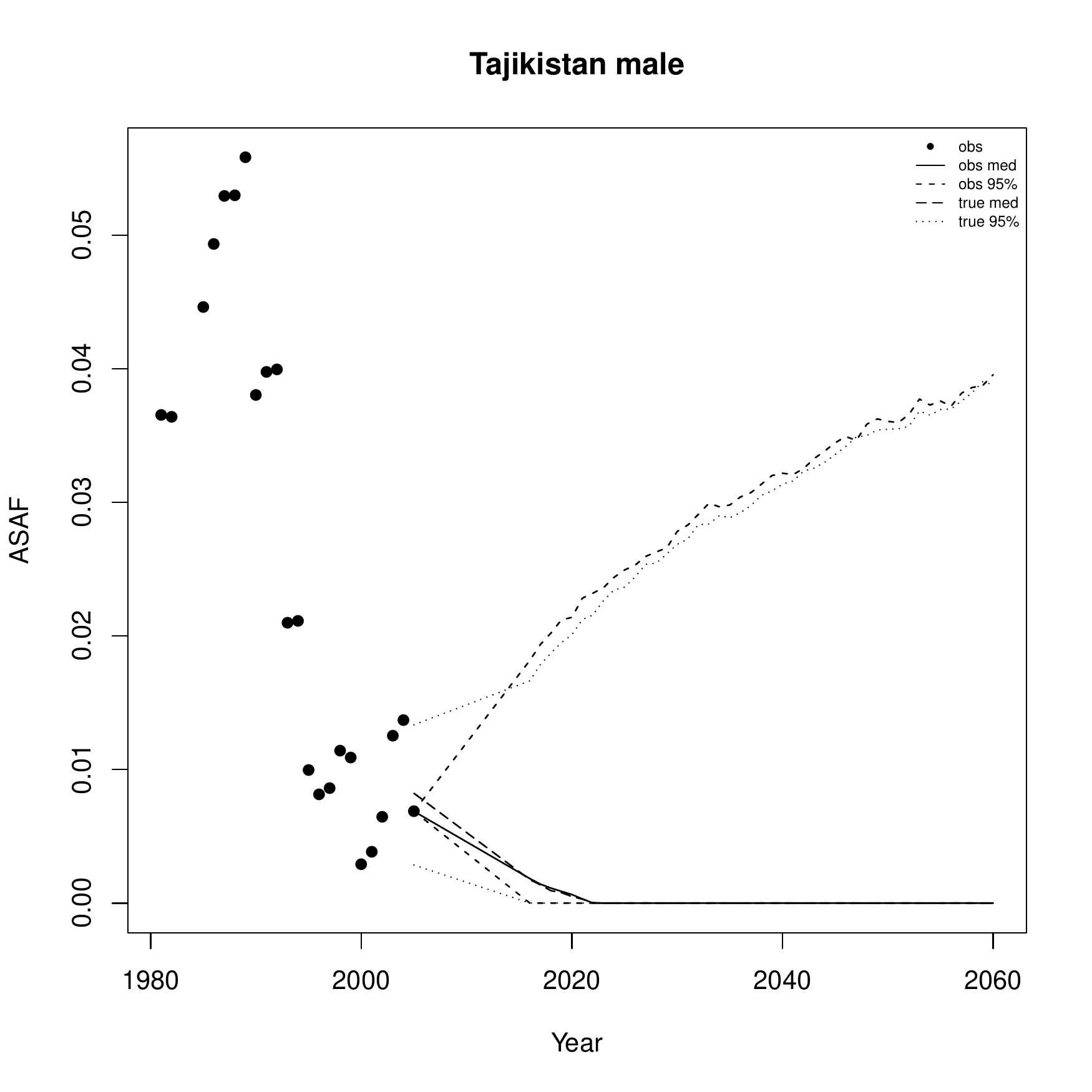}
		\includegraphics[scale=0.43]{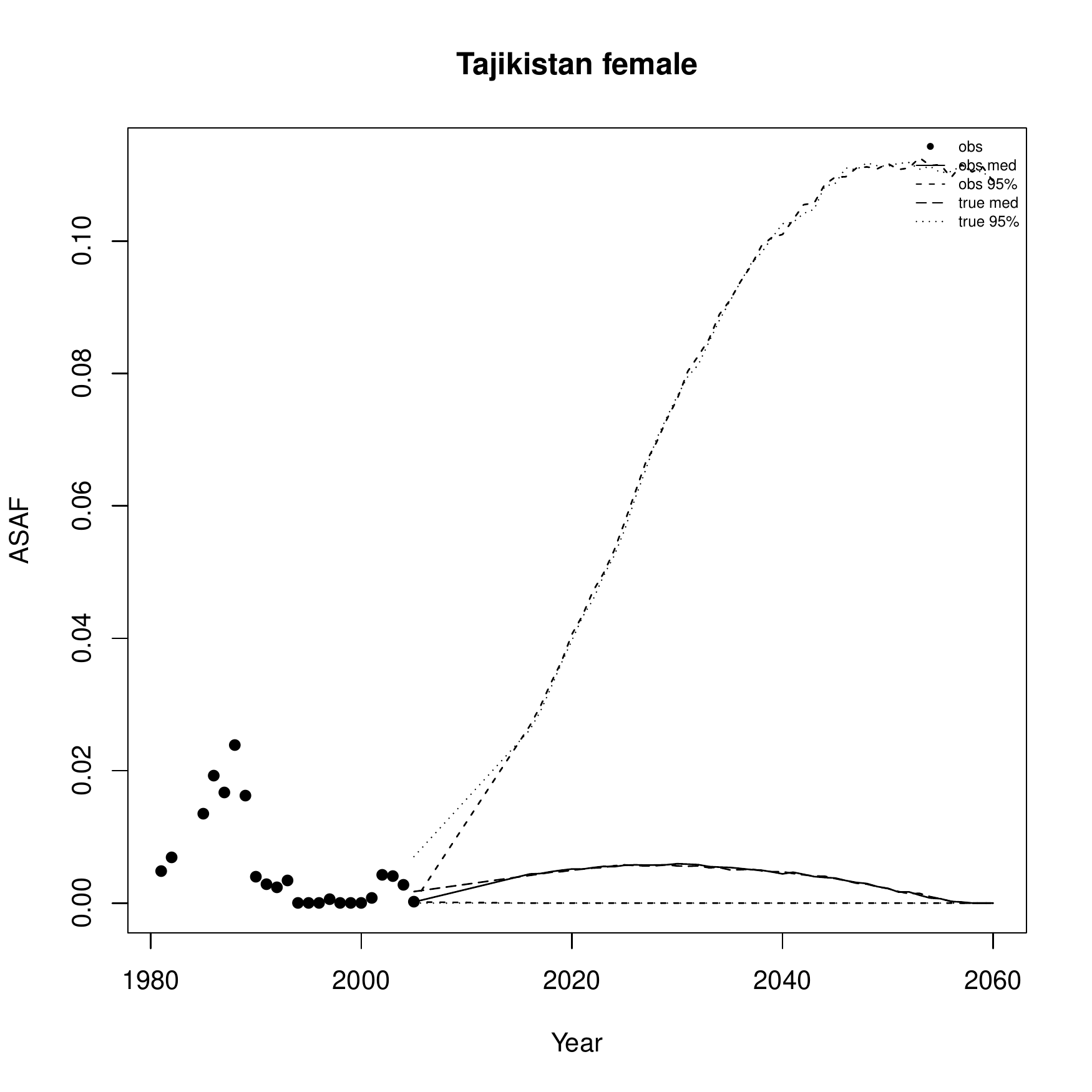}
		\includegraphics[scale=0.43]{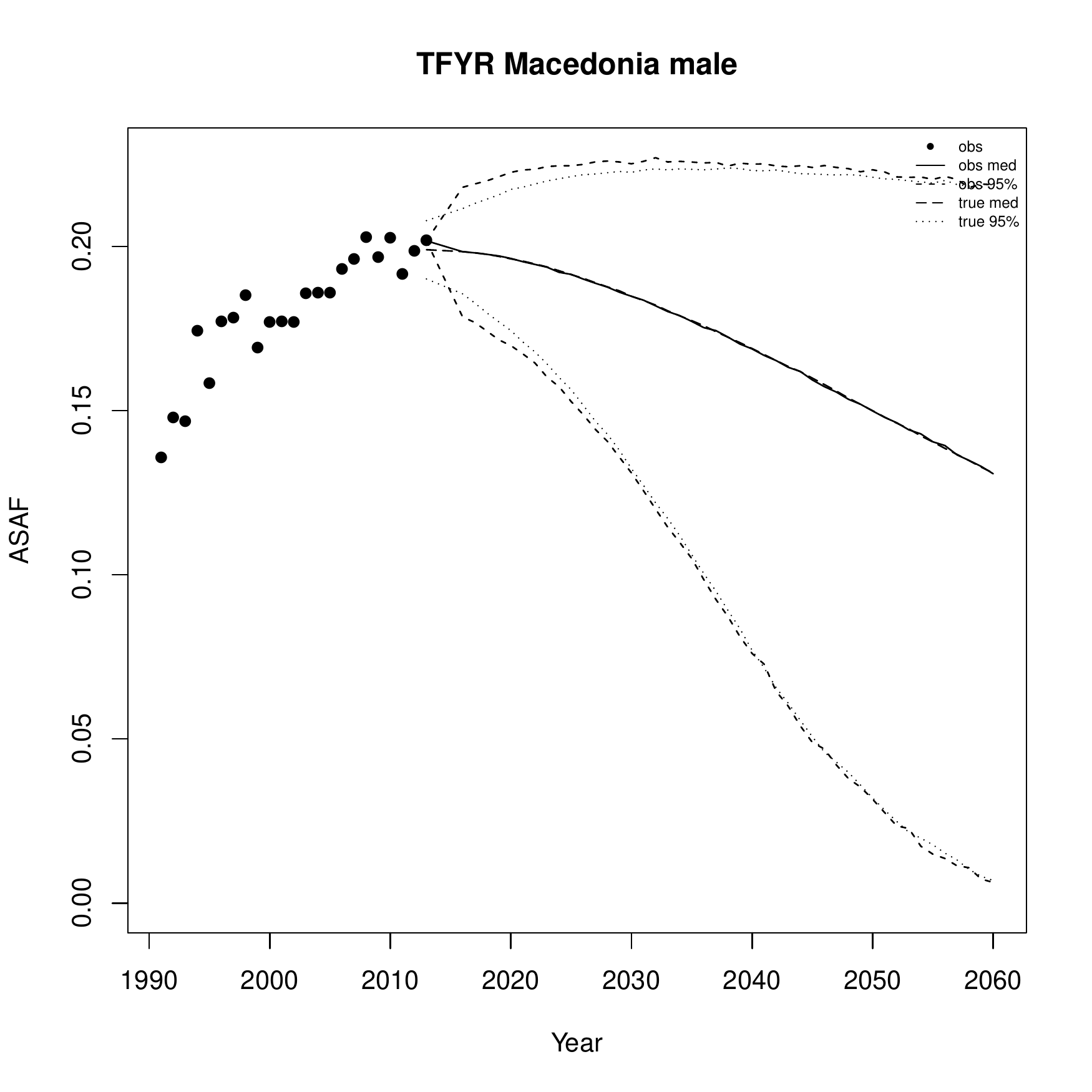}
		\includegraphics[scale=0.43]{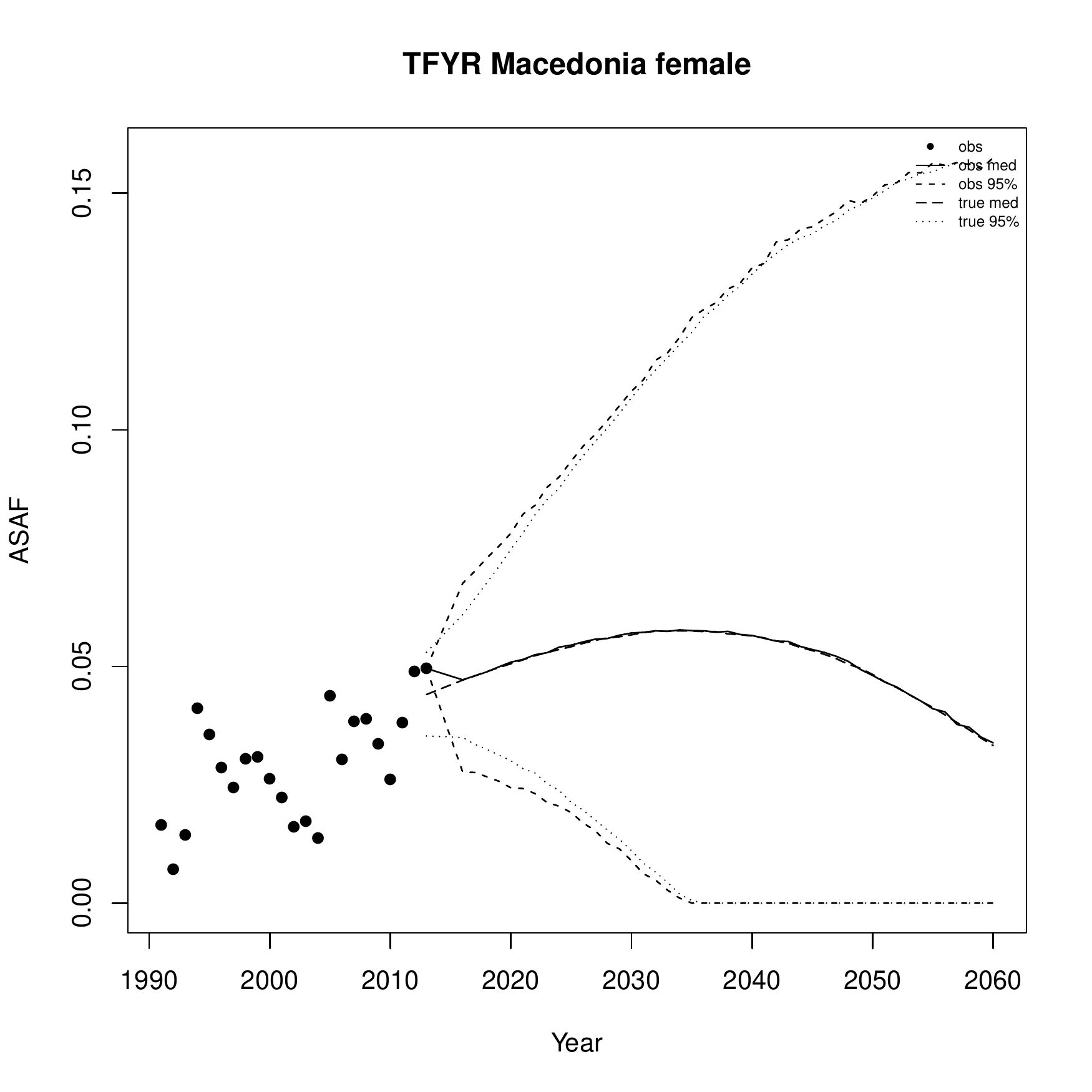}
		\includegraphics[scale=0.43]{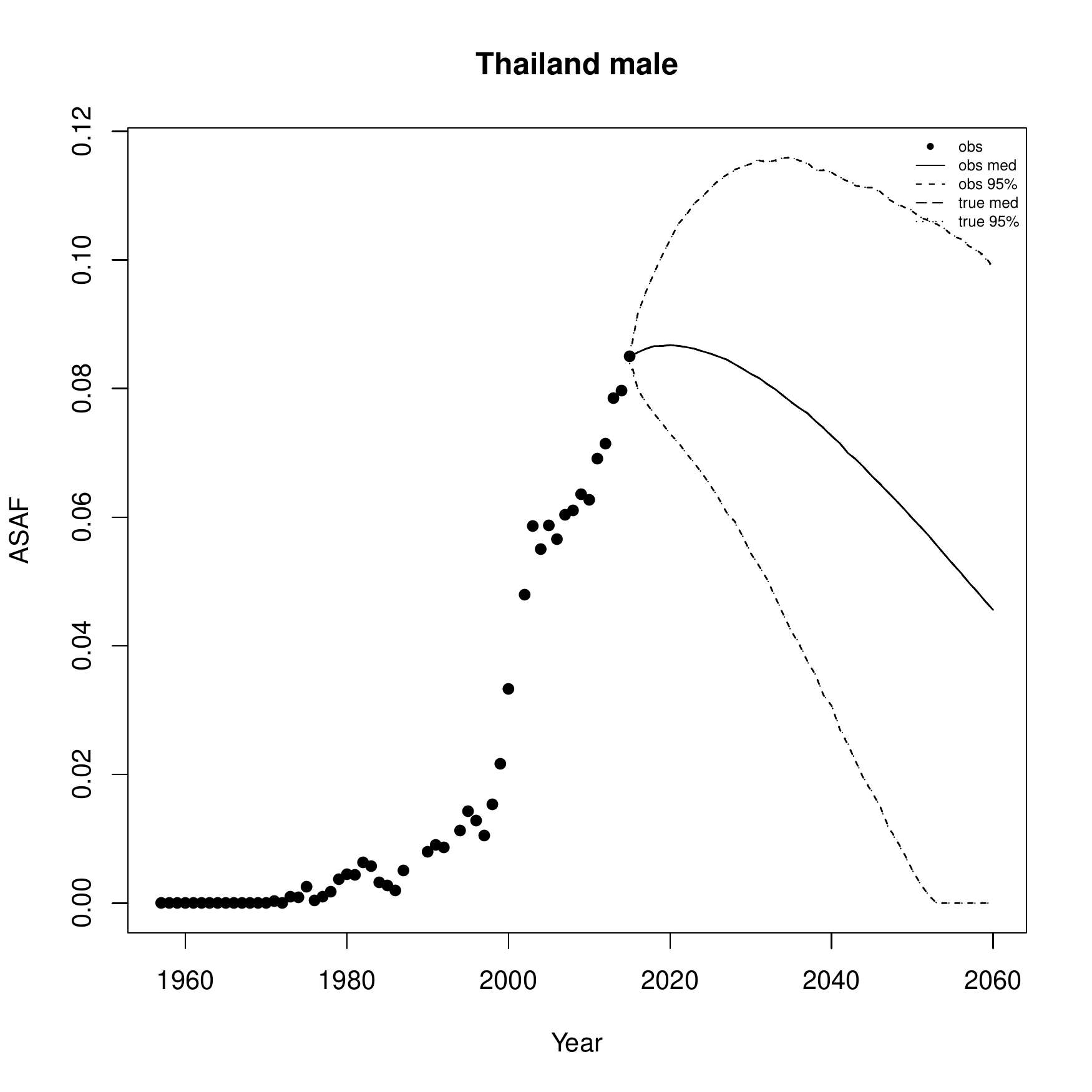}
		\includegraphics[scale=0.43]{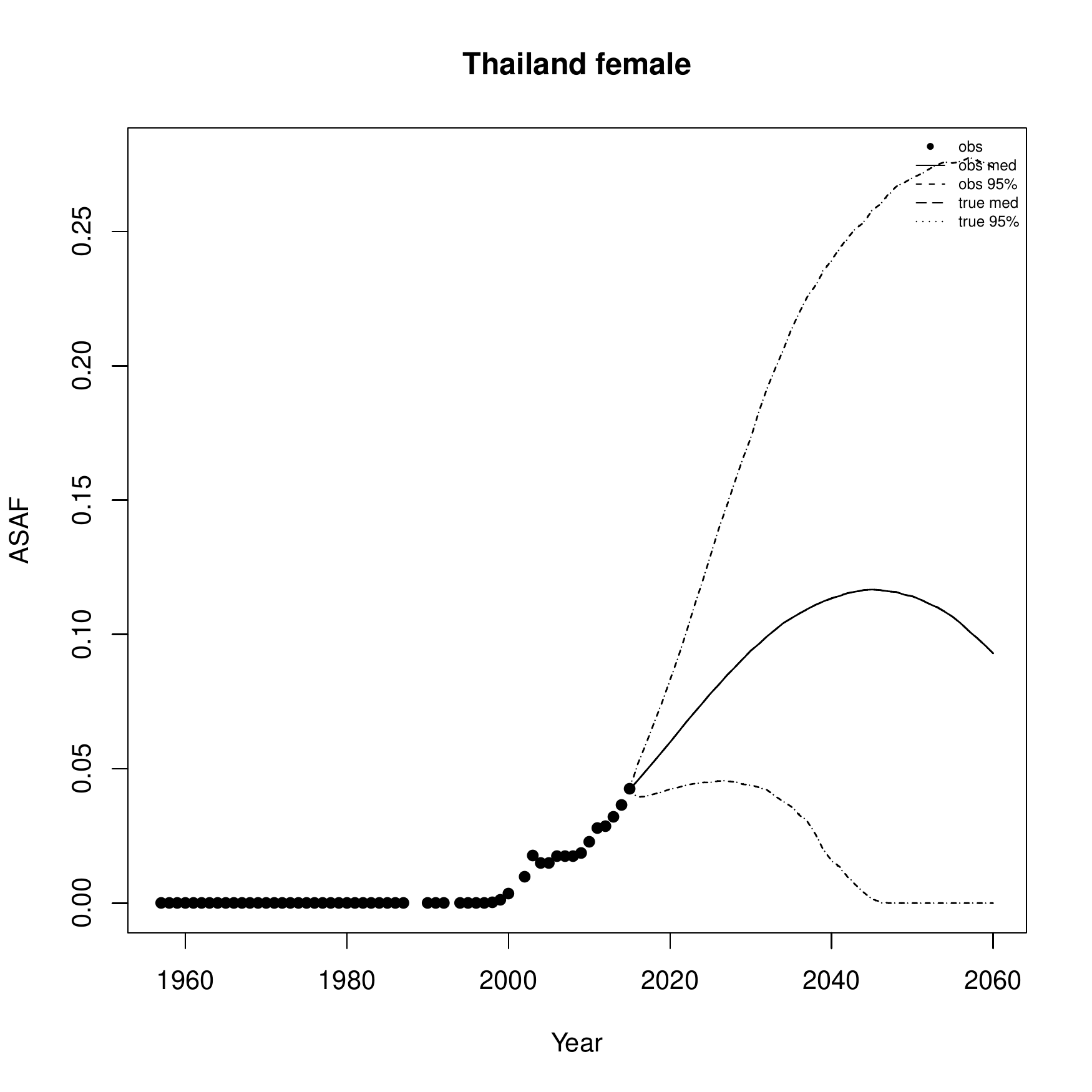}
	\end{center}
\end{figure}

\begin{figure}[H]
	\begin{center}
		\includegraphics[scale=0.43]{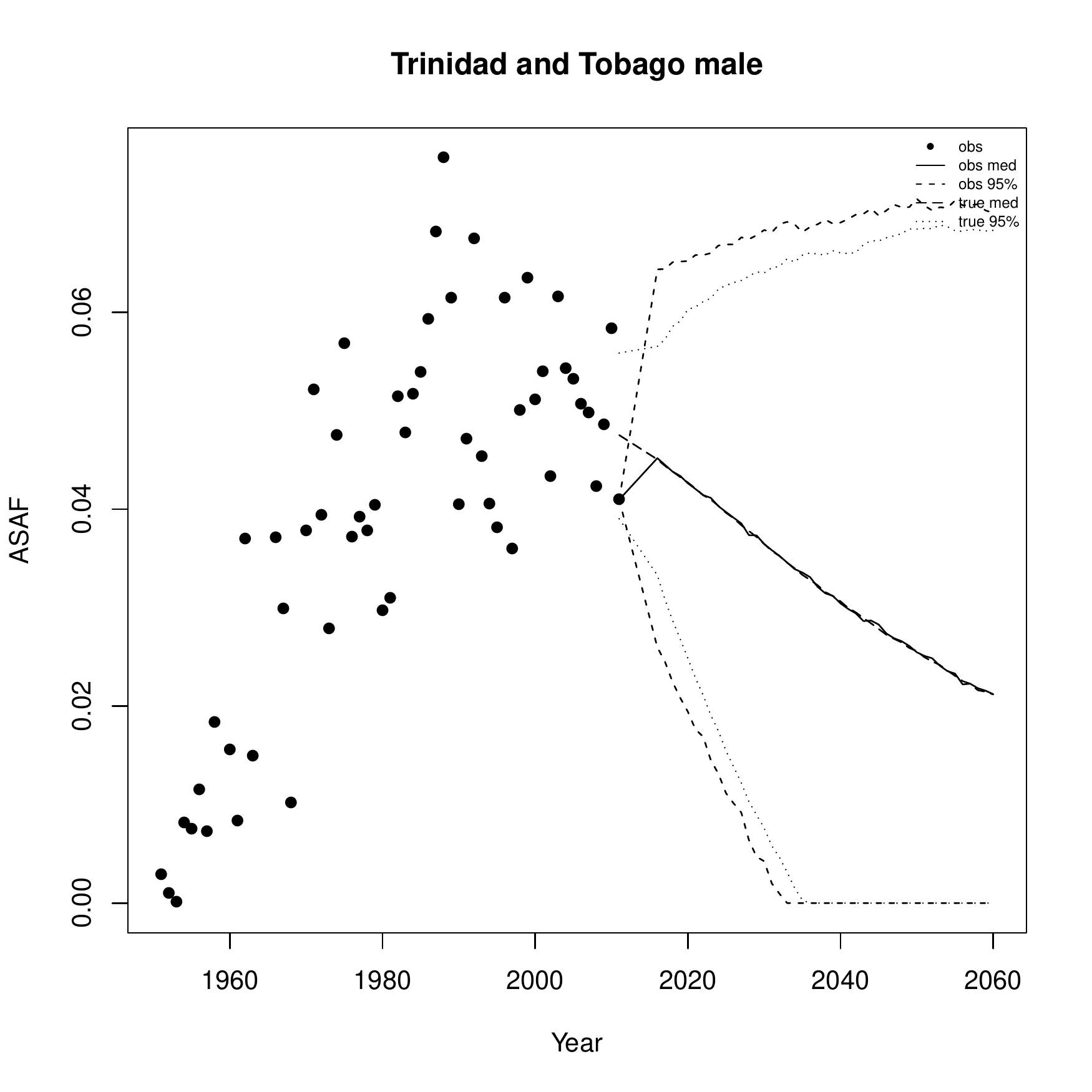}
		\includegraphics[scale=0.43]{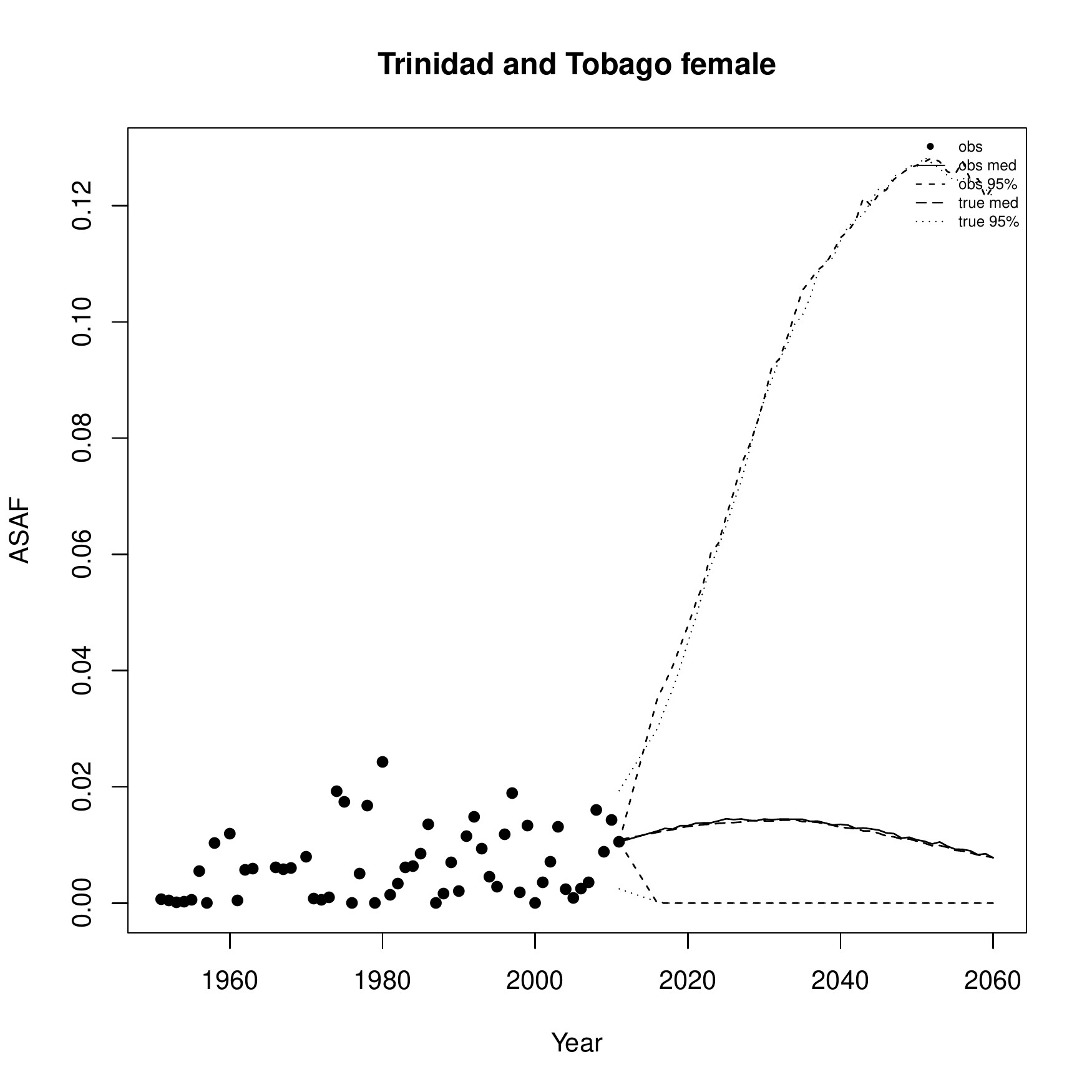}
		\includegraphics[scale=0.43]{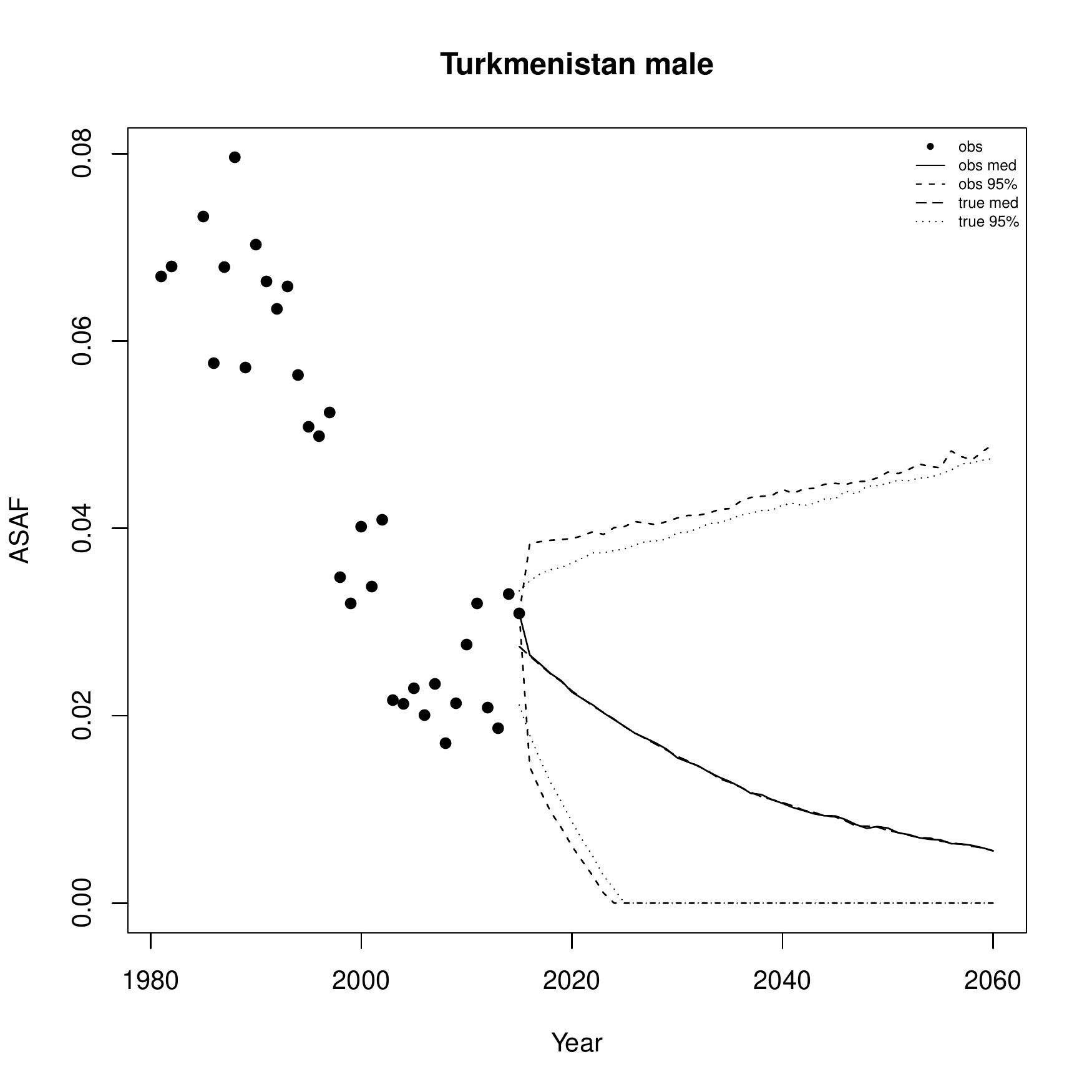}
		\includegraphics[scale=0.43]{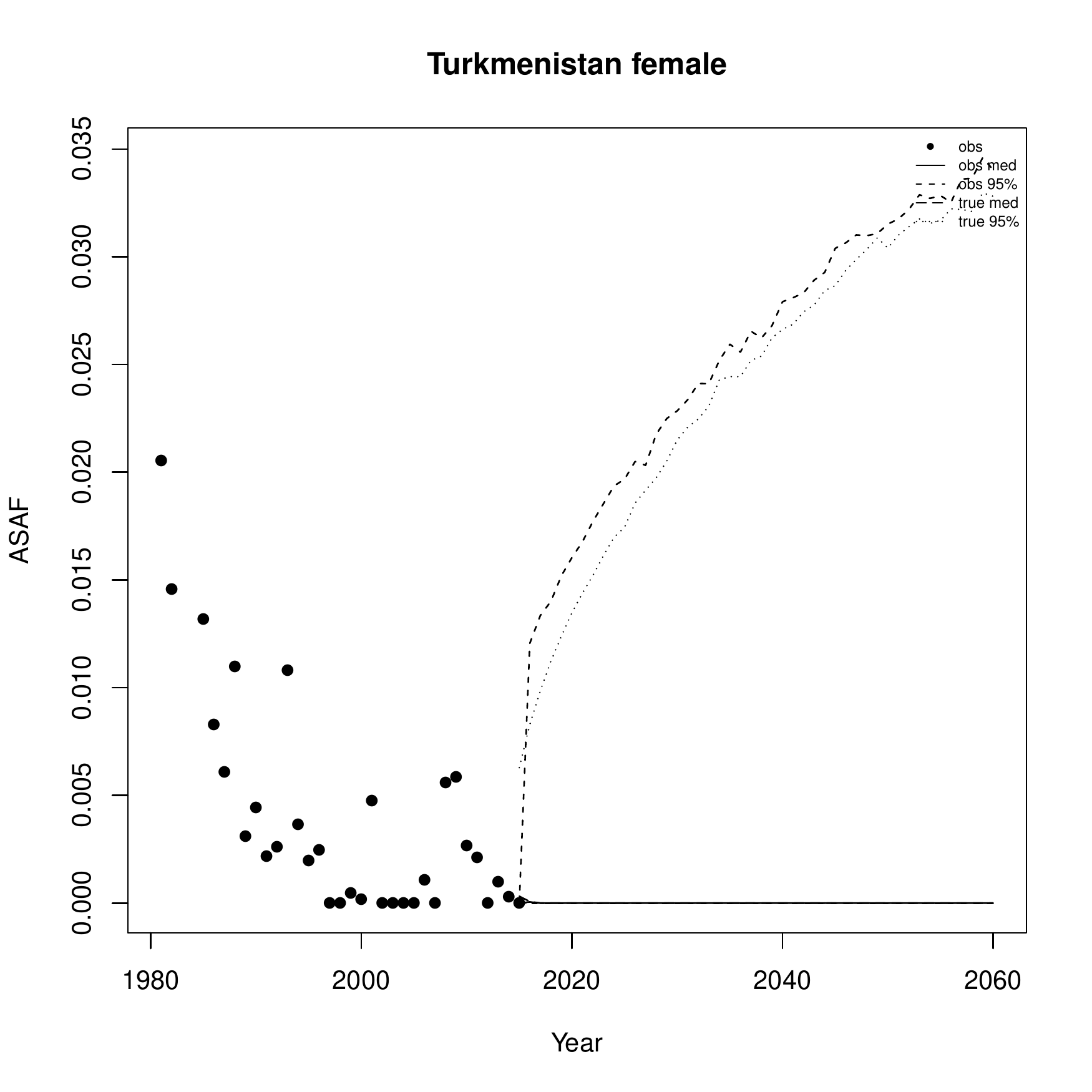}
		\includegraphics[scale=0.43]{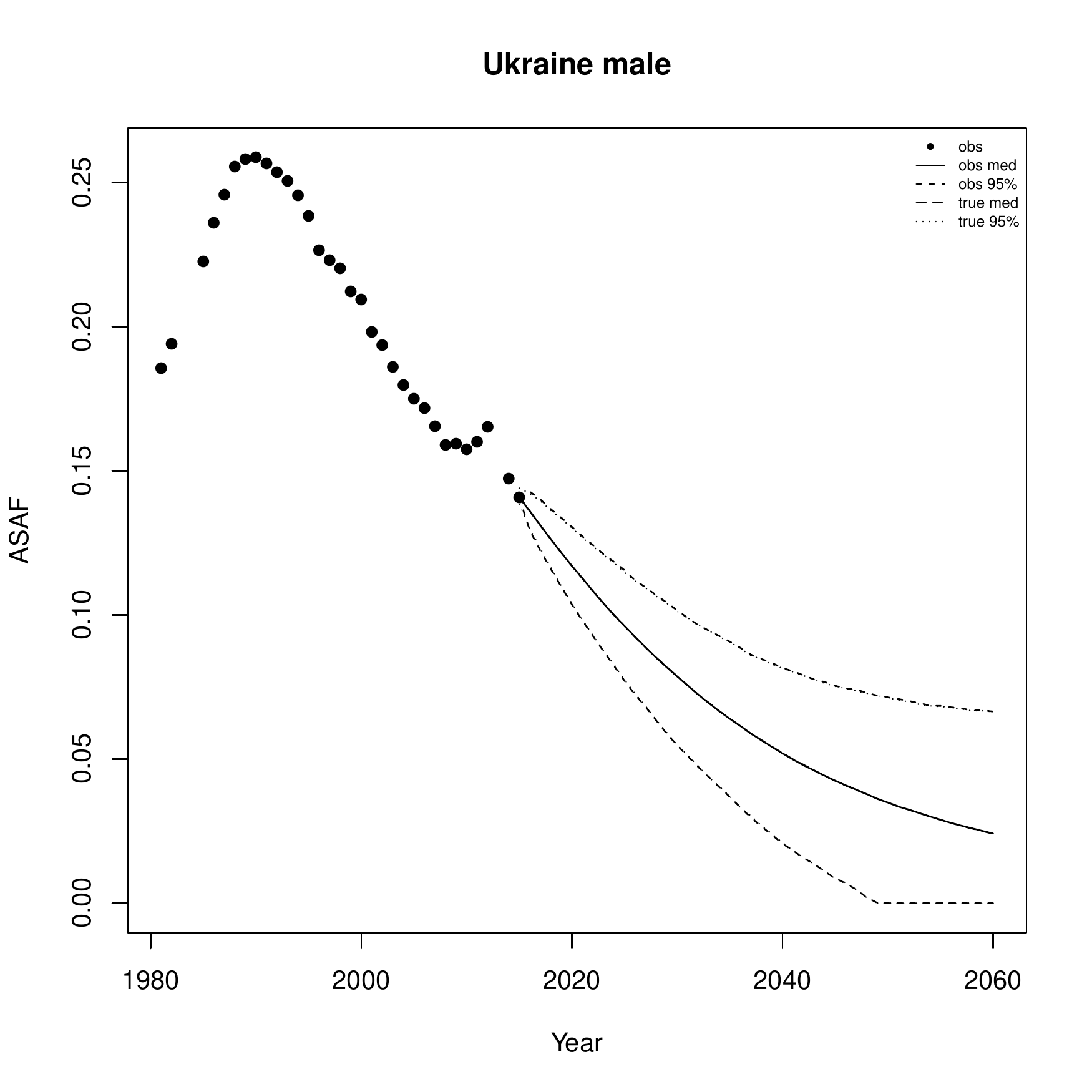}
		\includegraphics[scale=0.43]{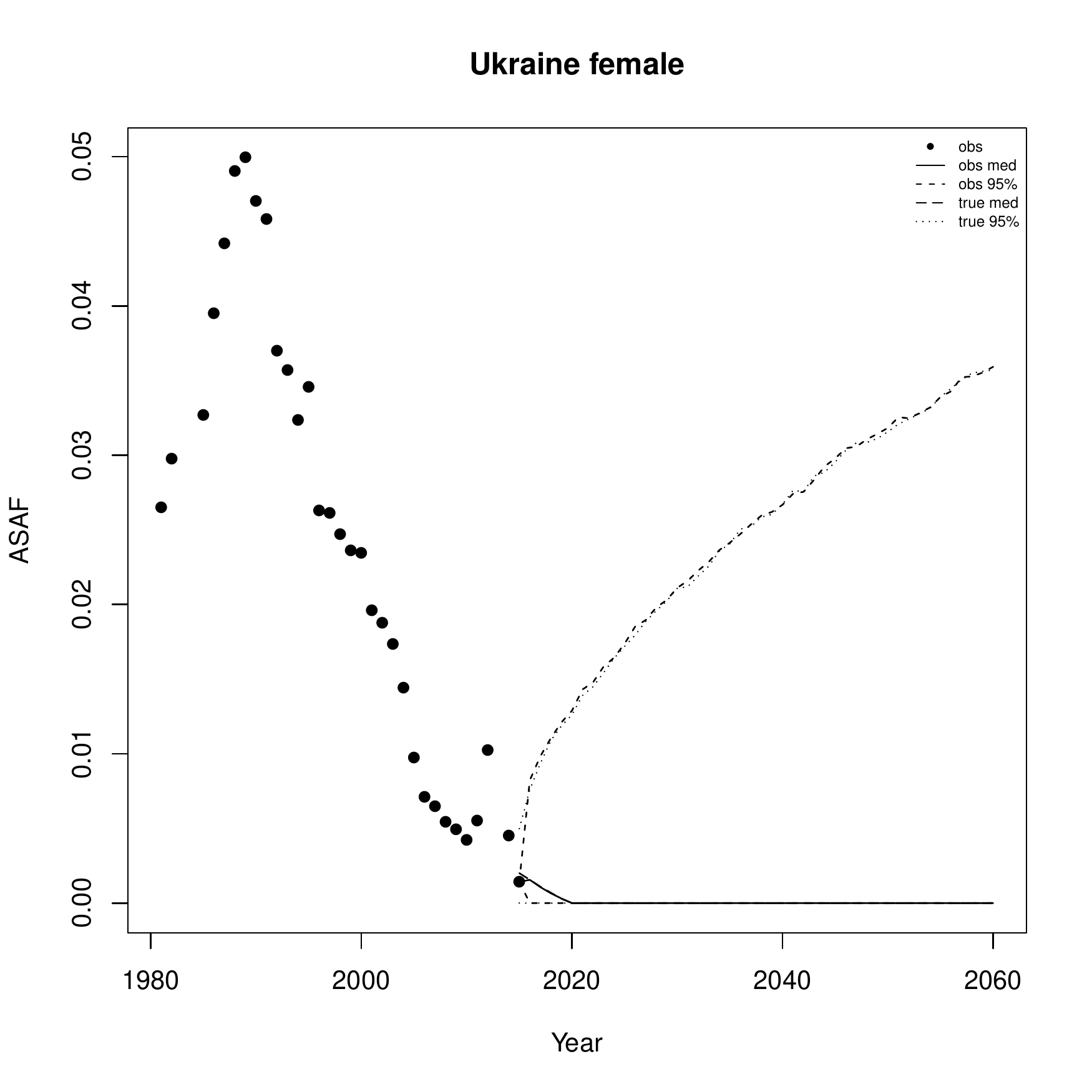}
	\end{center}
\end{figure}

\begin{figure}[H]
	\begin{center}
		\includegraphics[scale=0.43]{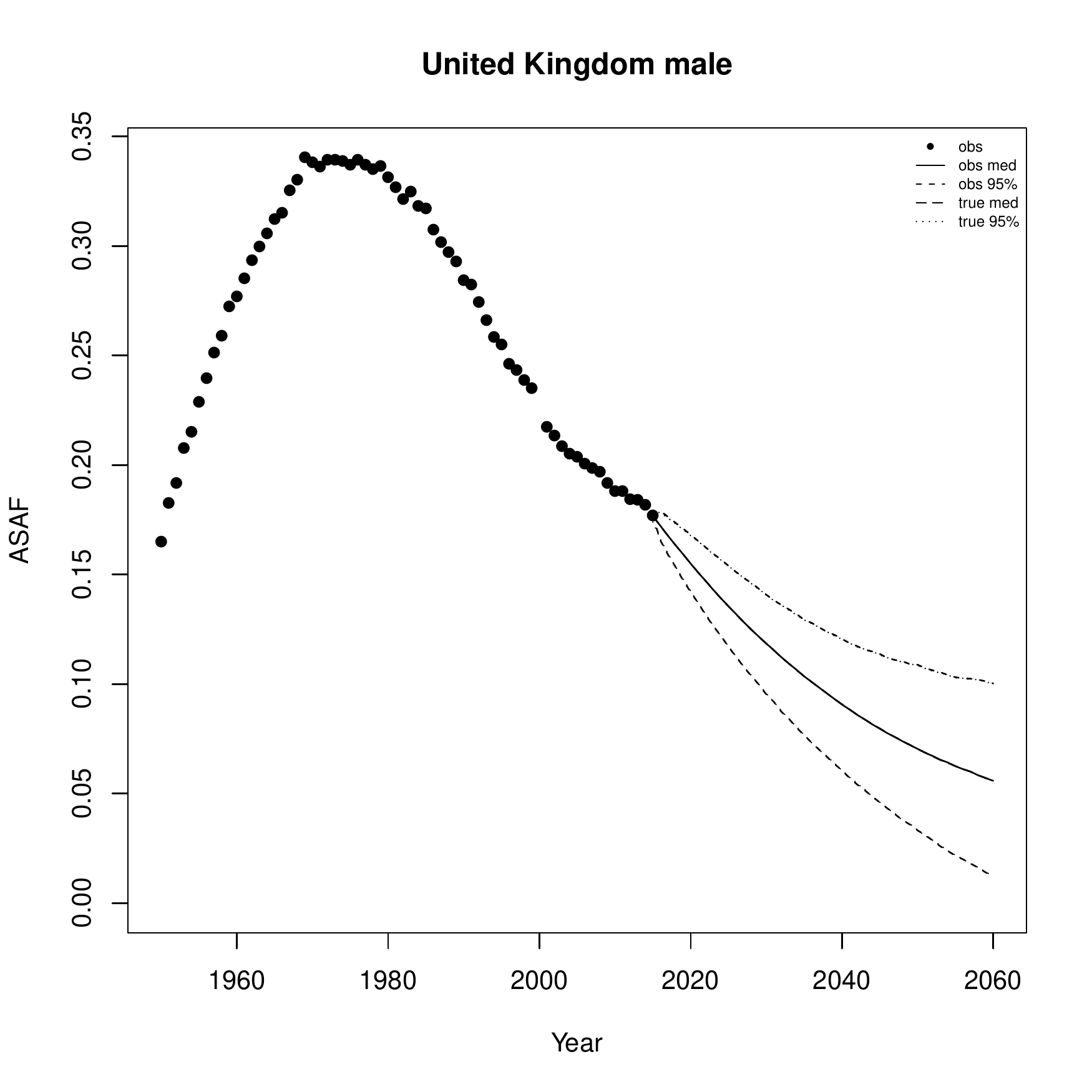}
		\includegraphics[scale=0.43]{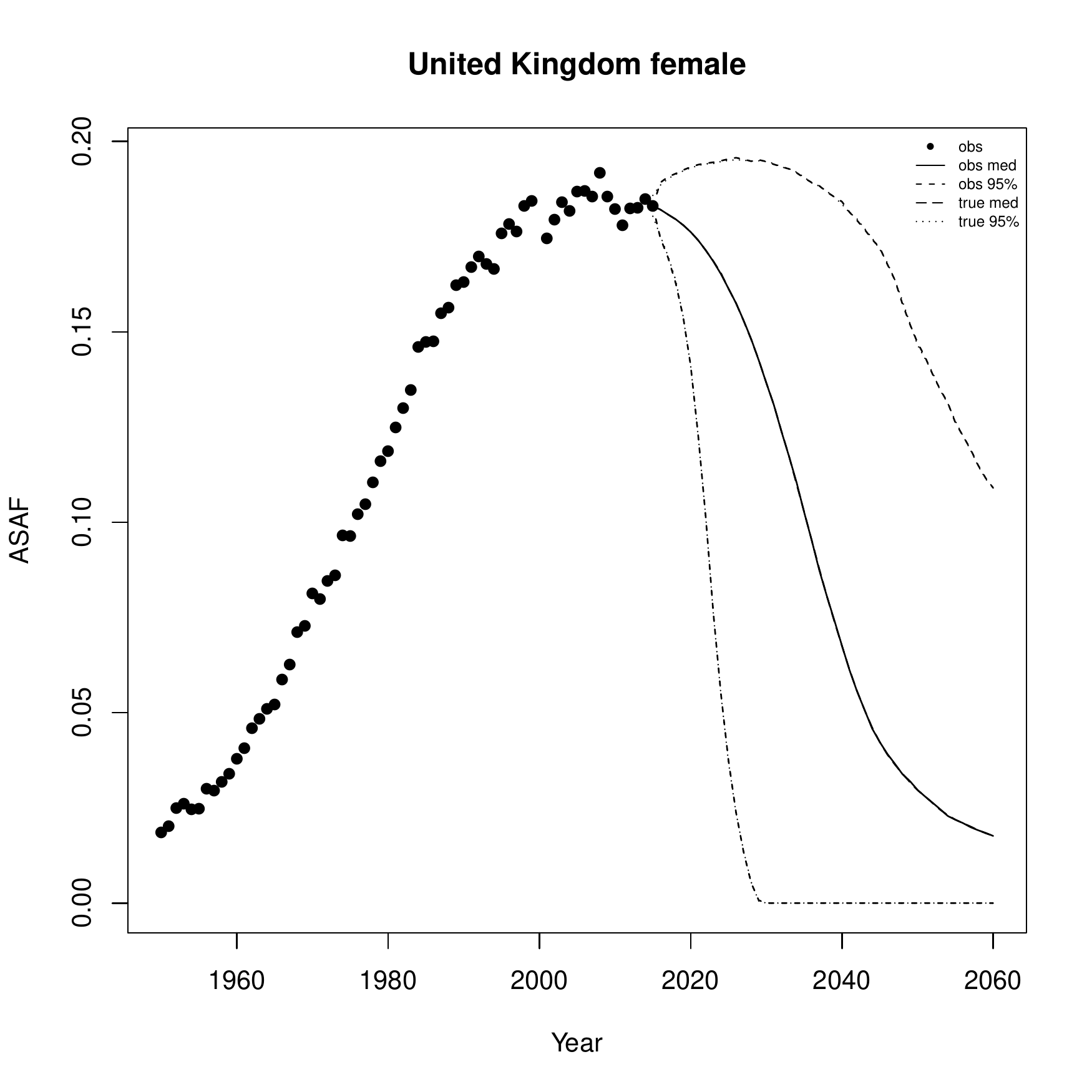}
		\includegraphics[scale=0.43]{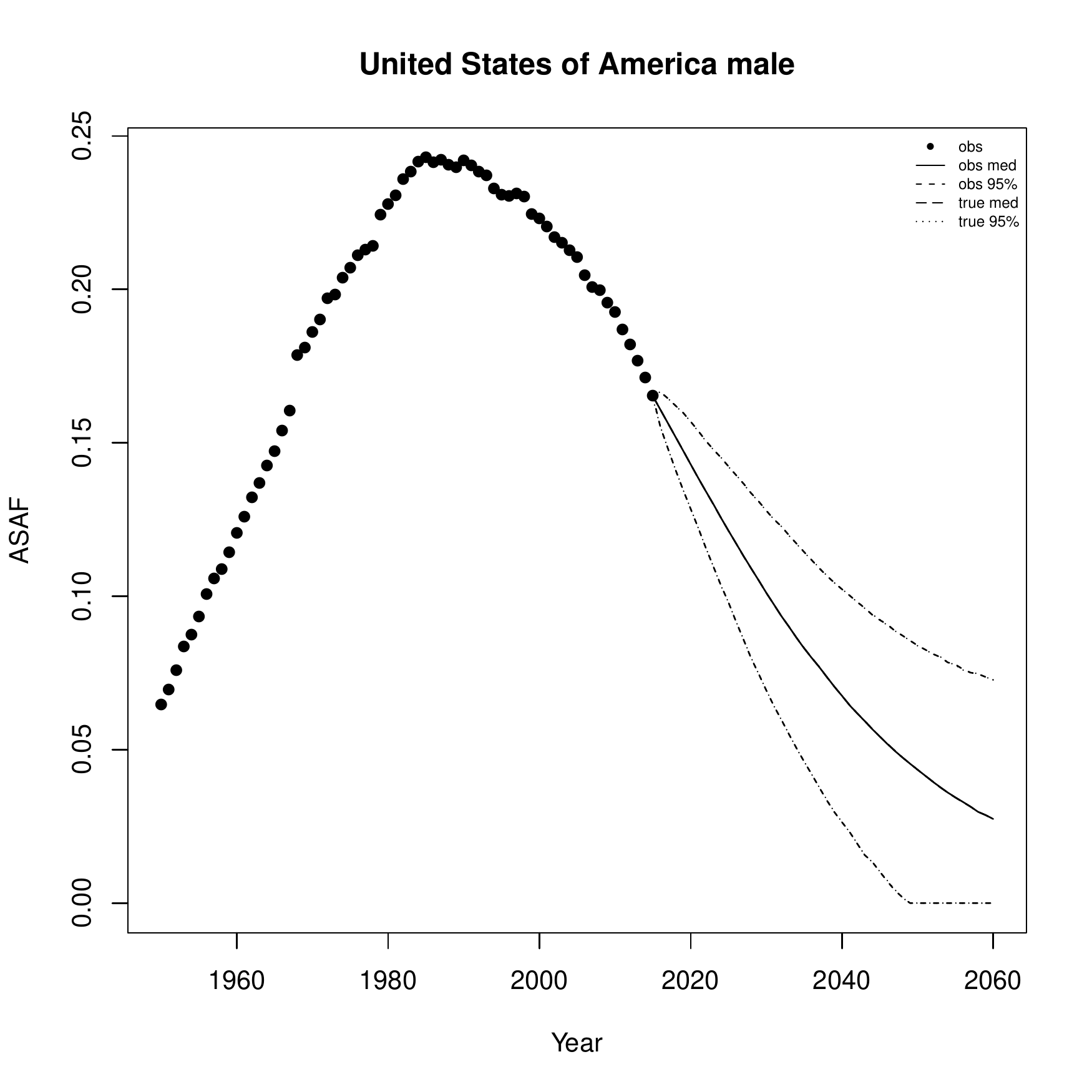}
		\includegraphics[scale=0.43]{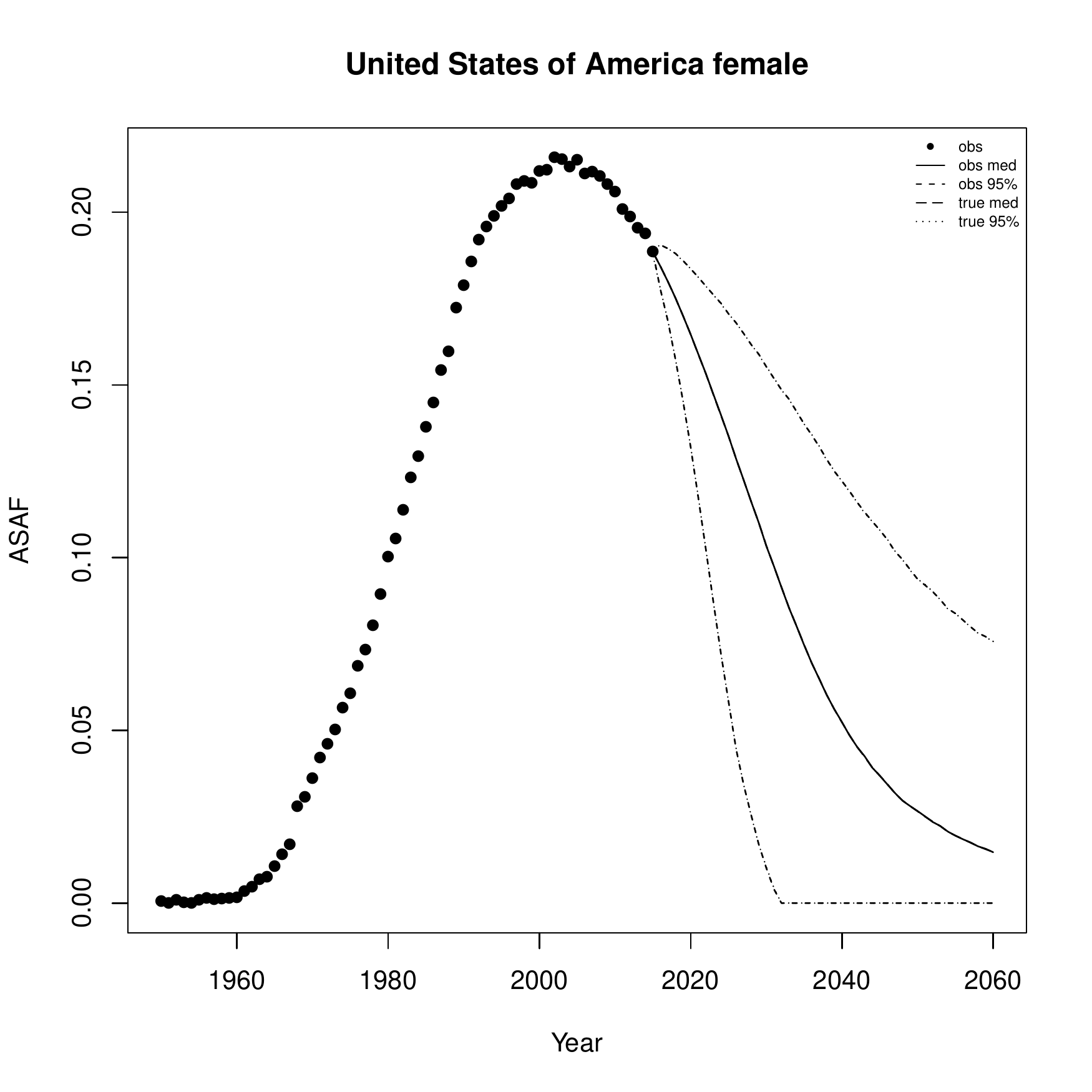}
		\includegraphics[scale=0.43]{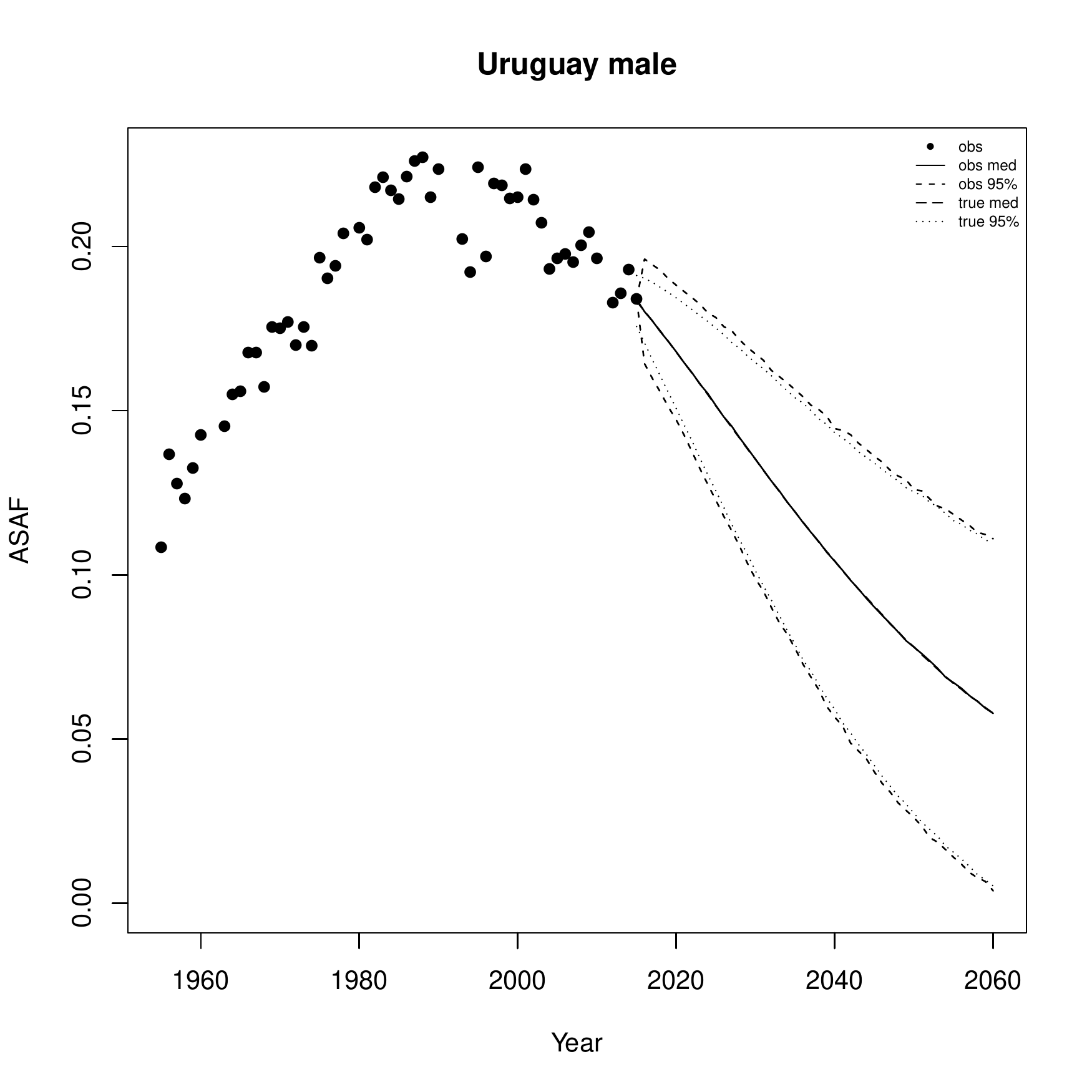}
		\includegraphics[scale=0.43]{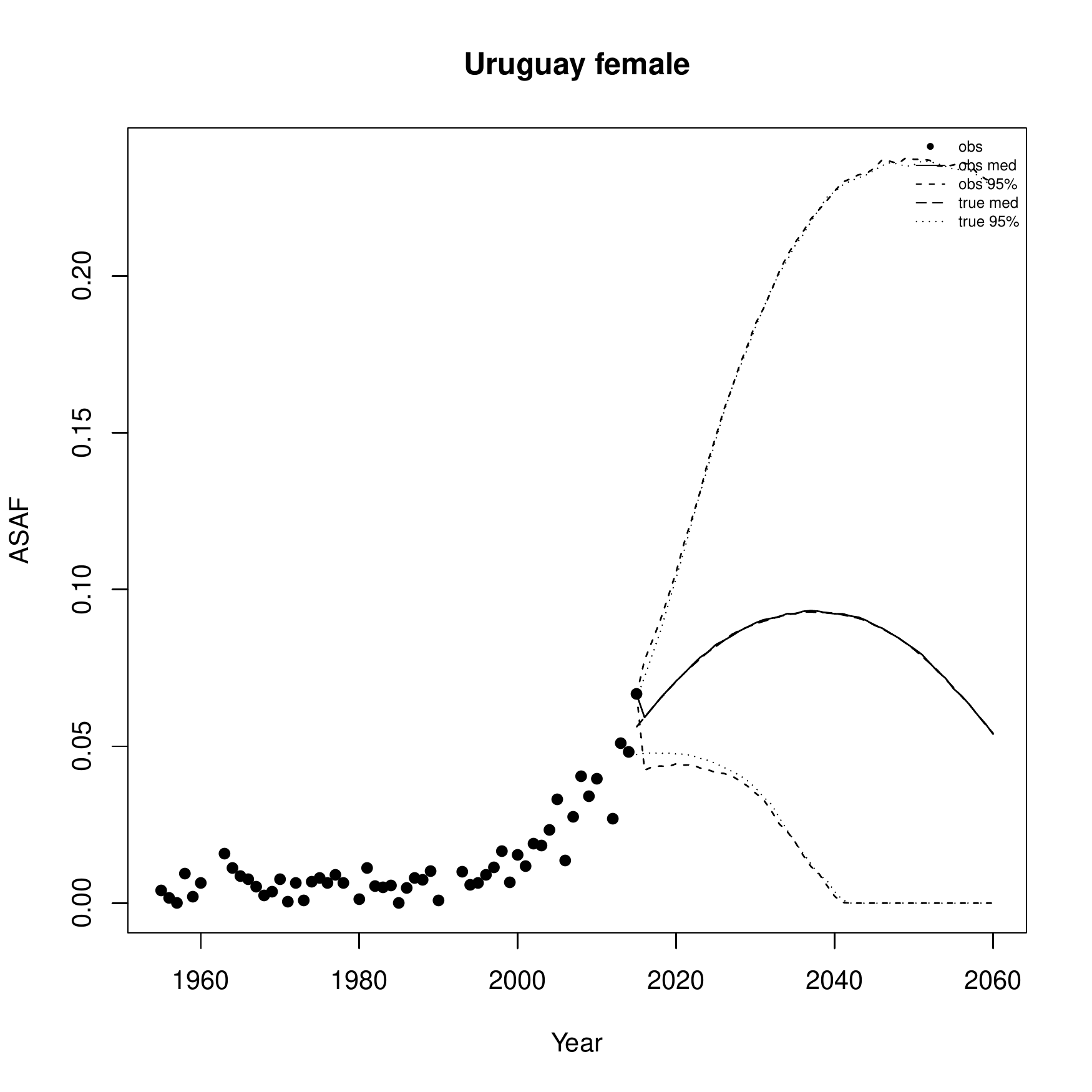}
	\end{center}
\end{figure}

\begin{figure}[H]
	\begin{center}
		\includegraphics[scale=0.43]{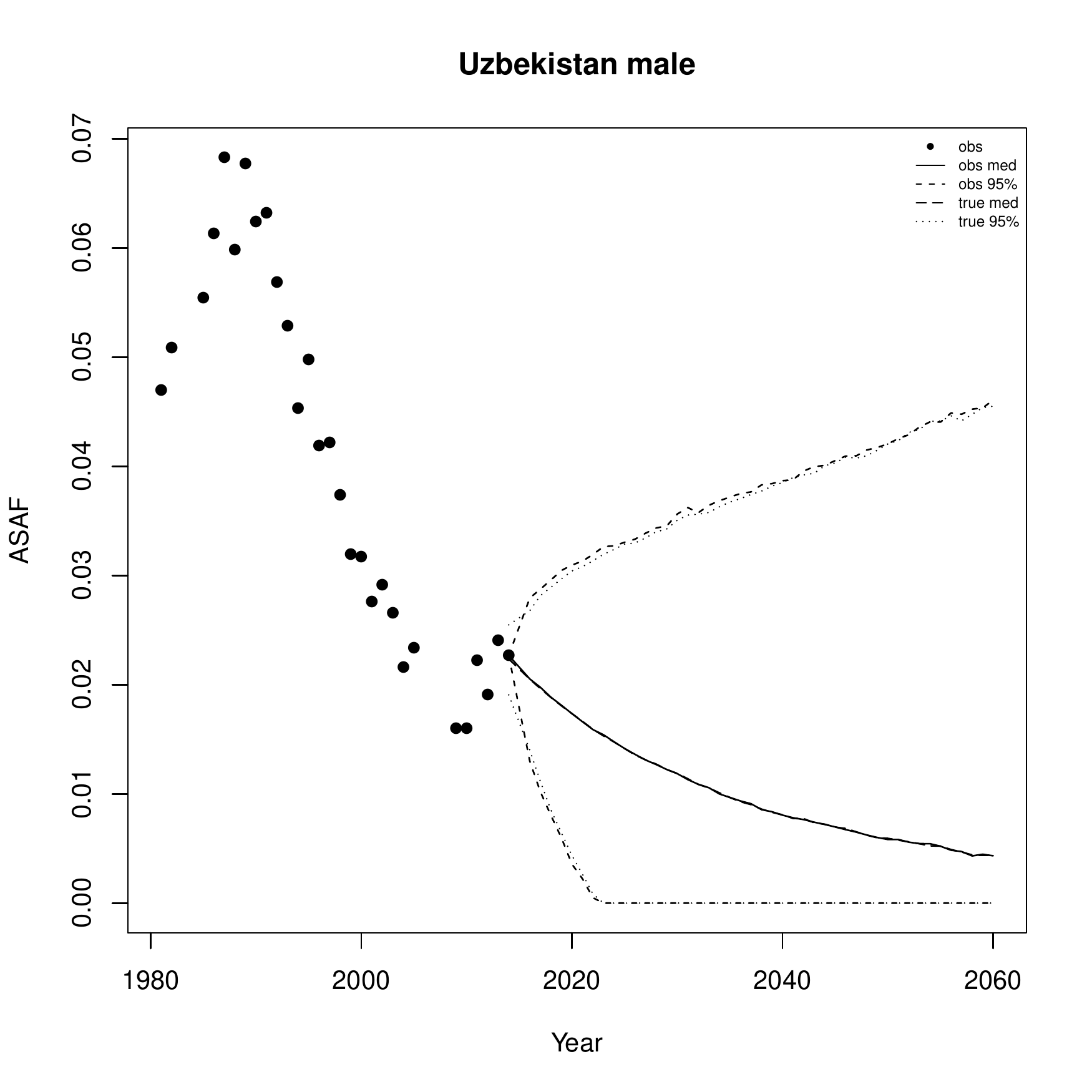}
		\includegraphics[scale=0.43]{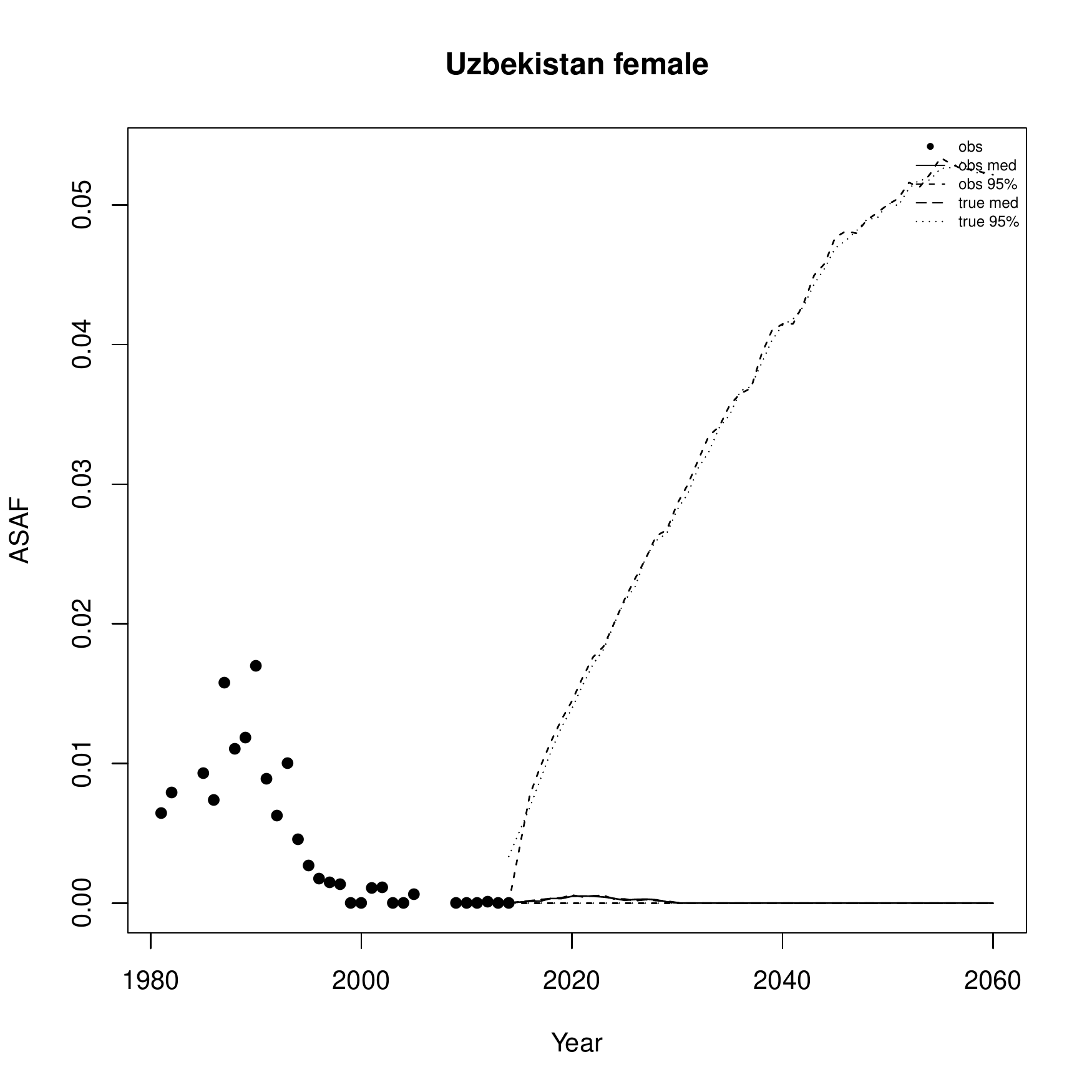}
		\includegraphics[scale=0.43]{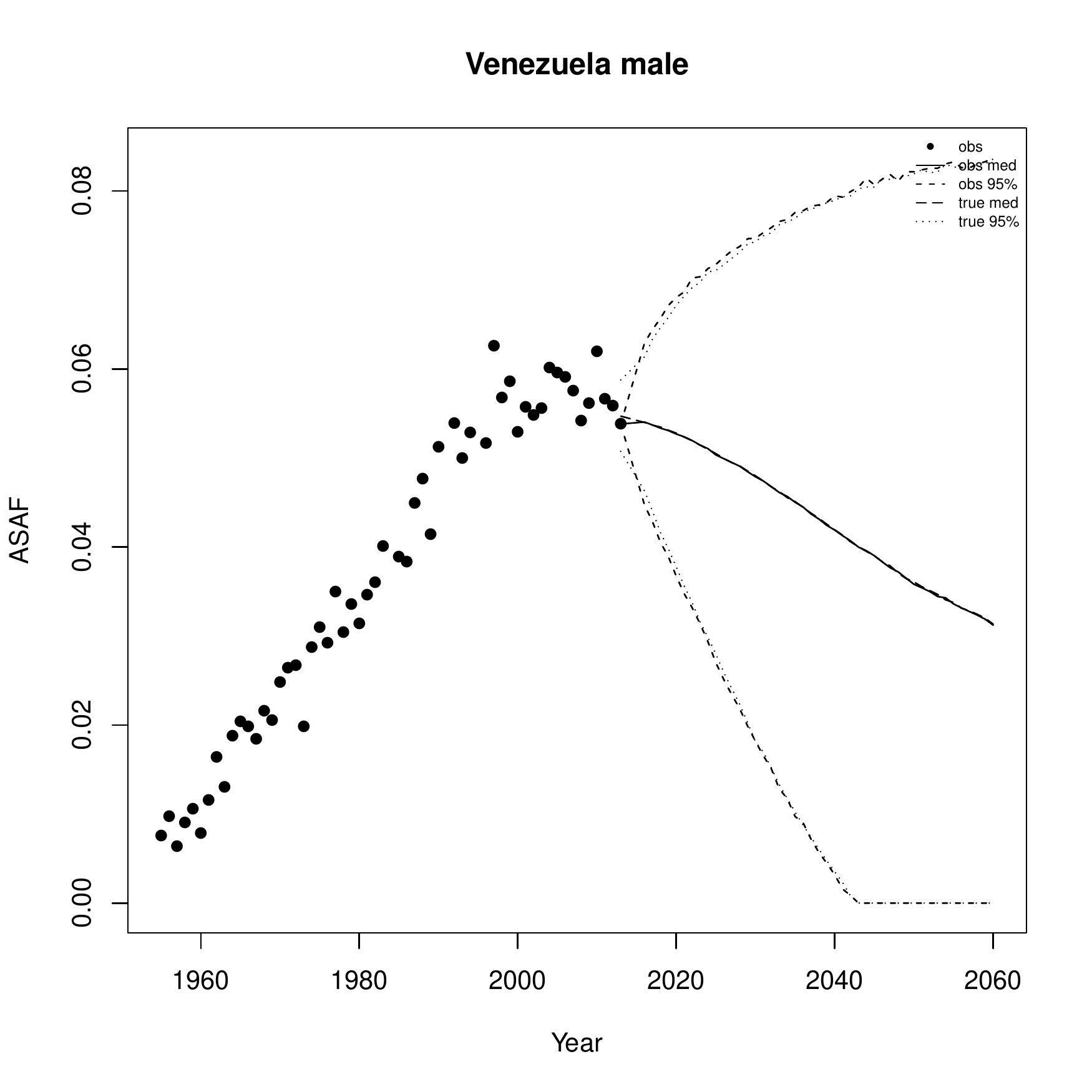}
		\includegraphics[scale=0.43]{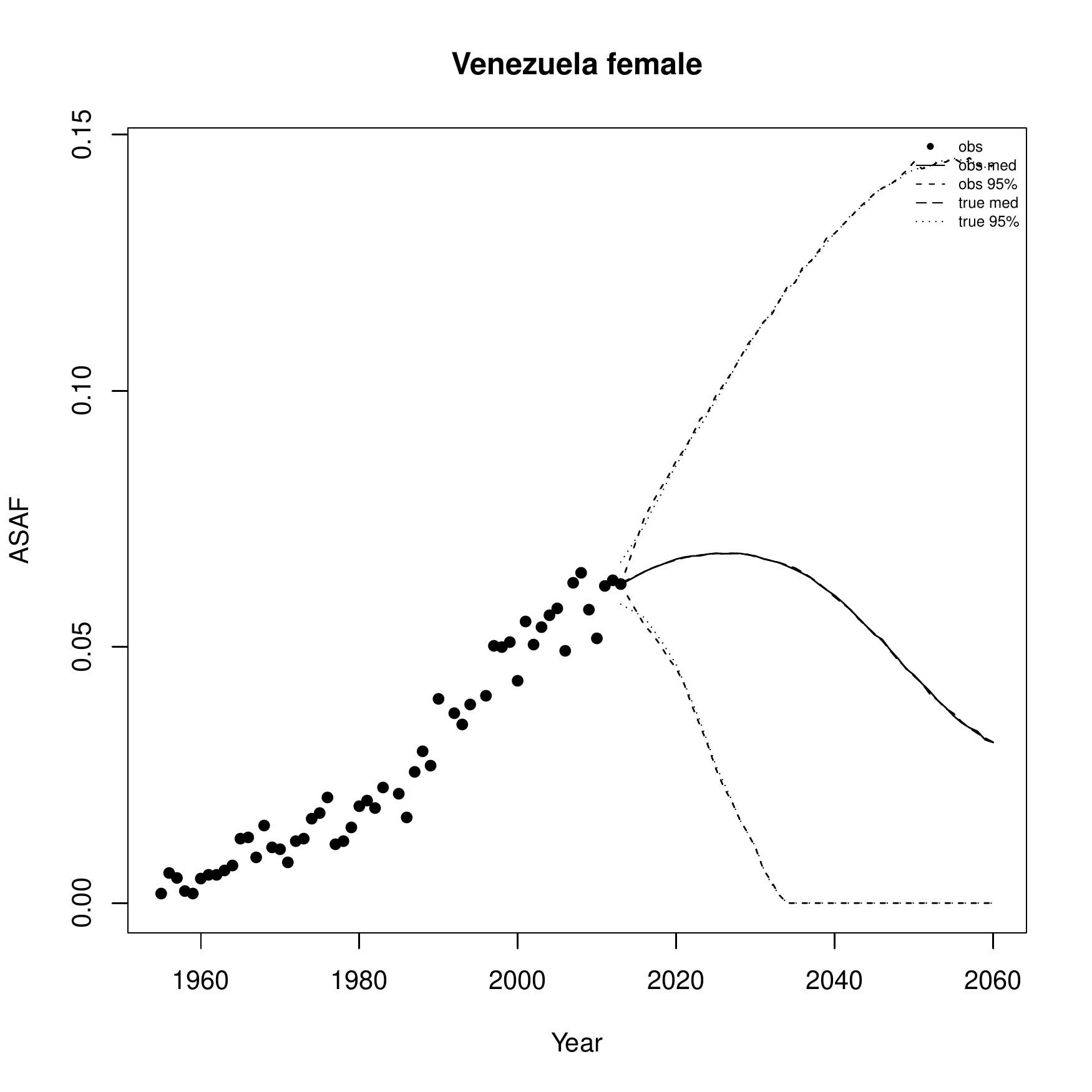}
	\end{center}
\end{figure}

\end{appendices}

\end{document}